\documentclass[10pt,journal,compsoc]{IEEEtran}
\ifCLASSOPTIONcompsoc
  \usepackage[nocompress]{cite}
\else
  \usepackage{cite}
\fi
\ifCLASSINFOpdf
\else
\fi

\usepackage{algorithm}
\usepackage{algorithmic}
\usepackage{alphalph}
\usepackage{amsmath}
\usepackage{amssymb}
\usepackage{amsthm}
\usepackage{array}
\usepackage{bbm}
\usepackage{bm}
\usepackage{caption}
\usepackage{cite}
\usepackage{color}
\usepackage{comment}
\usepackage{dsfont}
\usepackage{etoolbox}
\usepackage{fancyhdr}
\usepackage{graphicx}
\usepackage{multirow}
\usepackage{setspace}
\usepackage{stfloats}
\usepackage{subfigure}

\usepackage{url}

\newtheorem{theorem}{Theorem}
\newtheorem{lemma}{Lemma}
\newtheorem{corollary}{Corollary}

\newtheorem{proposition}{Proposition}

\newtheorem{problem}{Problem}
\newtheorem{observation}{Observation}

\begin{document}
%
\title{Pricing for Collaboration Between\\Online Apps and Offline Venues}
%
%
%
%

\author{Haoran~Yu,~\IEEEmembership{Member,~IEEE,}~George~Iosifidis,~Biying~Shou,~and~Jianwei~Huang,~\IEEEmembership{Fellow,~IEEE}
\IEEEcompsocitemizethanks{\IEEEcompsocthanksitem H. Yu is with the Department of Electrical and Computer Engineering, Northwestern University, USA. E-mail: haoran.yu@northwestern.edu. G. Iosifidis is with the School of Computer Science and Statistics, Trinity College Dublin, Ireland. E-mail: iosifidg@tcd.ie. B. Shou is with the Department of Management Sciences, City University of Hong Kong, Hong Kong. E-mail: biying.shou@cityu.edu.hk. J. Huang is with the Department of Information Engineering, The Chinese University of Hong Kong, Hong Kong, {{and the School of Science and Engineering, The Chinese University of Hong Kong, Shenzhen.}} E-mail: jwhuang@ie.cuhk.edu.hk.
\IEEEcompsocthanksitem {{The work of G. Iosifidis is supported by a Research Grant from the Science Foundation Ireland (SFI) under Grant 17/CDA/4760. The work of B. Shou is supported by the Hong Kong RGC General Research Fund (CityU 11527316). The work of J. Huang is supported by the General Research Funds (Project Number CUHK 14219016) established under the University Grant Committee of the Hong Kong Special Administrative Region, China.}}
}}

\IEEEtitleabstractindextext{%
\begin{abstract}
An increasing number of mobile applications (abbrev. apps), like Pokemon Go and Snapchat, {reward the users who physically visit some locations tagged as POIs (places-of-interest) by the apps.} We study the novel POI-based collaboration between apps and venues (e.g., restaurants). {On the one hand, an app charges a venue and tags the venue as a POI.} The POI tag motivates users to visit the venue, which potentially increases the venue's sales. {On the other hand, the venue can invest in the app-related infrastructure,} which enables more users to use the app and further benefits the app's business. The apps' existing POI tariffs cannot fully incentivize the venue's infrastructure investment, and hence cannot lead to the most effective app-venue collaboration. 
We design an optimal two-part tariff, {which charges the venue for becoming a POI, and subsidizes the venue every time a user interacts with the POI.} The subsidy design efficiently incentivizes the venue's infrastructure investment, and we prove that our tariff achieves the highest app's revenue among a general class of tariffs. Furthermore, we derive some counter-intuitive guidelines for the POI-based collaboration. For example, a bandwidth-consuming app should collaborate with a low-quality venue (users have low utilities when consuming the venue's products).
\end{abstract}

}


\newpage
\setcounter{page}{1}
\maketitle

\thispagestyle{empty}

\IEEEdisplaynontitleabstractindextext 


%
\IEEEpeerreviewmaketitle


\section{Introduction}

\subsection{Motivations}

{Many popular mobile applications (abbrev. apps), especially the augmented reality apps  \cite{huang2013mobile},} integrate users' digital experience with the real world. {For example, {Pokemon Go}, one of the most popular mobile games in 2016,} tags some real-world locations as ``PokeStops'' or ``Gyms''. 
When visiting these locations physically, users can collect game items or participate in ``battles'' in the game \cite{mcdonald}. {{Snapchat}, a popular image messaging app, provides various image filters, including ``Geofilters'',} which are unlocked only when users visit the specified real-world locations \cite{snapchat}. 
Many other apps, such as Ingress \cite{ingress}, Snatch \cite{snatch}, and Jurassic World Alive \cite{jurassic}, apply similar approaches to integrate users' digital experience and physical activities. {We use POIs (places-of-interest) to refer to the real-world locations where users can obtain rewards or unlock some features of the apps.}


When the locations are venues such as restaurants and cafes, the POI tags have the potential to benefit both the apps and the venues. On the one hand, the venues' infrastructure (e.g., smartphone chargers and Wi-Fi networks) enhances the users' experience of using the apps, and hence benefits the apps' businesses. For example, {many apps (especially the augmented reality apps like Pokemon Go \cite{battery}) drain the smartphones' batteries quickly, and some apps are data-hungry (e.g., Jurassic World Alive consumes more than 100MB per day for a regular player). The smartphone chargers and Wi-Fi networks at the venues alleviate users' needs of reducing the app usage because of battery or data usage concern.} 
On the other hand, the POI tags significantly increase the number of the venues' visitors, who might purchase the venues' products and increase the venues' sales. 
This explains the increasingly popular collaboration between apps (\emph{online businesses}) and venues (\emph{offline businesses}). 
{In 2016,} Pokemon Go collaborated with Sprint and McDonald's, tagging 10,500 Sprint stores in the U.S. and 3,000 McDonald's restaurants in Japan as POIs \cite{Sprint,mcdonald}. 
In particular, Sprint stores offered Pokemon Go players free smartphone charging stations \cite{Sprint}. 
It was estimated that each of the McDonald's restaurants that became POIs attracted up to 2,000 game players per day \cite{mcdonald}, and the POI tags increased some stores' sales by 100\% \cite{onehundred}. 
{In 2017,} Wendy's (a restaurant chain) made its ``Geofilters'' in Snapchat, which drove 42,000 additional visitors within a week \cite{snapchat}. {Yinyangshi tagged over 5,000 KFC restaurants in China as POIs \cite{yinyangshi}, and similarly Snatch partnered with Mitchells \& Butlers pubs in the U.K. \cite{snatch}. In 2018, Jurassic World Alive established the POI-based collaboration with Walmart and AMC Theaters \cite{jurassic}. Some augmented reality apps that will be released soon may plan for a similar partnership with venues \cite{2018ARgames}.
}


{All the aforementioned apps have augmented reality features.} It was estimated that the augmented reality and virtual reality market's worldwide revenue will be nearly \$215 billion in 2021 \cite{ARmarket}. Hence, the POI-based collaborations can potentially create substantial revenues for the apps and venues. However, as we show in the following, the apps and venues do not have fully aligned interests in attracting the users, which makes it difficult for the apps and venues to completely realize the collaborations' potential.

\begin{figure}[t]
  \centering
  \includegraphics[scale=0.52]{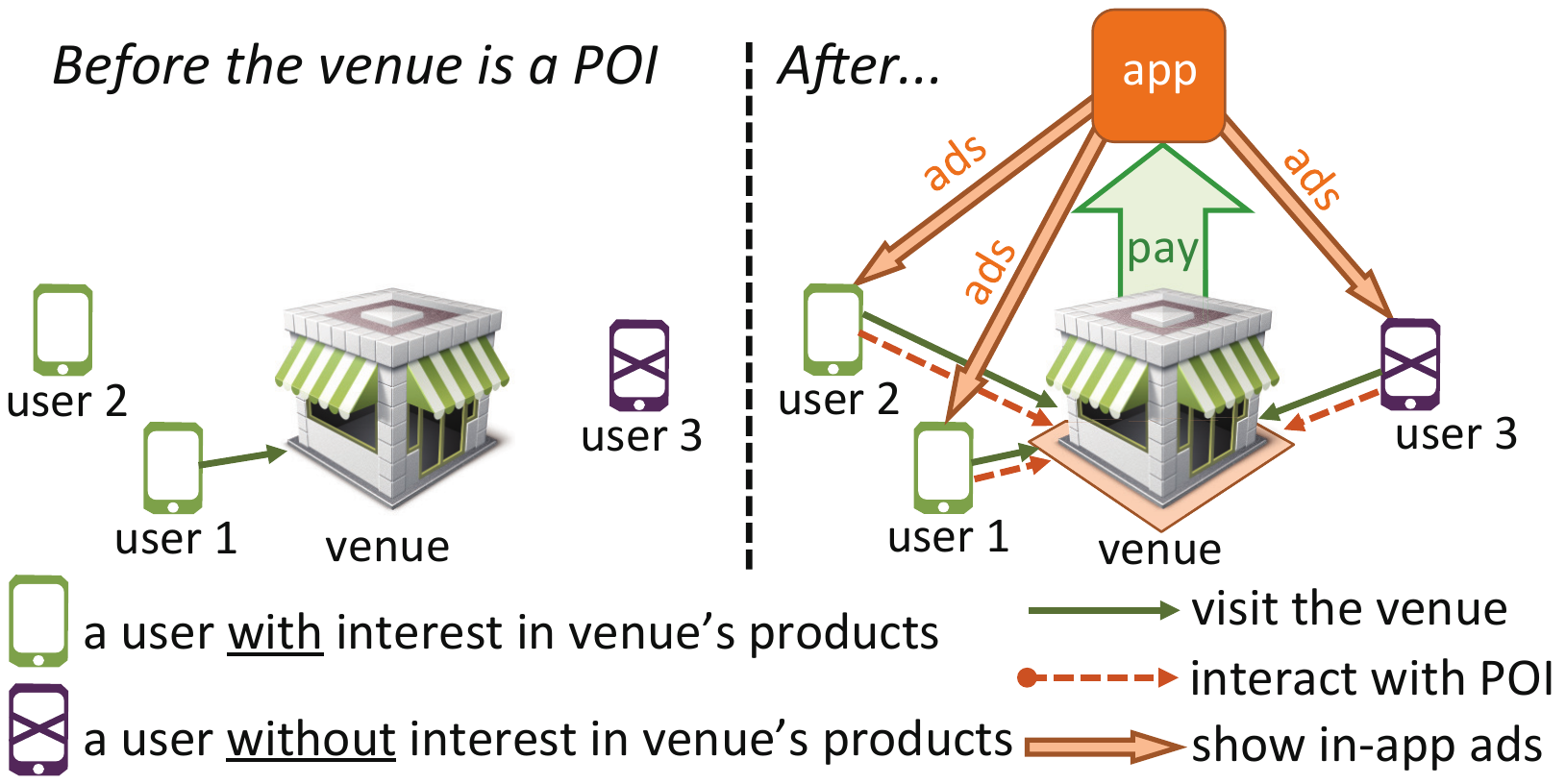}\\
  \caption{An example of POI-based collaboration (all users use the app).}
  \label{fig:system}
  \vspace{-0.45cm}
\end{figure}

In Fig. \ref{fig:system}, we illustrate the POI-based collaboration. {As shown in the abovementioned examples, an app usually collaborates with a store/restaurant chain (e.g., Sprint, McDonald's, and KFC).} 
In order to avoid internal competition, the venues in a chain are typically strategically spaced out. Therefore, we approximate the collaboration between the app and the chain by the collaboration between the app and a representative venue of the chain. 
In Fig. \ref{fig:system}, when the venue is not a POI, only the nearby users with interests in the venue's products (e.g., user $1$) visit the venue. 
{After the venue pays the app and becomes a POI, more users (including those without interests in the venue's products) visit the venue to interact with the POI (e.g., participate in the ``battles'' held at the POI). The number of these visitors depends on the venue's investment in the app-related infrastructure.} Moreover, the app can display location-dependent in-app advertisements to these visitors to get additional advertising revenue {(for an app that does not show any in-app advertisement, the corresponding advertising revenue is $0$ in our model)}. 
As Fig. \ref{fig:system} shows, the app and venue do not have fully aligned interests in terms of attracting the users. {The app delivers the advertisements to all users interacting with the POI (e.g., users $1\sim3$), and hence benefits from a high investment in the app-related infrastructure. 
The venue only gains profits from the users with interests in its products (e.g., users $1$ and $2$), and may not choose a high investment level.} A low investment level will reduce the number of users interacting with the POI. 

The key challenge is to design the app's optimal tariff scheme (which charges the venue for becoming a POI) to incentivize the venue's investment in the app-related infrastructure. The tariff schemes commonly used by the apps include the \emph{per-player-only tariff} and \emph{lump-sum-only tariff}. 
In the \emph{per-player-only tariff} scheme, the apps charge {a venue} according to the number of users playing the apps at the venue. {For example, Pokemon Go charges {a venue} up to \$0.50 per game player \cite{mcdonald}. 
In the \emph{lump-sum-only tariff} scheme, the apps (e.g., Snapchat) charge a venue a lump-sum fee, which is independent of the number of players at the venue.} 
We will show that these existing tariff schemes are not able to fully incentivize the venues' investments and achieve desirable users' experience. 
This motivates us to design a tariff scheme that induces high infrastructure investments at the venues and increases the number of users interacting with the POIs.

\subsection{Surveys}
To have a complete picture of the problem, we need to understand the POI-based collaboration's impact from the users' perspective. 
{There are several existing market surveys (such as \cite{SlantMarketing} and \cite{ClickZ}) about Pokemon Go players' engagements with the venues like restaurants, cafes, and bars. 
For example, in SLANT's survey, $71$\% of the $500$ respondents had visited these venues because of the POI features, and $51$\% of the respondents had visited at least one venue \emph{for the first time} because of Pokemon Go \cite{SlantMarketing}. 
These data reveal the venues' potential benefits from becoming POIs.}

{Because there is no prior survey about the dependence of users' experience on the POIs' infrastructure, we conducted a new survey involving $103$ Pokemon Go players in North America, Europe, and Asia.} $81$\% of the surveyed players stated that the infrastructure, including Wi-Fi networks, smartphone chargers, and air conditioners, could enhance their game experience. This data reveals that the app-related infrastructure at the POIs is important for the players.

Our survey {also} reveals the existence of both the \emph{network effect} and \emph{congestion effect} among the players. The \emph{network effect} means that when many players interact with the POI, each player's experience might improve, {as the players can share the app's information and play the app together}. The \emph{congestion effect} arises when the players compete for the limited infrastructure at the POI (e.g., Wi-Fi network), and this might deteriorate each player's experience. In our survey, $64$\% of the players stated that their game experience could be improved if there are nearby players playing the game (i.e., \emph{network effect}), and $59$\% of the players thought that the Wi-Fi speeds at the POIs affected their game experience (i.e., \emph{congestion effect}). More details of the survey can be found in Section \ref{SM:survey} of the supplemental material.

\subsection{Our Contributions}
In this work, we build a detailed model to capture and analyze the strategic interactions among an app, a venue, and users. {{Inspired by}} the two common tariffs in the market (i.e., lump-sum-only tariff and per-player-only tariff), {we design a \emph{two-part tariff}, under which the app charges the venue based on a lump-sum fee and a per-player charge.} {{We will show that the two-part tariff enjoys the advantages of both the lump-sum-only tariff and the per-player-only tariff. The two-part tariff has been applied to many commercial practices. For example, an amusement park may charge an admission fee and also a price for each ride that a user takes in the park. A shopping club can charge an annual membership fee and meanwhile charge for each of a customer's actual purchases.}}

Considering the inherent leader-follower relations among the app, venue, and users, {we model the problem by a three-stage Stackelberg game: in Stage I, the app announces the two-part tariff; in Stage II, the venue decides whether to become a POI and how much to invest in the infrastructure; in Stage III, the users decide whether to visit the venue and whether to interact with the POI. 
The game's analysis is very interesting and challenging because of the coexistence of the \emph{network effect} and \emph{congestion effect}.} 

Our first main result is the design of a \emph{charge-with-subsidy} tariff scheme, which achieves the highest app's revenue among a general class of tariff schemes (i.e., those charge the venue according to the venue's choices, the fraction of users consuming the offline products, and the fraction of users interacting with the POI). Specifically, we show that the app's optimal two-part tariff includes a \emph{positive} lump-sum fee and a \emph{negative} per-player charge, which implies that the app should first \emph{charge} the venue for becoming a POI, and then \emph{subsidize} the venue every time a user interacts with the POI. 
Furthermore, the amount of the per-player subsidy should equal the app's \emph{unit advertising revenue}, which is the app's revenue from showing the advertisements to one user. 
We also study its implementation in the case where the app does not know the \emph{unit advertising revenue} when it announces the tariff, and prove that a risk-averse app should choose a smaller lump-sum fee and a larger per-player charge (i.e., a smaller per-player subsidy).

Our second main result is the provision of counter-intuitive guidelines for the app regarding the type of venues to collaborate with. We analytically study the influences of the app's features (such as the congestion effect factor, network effect factor, and unit advertising revenue) and venue's characteristics (such as the venue's quality, venue's popularity, and population size) on the app's revenue, and obtain several important results. First, when the app's congestion effect factor is large,{\footnote{For example, a bandwidth-consuming app has a large congestion effect factor, since the users will easily experience the network congestion if they use the app in a low-speed Wi-Fi network.}} the app should collaborate with a low-quality venue, whose offline products induce low utilities to the users. Second, when the unit advertising revenue is small, the app may avoid collaborating with a popular venue (whose products are liked by a large fraction of users) or a venue in a busy area (where the number of users is large). 
One key reason behind these counter-intuitive results is that if a venue is high-quality, popular, or in a busy area, it already attracts many visitors before becoming a POI. After it becomes a POI, the initial visitors induce large congestion to the new visitors in terms of playing the app. This potentially decreases the number of new visitors, and further reduces the venue's willingness to become a POI. We analytically derive the conditions under which this negative impact dominates. In this case, the app should avoid collaborating with a venue which is high-quality, popular, or in a busy area.

{We summarize our major contributions as follows:}
\begin{itemize}
\item \emph{Analytical Study of POI-Based Collaboration between Online Apps and Offline Venues}: {Motivated by our survey, we model the interactions among the app, venue, and users as a three-stage game, and characterize their equilibrium strategies.} The analysis is particularly challenging because of the coexistence of network effect and congestion effect.  
\item {\emph{Design of Optimal Two-Part Tariff}: We design the optimal two-part tariff for the app and show the \emph{charge-with-subsidy} structure.} We also consider the tariff design under the uncertainty about the unit advertising revenue, which makes our tariff scheme robust for implementation.
\item \emph{Analysis of Parameters' Influences}: We provide counter-intuitive guidelines for the collaboration via studying the influences of the app's and venue's characteristics on the app's revenue. For example, we show that a bandwidth-consuming app should collaborate with a low-quality venue.
\item \emph{Comparison with State-of-the-Art Schemes}: We analytically prove that our two-part tariff achieves the highest app's revenue among a general class of schemes. We also numerically show our scheme's performance improvement over the current market practices (i.e., {lump-sum-only tariff} and {per-player-only tariff}). 
\end{itemize}

\section{Literature Review}\label{sec:related} 

\subsection{Cooperation of Online and Offline Businesses}
There are few references studying the cooperation between online and offline businesses. Berger \emph{et al.} in \cite{berger2006optimal} investigated the cooperative advertising between a manufacturer who has an online channel and a retailer who has an offline channel. Yu \emph{et al.} in \cite{yu2017public} studied a situation where the online advertisers sponsor the venues' public Wi-Fi services, and deliver mobile advertisements to the venues' visitors. Our work is closely related to \cite{pamuru2017impact} and \cite{colley2017geography}, which conducted \emph{empirical} studies of Pokemon Go's impact on the offline businesses. {Pamuru \emph{et al.} in \cite{pamuru2017impact} collected consumers' reviews of 2,032 restaurants in Houston, and analyzed the correlation between the reviews and whether the restaurants are covered by the POIs (``PokeStops''). Colley \emph{et al.} in \cite{colley2017geography} surveyed $375$ Pokemon Go players, and showed that $46$\% of the players had purchased the venues' offline products because of the POIs. Different from \cite{pamuru2017impact} and \cite{colley2017geography},} our work surveys the impact of the venues' infrastructure on Pokemon Go players' game experience, and {provides the first model and analysis for the cooperation between online apps and offline businesses.}

\subsection{Two-Part Tariffs}
{Since the studies in \cite{oi1971disneyland} and \cite{lewis1941two}, there have been many references analyzing the two-part tariffs and their applications.} 
For example, references \cite{cachon2010competing} and \cite{shin2010firms} analyzed the two-part tariff contracts for the manufacturers or suppliers to coordinate the channels. 
References \cite{zhang2017contract} and \cite{fibich2017optimal} designed the two-part tariffs (or three-part tariffs, which additionally consider free units of service) for the service providers to extract the consumer surplus. 
We consider a two-part tariff scheme, because it can induce the buyer's efficient decision (e.g., welfare-maximizing decision). 
This is particularly useful in the POI-based collaboration, where the app (seller) induces the venue's (buyer's) efficient investment. 
{{Our two-part tariff design is different from those in \cite{oi1971disneyland}--\!\cite{fibich2017optimal}. First, the schemes proposed in \cite{oi1971disneyland}--\!\cite{fibich2017optimal} include positive per-unit charges, while our optimal two-part tariff includes a \emph{negative} per-player charge.}} This is because the investment cost is paid by the venue rather than the app, and the app needs to subsidize the venue's investment via a {negative} per-player charge. {{Second, we consider both the network effect and the congestion effect among the users, which significantly complicates the optimal design of the lump-sum fee. The optimal lump-sum fee's complicated dependence on the system parameters leads to some counter-intuitive guidelines for the POI-based collaboration (e.g., a bandwidth-consuming app should collaborate with a low-quality venue).}}


\subsection{Congestion Effect and Network Effect}
{{Some prior studies analyzed mobile services by jointly considering the congestion effect and network effect among users \cite{zhang2018motivating,gong2015network,xiong2017economic,bai2016optimal}. For example, Zhang \emph{et al.} in \cite{zhang2018motivating} studied mobile caching users, who pre-cache contents and disseminate the contents to users requesting them. The authors considered the delay caused by serving a large number of content requests and also the social connection among the users. Gong \emph{et al.} in \cite{gong2015network} analyzed a wireless provider's data pricing by jointly considering the congestion effect in the physical wireless domain and the social relation among users. In these studies, the congestion levels are determined by the congestion effect factors and users' demand, and service providers cannot alleviate the congestion by investing in the service-related infrastructure. In contrast, our work focuses on studying the venue's investment in the app-related infrastructure and how the app incentivizes the venue to invest, which is the contribution of our work.}}

\section{Model}\label{sec:model}
As explained above, an app usually collaborates with a store/restaurant chain (e.g., Sprint, McDonald's, and KFC). Because the venues in a chain are typically strategically spaced out to avoid internal competitions, we focus on the interaction between an app and a chain's representative venue. 
Our study serves as a first step towards understanding the more general scenario where an app interacts with multiple venues of different owners in the same area. We will briefly discuss the challenges of analyzing this scenario in Section \ref{sec:conclusion}. 

In the following, {we introduce the strategies of the app, the representative venue, and the users, and formulate their interactions as a three-stage game.} 

\subsection{App's Two-Part Tariff}\label{subsec:app}
{Since most popular apps (e.g., Pokemon Go, Snapchat, and Snatch) are free to users, we {{consider an app that does not charge the users.}} In our model, the app only decides the two-part tariff. When the venue becomes a POI, its payment to the app contains: (i) the lump-sum fee $l\in{\mathbb R}$, and (ii) the product between the per-player charge $p\in{\mathbb R}$ and the number of users interacting with the POI. Note that both $l$ and $p$ can be negative, in which case the venue \emph{receives} a payment from the app. 
The app's revenue has two components: (i) the venue's payment, and (ii) the advertising revenue from the in-app advertisements.}

\subsection{Venue's POI and Investment Choices}
We use $r\in\left\{0,1\right\}$ to denote the venue's choice to become a POI ($r=1$) or not ($r=0$). Moreover, we use $I\ge0$ to denote the venue's investment level on the app-related infrastructure, e.g., smartphone chargers and Wi-Fi networks. Note that cellular technologies (e.g., LTE technology) suffer from building penetration loss and may have poor indoor performance \cite{andersen1995propagation}. Hence, it is necessary for the venue to offer high-quality Wi-Fi service, which guarantees users' wireless connection and enhances users' game experience. Some app-related infrastructure might be initially available at the venue. Let parameter $\!I_0\ge0$ denote the \emph{initial investment level}. Accordingly, $I_0+I$ is the \emph{total investment level}. 

\subsection{Users' Types, Decisions, and Payoffs}
{We consider a continuum of users who use the app and seek to interact with a POI. We denote the mass of users by $N$. We assume that the number of users using the app is relatively small, compared with the number of users who do not use the app. In this case, the users who do not use the app are not affected by whether the venue is a POI, and they are not considered in our model.}

\subsubsection{User Type}
Each user is characterized by attributes $\omega$ and $c$. The first attribute $\omega\in\left\{0,1\right\}$ indicates whether the user has an intrinsic interest in consuming the venue's \emph{offline} products. 
We assume that $\eta N$ users have $\omega=1$ (will consume the offline products when visiting the venue), and the remaining $\left(1-\eta\right) N$ users have $\omega=0$. Hence, parameter $\eta\in\left[0,1\right]$ represents the \emph{venue's popularity}. 
The second attribute $c$ denotes the user's transportation cost for visiting the venue, and we assume that $c$ is uniformly distributed in $\left[0,c_{\max}\right]$ \cite{hotelling1929stability,dewenter2012file,rasch2013piracy}. 
{{The app, venue, and users only know the value of $\eta$ and the uniform distribution of $c$, and do not know the actual attributes of each user.}}

\subsubsection{User Decision and Payoff}
{We denote a user's decision by $d$, which has three possible values: $d=0$ (do not visit the venue), $d=1$ (visit the venue but do not interact with the POI), and $d=2$ (visit the venue and interact with the POI).} Before computing a user's payoffs under different decisions $d$, we introduce the following notations:
\begin{itemize}
\item Parameter $U>0$ denotes the utility of a user with $\omega=1$ when it consumes the offline products;
\item Parameter $V>0$ denotes a user's base utility (without considering the network effect and congestion effect) of interacting with the POI;
\item Parameters $\theta\ge0$ and $\delta>0$ denote the \emph{network effect factor} and \emph{congestion effect factor}, respectively; 
\item {Function ${\bar y}\left(r,I\right)\in\left[0,1\right]$ denotes the fraction of users choosing $d=2$ (i.e., interacting with the POI), given the venue's choices $r$ and $I$. The ${\bar y}\left(r,I\right)$ depends on all users' equilibrium decisions, and will be computed in Section \ref{sec:stageIII}.}
\end{itemize}

{A type-$\left(\omega,c\right)$ user's payoff under the venue's choices $r$ and $I$ is
\begin{align}
\nonumber
&\Pi^{\rm user}\left(\omega,c,d,r,I\right)=  \\
& \!\!\!\!\left\{ {\begin{array}{*{20}{l}}
{\!\!0,}&{\!\!{\rm if~}d=0,}\\
{\!\!U\omega-c,}&{\!\!{\rm if~}d=1,}\\
{\!\!U\omega-c+V {+ \theta {\bar y}\left(r,I\right)N}-  \frac{\delta}{I+I_0}{\bar y}\left(r,I\right)N,}&{\!\!{\rm if~}d=2.}\\
\end{array}} \right.\label{equ:userpayoff}
\end{align}

When $d=0$, the user's payoff is $0$.{\footnote{{Even if the users do not interact with the POI (i.e., $d=0$ or $1$), they might still use the app. However, in this case, the app's usage will be much smaller than that when the users interact with the POI. Furthermore, the users who do not interact with the POI might use the app at different locations. Therefore, we do not consider the congestion effect and network effect among these users. Without loss of generality, we normalize these users' utilities of using the app to $0$ in (\ref{equ:userpayoff}).}}} When $d=1$, the user's payoff is the difference between $U\omega$ and the transportation cost $c$. Recall that the user consumes the offline products during its visit if and only if $\omega=1$} (i.e., it has an intrinsic interest in the venue's products).

Compared with $d=1$, the user's payoff under $d=2$ is additionally affected by the base utility of interacting with the POI (i.e., parameter $V$), network effect, and congestion effect. Specifically, the term $\theta {\bar y}\left(r,I\right)N$ corresponds to the network effect, which increases with the number of users interacting with the POI  \cite{viswanathan2005competing,dewenter2012file,rasch2013piracy}. Moreover, the term $-{\delta}{\bar y}\left(r,I\right)N/\left({I+I_0}\right)$ corresponds to the congestion effect of sharing the app-related infrastructure. The congestion level ${\delta}{\bar y}\left(r,I\right)N/\left({I+I_0}\right)$ increases with the number of users interacting with the POI,{\footnote{{In our future work, we can study the case where the users who do not interact with the POI also use the venue's infrastructure (e.g., Wi-Fi networks) and cause congestion. We provide some detailed discussions in Section \ref{SM:model:congestion} of our supplemental material.}}} and decreases with the total investment level $I+I_0$.{\footnote{References \cite{liu2016impact} and \cite{dipalantino2011competition} used similar investment models, which capture the fact that the marginal reduction in the congestion level decreases with the investment. References \cite{nguyen2016cost} and \cite{christodoulou2005price} also considered similar linear congestion costs.}} As we can see in Section \ref{sec:stageIII}, when $I+I_0$ approximates $0$, we have ${\bar y}\left(r,I\right)=0$. This implies that no user will interact with the POI at the equilibrium when there is no app-related infrastructure (e.g., no wireless network). 

\subsubsection{Fractions ${\bar x}\left(r,I\right)$ and ${\bar y}\left(r,I\right)$}\label{subsubsec:fractions}
We use function ${\bar x}\left(r,I\right)\in\left[0,1\right]$ to denote the fraction of users that have $\omega=1$ and visit the venue (i.e., choose $d=1$ or $2$) under the venue's choices $r$ and $I$, and we will compute ${\bar x}\left(r,I\right)$ in Section \ref{sec:stageIII}. 
Function ${\bar x}\left(r,I\right)$ corresponds to the fraction of users consuming the venue's \emph{offline} products, hence the venue wants to increase ${\bar x}\left(r,I\right)$. Recall that ${\bar y}\left(r,I\right)$ is the fraction of users interacting with the POI (i.e., choosing $d=2$) at the equilibrium, hence the app wants to increase ${\bar y}\left(r,I\right)$. 
The difference between ${\bar x}\left(r,I\right)$ and ${\bar y}\left(r,I\right)$ reveals that the venue and app have overlapping but not fully aligned interests in attracting the users.

\subsection{Three-Stage Stackelberg Game}\label{subsec:stackelberg}
\begin{figure}[t]
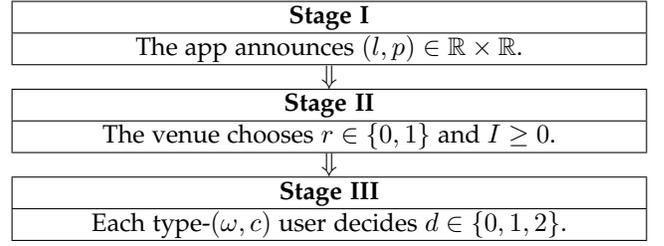

\centering
\begin{tabular}{|p{8cm}<{\centering}|}
\hline
{{\bf{Stage I}}}\\
\hline
{The app announces $\left(l,p\right)\in{\mathbb R}\times {\mathbb R}$.}\\
\hline
\end{tabular}
\centerline{$\Downarrow$}
\begin{tabular}{|p{8cm}<{\centering}|}
\hline
{\bf{Stage II}}\\
\hline
{The venue chooses $r\in\left\{0,1\right\}$ and $I\ge0$.}\\
\hline
\end{tabular}
\centerline{$\Downarrow$}
\begin{tabular}{|p{8cm}<{\centering}|}
\hline
{\bf{Stage III}}\\
\hline
{Each type-$\left(\omega,c\right)$ user decides $d\in\left\{0,1,2\right\}$.}\\
\hline
\end{tabular}
\caption{Three-Stage Game.}\label{fig:threestage}
\end{figure}

We formulate the interactions among the app, venue, and users by a three-stage Stackelberg game, as illustrated in Fig. \ref{fig:threestage}. Since the app has the market power and decides whether to tag the venue as a POI, we assume that the app is the leader and first-mover in the game. {{In Section \ref{subsubsec:revenuepay}, we will discuss the case where the app and the venue negotiate the two-part tariff via bargaining.}}

We assume that the users' maximum transportation cost $c_{\max}$ is large so that $c_{\max}>U+V+\theta N$. This captures a general case where some users are located far from the venue and will not visit it even if it becomes a POI. References \cite{dewenter2012file} and \cite{rasch2013piracy} considered similar cases, i.e., the range of users' transportation costs is large so that a firm cannot attract all users in the market. {{We summarize our paper's key notations in Section \ref{SM:notation} of our supplemental material.}}

\section{Three-Stage Game Analysis}\label{sec:analyzegame}
In this section, we analyze the three-stage game by backward induction. Because of page limit, we leave all detailed proofs in our supplemental material. 

\subsection{Stage III: Users' Decisions}\label{sec:stageIII}
Given the app's tariff $\left(l,p\right)$ in Stage I and the venue's choices of $r$ and $I$ in Stage II, each type-$\left(\omega,c\right)$ user solves the following problem in Stage III.
\begin{problem}
A type-$\left(\omega,c\right)$ user decides $d^*$ by solving
\begin{align}
& \max \Pi^{\rm user}\left(\omega,c,d,r,I\right)\\
& {\rm~var.~} d\in \left\{ {\begin{array}{*{20}{l}}
{\left\{0,1\right\},}&{{\rm~if~}r=0,}\\
{\left\{0,1,2\right\},}&{{\rm~if~}r=1,}
\end{array}} \right.\label{equ:useropt:var}
\end{align}
where the payoff function $\Pi^{\rm user}\left(\omega,c,d,r,I\right)$ is given in (\ref{equ:userpayoff}). 
\end{problem}

\begin{figure}[t]
  \centering
  \includegraphics[scale=0.35]{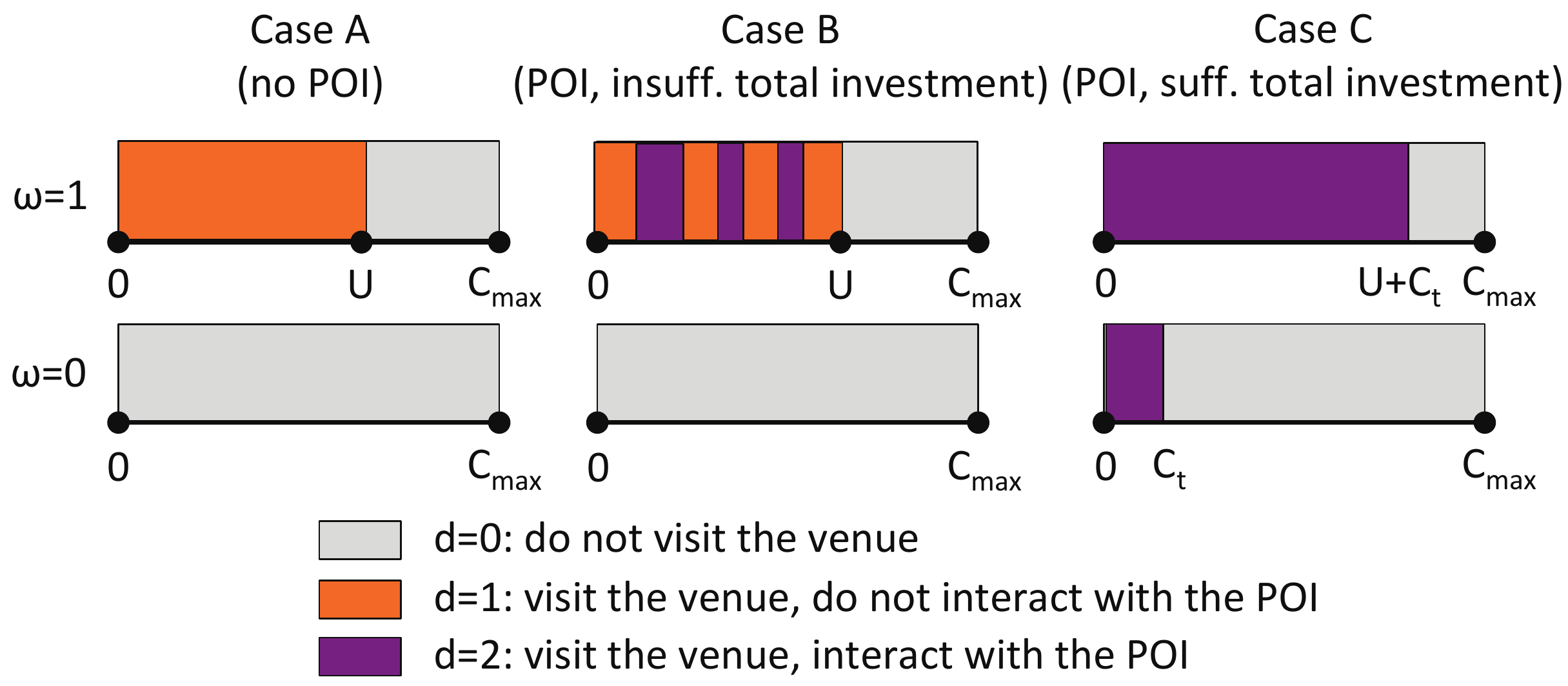}\\
  \caption{Users' equilibrium decisions in Stage III. We use three colors to indicate the equilibrium decisions of the users with different $\omega$ and $c$. In Case B, users have infinitely many equilibria, with one example illustrated here.}
  \label{fig:stageIII}
\end{figure}

Here, (\ref{equ:useropt:var}) implies that the user can interact with the POI if and only if the venue is a POI. Based on the venue's choices of $r$ and $I$ (i.e., whether the venue becomes a POI and how much to invest), we show the users' optimal decisions in the following three cases and illustrate them in Fig. \ref{fig:stageIII}. 

\subsubsection{Case A: No POI}
In Proposition \ref{proposition:stageIII:a}, we will show that when the venue is not a POI, only the users with intrinsic interests in the offline products (i.e., $\omega=1$) and small transportation costs (i.e., $c<U$) will visit the venue. We also derive ${\bar x}\left(r,I\right)$ and ${\bar y}\left(r,I\right)$ in this case. 

\begin{proposition}\label{proposition:stageIII:a}
When $r=0$, the unique equilibrium at Stage III is
\begin{align}
d^*\left(\omega,c,r,I\right)= \left\{ {\begin{array}{*{20}{l}}
{1,}&{{\rm~if~}c\in\left[0,U\omega\right),}\\
{0,}&{{\rm~if~}c\in\left[U\omega,c_{\max}\right],}
\end{array}} \right.\label{equ:user:caseA}
\end{align}
where $\omega\in\left\{0,1\right\}$ and $c\in\left[0,c_{\max}\right]$.{\footnote{{\label{footnote:boundary}At the equilibrium, the user whose $\omega$ and $c$ satisfy $c=U\omega$ has the same payoff under choices $d=0$ and $d=1$. This user's decision does not affect the computation of ${\bar x}\left(r,I\right)$ (and the analysis of Stages II and I), because $c$ follows a continuous distribution and the probability for a user to have $c=U\omega$ is zero. Without affecting the analysis, we assume that such a user always chooses $d=0$ to simplify the presentation. Similar assumptions are made in Propositions \ref{proposition:stageIII:b} and \ref{proposition:stageIII:c}.}}} 
Moreover, ${\bar x}\left(r,I\right)=\eta {U}/{c_{\max}}$ and ${\bar y}\left(r,I\right)=0$.
\end{proposition}

\subsubsection{Case B: POI with Insufficient Total Investment} In Proposition \ref{proposition:stageIII:b}, we will show that even after becoming a POI, the venue cannot attract new visitors (compared with Case A) if its total investment $I+I_0\le \frac{\delta}{\theta+\frac{Vc_{\max}}{\eta UN}}$. 
We define $I_{\rm th}\triangleq \frac{\delta}{\theta+\frac{Vc_{\max}}{\eta UN}}$ as the \emph{threshold investment level}. Moreover, we say the total investment is \emph{insufficient} if $I+I_0\le I_{\rm th}$, and it is \emph{sufficient} otherwise. 

\begin{proposition}\label{proposition:stageIII:b}
When $r=1$ and $I+I_0\le I_{\rm th}$, the unique form of equilibrium at Stage III is
\begin{align}
d^*\left(\omega,c,r,I\right)= \left\{ {\begin{array}{*{20}{l}}
{2,}&{{\rm~if~}c\in{\hat{\cal C}}{\rm~and~}\omega=1,}\\
{1,}&{{\rm~if~}c\in[0,U\omega)\setminus {\hat{\cal C}},}\\
{0,}&{{\rm~if~}c\in\left[U\omega,c_{\max}\right],}\\
\end{array}} \right.\label{equ:user:caseB}
\end{align}
where $\omega\in\left\{0,1\right\}$, $c\in\left[0,c_{\max}\right]$, and ${\hat{\cal C}}\subseteq \left[0,U\right)$ can be any set that satisfies $\eta\int_{0}^{U}\frac{1}{{c_{\max}}}{{\boldsymbol 1}_{\left\{c\in{\hat{\cal C}}\right\}}} dc={\frac{V}{\left(\frac{\delta}{I+I_0}-\theta\right) N}}$.{\footnote{Here, ${\boldsymbol 1}_{\left\{\cdot\right\}}$ is the indicator function, which equals $1$ if the event in the braces is true, and equals $0$ otherwise.}} Moreover, ${\bar x}\left(r,I\right)=\eta \frac{U}{c_{\max}}$ and ${\bar y}\left(r,I\right)={\frac{V}{\left(\frac{\delta}{I+I_0}-\theta\right) N}}$.
\end{proposition}
 
In Case B, the app-related infrastructure simply enables some of the initial visitors (i.e., the visitors to the venue when the venue is not a POI) to interact with the POI. We use $\hat{\cal C}$ to denote the set of these visitors' transportation costs in (\ref{equ:user:caseB}). There are infinitely many ${\hat {\cal C}}$ satisfying Proposition \ref{proposition:stageIII:b}, and hence the users have infinitely many equilibria in Case B. Proposition \ref{proposition:stageIII:b} shows that all the equilibria lead to the same ${\bar x}\left(r,I\right)$ and ${\bar y}\left(r,I\right)$, so the existence of infinitely many equilibria does not affect the analysis of Stages II and I. Note that $\hat{\cal C}$ need not be an interval. This is because each initial visitor has the incentive to interact with the POI until the fraction of visitors interacting with the POI reaches ${\frac{V}{\left(\frac{\delta}{I+I_0}-\theta\right) N}}$. We show one example of $\hat{\cal C}$ in Fig. \ref{fig:stageIII}, where the set of transportation costs of the initial visitors who interact with the POI consists of three intervals (i.e., the purple intervals). The total ``length'' of these purple intervals is ${\frac{V c_{\max}}{\left(\frac{\delta}{I+I_0}-\theta\right) \eta N}}$, which is no greater than $U$ in Case B.

\subsubsection{Case C: POI with Sufficient Total Investment} In Proposition \ref{proposition:stageIII:c}, we will show that after becoming a POI, the venue can attract new visitors (compared with Case A) if its total investment is sufficient. As we can see in Fig. \ref{fig:stageIII}, {the new visitors include users without intrinsic interests in the offline products (i.e., $\omega=0$). 

\begin{proposition}\label{proposition:stageIII:c}
When $r=1$ and $I+I_0>I_{\rm th}$, the unique equilibrium at Stage III is
\begin{align}
d^*\left(\omega,c,r,I\right)= \left\{ {\begin{array}{*{20}{l}}
{2,}&{{\rm~if~}c\in\left[0,U\omega+c_t\right),}\\
{0,}&{{\rm~if~}c\in\left[U\omega+c_t,c_{\max}\right],}\\
\end{array}} \right.
\end{align}
where $\omega\in\left\{0,1\right\}$, $c\in\left[0,c_{\max}\right]$, and $c_t \triangleq {\frac{Vc_{\max}\left(I+I_0\right)-\eta UN\delta+\eta UN\theta \left(I+I_0\right)}{c_{\max}\left(I+I_0\right)+N\delta -N\theta\left(I+I_0\right)}}$. Moreover, ${\bar x}\left(r,I\right)=\eta\frac{U+c_t}{c_{\max}}$ and ${\bar y}\left(r,I\right)=\frac{\eta U+c_t}{c_{\max}}$.
\end{proposition}}

\subsection{Stage II: Venue's POI and Investment Choices}\label{sec:stageII}
{In Stage II, the venue solves the following problem by responding to the app's tariff $\left(l,p\right)$ in Stage I and anticipating the users' decisions $d^*\left(\omega,c,r,I\right)$ in Stage III.

\begin{problem}\label{problem:venue}
The venue makes the POI choice $r^*$ and investment choice $I^*$ by solving
\begin{align}
& \max \Pi^{\rm venue}\left(r,I,l,p\right)\triangleq b N {\bar x}\left(r,I\right)-k I- r \left(l+pN {\bar y}\left(r,I\right)\right) \label{equ:venueopt:obj}\\
& {\rm~var.~} r\in\left\{0,1\right\}, I\ge 0.
\end{align}
Here, $b>0$ is the venue's profit due to one user's consumption of the offline products, and $k>0$ denotes the unit investment cost.
\end{problem}
In (\ref{equ:venueopt:obj}), $\Pi^{\rm venue}\left(r,I,l,p\right)$ is the venue's payoff. The term $b N {\bar x}\left(r,I\right)$ is the venue's aggregate profit due to its offline products' sales, the term $kI$ is the investment cost \cite{liu2016impact}, and the term $r \left(l+pN {\bar y}\left(r,I\right)\right)$ is the payment to the app under the two-part tariff. Recall that ${\bar x}\left(r,I\right)$ and ${\bar y}\left(r,I\right)$ are the fractions of users consuming the offline products and interacting with the POI, respectively, and they are given in Propositions \ref{proposition:stageIII:a}, \ref{proposition:stageIII:b}, and \ref{proposition:stageIII:c} in Section \ref{sec:stageIII}.} The fact that ${\bar x}\left(r,I\right)$ and ${\bar y}\left(r,I\right)$ have different and possibly complicated expressions under different values of $r$ and $I$ makes the analysis of Problem \ref{problem:venue} quite challenging.

Next, we analyze three situations with different $I_0$ and $\delta$, and derive the venue's corresponding optimal choices.

\subsubsection{Situation I: Small Initial Investment and Large Congestion Effect}\label{subsec:situI}

\begin{figure}[t]
  \centering
  \includegraphics[scale=0.43]{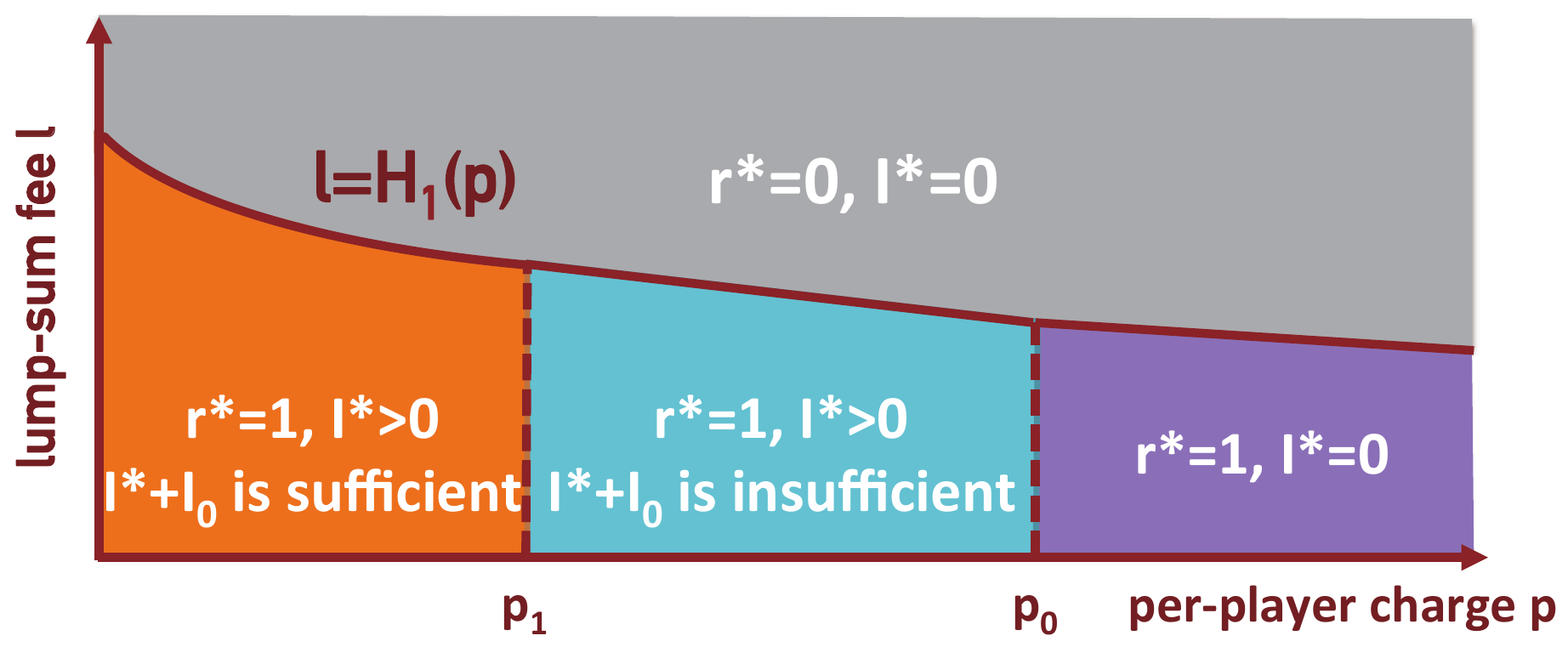}\\
  \caption{Situation I of Venue's POI Decision $r^*$ and Investment $I^*$ in Stage II.}
  \label{fig:stageII}
\end{figure}

In Proposition \ref{proposition:stageII:situationI}, we will show that when the initial investment $I_0\le I_{\rm th}$ and the congestion effect factor $\delta>\delta_{\rm th}\triangleq \frac{\left(Vc_{\max}+\theta\eta UN\right)\left(b\eta \left(Vc_{\max}+\theta\eta U N\right)-\theta I_0 c_{\max}k \right)}{k c_{\max} \eta U \left(c_{\max}-\theta N\right)}$, the venue's chosen investment level $I^*\left(l,p\right)$ may be positive and meanwhile lead to an insufficient total investment (i.e., $I^*\left(l,p\right)+I_0\le I_{\rm th}$). We call $\delta_{\rm th}$ as the \emph{threshold congestion effect factor}. 
We illustrate Proposition \ref{proposition:stageII:situationI} in Fig. \ref{fig:stageII} (the illustrations of Propositions \ref{proposition:stageII:situationII} and \ref{proposition:stageII:situationIII} are provided in Section \ref{SM:illustration:pro5} of the supplemental material).

\begin{figure*}
{{
\begin{align}
H_1\left(p\right)\triangleq  \left\{ {\begin{array}{*{20}{l}}
{-\frac{N}{c_{\max}}b\eta^2U+\frac{N}{c_{\max}-N\theta}\left(\sqrt{\left(V+\eta U\right)\left(b\eta -p\right)}-\sqrt{\delta k}\right)^2+kI_0,}&{{\rm if~}p< p_1,}\\
{-pN\eta \frac{U}{c_{\max}}-k I_{\rm th}+k I_0,}&{{\rm if~}p_1\le p\le p_0,}\\
{-\frac{V}{\frac{\delta}{I_0}-\theta}p,}&{{\rm if~}p>p_0.}\\
\end{array}} \right.\label{equ:H1}
\end{align}}}
\vspace{-0.5cm}

{{
\begin{align}
H_2\left(p\right)\triangleq \left\{ {\begin{array}{*{20}{l}}
{-\frac{N}{c_{\max}}b\eta^2U+\frac{N}{c_{\max}-N\theta}\left(\sqrt{\left(V+\eta U\right)\left(b\eta -p\right)}-\sqrt{\delta k}\right)^2+kI_0,}&{{\rm if~}p< p_2,}\\
{-\frac{V}{\frac{\delta}{I_0}-\theta}p,}&{{\rm if~}p\ge p_2.}\\
\end{array}} \right.\label{equ:H2}
\end{align}}}

\vspace{-0.5cm}
\hrulefill

\end{figure*}

\begin{proposition}\label{proposition:stageII:situationI}
{When {$I_0\le I_{\rm th}$} and {$\delta>\delta_{\rm th}$}, the venue's optimal choices are 
{\small
\begin{align}
\nonumber
& \!\left(r^*\left(l,p\right),I^*\left(l,p\right)\right)\!=\!\\
& \!\!\!\left\{ {\begin{array}{{l}{l}}
{\!\!\!\left(0,0\right),}&{\!\!\!\!\!{\rm if~}l\!>\!H_1\!\left(p \right)\!,}\\
{\!\!\!\!\!\!\left(\!1,\!\frac{N}{{c_{\max}\!-\!N\theta}}\!\sqrt{\!\frac{\delta\left(V+\eta U\right)\left(b\eta -p\right)}{k}}\!-\! \!\frac{\delta N}{{c_{\max}-N\theta}} {-\!I_0}\!\!\right)\!\!,}&{\!\!\!\!\!{\rm if~}l\!\le\! H_1\!\left(p \right)\!,p\!<\! p_1,}\\
{\!\!\!\left(1,I_{\rm th}-I_0\right),}&{\!\!\!\!\!{\rm if~}l\!\le\! H_1\!\left(p \right)\!,p_1\!\le \!p\!\le\! p_0,}\\
{\!\!\!\left(1,0\right),}&{\!\!\!\!\!{\rm if~}l\!\le\! H_1\!\left(p \right)\!,p\!>\!p_0,}
\end{array}} \right.\label{equ:venue:situationI}
\end{align}}
where $p_0\triangleq -k\frac{\left(\delta-\theta I_0\right)c_{\max}}{Vc_{\max}+\theta\eta UN}<0$, $p_1 \triangleq b\eta-\frac{\delta k \left(V+\eta U\right)c_{\max}^2}{\left(Vc_{\max}+\theta\eta UN\right)^2}$,} and function $H_1\left(p\right)$ defined in (\ref{equ:H1}) are used to describe different conditions of the app's tariff $\left(l,p\right)$ (in Fig. \ref{fig:stageII}, we can divide the $\left(p,l\right)$-plane into four regions based on $p_0$, $p_1$, and $H_1\left(p\right)$). 
\end{proposition}
{First, we see that the venue will become a POI (i.e., $r^*\left(l,p\right)=1$) if and only if $l$ and $p$ satisfy $l\le H_1\left(p \right)$ (i.e., the orange, blue, and purple parts in Fig. \ref{fig:stageII}). This implies that $H_1\left(p\right)$ reflects the maximum lump-sum fee under which the venue will be a POI in Situation I, given the per-player charge $p$.} 
We can show that $H_1\left(p\right)$ is convexly decreasing in $p$. Intuitively, when the app increases $p$, it has to reduce $l$ to ensure that the venue becomes a POI.

{Second, we discuss the venue's investment $I^*\left(l,p\right)$. When $l>H_1\left(p \right)$, the venue does not become a POI, and hence chooses $I^*\left(l,p\right)=0$. When $l\le H_1\left(p \right)$, $I^*\left(l,p\right)$ is independent of $l$, and is decreasing in $p$. Specifically, $I^*\left(l,p\right)$ has three different expressions based on the value of $p$: (a) when $p< p_1$, the venue chooses $I^*\left(l,p\right)=\frac{N}{{c_{\max}-N\theta}}\sqrt{\frac{\delta\left(V+\eta U\right)\left(b\eta -p\right)}{k}}-\delta \frac{N}{{c_{\max}-N\theta}} {-I_0}>I_{\rm th}-I_0$; (b) when $p_1\le p\le p_0$ (we can prove that $p_1<p_0$ in Situation I), the venue chooses $I^*\left(l,p\right)=I_{\rm th}-I_0$. Since $I^*\left(l,p\right)+I_0\le I_{\rm th}$, the total investment is insufficient, and cannot attract new visitors to the venue.} The reason the venue still chooses a positive investment level in this case is that the per-player charge $p$ is negative (i.e., $p\le p_0<0$). Hence, the app actually incentivizes the venue to invest by charging a negative $p$ (i.e., providing a subsidy); {(c) when $p>p_0$, the per-player charge is large, and the venue does not invest.}

\subsubsection{Situation II: Small Initial Investment and Small Congestion Effect} We will show that when {$I_0\le I_{\rm th}$} and {$\delta\le\delta_{\rm th}$}, if the venue's chosen investment level $I^*\left(l,p\right)$ is positive, it will always lead to a sufficient total investment (i.e., $I^*\left(l,p\right)+I_0> I_{\rm th}$) and hence attract new visitors. {This is different from Situation I, because the congestion effect factor $\delta$ in Situation II is smaller than that in Situation I, which makes it easier for the venue to attract new visitors.} 

We first use Lemma \ref{lemma:p2unique} to introduce $p_2$ (which will be used to describe different conditions of the app's tariff $\left(l,p\right)$ in Proposition \ref{proposition:stageII:situationII}), and then show Proposition \ref{proposition:stageII:situationII}.

\begin{lemma}\label{lemma:p2unique}
When {$I_0\le I_{\rm th}$} and {$\delta\le\delta_{\rm th}$}, there is a unique $p\in\left[p_0,p_1\right]$ that satisfies $\frac{N}{c_{\max}-N\theta}\left(\sqrt{\left(V+\eta U\right)\left(b\eta -p\right)}-\sqrt{\delta k}\right)^2-\frac{N}{c_{\max}}b\eta^2U+kI_0+\frac{V}{\frac{\delta}{I_0}-\theta}p=0$, and we denote it by $p_2$. 
\end{lemma}

\begin{figure*}

{{
\begin{align}
H_3\left(p\right)\triangleq \left\{ {\begin{array}{*{20}{l}}
{-\frac{N}{c_{\max}}b\eta^2U+\frac{N}{c_{\max}-N\theta}\left(\sqrt{\left(V+\eta U\right)\left(b\eta -p\right)}-\sqrt{\delta k}\right)^2+kI_0,}&{{\rm if~}p< p_3,}\\
{{\left(b\eta-p\right)N {\frac{Vc_{\max}I_0-\eta UN\delta+\eta UN\theta I_0}{c_{\max}^2I_0+c_{\max}N\delta -c_{\max}N\theta I_0}}-pN\eta\frac{U}{c_{\max}}},}&{{\rm if~}p\ge p_3.}
\end{array}} \right.\label{equ:H3}
\end{align}}}

\vspace{-0.5cm}
\hrulefill

\end{figure*}

{
\begin{proposition}\label{proposition:stageII:situationII}
When {$I_0\le I_{\rm th}$} and {$\delta\le\delta_{\rm th}$}, the venue's optimal choices are 
{\small
\begin{align}
\nonumber
& \left(r^*\left(l,p\right),I^*\left(l,p\right)\right)= \\
& \!\!\!\left\{ {\begin{array}{{l}{l}}
{\!\!\!\!\left(0,0\right),}&{\!\!\!\!\!{\rm if~}\!l\!>\!H_2\left(p \right)\!,}\\
{\!\!\!\!\!\left(1,\!\frac{N}{{c_{\max}-N\theta}}\!\sqrt{\frac{\delta\left(V+\eta U\right)\left(b\eta -p\right)}{k}}\!-\! \frac{\delta N}{{c_{\max}-N\theta}} {-I_0}\!\!\right)\!\!,}&{\!\!\!\!\!{\rm if~}\!l\!\le\! H_2\!\left(p \right)\!,p\!<\!p_2,}\\
{\!\!\!\!\left(1,0\right),}&{\!\!\!\!\!{\rm if~}\!l\!\le \!H_2\!\left(p \right)\!,p\!\ge\! p_2,}
\end{array}} \right.\label{equ:venue:situationII}
\end{align}}
where $H_2\left(p\right)$ is defined in (\ref{equ:H2}).

\end{proposition}
First, the venue becomes a POI if and only if $l\le H_2\left(p \right)$.} Similar to $H_1\left(p\right)$ in Situation I, $H_2\left(p\right)$ shows, for a given $p$, the maximum lump-sum fee under which the venue will be a POI in Situation II. 
{Second, when $l\le H_2\left(p \right)$, the venue's optimal investment level $I^*\left(l,p\right)$ has two different expressions: (a) when $p< p_2$, the venue achieves a sufficient total investment and attracts new visitors; (b) when $p\ge p_2$, the venue does not invest because of the large per-player charge.} 



\subsubsection{Situation III: Large Initial Investment} In Proposition \ref{proposition:stageII:situationIII}, we will see that when $I_0>I_{\rm th}$, {the venue's total investment is always sufficient, regardless of its chosen investment level $I^*\left(l,p\right)$. In this situation, as long as the venue becomes a POI, it attracts new visitors.}

\begin{proposition}\label{proposition:stageII:situationIII}
{When {$I_0>I_{\rm th}$}, the venue's optimal choices are 
{\small
\begin{align}
\nonumber 
&\left(r^*\left(l,p\right),I^*\left(l,p\right)\right)=\\
& \!\!\!\left\{ {\begin{array}{{l}{l}}
{\!\!\!\!\left(0,0\right),}&{\!\!\!\!\!{\rm if~}\!l\!>\!H_3\!\left(p \right)\!,}\\
{\!\!\!\!\!\left(1,\!\frac{N}{{c_{\max}\!-\!N\theta}}\!\sqrt{\frac{\delta\left(V+\eta U\right)\left(b\eta -p\right)}{k}}\!-\! \frac{\delta N}{{c_{\max}-N\theta}} {-I_0}\!\right)\!,}&{\!\!\!\!\!{\rm if~}\!l\!\le\! H_3\!\left(p \right)\!,p\!<\!p_3,}\\
{\!\!\!\!\left(1,0\right),}&{\!\!\!\!\!{\rm if~}\!l\!\le \!H_3\!\left(p \right)\!,p\!\ge\! p_3,}
\end{array}} \right.\label{equ:venue:situationIII}
\end{align}}
where $p_3\triangleq b\eta-\frac{k\left(\left(c_{\max}-\theta N\right)I_0+\delta N\right)^2}{\delta\left(V+\eta U\right)N^2}$} and $H_3\left(p\right)$ defined in (\ref{equ:H3}) are used to describe different conditions of the app's tariff $\left(l,p\right)$.
\end{proposition}

\subsection{Stage I: App's Two-Part Tariff}\label{sec:stageI}
\subsubsection{Problem Formulation}
{In Stage I, the app solves Problem \ref{problem:app}, anticipating the venue's and users' decisions in Stages II and III, respectively. 
\begin{problem}\label{problem:app}
The app determines $\left(l^*,p^*\right)$ by solving
\begin{align}
\nonumber
& \max R^{\rm app}\left(l,p\right)\triangleq {r^*}\left(l,p\right)\Bigl(l+p N{\bar y}\left(r^*\left(l,p\right),I^*\left(l,p\right)\right)\Bigr)\\
&{~~~~~~~~~~~~~~~~~~~~~~~~~~~~}+\phi N{\bar y}\left(r^*\left(l,p\right),I^*\left(l,p\right)\right) \label{equ:appopt:obj}\\
& {\rm var.~~~} l,p\in{\mathbb R}.
\end{align}
Here, $\!\phi\ge0$ is the unit advertising revenue, representing the app's advertising revenue because of a user's interaction with the POI.
\end{problem}
In (\ref{equ:appopt:obj}), $R^{\rm app}\left(l,p\right)$ is the app's revenue, which has two components: the venue's payment based on the two-part tariff, and the app's advertising revenue. 
Function ${\bar y}\left(r,I\right)$ is given in Propositions \ref{proposition:stageIII:a}, \ref{proposition:stageIII:b}, and \ref{proposition:stageIII:c}. Functions $r^*\left(l,p\right)$ and $I^*\left(l,p\right)$ are given in Propositions \ref{proposition:stageII:situationI}, \ref{proposition:stageII:situationII}, and \ref{proposition:stageII:situationIII}.} 

We can easily extend our model to consider the case where the app has a non-negligible cost $\xi>0$ of supporting a user to interact with the POI (e.g., coordinating {the ``battles'' between this user and other users}) by replacing $\phi$ in (\ref{equ:appopt:obj}) with $\phi-\xi$. {{Furthermore, we focus on the monetary payment between the app and venue in this work. Our framework can be extended to analyze the app-venue collaboration with non-monetary rewards, and we provide the detailed discussions in Section \ref{SM:model:nonmonetary} of the supplemental material.}}

\subsubsection{Optimal Two-Part Tariff}
{We show the app's optimal two-part tariff in Theorem \ref{theorem:optimaltariff}.

\begin{theorem}\label{theorem:optimaltariff}
The app's optimal two-part tariff is 
\begin{align}
p^*=-\phi,{~}l^*={\tilde H}\left(-\phi\right),\label{equ:optimalwp}
\end{align}
where function ${\tilde H}\left(p\right), p\in{\mathbb R},$ is defined as
\begin{align}
& {\tilde H}\left(p\right) \triangleq \left\{ {\begin{array}{*{20}{l}}
{H_1\left(p\right),}&{{\rm if~}{I_0\le I_{\rm th}}{~\rm and~}{\delta>\delta_{\rm th}},}\\
{H_2\left(p\right),}&{{\rm if~}{I_0\le I_{\rm th}}{~\rm and~}{\delta\le\delta_{\rm th}},}\\
{H_3\left(p\right),}&{{\rm if~}{I_0>I_{\rm th}}.}
\end{array}} \right.\label{equ:defineH}
\end{align}
The per-player charge $p^*\le0$ and the lump-sum fee $l^*\ge0$.
\end{theorem}
From Propositions \ref{proposition:stageII:situationI}, \ref{proposition:stageII:situationII}, and \ref{proposition:stageII:situationIII}, ${\tilde H}\left(p\right)$ is the maximum lump-sum fee under which the venue will be a POI, given the $p$. {{In practice, the app simply needs to compute its two-part tariff based on (\ref{equ:optimalwp}) and (\ref{equ:defineH}), and charge the venue accordingly. The computational complexity is $O\left(1\right)$, which does not increase with the system scale (e.g., the number of users).}}

We first discuss the intuitions behind Theorem \ref{theorem:optimaltariff}. With $p^*\le0$, the app \emph{pays} the venue based on the number of users interacting with the POI. This incentivizes the venue to invest in the app-related infrastructure, which attracts more users to interact with the POI.} When $p^*=-\phi$, we can prove that the venue's investment level in response to $p^*$ will also maximize the summation of the app's revenue and the venue's payoff. {Meanwhile, the app sets $l^*={\tilde H}\left(-\phi\right)$, which is the maximum lump-sum fee the venue will accept under $p^*=-\phi$. With $l^*={\tilde H}\left(-\phi\right)$, the app extracts all the venue's surplus. Hence, we can see that $p^*$ and $l^*$ maximize the app's revenue. 

Theorem \ref{theorem:optimaltariff} leads to the following practical insights. The app should announce a \emph{charge-with-subsidy} scheme to the venue: (i) in order to become a POI, the venue needs to pay the app ${\tilde H}\left(-\phi\right)$; (ii) every time a user interacts with the POI, the app pays the venue $\phi$ (unit advertising revenue).} 

{{From Theorem \ref{theorem:optimaltariff}, we can also see the challenge of considering the congestion effect and network effect. Based on (\ref{equ:optimalwp}), (\ref{equ:defineH}), (\ref{equ:H1}), (\ref{equ:H2}), and (\ref{equ:H3}), the optimal lump-sum fee $l^*$ has different concrete expressions under different parameter settings, and the corresponding thresholds (such as $\delta_{\rm th}$, $I_{\rm th}$, $p_0$, $p_1$, $p_2$, and $p_3$) have complicated expressions. If there is no congestion effect, it is equivalent to assuming that $I_0$ goes to infinity. We can prove that in this case, $l^*$ only has one possible expression, and the analysis of the venue's and users' strategies can be significantly simplified. Furthermore, if there is no network effect, the expressions of those thresholds will be much simpler. For example, when $\theta=0$, the value of $\delta_{\rm th}$ becomes $\frac{bV^2}{kU}$.}}

\subsubsection{Two-Part Tariff's Performance}\label{subsubsec:performancetwopart}
Next, we show that our two-part tariff scheme is optimal among the class of tariff schemes that charge the venue according to the venue's choices $r$ and $I$, the fraction of users consuming the venue's products (i.e., ${\bar x}\left(r,I\right)$), and the fraction of users interacting with the POI (i.e., ${\bar y}\left(r,I\right)$). Intuitively, when maximizing the app's revenue, we can focus on this class of tariff schemes and do not need to consider other tariff schemes (e.g., charge the venue according to a particular user's visiting decision), because the app and venue are only interested in the fractions of users consuming their products or using their services. 

Note that the values of ${\bar x}\left(r,I\right)$ and ${\bar y}\left(r,I\right)$ are determined by $r$ and $I$, as shown in Propositions \ref{proposition:stageIII:a}, \ref{proposition:stageIII:b}, and \ref{proposition:stageIII:c}. Therefore, we call the abovementioned class of tariff schemes as $\left(r,I\right)$\emph{-dependent tariff schemes}, and the venue's payment to the app under any such a scheme can be represented by function $T\left(r,I\right):\left\{0,1\right\}\times\left[0,\infty\right)\rightarrow {\mathbb R}$.
For example, our optimal two-part tariff scheme in Theorem \ref{theorem:optimaltariff} is an $\left(r,I\right)${-dependent tariff scheme}, and can be represented by $T\left(r,I\right)\!=\!r\!\left({\tilde H}\left(-\phi\right)-\phi N {\bar y}\left(r,I\right) \right)$. Once the venue becomes a POI (i.e., $r=1$), its payment to the app includes the lump-sum fee (i.e., ${\tilde H}\left(-\phi\right)$) as well as the product between the per-player charge (i.e., $-\phi$) and the number of users interacting with the POI (i.e., $N{\bar y}\left(r,I\right)$). 
Note that the two state-of-the-art schemes, i.e., per-player-only and lump-sum-only tariff schemes, are also $\left(r,I\right)$-dependent tariff schemes.

We say an $\left(r,I\right)${-dependent tariff scheme} is \emph{feasible} if and only if $T\left(0,I\right)=0$, i.e., the venue need not pay the app when the venue does not become a POI. We introduce the following theorem. 
\begin{theorem}\label{theorem:optimalforall}
Our optimal two-part tariff scheme $T\left(r,I\right)=r\left({\tilde H}\left(-\phi\right)-\phi N {\bar y}\left(r,I\right) \right)$ achieves the highest app's revenue among all feasible $\left(r,I\right)${-dependent tariff schemes}.
\end{theorem}
As explained before, the venue's choices under our optimal two-part tariff maximize the summation of the app's revenue and the venue's payoff. Our optimal two-part tariff also extracts all the venue's surplus, which ensures our tariff's optimality among all feasible $\left(r,I\right)${-dependent tariffs}.

\subsubsection{App's Revenue and Venue's Payoff}\label{subsubsec:revenuepay}
{Under $l^*$ and $p^*$, the app's revenue and the venue's payoff are given in the following corollary. 
\begin{corollary}\label{corollary:optimalrevenue}
Under $l^*$ and $p^*$, we have
\begin{align}
& R^{\rm app}\left(l^*,p^*\right)={\tilde H}\left(-\phi\right)\ge0,\label{optimal:app}\\
& \Pi^{\rm venue}\left(r^*\left(l^*,p^*\right),I^*\left(l^*,p^*\right),l^*,p^*\right)=b N \eta \frac{U}{c_{\max}}.\label{optimal:venue}
\end{align}
\end{corollary}
Based on (\ref{equ:appopt:obj}) and $p^*=-\phi$, the app's payment to the venue due to the negative per-player charge cancels out the app's total advertising revenue. Hence, the app's optimal revenue equals its lump-sum fee, i.e., $R^{\rm app}\left(l^*,p^*\right)=l^*={\tilde H}\left(-\phi\right)$. 

From (\ref{optimal:venue}), we see that the venue's payoff under the app's optimal two-part tariff is $b N \eta {U}/{c_{\max}}$, which equals the venue's payoff when it does not become a POI. This is because we assume that the app has the market power. In this case, the app can extract all the venue's surplus via the tariff.} 
In Section \ref{SM:bargaining} of the supplemental material, we have studied a more general bargaining-based negotiation model between the app and venue in Stage I. It is important to note that the bargaining formulation only changes the profit {split} between the app and venue, and does not affect the venue's choices in Stage II and the users' decisions in Stage III. {{Under the bargaining model, the per-player charge will still be $-\phi$, and the lump-sum fee will be the product between ${\tilde H}\left(-\phi\right)$ and a parameter capturing the app's bargaining power.}} Moreover, the venue's payoff increases with its bargaining power and can be higher than $b N \eta {U}/{c_{\max}}$. When both the app and venue have positive bargaining power, the POI-based collaboration leads to a win-win situation for them. 

\section{Guidelines for Collaboration}\label{sec:sensitivity}
In this section, we analyze the influences of the venue's quality $U$, venue's popularity $\eta$, and population size $N$ on the app's optimal revenue. 
As shown by (\ref{equ:defineH}), (\ref{equ:H1}), (\ref{equ:H2}), and (\ref{equ:H3}), the dependence of ${\tilde H}\left(p\right)$'s expression on these parameters is very complicated. The thresholds, such as $I_{\rm th}$, $\delta_{\rm th}$, $p_0$, $p_1$, $p_2$, and $p_3$, are also affected by these parameters. These make the analysis very challenging. The inherent reason behind the complicated dependence is that each of these parameters has different impacts on different components of the app's revenue, as explained in the following subsections. The influences of the other parameters are intuitive (e.g., the app's optimal revenue increases with the network effect factor $\theta$), and hence the corresponding analysis is omitted.

\subsection{Influence of Venue's Quality $U$}\label{subsec:influence:quality}
Recall that if a user has an intrinsic interest in the venue's products, parameter $U$ captures the user's utility of consuming the products. Hence, $U$ reflects the venue's quality. Next, we discuss $U$'s impact on the two components of the app's optimal revenue $R^{\rm app}\left(l^*,p^*\right)$: the advertising revenue and the venue's payment. First, the advertising revenue increases with $U$. When $U\!$ increases, {the venue attracts more users with intrinsic interests in the offline products, which increases the number of users interacting with the POI. This enables the app to obtain a higher advertising revenue.} Second, the impact of $U$ on the venue's payment depends on the relative intensity between the \emph{network effect} and \emph{congestion effect}. When $U$ increases, there are more initial visitors (who visit the venue before it becomes a POI). After the venue becomes a POI and makes sufficient investment, the initial visitors generate a larger \emph{network effect} and a larger \emph{congestion effect} to the new visitors. If the {network effect} dominates over the {congestion effect}, the number of new visitors to the venue increases with $U$, and hence the venue's payment increases with $U$; otherwise, the number of new visitors decreases with $U$, and the venue's payment decreases with $U$. 

Parameter $U$'s impact on $R^{\rm app}\left(l^*,p^*\right)$ is jointly determined by $U$'s impact on the advertising revenue and its impact on the venue's payment. Next, we introduce Propositions \ref{proposition:impactU:smalldelta} and \ref{proposition:impactU:largedelta}.

{\begin{proposition}\label{proposition:impactU:smalldelta}
When $\delta \le \frac{\phi c_{\max} I_0}{b\eta N}+\theta I_0$, $R^{\rm app}\left(l^*,p^*\right)$ increases with $U\in\left(0,\infty\right)$. 
\end{proposition}}

\begin{figure*}[t]
  \centering
  \subfigure[Case $\delta \le \frac{\phi c_{\max} I_0}{b\eta N}+\theta I_0$.]{
    \label{fig:simu:pro7}
    \includegraphics[scale=0.36]{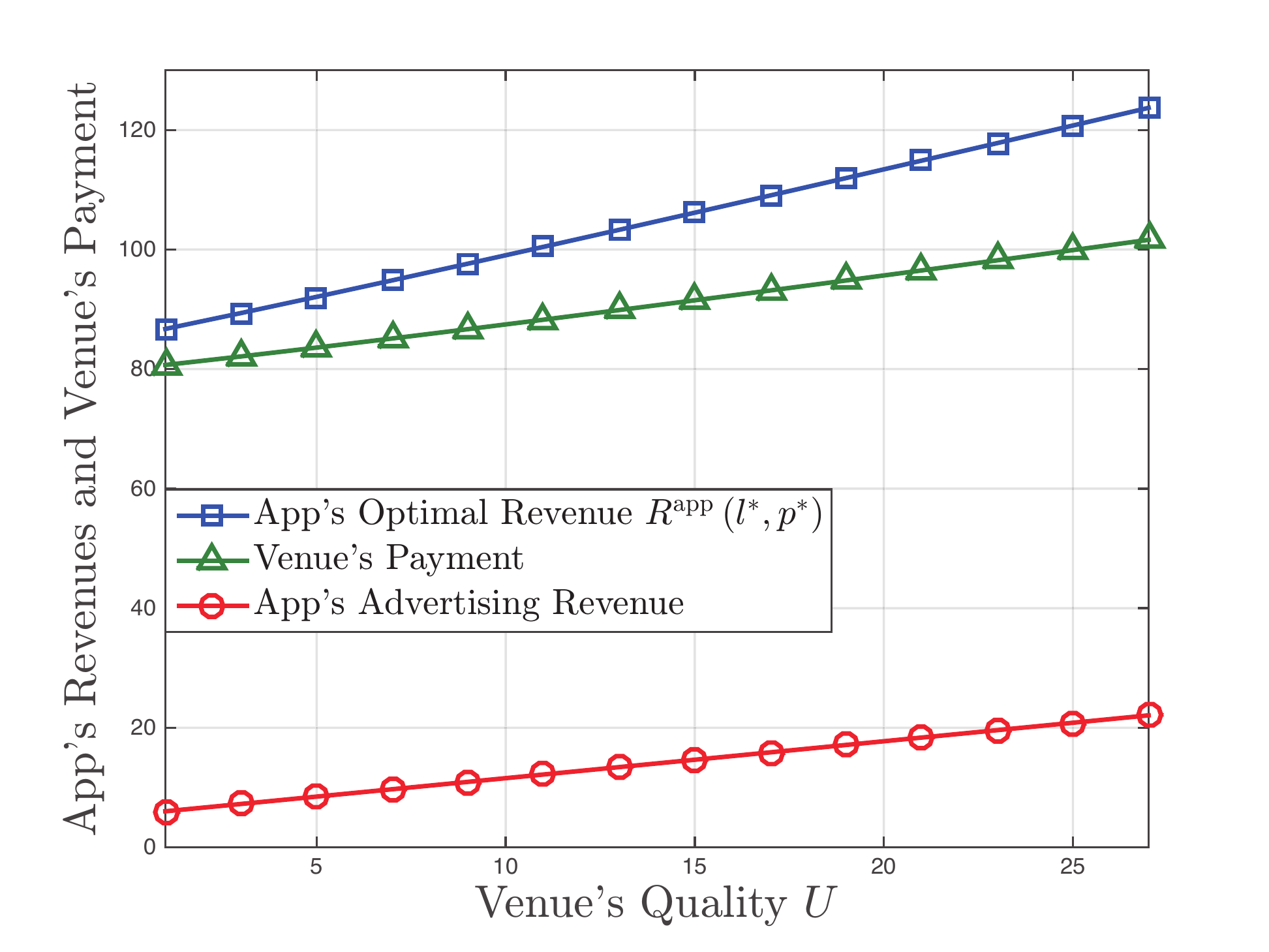}}
  \subfigure[Case $\delta > \frac{\phi c_{\max} I_0}{b\eta N}+\theta I_0$.]{
    \label{fig:simu:pro8}
    \includegraphics[scale=0.36]{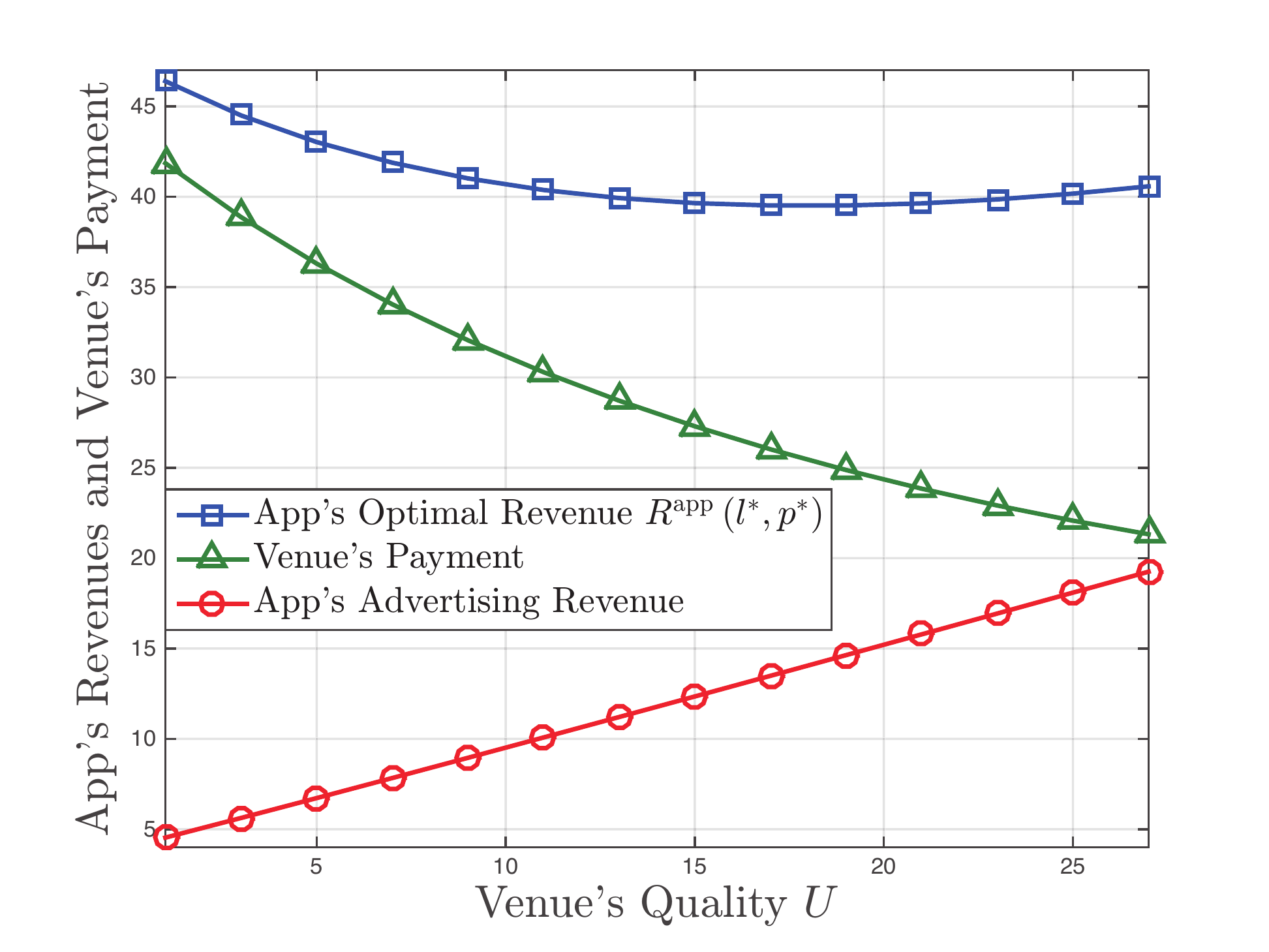}}
  \caption{Illustrations of Propositions \ref{proposition:impactU:smalldelta} and \ref{proposition:impactU:largedelta}.}
\end{figure*}

When $\delta \le \frac{\phi c_{\max} I_0}{b\eta N}+\theta I_0$, either the unit advertising revenue $\phi$ or the network effect factor $\theta$ is large. If $\phi$ is large, the app's optimal revenue mainly consists of the advertising revenue, which increases with $U$. If $\theta$ is large, the network effect is large. Based on our prior discussion, in both cases, the app's optimal revenue increases with $U$. We illustrate Proposition \ref{proposition:impactU:smalldelta} in Fig. \ref{fig:simu:pro7}. We choose $N=400$, $c_{\max}=36$, $V=4.5$, $I_0=1$, $k=2$, $b=3$, $\eta=0.5$, $\theta=0.01$, $\phi=0.1$, and $\delta=0.005$. It is easy to verify that $\delta \le \frac{\phi c_{\max} I_0}{b\eta N}+\theta I_0$. We observe that $R^{\rm app}\left(l^*,p^*\right)$ as well as its two components (venue's payment and app's advertising revenue) increase with $U$.

{\begin{proposition}\label{proposition:impactU:largedelta}
When $\delta > \frac{\phi c_{\max} I_0}{b\eta N}+\theta I_0$, $R^{\rm app}\left(l^*,p^*\right)$ decreases with $U$ for $U\in\left(0,\frac{\delta k\left(b\eta+\phi\right)}{\eta \left(\frac{N\theta b \eta }{c_{\max}}+\phi\right)^2}-\frac{V}{\eta}\right]$, and increases with $U$ for $U\in\left[\frac{\delta k\left(b\eta+\phi\right)}{\eta \left(\frac{N\theta b \eta }{c_{\max}}+\phi\right)^2}-\frac{V}{\eta},\infty\right)$. 
\end{proposition}}

When $\delta > \frac{\phi c_{\max} I_0}{b\eta N}+\theta I_0$, both $\phi$ and $\theta$ are small. Hence, the app's optimal revenue mainly consists of the venue's payment, and the network effect is small. When $U$ increases, the venue's payment first decreases because of the congestion effect. However, the marginal decrease in the venue's payment decreases with $U$.{\footnote{According to Theorem \ref{theorem:optimaltariff} and Propositions \ref{proposition:stageII:situationI}, \ref{proposition:stageII:situationII}, and \ref{proposition:stageII:situationIII}, we can easily prove that if the venue invests in the infrastructure at the equilibrium, i.e., $I^*\left(l^*,p^*\right)>0$, the investment level $I^*\left(l^*,p^*\right)$ concavely increases with $U$. This implies that the marginal increase in the venue's total investment cost $k I^*\left(l^*,p^*\right)$ decreases with $U$. As a result, the marginal decreases in both the venue's willingness to become a POI and its payment to the app decrease with $U$.}} Therefore, the app's optimal revenue first decreases and then increases with $U$. 
We illustrate Proposition \ref{proposition:impactU:largedelta} in Fig. \ref{fig:simu:pro8}.  We let $\delta=0.3$, and the other parameters are the same as in Fig. \ref{fig:simu:pro7}. It is easy to verify that $\delta > \frac{\phi c_{\max} I_0}{b\eta N}+\theta I_0$. We can observe that the venue's payment convexly decreases with $U$, and the app's advertising revenue increases with $U$. Moreover, we can see that $R^{\rm app}\left(l^*,p^*\right)$ first decreases and then increases with $U$. 

{Based on our assumption in Section \ref{subsec:stackelberg}, $U$ is upper-bounded by $c_{\max}-V-\theta N$. If $\delta$ is very large such that $\frac{\delta k\left(b\eta+\phi\right)}{\eta \left(\frac{N\theta b \eta }{c_{\max}}+\phi\right)^2}-\frac{V}{\eta}$ exceeds this upper bound, the app's optimal revenue will decrease with $U$ for $U\in\left(0,c_{\max}-V-\theta N\right)$.} We formally show this result in Corollary \ref{corollary:impactU}.

\begin{corollary}\label{corollary:impactU}
When $\!\delta\!>\! \delta_1\triangleq \max\left\{\frac{\phi c_{\max} I_0}{b\eta N}+\theta I_0,\right.\left.\frac{\left(\frac{N\theta b \eta }{c_{\max}}+\phi\right)^2\left(\eta c_{\max}+\left(1-\eta\right)V-\eta \theta N\right)}{k\left(b\eta+\phi\right)}\right\}$, $R^{\rm app}\left(l^*,p^*\right)$ decreases with $U$ for $U\in\left(0,c_{\max}-V-\theta N\right)$.
\end{corollary}

Based on Proposition \ref{proposition:impactU:smalldelta} and Corollary \ref{corollary:impactU}, if the app has a small congestion effect factor and a large network effect factor, it achieves a high revenue when cooperating with a high-quality venue. If the app's congestion effect factor is very large, it achieves a high revenue when cooperating with a low-quality venue, which is a surprising result. 

In Table \ref{table:1}, we summarize these insights (and the other insights derived in this section), which provide guidelines for the app to select the optimal venue to collaborate with. 

\subsection{Influence of Venue's Popularity $\eta$}\label{subsec:impacteta}
The venue's popularity $\eta\in\left[0,1\right]$ reflects the fraction of users with intrinsic interests in the offline products. In order to understand the impact of $\eta$ on the app's revenue, we again examine its impacts on the advertising revenue and the venue's payment. Compared with $U$, parameter $\eta$'s impact on the advertising revenue is similar, i.e., the advertising revenue increases with $\eta$, but $\eta$'s impact on the venue's payment is more complicated. The impact of $\eta$ on the venue's payment depends not only on the \emph{network effect} and \emph{congestion effect}, but also on the following \emph{alignment effect}. As mentioned in Section \ref{subsubsec:fractions}, the app can gain revenue by delivering advertisements to all types of users, and the venue can only sell its products to the users with intrinsic interests in the venue. When $\eta$ increases, the app and venue have more \emph{aligned} interests in attracting the users, and the venue is more willing to be a POI, which potentially increases the venue's payment. 

In Proposition \ref{proposition:impacteta:largephi}, we show $\eta$'s impact under a large unit advertising revenue $\phi$.

\begin{proposition}\label{proposition:impacteta:largephi}
When $\phi\ge{bV}/{U}$, $R^{\rm app}\left(l^*,p^*\right)$ increases with $\eta\in\left(0,\infty\right)$.
\end{proposition} 

Recall that $b$ is the venue's profit due to one user's consumption of offline products. When $\phi\ge{bV}/{U}$, the unit advertising revenue is large and the venue does not have a strong incentive to become a POI because of the small $b$. In this case, the app's revenue mainly consists of the advertising revenue rather than the venue's payment. Since the advertising revenue increases with $\eta$, the app's revenue increases with $\eta$. 

Next, we use Lemma \ref{lemma:delta1} to introduce $\delta_2$, and use Proposition \ref{proposition:eta:no} to show that when $\phi\!<\!{bV}/{U}$ and $\delta\!>\!\delta_2$, surprisingly, the app's revenue may decrease with $\eta$. 

\begin{lemma}\label{lemma:delta1}
There is a unique $\delta\in\left(\theta I_0,\infty\right)$ that satisfies $\delta^2-2\theta I_0\delta -\frac{bV^2}{kU} \delta+\theta^2 I_0^2=0$, and we denote it by $\delta_2$. 
\end{lemma}

\begin{proposition}\label{proposition:eta:no}
When $\phi<\frac{bV}{U}$ and $\delta>\delta_2$, $R^{\rm app}\left(l^*,p^*\right)$ decreases with $\eta$ for $\eta\in\left(\eta_A,\eta_B\right)$, where $\eta_A \triangleq \frac{\left(bV+\phi U\right)c_{\max}I_0}{2bUN\left(\delta-\theta I_0\right)}$, $\eta_B \triangleq \frac{Vc_{\max}I_0}{UN\left(\delta-\theta I_0\right)}$, and $\eta_A < \eta_B$.
\end{proposition}

\begin{table*}[t]
\centering
\caption{Influences of System Parameters.}\label{table:1}
{\begin{tabular}{|llp{9.2cm}<{}|}
\hline
{{Parameter Conditions}} & {App's Revenue} & {Guidelines for Collaboration}\\
\hline
{{$U$$\uparrow$ and $\delta \le \frac{\phi c_{\max} I_0}{b\eta N}+\theta I_0$}} & {always $\uparrow$} & {If an app has a small congestion effect factor (e.g., it is bandwidth-efficient) and a large network effect factor (e.g., it encourages interactions among users), it should collaborate}\\
{{$U$$\uparrow$ and $\delta > \delta_1$}} & {always $\downarrow$} & {with a high-quality venue; if the congestion effect factor is very large, the app should collaborate with a low-quality venue.}\\
\hline
{{$\eta$$\uparrow$ and $\phi \ge \frac{bV}{U}$}} & {always $\uparrow$} & {If the unit advertising revenue is large, an app should collaborate with a popular venue; otherwise, it may avoid collab-}\\
{{$\eta$$\uparrow$ and $\phi < \frac{bV}{U}$}} & {may $\downarrow$} & {orating with a popular venue.}\\
\hline
{{$N$$\uparrow$ and $\phi \ge \frac{bV}{U}$}} & {always $\uparrow$} & {If the unit advertising revenue is large, an app should collaborate with a venue in a busy area; otherwise, it may avoid}\\
{{$N$$\uparrow$ and $\phi < \frac{bV}{U}$}} & {may $\downarrow$} & {collaborating with a venue in a busy area.}\\
\hline
\end{tabular}}
\end{table*}

If $\phi<{bV}/{U}$, the unit advertising revenue is small and the venue has a strong incentive to be a POI because of the large $b$. In this case, the app's revenue mainly consists of the venue's payment rather than the advertising revenue. If $\delta>\delta_2$, the congestion effect is large, and may dominate over the network effect and alignment effect. Hence, we show that when $\phi<{bV}/{U}$ and $\delta>\delta_2$, $R^{\rm app}\left(l^*,p^*\right)$ decreases with $\eta$ for $\eta\in\left(\eta_A,\eta_B\right)$. 
When $\phi<{bV}/{U}$ and $\delta\le\delta_2$, the analytical study of $\eta$'s impact on $R^{\rm app}\left(l^*,p^*\right)$ is more challenging because of the more complicated comparison between the advertising revenue and venue's payment (affected by the network effect, congestion effect, and alignment effect). In Section \ref{SM:numerical:popularity} of the supplemental material, we numerically show that when $\phi<{bV}/{U}$ and $\delta\le\delta_2$, $R^{\rm app}\left(l^*,p^*\right)$ may also decrease with $\eta$. 
The main insights about $\eta$ are summarized in Table \ref{table:1}.

\subsection{Influence of Population Size $N$}\label{subsec:impactN}
Recall that $N$ captures the population size. A large $N$ implies that the venue is located in a busy area. Compared with $U$, parameter $N$ has a similar impact on the advertising revenue, i.e., the advertising revenue increases with $N$, but has a more complicated impact on the venue's payment. 
Specifically, the impact of $N$ on the venue's payment depends not only on the network effect and congestion effect, but also on the following \emph{proximity effect}. When $N$ increases, there are more users who are close to the venue. In this case, it is easier for the venue to attract a large number of visitors, since the average transportation cost of these visitors decreases. Hence, the venue will be more willing to be a POI, which potentially increases the venue's payment. 

In Proposition \ref{proposition:N:largephi}, we show $N$'s impact under a large unit advertising revenue $\phi$.

\begin{proposition}\label{proposition:N:largephi}
When $\phi\ge{bV}/{U}$, $R^{\rm app}\left(l^*,p^*\right)$ increases with $N\in\left(0,\infty\right)$.
\end{proposition}

The condition $\phi\ge{bV}/{U}$ implies that the app's revenue mainly consists of the advertising revenue (affected by $\phi$) rather than the venue's payment (affected by $b$). Since the advertising revenue increases with $N$, the app's revenue increases with $N$. 

Next, we use Lemma \ref{lemma:delta:2} and Lemma \ref{lemma:N:1} to introduce $\delta_3$ and $N_A$, respectively. In Proposition \ref{proposition:N:no}, we show that when $\phi<{bV}/{U}$ and $\delta>\delta_3$, surprisingly, the app's revenue may decrease with $N$.

\begin{lemma}\label{lemma:delta:2}
There is a unique $\delta\in\left(\theta I_0,\infty \right)$ that satisfies $\delta^2-2\theta I_0 \delta -\frac{V^2\left(b\eta+\phi\right)}{k\left(V+\eta U\right)}\delta + \theta^2 I_0^2=0$, and we denote it by $\delta_3$. 
\end{lemma}

\begin{lemma}\label{lemma:N:1}
When $\phi<\frac{bV}{U}$ and $\delta>\delta_3$, there is a unique $N\in\left(0,\frac{Vc_{\max}I_0}{\eta U\left(\delta-\theta I_0\right)}\right)$ that satisfies 
\begin{multline}
-b \eta^2 c_{\max} U \left(\delta-\theta I_0\right)^2 N^2 +2b \eta^2 c_{\max}^2 I_0 U\left(\theta I_0-\delta\right) N \\
+b\eta c_{\max}^3 I_0^2 V+\phi \left(\eta U+V\right) c_{\max}^3 I_0^2=0.
\end{multline}
We denote this $N$ by $N_A$.
\end{lemma}

\begin{proposition}\label{proposition:N:no}
When $\phi<\frac{bV}{U}$ and $\delta>\delta_3$, $R^{\rm app}\left(l^*,p^*\right)$ decreases with $N$ for $N\in\left(N_A,N_B\right)$, where $N_B \triangleq \frac{Vc_{\max}I_0}{\eta U\left(\delta-\theta I_0\right)}$.
\end{proposition}

When $\phi<{bV}/{U}$ and $\delta>\delta_3$, the venue has a strong incentive to be a POI because of the large $b$, and hence the app's revenue mainly consists of the venue's payment. The congestion effect is large, and may dominate over the network effect and proximity effect. Therefore, $R^{\rm app}\left(l^*,p^*\right)$ decreases with $N$ for $N\in\left(N_A,N_B\right)$. 
When $\phi<{bV}/{U}$ and $\delta\le\delta_3$, the analytical study of $N$'s impact on $R^{\rm app}\left(l^*,p^*\right)$ is more challenging because of the more complicated comparison between the advertising revenue and venue's payment (affected by the network effect, congestion effect, and proximity effect). In Section \ref{SM:numerical:populationsize} of the supplemental material, we use numerical results to show that when $\phi<{bV}/{U}$ and $\delta\le\delta_3$, $R^{\rm app}\left(l^*,p^*\right)$ may also decrease with $N$. The main insights about $N$ are summarized in Table \ref{table:1}.

  \begin{figure*}[t]
  \centering
  \subfigure[Comparison Under Different $\delta$.]{
    \label{fig:simu:1}
    \includegraphics[scale=0.299]{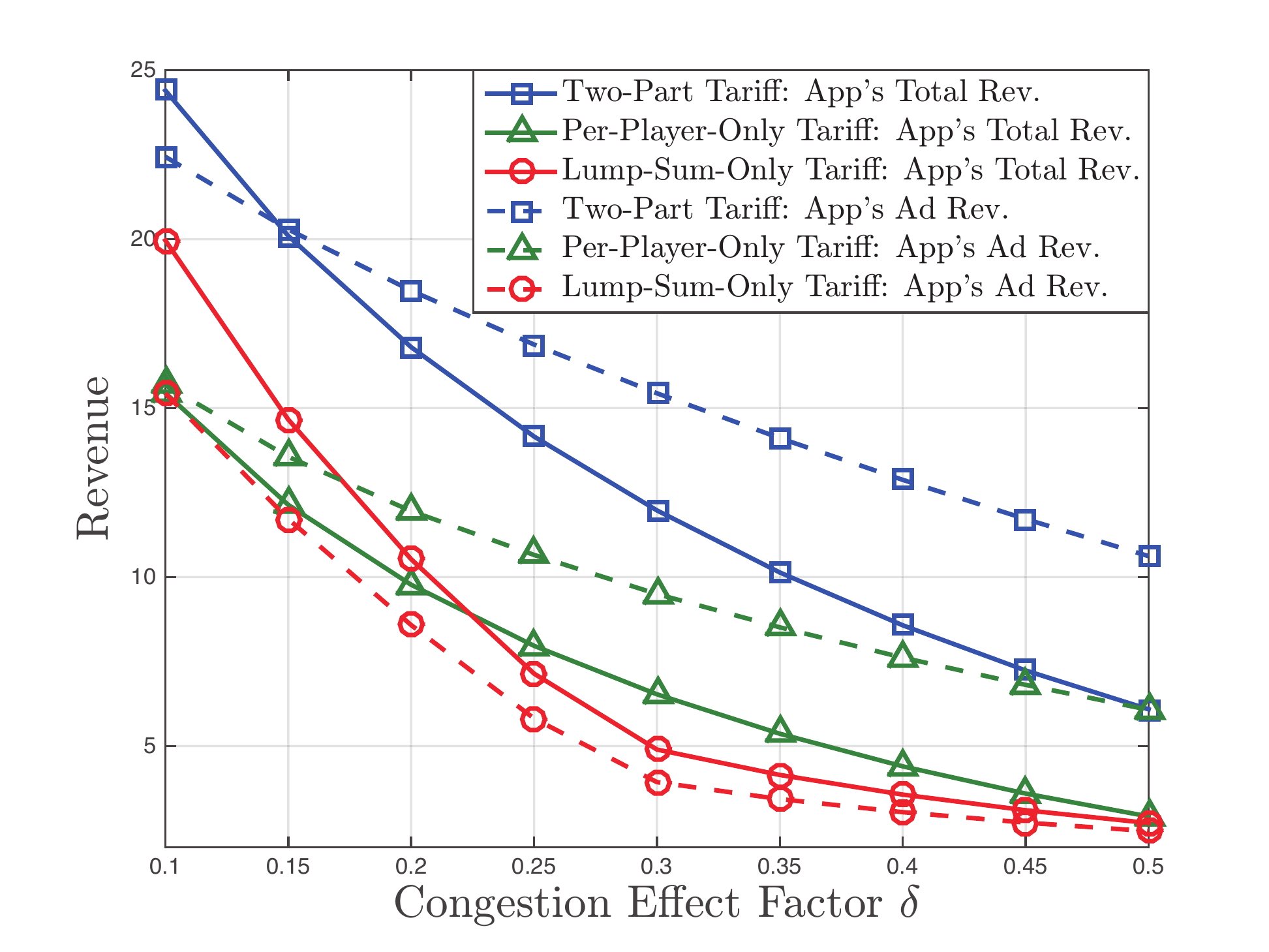}}
  \subfigure[Comparison Under Different $\theta$.]{
    \label{fig:simu:2}
    \includegraphics[scale=0.299]{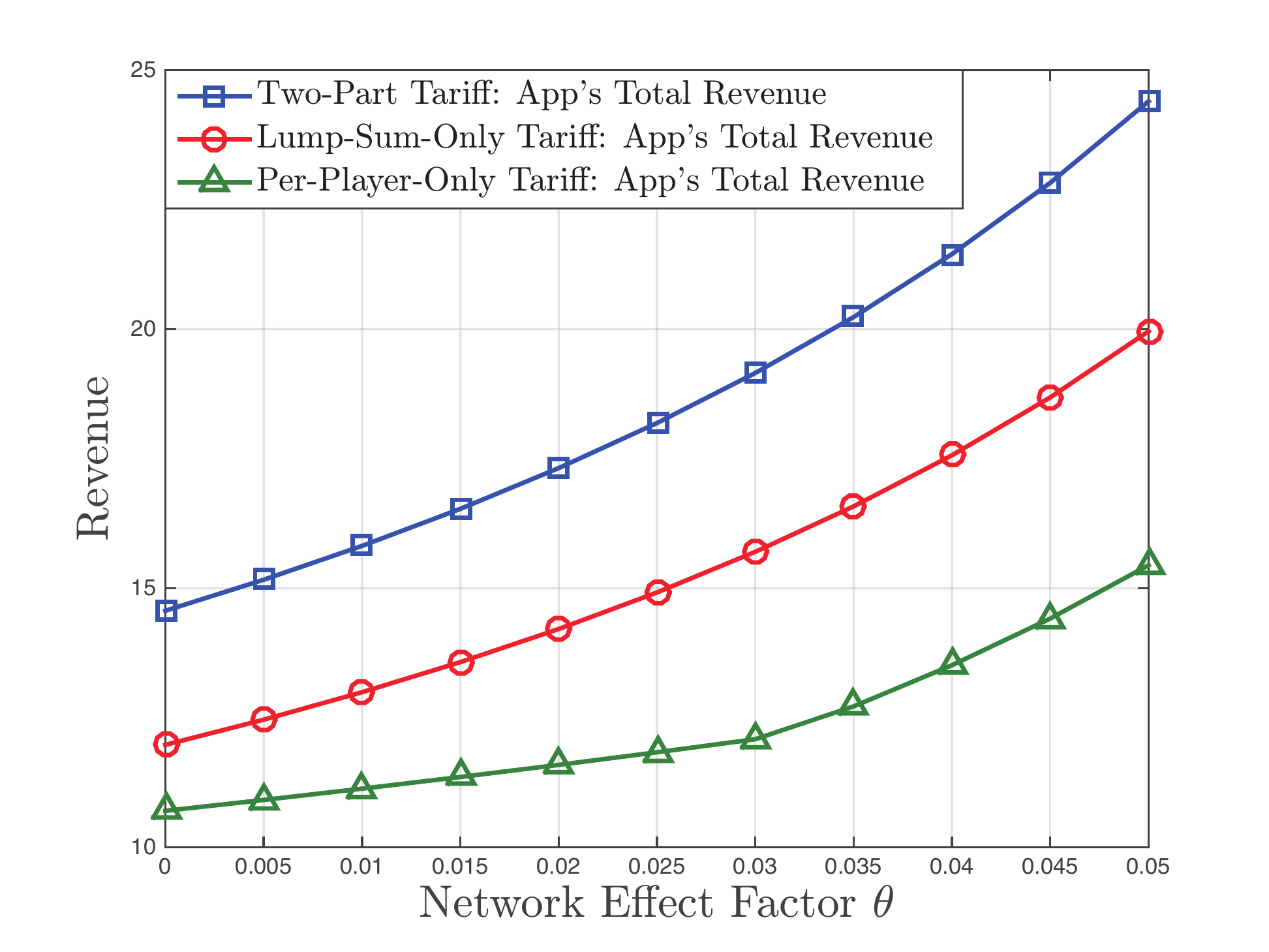}}
  \subfigure[Comparison Under Different $\phi$.]{
    \label{fig:simu:3}
    \includegraphics[scale=0.299]{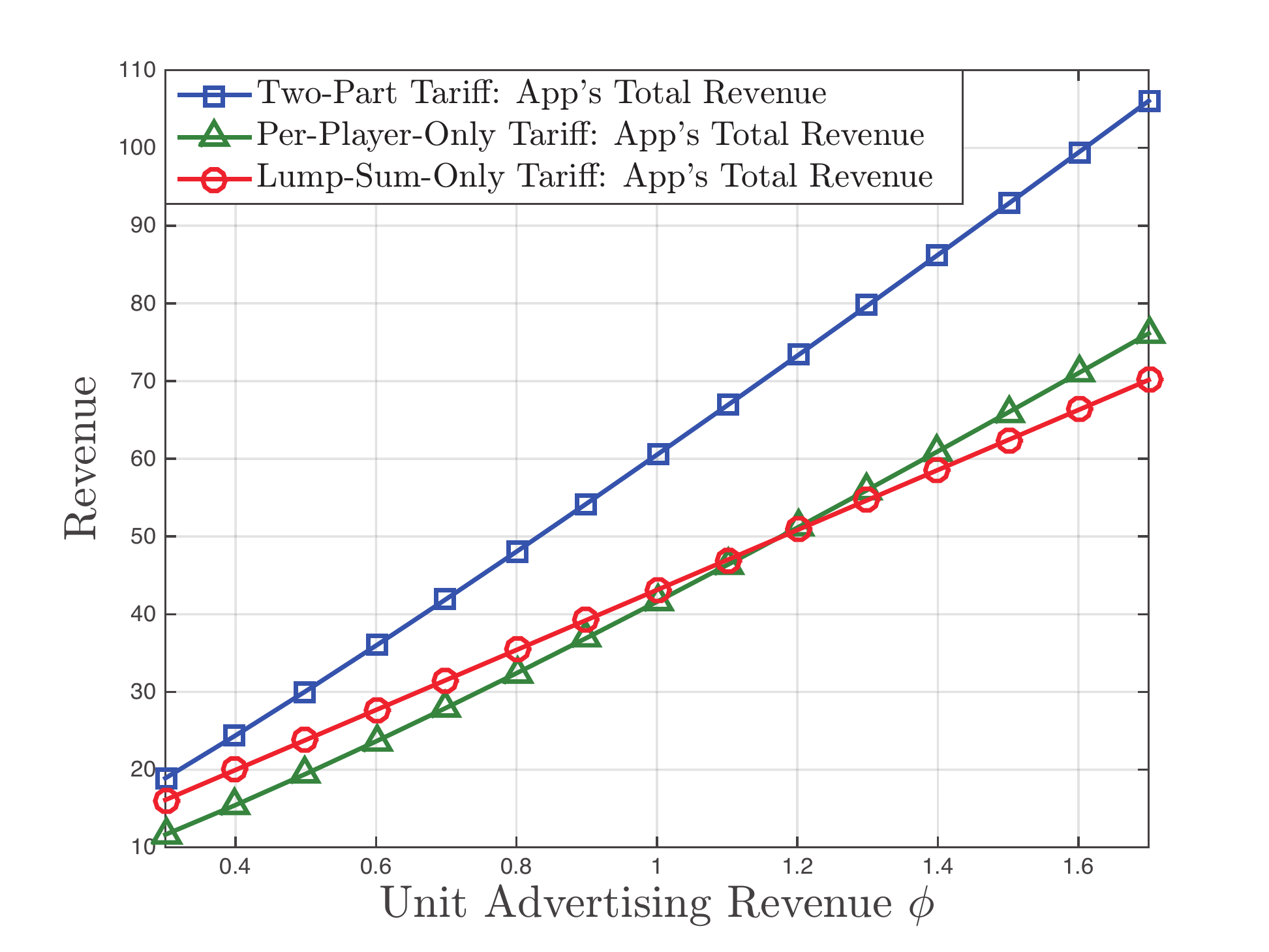}}    
  \caption{Comparison Between Tariffs.}
  \vspace{-0.3cm}
  \end{figure*}

\section{Two-Part Tariff Under Uncertainty}\label{sec:uncertainty}
In Section \ref{sec:stageI}, we assume that \emph{before} the venue becomes a POI, the app knows the exact unit advertising revenue $\phi$ and hence is able to set the optimal tariff $\left(l^*,p^*\right)=({\tilde H}\left(-\phi\right),-\phi)$. 
However, the assumption may not always hold in practice. For example, if the advertisers pay the app based on the click-through rates of their advertisements, the app will know the exact unit advertisement revenue \emph{only after} the venue becomes a POI and the users interact with the POI. 
Therefore, the app will be uncertain about $\phi$ when designing the tariff in Stage I.

In this section, we relax the assumption and extend Problem \ref{problem:app} in Section \ref{sec:stageI} by considering an app which decides its tariff only with the probabilistic information of $\phi$. Meanwhile, we investigate the impact of the app's risk preference on the optimal two-part tariff. Under uncertainty about $\phi$, the app solves Problem \ref{problem:uncertainty} in Stage I (the uncertainty of $\phi$ does not affect Stages II and III).
\begin{problem}\label{problem:uncertainty}
The app determines $\left(l_u^*,p_u^*\right)$ by solving
\begin{align}
& \max {{\mathbb E}_{\phi}\left\{J\big(R^{\rm app}\left(l_u,p_u\right) \big)\right\}} \label{equ:uncertainty:obj}\\
& {\rm var.~~~} l_u,p_u\in{\mathbb R},
\end{align}
where $\phi\in\left[\phi_{\min},\phi_{\max}\right]$ ($0\le \phi_{\min} < \phi_{\max}$) is a random variable that follows a general distribution. Function $J\left(\cdot\right)$ is the app's utility, with $J'\left(z\right)\ge0$ for all $z\in{\mathbb R}$. 
\end{problem}
Here, $R^{\rm app}\left(\cdot,\cdot\right)$ is the app's revenue function defined in (\ref{equ:appopt:obj}) (it is also called the app's wealth in the expected utility theory). The utility function $J\left(\cdot\right)$ reflects the app's risk preference \cite{png2010buyer}. In (\ref{equ:uncertainty:obj}), the app maximizes its expected utility via the tariff, where the expectation is taken with respect to $\phi$. 
We discuss the app's optimal two-part tariff in Theorem \ref{theorem:uncertainty}.

\begin{theorem}\label{theorem:uncertainty}
We characterize the optimal two-part tariff under the uncertainty about $\phi$ as follows:
\begin{itemize}
\item Risk-neutral app ($J''\left(z\right)=0,z\in{\mathbb R}$): $p_u^*=-\mathbb{E}\left\{\phi\right\},l_u^*={\tilde H}\left(-\mathbb{E}\left\{\phi\right\}\right)$;
\item Risk-averse app ($J''\left(z\right)<0,z\in{\mathbb R}$): $-\mathbb{E}\left\{\phi\right\} \le p_u^* \le -\phi_{\min},l_u^*={\tilde H}\left(p_u^*\right)\le{\tilde H}\left(-\mathbb{E}\left\{\phi\right\}\right)$;
\item Risk-seeking app ($J''\left(z\right)>0,z\in{\mathbb R}$): $-\phi_{\max} \le p_u^*\le-\mathbb{E}\left\{\phi\right\}, l_u^*={\tilde H}\left(p_u^*\right) \ge {\tilde H}\left(-\mathbb{E}\left\{\phi\right\}\right)$.
\end{itemize}
\end{theorem}

A risk-neutral app's optimal tariff (in the first bullet) is similar to the one in the complete information case, by replacing $\phi$ in (\ref{equ:optimalwp}) with $\mathbb{E}\left\{\phi\right\}$. Compared with a risk-neutral app, a risk-averse app (with a concave utility function $J\left(\cdot\right)$) should set a higher per-player charge $p_u^*$ and a lower lump-sum fee $l_u^*$. This strategy reduces the risk faced by the app. First, when the app increases $p_u^*$ ($p_u^*\le0 $), it provides a smaller subsidy for the venue to invest in the infrastructure. As a result, the investment level decreases, and the fraction of users interacting with the POI also decreases. According to the app's revenue defined in (\ref{equ:appopt:obj}), this reduces the difference between the app's revenues under different $\phi$, and hence reduces the risk faced by the app. Second, when the app increases $p_u^*$, it has to decrease $l_u^*$ to motivate the venue to become a POI.

Compared with a risk-neutral app, a risk-seeking app should set a lower per-player charge $p_u^*$ (hence a larger subsidy) and a higher lump-sum fee $l_u^*$ to increase the risk faced by the app. The detailed explanations are opposite to those for a risk-averse app.

In Section \ref{SM:riskpreference} of the supplemental material, we provide a numerical approach to compute $p_u^*$ and $l_u^*$ for the risk-averse and risk-seeking apps (in the second and third bullets), and numerically investigate the impacts of the degrees of app's risk aversion and risk seeking on the $p_u^*$ and $l_u^*$.

{{We can extend our work to consider other incomplete information cases. For example, the app only knows the probability distribution of $U$. In this case, when the two-part tariff is fixed, the users' and venue's equilibrium strategies under any given $U$ are still characterized by Propositions \ref{proposition:stageIII:a}-\ref{proposition:stageII:situationIII}. Hence, for a fixed two-part tariff, the app can compute its expected revenue based on the users' and venue's strategies and the probability distribution of $U$. Then, the app can decide its optimal two-part tariff by choosing the one that maximizes its expected revenue.}}

\section{Numerical Results}\label{sec:numerical}
In this section, we compare our two-part tariff scheme with {two state-of-the-art} tariff schemes: the \emph{lump-sum-only tariff} (e.g., used by Snapchat), where the app charges the venue based on the lump-sum fee $l_{\rm only}^*={{\arg\max}_{l\in{\mathbb R}}}R^{\rm app}\left(l,0\right)$; the \emph{per-player-only tariff} (e.g., used by Pokemon Go), where the app charges the venue based on the per-player charge $p_{\rm only}^*={{\arg\max}_{p\in{\mathbb R}}}R^{\rm app}\left(0,p\right)$.{\footnote{{{Another possible tariff scheme is the usage-based tariff, where the app charges the venue based on the users' overall usage of the app at the venue. It is reasonable to assume that if a user interacts with the POI, its usage of the app at the venue is a random variable that is independent of the user's attributes $\omega$ and $c$. In this case, the users' overall usage of the app at the venue is proportional to the number of users interacting with the POI at the venue. Hence, the performance of the usage-based tariff is the same as that of the per-player-only tariff.}}}} 

{{We will answer the following question: \emph{how does our two-part tariff's performance improvement over the two state-of-the-art tariffs change with system parameters (e.g., $\delta$, $\theta$, and $\phi$)?} The answer can help the apps that currently use lump-sum-only tariff or per-player-only tariff understand whether it is worth switching to the two-part tariff.}}

\subsection{Impact of Congestion Effect Factor $\delta$} 
In Fig. \ref{fig:simu:1}, we compare the three schemes under different congestion effect factor $\delta$. We choose $N=200$, $c_{\max}=24$, $U=3$, $V=5$, $I_0=0.6$, $k=3$, $b=1$, $\eta=0.2$, $\theta=0.05$, and $\phi=0.4$. We change $\delta$ from $0.1$ to $0.5$, and plot the app's total revenues $R^{\rm app}$ (solid curves) and advertising revenues (dash curves) under different schemes with respect to $\delta$. 
{{Since there is no randomness in the experiment, we only need one simulation run to get the plot.}}

First, we observe that the two-part tariff always achieves the highest app's total revenue (solid blue curve), compared to the per-player-only tariff and the lump-sum-only tariff schemes, which is consistent with Theorem \ref{theorem:optimalforall}. 
{For example, the two-part tariff improves the app's total revenue over the per-player-only tariff by at least $55\%$ for all $\delta$'s values shown in Fig. \ref{fig:simu:1}.
Second, the two-part tariff always achieves the highest app's advertising revenue (dash blue curve), which implies that it also achieves the highest number of users interacting with the POI. This is because the two-part tariff has the lowest per-player charge, and can best incentivize the venue to invest in the app-related infrastructure and relieve the congestion.} 

{When $\delta$ is medium (e.g., $0.2\le\delta\le0.35$), the two-part tariff significantly improves the app's total revenue compared with the lump-sum-only tariff. To understand this, note that the solid blue curve could be below the dash blue curve under the two-part tariff. This means that the app pays the venue to incentivize the investment. 
Under the lump-sum-only tariff, the app cannot incentivize investment by paying the venue. Hence, when $\delta$ is medium, the two-part tariff relieves the congestion, and significantly outperforms the lump-sum-only tariff.} 
When $\delta$ is small (e.g., $\delta<0.2$), the congestion does not heavily reduce the users' payoffs, so whether the venue is incentivized to invest does not strongly affect the number of users interacting with the POI; when $\delta$ is large (e.g., $\delta>0.35$), the congestion cannot be efficiently relieved even with the investment. In both cases, the gap between the app's total revenues under the two-part tariff and lump-sum-only tariff is smaller.

\subsection{Impact of Network Effect Factor $\theta$}
{In Fig. \ref{fig:simu:2}, we compare the three tariff schemes under different network effect factor $\theta$ (from $0$ to $0.05$). We let $\delta=0.1$, and the other parameters are the same as in Fig. \ref{fig:simu:1}.} 
When $\theta=0$, the two-part tariff improves the app's total revenue over the per-player-only tariff by $36\%$. When $\theta=0.05$, this improvement becomes more significant, i.e., $58\%$. {This is because a large network effect factor enables the POI to attract many visitors, and the venue is willing to pay the app for becoming a POI. 
Under the two-part tariff, the app can set a large lump-sum fee to receive a high venue's payment. 
Under the per-player-only tariff, the app cannot set a large per-player charge to obtain a high venue's payment, since this will reduce the venue's investment, the number of users interacting with the POI, and the app's advertising revenue.} 

\subsection{Impact of Unit Advertising Revenue $\phi$}
In Fig. \ref{fig:simu:3}, we compare the three tariff schemes under different unit advertising revenue $\phi$ (from $0.3$ to $1.7$). We choose $\delta=0.1$, and the other parameters are the same as in Fig. \ref{fig:simu:1}. 
We can observe that when $\phi$ is large, the two-part tariff significantly outperforms the other two schemes. For example, when $\phi=1.7$, the two-part tariff improves the app's total revenue over the per-player-only tariff and lump-sum-only tariff by $39\%$ and $51\%$, respectively. This is because the two-part tariff best incentivizes the venue's investment, and hence results in the highest number of users interacting with the POI. When $\phi$ is large, the two-part tariff achieves a much higher app's total revenue than the other two schemes.

{{We summarize the key insights obtained from the numerical results as follows.

\begin{observation}
If the congestion effect factor is medium or the unit ad revenue is large, our tariff significantly outperforms the lump-sum-only tariff. If the network effect factor or the unit ad revenue is large, our tariff significantly outperforms the per-player-only tariff.
\end{observation}

These insights are robust to the change in the parameter settings. In Section \ref{SM:numerical:result} of our supplemental material, we provide more numerical results, and show that these insights hold under different parameter settings.
}}

\section{Conclusion}\label{sec:conclusion}
The economics of the online apps (especially the augmented reality apps) and offline venues' collaboration is a fast-emerging business area. The POI-based collaboration is increasingly popular, but there are no prior analytical studies to investigate the collaboration schemes and characterize the equilibria. 
We designed a \emph{charge-with-subsidy} tariff scheme, which achieves the highest app's revenue among all feasible $\left(r,I\right)${-dependent tariff schemes}. Our tariff scheme also significantly improves the users' engagements with the venues, {compared with the state-of-the-art tariff schemes.} Moreover, we provided some counter-intuitive guidelines for the collaboration. For example, a bandwidth-consuming app should collaborate with a low-quality venue, and an app with a small unit advertising revenue may avoid collaborating with a venue that is popular or in a busy area.

Our work opens up exciting directions for future works. Since most apps currently collaborate with store/restaurant chains, we considered the collaboration between an app and a store/restaurant chain's representative venue in this paper. For future research, it is interesting to study the collaboration between an app and multiple venues owned by different entities in the same area. This extension will be challenging. The users need to decide which venues to visit, by comparing both the qualities of the venues' offline products and the venues' investment levels on the app-related infrastructure (related to the qualities of the online products). 
{{
Moreover, when there are multiple apps, the problem's analysis will be more challenging. If an app pays a venue to incentivize its investment in the infrastructure, this potentially benefits other apps. This is because the users who choose to play other apps can also use this infrastructure. Intuitively, when there are multiple apps, the users will be better off, and more users can play apps.}}

\bibliographystyle{IEEEtran}
\bibliography{tmc}

\begin{IEEEbiography}
[{\includegraphics[width=1in,height=1.25in,clip,keepaspectratio]{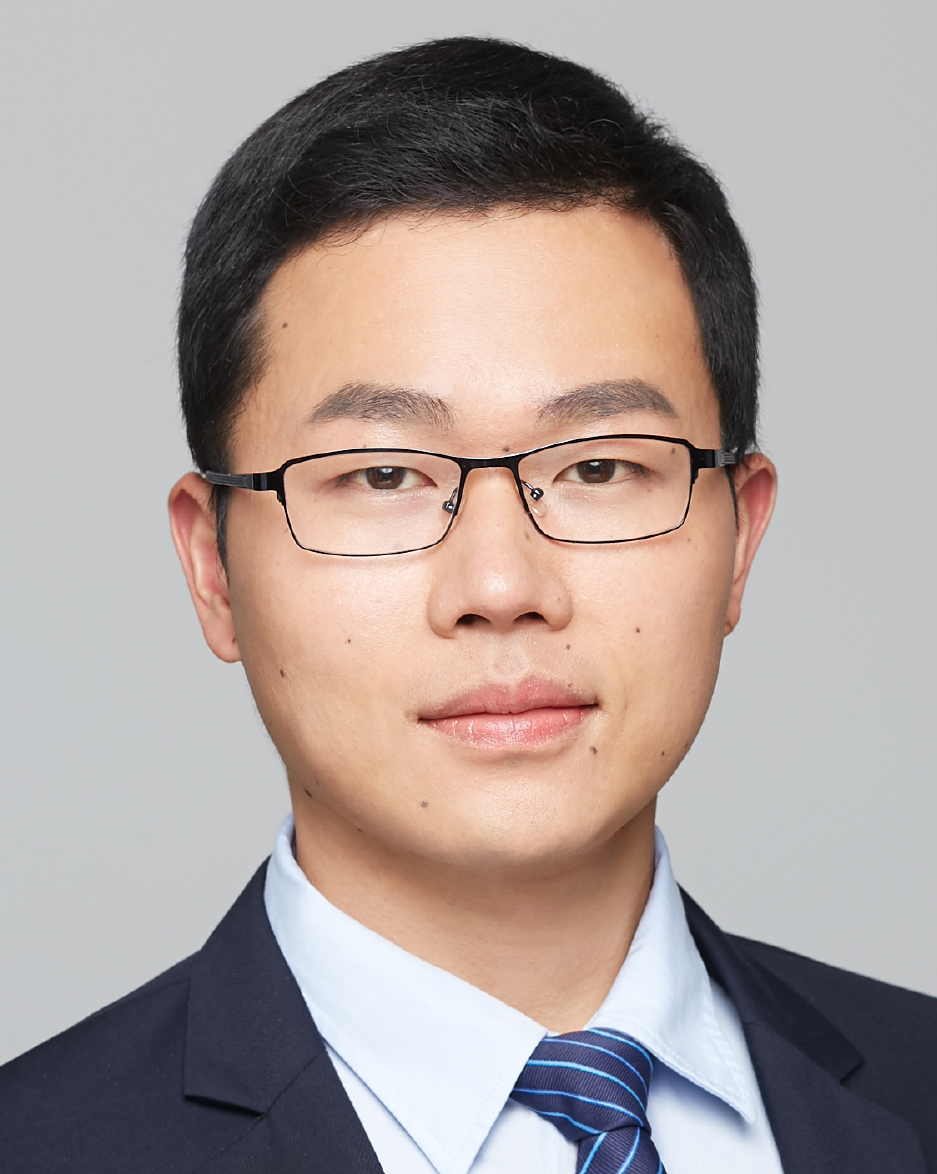}}]
{Haoran Yu} (S'14-M'17) is a Post-Doctoral Fellow in the Department of Electrical and Computer Engineering at Northwestern University. He received the Ph.D. degree from the Chinese University of Hong Kong in 2016. He was a Visiting Student in the Yale Institute for Network Science and the Department of Electrical Engineering at Yale University during 2015-2016. His research interests lie in the field of network economics, with current emphasis on economics of Wi-Fi networks, location-based services, and business models of mobile advertising. He was awarded the Global Scholarship Programme for Research Excellence by the Chinese University of Hong Kong. His paper in {\it IEEE INFOCOM 2016} was selected as a Best Paper Award finalist and one of top 5 papers from over 1600 submissions.
\end{IEEEbiography}

\vspace{-1cm}

\begin{IEEEbiography}
[{\includegraphics[width=1in,height=1.25in,clip,keepaspectratio]{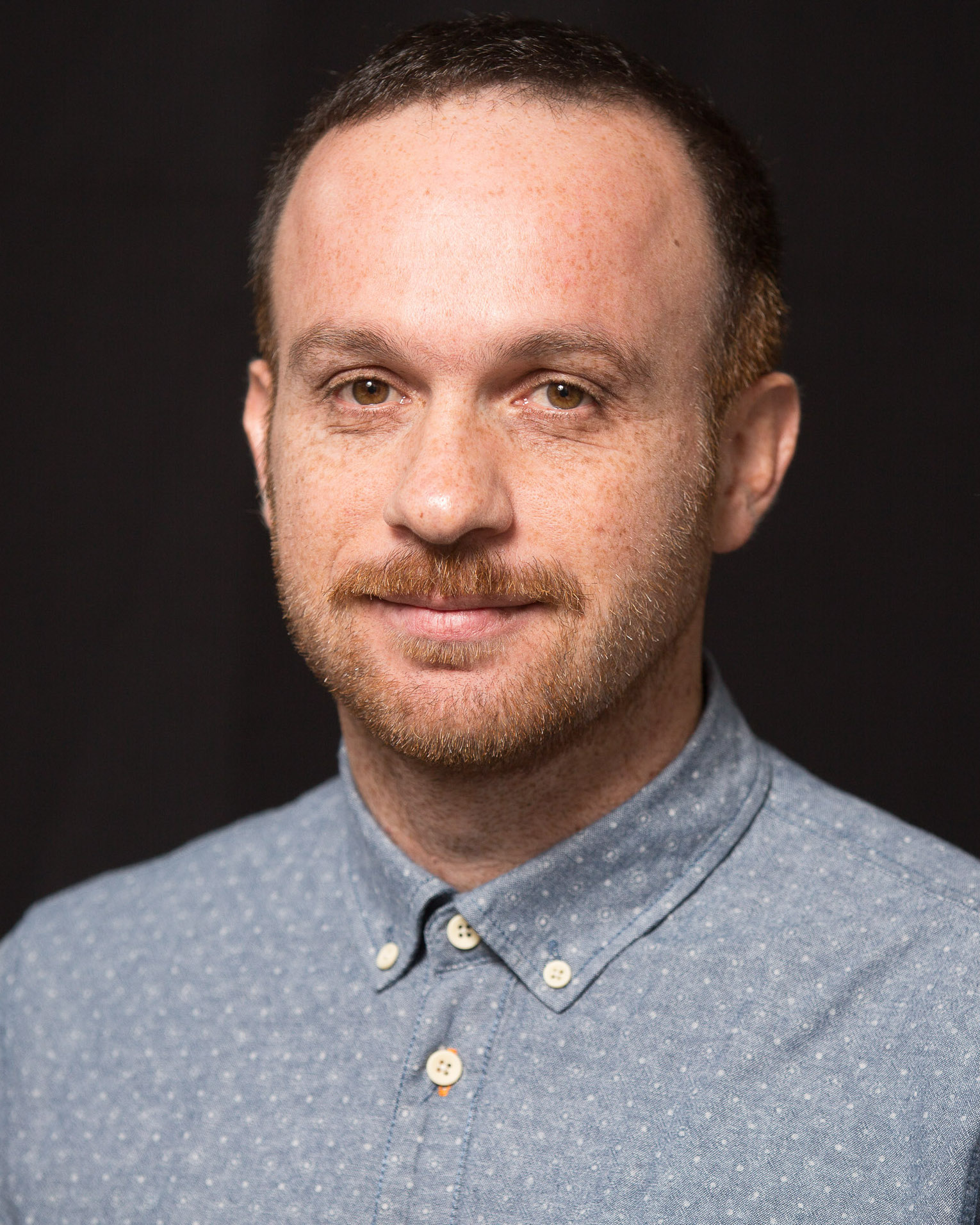}}]
{George Iosifidis} is the Ussher Assistant Professor in Future Networks, at Trinity College Dublin, University of Dublin, Ireland. He received a Diploma in Electronics and Communications, from the Greek Air Force Academy (Athens, 2000) and a PhD degree from the Department of Electrical and Computer Engineering, University of Thessaly in 2012. He was a Postdoctoral researcher (’12-’14) at CERTH-ITI in Greece, and Postdoctoral/Associate research scientist at Yale University (’14-’17). He is a co-recipient of the best paper awards in {\it WiOPT 2013} and {\it IEEE INFOCOM 2017} conferences, served as a guest editor for the IEEE Journal on Selected Areas in Communications,  and has received an SFI Career Development Award in 2018. 
\end{IEEEbiography}

\vspace{-1cm}

\begin{IEEEbiography}
[{\includegraphics[width=1in,height=1.25in,clip,keepaspectratio]{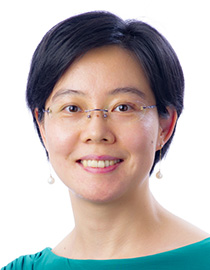}}]
{Biying Shou} is an Associate Professor of Management Sciences at City University of Hong Kong. She received B.E. in Management Information Systems from Tsinghua University and Ph.D. in Industrial Engineering and Management Sciences from Northwestern University. Her main research interests include operations and supply chain management, operations-marketing interface, and network economics. Her papers have been published in leading journals including Operations Research, Production and Operations Management, IEEE Transactions on Mobile Computing, and others. She also had work and consulting experience in the telecommunications, automobile, and retailing industries. 
\end{IEEEbiography}

\vspace{-1cm}

\begin{IEEEbiography}
[{\includegraphics[width=1.03in,height=1.25in,clip,keepaspectratio]{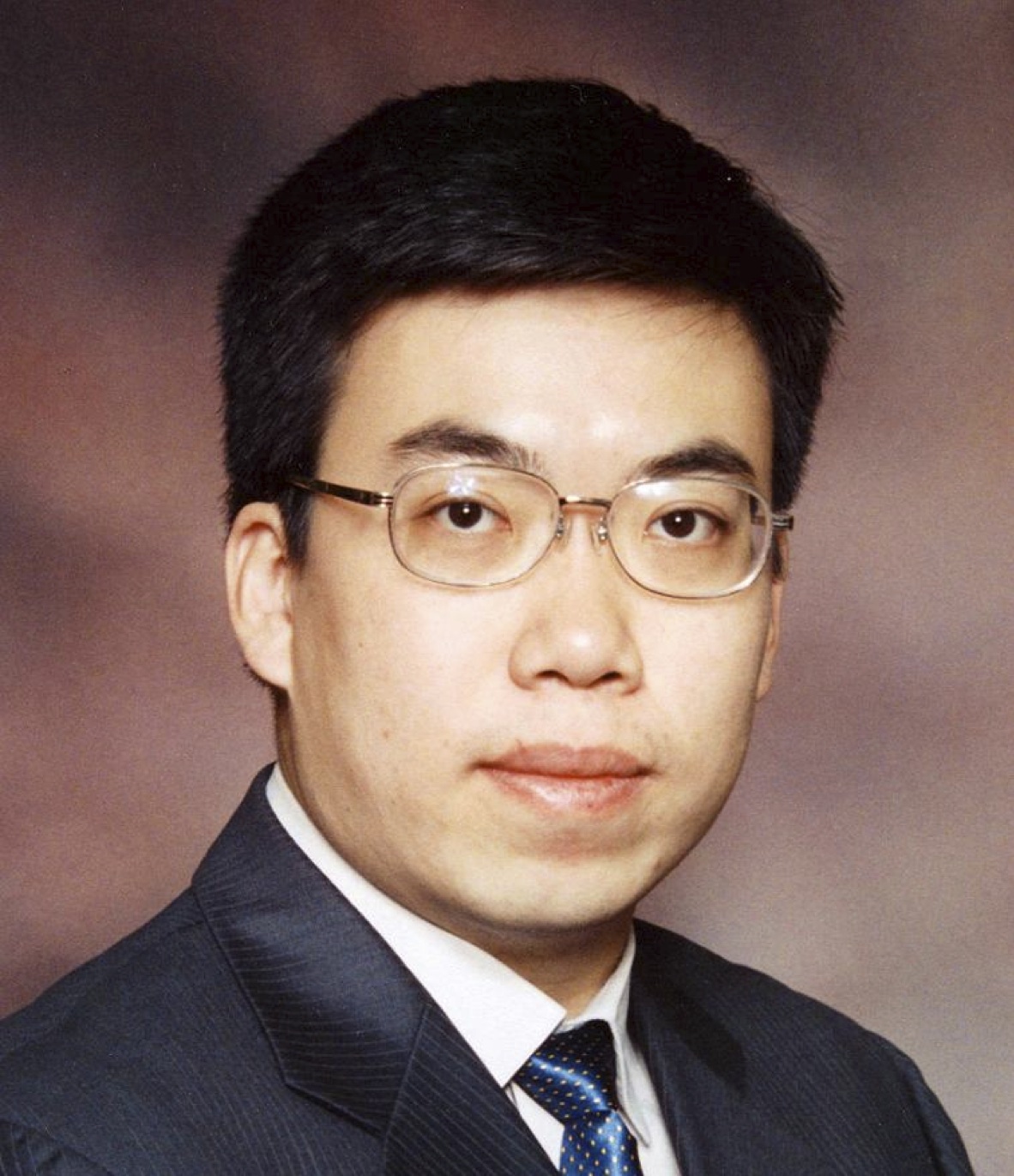}}]
{Jianwei Huang} is a Presidential Chair Professor and the Associate Dean of the School of Science and Engineering, The Chinese University of Hong Kong, Shenzhen. He is also a Professor in the Department of Information Engineering, The Chinese University of Hong Kong. He is the co-author of 9 Best Paper Awards, including IEEE Marconi Prize Paper Award in Wireless Communications 2011. He has co-authored six books, including the textbook on ``Wireless Network Pricing''. He has served as the Chair of IEEE ComSoc Cognitive Network Technical Committee and Multimedia Communications Technical Committee. He has been an IEEE Fellow, an IEEE ComSoc Distinguished Lecturer, and a Clarivate Analytics Highly Cited Researcher. More information at http://jianwei.ie.cuhk.edu.hk/.
\end{IEEEbiography}

%







\newpage
\setcounter{page}{1}

\begin{centering}
{\Large{\bf{Supplemental Material}}}\\
\end{centering}


\vspace{0.5cm}


\begin{centering}
{{\large{\bf{Outline}}}}\\
\end{centering}

\vspace{0.2cm}

\appendices




{\bf{Part I: Our Survey}}

\emph{Section \ref{SM:survey}. Details of Our Survey}

{\bf{Part II: Proofs and Illustrations}}

\emph{Section \ref{SM:notation}. Notation Table}
\vspace{0.05cm}

\emph{Section \ref{SM:proposition1}. Proof of Proposition \ref{proposition:stageIII:a} in Section \ref{sec:stageIII}}
\vspace{0.05cm}

\emph{Section \ref{SM:proposition2}. Proof of Proposition \ref{proposition:stageIII:b} in Section \ref{sec:stageIII}}
\vspace{0.05cm}

\emph{Section \ref{SM:proposition3}. Proof of Proposition \ref{proposition:stageIII:c} in Section \ref{sec:stageIII}}
\vspace{0.05cm}

\emph{Section \ref{SM:proposition4}. Proof of Proposition \ref{proposition:stageII:situationI} in Section \ref{sec:stageII}}
\vspace{0.05cm}

\emph{Section \ref{SM:uniquep2}. Proof of Lemma \ref{lemma:p2unique} in Section \ref{sec:stageII}}
\vspace{0.05cm}

\emph{Section \ref{SM:illustration:pro5}. \!Illustrations of Propositions \ref{proposition:stageII:situationII} and \ref{proposition:stageII:situationIII} in Section \ref{sec:stageII}}
\vspace{0.05cm}

\emph{Section \ref{SM:proposition5}. Proof of Proposition \ref{proposition:stageII:situationII} in Section \ref{sec:stageII}}
\vspace{0.05cm}

\emph{Section \ref{SM:proposition6}. Proof of Proposition \ref{proposition:stageII:situationIII} in Section \ref{sec:stageII}}
\vspace{0.05cm}

\emph{Section \ref{SM:theorem:O2}. Proof of Theorem \ref{theorem:optimaltariff} in Section \ref{sec:stageI}}
\vspace{0.05cm}

\emph{Section \ref{SM:theorem2}. Proof of Theorem \ref{theorem:optimalforall} in Section \ref{sec:stageI}}
\vspace{0.05cm}

\emph{Section \ref{SM:corollary1}. Proof of Corollary \ref{corollary:optimalrevenue} in Section \ref{sec:stageI}}
\vspace{0.05cm}

\emph{Section \ref{SM:proposition78}. Proofs of Propositions \ref{proposition:impactU:smalldelta} and \ref{proposition:impactU:largedelta} in Section \ref{subsec:influence:quality}}
\vspace{0.05cm}

\emph{Section \ref{SM:corollary:impactU}. Proof of Corollary \ref{corollary:impactU} in Section \ref{subsec:influence:quality}}
\vspace{0.05cm}

\emph{Section \ref{SM:proposition9}. Proof of Proposition \ref{proposition:impacteta:largephi} in Section \ref{subsec:impacteta}}
\vspace{0.05cm}

\emph{Section \ref{SM:lemma2}. Proof of Lemma \ref{lemma:delta1} in Section \ref{subsec:impacteta}}\vspace{0.05cm}

\emph{Section \ref{SM:proposition10}. Proof of Proposition \ref{proposition:eta:no} in Section \ref{subsec:impacteta}}\vspace{0.05cm}

\emph{Section \ref{SM:proposition11}. Proof of Proposition \ref{proposition:N:largephi} in Section \ref{subsec:impactN}}\vspace{0.05cm}

\emph{Section \ref{SM:lemma3}. Proof of Lemma \ref{lemma:delta:2} in Section \ref{subsec:impactN}}\vspace{0.05cm}

\emph{Section \ref{SM:lemma4}. Proof of Lemma \ref{lemma:N:1} in Section \ref{subsec:impactN}}\vspace{0.05cm}

\emph{Section \ref{SM:proposition12}. Proof of Proposition \ref{proposition:N:no} in Section \ref{subsec:impactN}}\vspace{0.05cm}

\emph{Section \ref{SM:theoremuncertainty}. Proof of Theorem \ref{theorem:uncertainty} in Section \ref{sec:uncertainty}}
\vspace{0.05cm}


{\bf{Part III: Supplemental Results}}

\emph{Section \ref{SM:model:congestion}. Modeling Congestion in General Infrastructure}
\vspace{0.05cm}

\emph{Section \ref{SM:model:nonmonetary}. Modeling Non-Monetary Rewards}
\vspace{0.05cm}

\emph{Section \ref{SM:bargaining}. Bargaining Between App and Venue}
\vspace{0.05cm}


\emph{Section \ref{SM:numerical:popularity}. Influence of $\eta$ When $\phi<{bV}/{U}$ and $\delta\le\delta_2$}
\vspace{0.05cm}

\emph{Section \ref{SM:numerical:populationsize}. Influence of $N$ When $\phi<{bV}/{U}$ and $\delta\le\delta_3$}
\vspace{0.05cm}

\emph{Section \ref{SM:riskpreference}. Numerical Computation of $p_u^*$ and $l_u^*$ and Impact of App's Risk Preference}
\vspace{0.05cm}

\emph{Section \ref{SM:numerical:result}. More Numerical Results}
\vspace{0.05cm}

\vspace{-0.3cm}


\section{Details of Our Survey}\label{SM:survey}

In this section, we show the illustrations for the key results of our survey. 

\subsection{Respondents' Basic Information} 
\begin{figure}[h]
  \centering
  \includegraphics[scale=0.85]{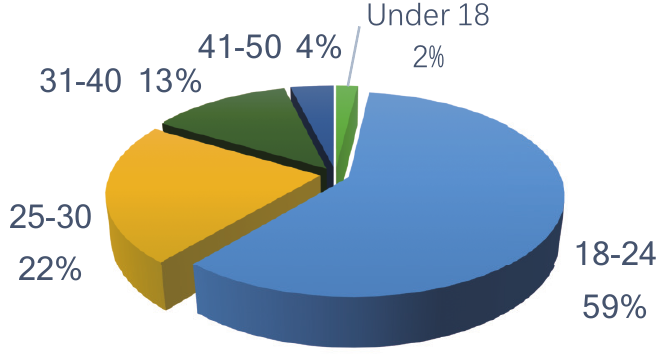}\\
  \caption{Ages of Respondents.}
  \label{SM:survey:age}
\end{figure}

In Fig. \ref{SM:survey:age}, we show the distribution of the respondents' ages. $81$\% of the respondents are between $18$ and $30$ years old.

\begin{figure}[h]
  \centering
  \includegraphics[scale=0.8]{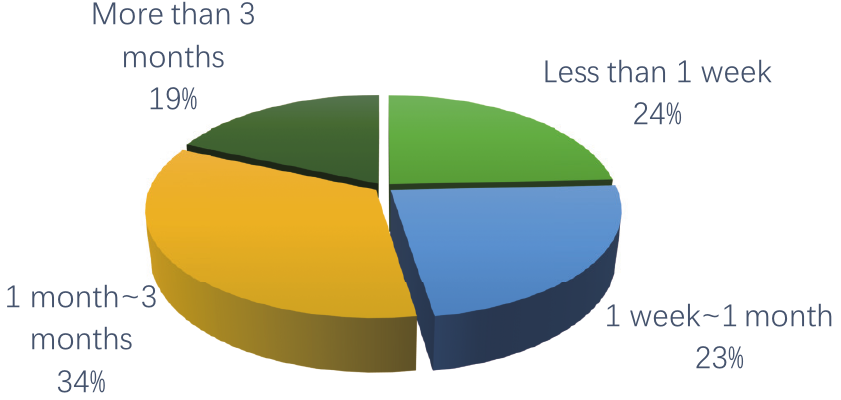}\\
  \caption{Duration of Respondents' Game Experience.}
  \label{SM:survey:duration}
\end{figure}

In Fig. \ref{SM:survey:duration}, we show the distribution of the duration of respondents' game experience. We can see that $53$\% of the respondents have played Pokemon Go for more than one month. 

\subsection{Impact of POIs' Infrastructure} 
\begin{figure}[h]
  \centering
  \includegraphics[scale=0.32]{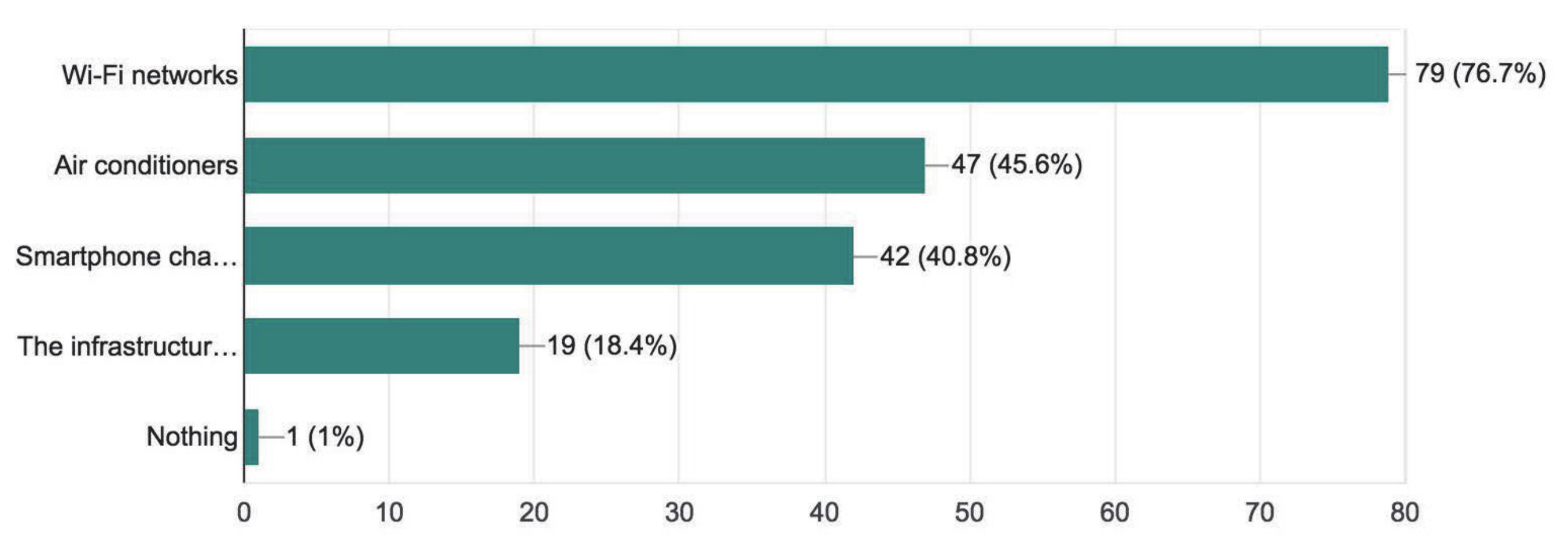}\\
  \caption{Infrastructure That Enhances Game Experience.}
  \label{SM:survey:type}
\end{figure}

We asked \emph{``What types of infrastructure at the sponsored PokeStops/Gyms might enhance your experience of playing Pokemon Go''}, and illustrate the responses in Fig. \ref{SM:survey:type}. Note that this is a multi-choice question. 

$76.7$\% of the respondents thought that the Wi-Fi networks at the sponsored PokeStops/Gyms (i.e., POIs) might enhance their experience. $45.6$\% of the respondents and $40.8$\% of the respondents thought that the air conditioners and smartphone chargers might enhance their experience. $19.4$\% of the respondents (including the respondent who chose ``nothing'') answered that the infrastructure had negligible impacts on their game experience. That is to say, the infrastructure, such as Wi-Fi networks, smartphone chargers, and air conditioners, could enhance the game experience of 80.6$\% ($=100$\%$-19.4$\%$) of the respondents.  

\begin{figure}[h]
  \centering
  \includegraphics[scale=0.32]{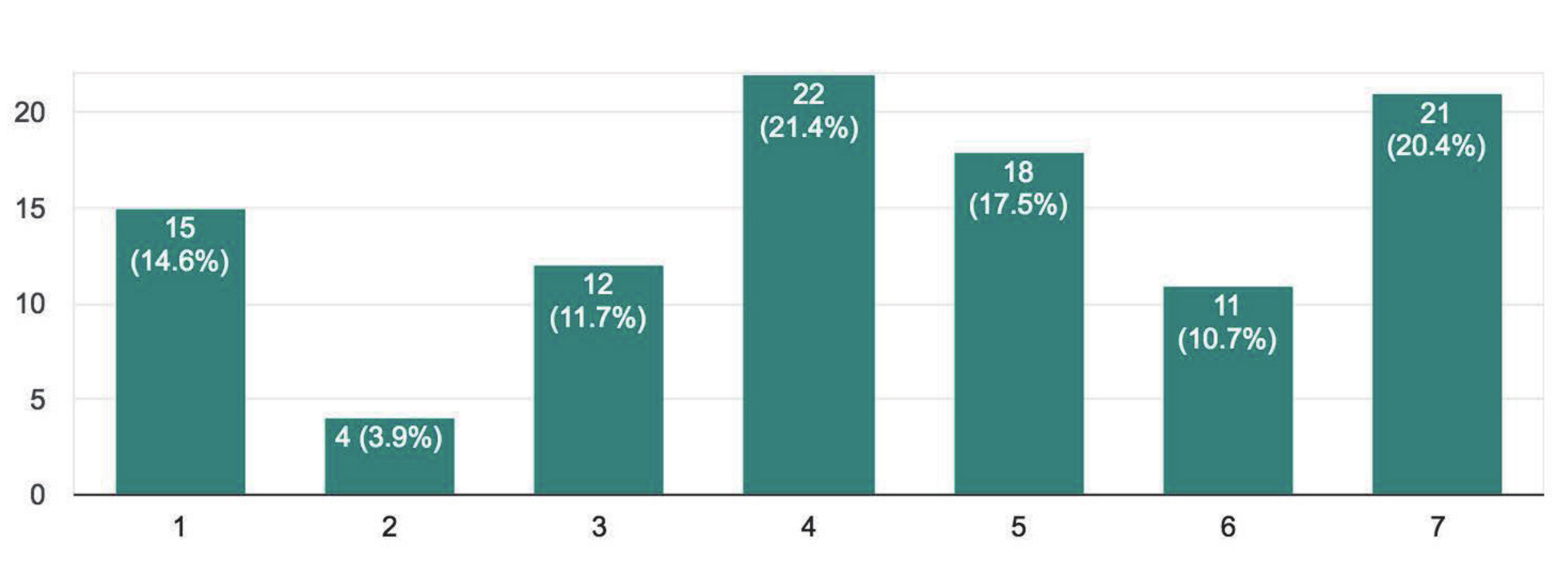}\\
  \caption{Respondents' Opinions Towards Infrastructure's Investment.}
  \label{SM:survey:whether}
\end{figure}

We asked \emph{``Do you think the sponsored PokeStops/Gyms could become more attractive to you through investing on their infrastructure (e.g., having faster Wi-Fi Internet, better air conditioning, or more smartphone chargers)''}, and invited the respondents to rate their degrees of agreements ($7$ means ``surely''). We illustrate the respondents' ratings in Fig. \ref{SM:survey:whether}. We can see that almost half of the respondents chose high values (i.e., $5$, $6$, or $7$), which implies the need for infrastructure's investment. 

\subsection{Externalities Among Players}

\begin{figure}[h]
  \centering
  \includegraphics[scale=0.32]{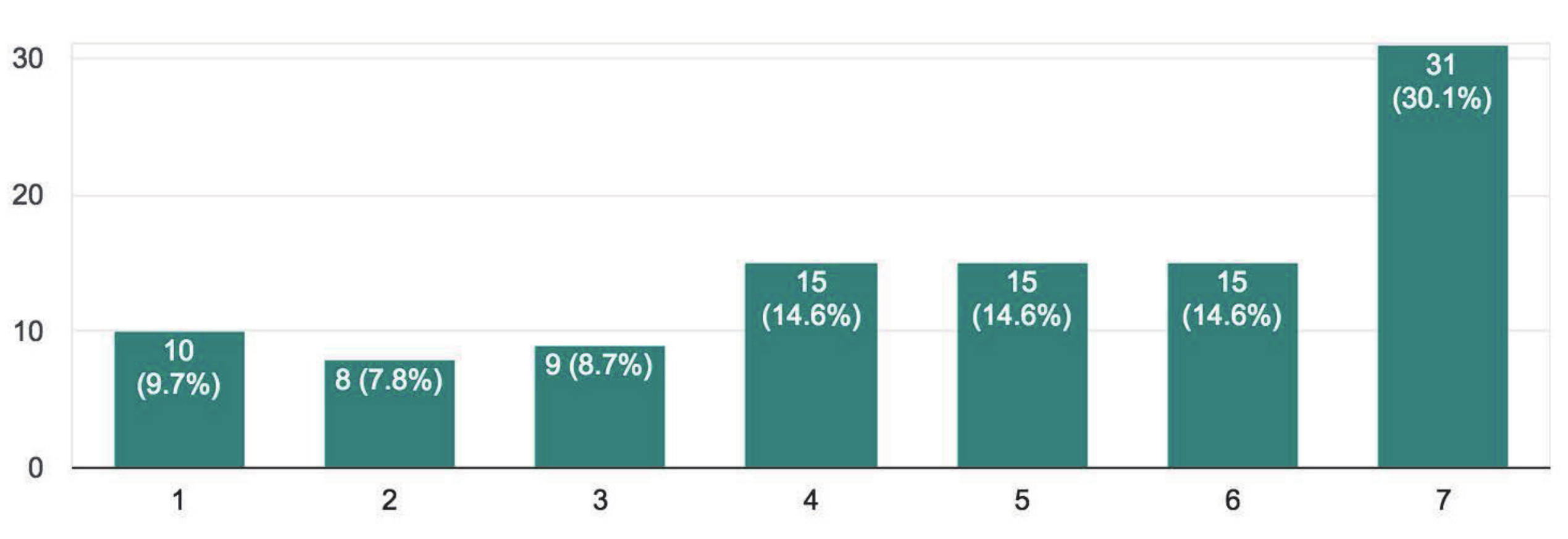}\\
  \caption{Impact of Wi-Fi Speed.}
  \label{SM:survey:congestion}
\end{figure}

We asked \emph{``Suppose you are using the Wi-Fi service at a sponsored PokeStop/Gym to play Pokemon Go, do you think the Wi-Fi speed will affect your experience of playing the game''}, and collected the respondents' ratings of their degrees of agreements ($7$ means ``surely''). We illustrate the responses in Fig. \ref{SM:survey:congestion}, and can find that $59.3$\% of the respondents chose high values (i.e., $5$, $6$, or $7$). In particular, among the fractions of the respondents choosing different values, the fraction of the respondents choosing $7$ is the largest one. 

\begin{figure}[h]
  \centering
  \includegraphics[scale=0.32]{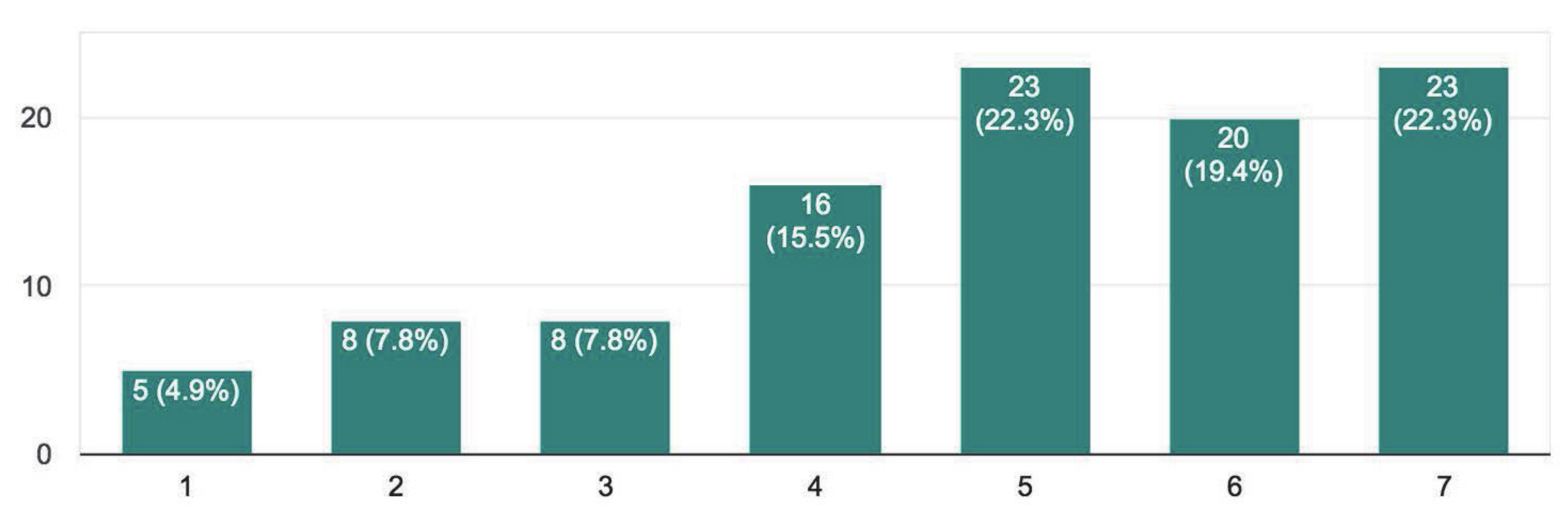}\\
  \caption{Impact of Nearby Players' Engagements in Pokemon Go.}
  \label{SM:survey:network}
\end{figure}

We asked \emph{``When you play Pokemon Go, if there are nearby people playing the game as well, will this enhance your game experience''}, and collected the respondents' ratings of their degrees of agreements ($7$ means ``surely''). We illustrate the responses in Fig. \ref{SM:survey:network}, and find that $64$\% of the respondents chose high values (i.e., $5$, $6$, or $7$). 

\section{Notation Table}\label{SM:notation}
We summarize our paper's key notations in Table \ref{table:notation}.

\begin{table}
\caption{Key Notations.}\label{table:notation}
\begin{tabular}{|p{2.3cm}|p{5.7cm}|}
\hline
\multicolumn{2}{|c|}{{\bf Decision Variables}}\\
\hline
{$d\in\left\{0,1,2\right\}$} & {User's visiting and interaction decision\!}\\
{$r\in\left\{0,1\right\}$} & {Venue's POI choice}\\
{$I\ge0$} & {Venue's investment choice}\\
{$l\in{\mathbb R}$} & {App's lump-sum fee}\\
{$p\in{\mathbb R}$} & {App's per-player charge}\\
\hline
\hline
\multicolumn{2}{|c|}{{\bf Parameters}}\\
\hline
{$N>0$} & {Mass of users}\\
{$c\in\left[0,c_{\max}\right]$} & {A user's attribute indicating its transportation cost}\\
{$\omega\in\left\{0,1\right\}$} & {A user's attribute indicating its interest in the venue's offline products}\\
{$\eta\in\left[0,1\right]$} & {Fraction of users with $\omega=1$ (\emph{reflect venue's popularity})}\\
{$U>0$} & {Utility of a user with $\omega=1$ when it consumes offline products (\emph{reflect venue's quality})}\\
{$V>0$} & {A user's base utility of interacting with the POI}\\
{$\theta\ge0$} & {Network effect factor}\\
{$\delta>0$} & {Congestion effect factor}\\
{$I_0\ge0$} & {Venue's initial investment level}\\
{$b>0$} & {Venue's profit due to one user's consumption of its products}\\
{$k>0$} & {Venue's unit investment cost}\\
{$\phi\ge0$} & {App's unit advertising revenue}\\
\hline
\hline
\multicolumn{2}{|c|}{{\bf Functions}}\\
\hline
{${\bar x}\left(r,I\right)\in\left[0,1\right]$} & {Fraction of users consuming venue's products (have $\omega=1$ and visit venue)}\\
{${\bar y}\left(r,I\right)\in\left[0,1\right]$} & {Fraction of users interacting with POI}\\
{$\Pi^{\rm user}\!\left(\omega,c,d,r,\!I\!\right)$} & {A type-$\left(\omega,c\right)$ user's payoff function}\\
{$\Pi^{\rm venue}\left(r,I,l,p\right)$} & {Venue's payoff function}\\
{$R^{\rm app}\left(l,p\right)$} & {App's revenue function}\\
{${\tilde H}\left(p\right)$} & {Maximum lump-sum fee under which venue becomes a POI, given per-player charge $p$}\\
\hline
\end{tabular}
\end{table}

\section{Proof of Proposition \ref{proposition:stageIII:a} in Section \ref{sec:stageIII}}\label{SM:proposition1}
\proof
According to (\ref{equ:useropt:var}), when $r=0$, a type-$\left(\omega,c\right)$ user can only choose $d=0$ or $d=1$. From (\ref{equ:userpayoff}), the user's payoff is $0$ under $d=0$, and is $U\omega-c$ under $d=1$. Based on endnote \ref{footnote:boundary}, the user chooses $d=1$ if and only if $c<U\omega$. 

Hence, when $r=0$, a type-$\left(\omega,c\right)$ user's optimal strategy is 
\begin{align}
d^*\left(\omega,c,r,I\right)= \left\{ {\begin{array}{*{20}{l}}
{1,}&{{\rm~if~}c\in\left[0,U\omega\right),}\\
{0,}&{{\rm~if~}c\in\left[U\omega,c_{\max}\right],}
\end{array}} \right.\label{SM:equ:proposition1}
\end{align}
where $\omega\in\left\{0,1\right\}$ and $c\in\left[0,c_{\max}\right]$. 

Recall that ${\bar x}\left(r,I\right)$ is the fraction of users who have $\omega=1$ and choose $d=1$ or $2$. Based on (\ref{SM:equ:proposition1}), we compute ${\bar x}\left(r,I\right)$ by
\begin{align}
{\bar x}\left(r,I\right)=\eta\int_{0}^{U} \frac{1}{c_{\max}} dc=\eta\frac{U}{c_{\max}}.
\end{align}
Recall that ${\bar y}\left(r,I\right)$ is the fraction of users who choose $d=2$. Based on (\ref{SM:equ:proposition1}), we have ${\bar y}\left(r,I\right)=0$.
\endproof

\section{Proof of Proposition \ref{proposition:stageIII:b} in Section \ref{sec:stageIII}}\label{SM:proposition2}
\proof
{\bf (Step 1)} we prove that $V + \theta {\bar y}\left(r,I\right)N-  \frac{\delta}{I+I_0}{\bar y}\left(r,I\right)N=0$ holds at the equilibrium (a user's net payoff of interacting with the POI at the equilibrium is zero). We prove it by contradiction, and assume that $V {+ \theta {\bar y}\left(r,I\right)N}-  \frac{\delta}{I+I_0}{\bar y}\left(r,I\right)N\ne0$.

First, we discuss the possibility of $V {+ \theta {\bar y}\left(r,I\right)N}-  \frac{\delta}{I+I_0}{\bar y}\left(r,I\right)N<0$.
When $V {+ \theta {\bar y}\left(r,I\right)N}-  \frac{\delta}{I+I_0}{\bar y}\left(r,I\right)N<0$, we can see that ${\bar y}\left(r,I\right)\ne0$. 
Moreover, the fraction ${\bar y}\left(r,I\right)$'s definition implies that ${\bar y}\left(r,I\right)\ge0$. 
Hence, ${\bar y}\left(r,I\right)>0$ (the fraction of users interacting with the POI is positive). Because $V {+ \theta {\bar y}\left(r,I\right)N}-  \frac{\delta}{I+I_0}{\bar y}\left(r,I\right)N<0$ (a user's net payoff of interacting with the POI at the equilibrium is negative), a user who interacts with the POI at the equilibrium can strictly improve its payoff by switching its strategy from $d=2$ to $d=1$. This violates the concept of equilibrium. Therefore, it is impossible that $V {+ \theta {\bar y}\left(r,I\right)N}-  \frac{\delta}{I+I_0}{\bar y}\left(r,I\right)N<0$.

Second, we discuss the possibility of $V {+ \theta {\bar y}\left(r,I\right)N}-\frac{\delta}{I+I_0}{\bar y}\left(r,I\right)N>0$. When $V {+ \theta {\bar y}\left(r,I\right)N}-\frac{\delta}{I+I_0}{\bar y}\left(r,I\right)N>0$ holds at the equilibrium, all the users with $c<U\omega$ can maximize their payoffs by interacting with the POI at the equilibrium based on (\ref{equ:userpayoff}). In other words, all the users with $c<U\omega$ choose $d=2$ at the equilibrium, and the fraction ${\bar y}\left(r,I\right)$ is no smaller than the fraction of users with $c<U\omega$. Recall that $c$ is uniformly distributed in $\left[0,c_{\max}\right]$. The fraction of users with $c<U\omega$ is $\eta \frac{U}{c_{\max}}$, and hence ${\bar y}\left(r,I\right)\ge \eta \frac{U}{c_{\max}}$. 
Recall that the conditions of Case B include $I+I_0\le I_{\rm th}=\frac{\delta}{\theta+\frac{Vc_{\max}}{\eta UN}}$. 
We can easily derive 
\begin{align}
\theta {\bar y}\left(r,I\right)N-\frac{\delta}{I+I_0}{\bar y}\left(r,I\right)N\le - \frac{V c_{\max}}{\eta U} {\bar y}\left(r,I\right).
\end{align}
Considering ${\bar y}\left(r,I\right)\ge \eta \frac{U}{c_{\max}}$, we have
\begin{align}
\theta {\bar y}\left(r,I\right)N-\frac{\delta}{I+I_0}{\bar y}\left(r,I\right)N\le -V.
\end{align}
This contradicts with $V {+ \theta {\bar y}\left(r,I\right)N}-\frac{\delta}{I+I_0}{\bar y}\left(r,I\right)N>0$. Therefore, it is impossible that $V {+ \theta {\bar y}\left(r,I\right)N}-  \frac{\delta}{I+I_0}{\bar y}\left(r,I\right)N>0$.

Combing the above analysis, we conclude that $V {+ \theta {\bar y}\left(r,I\right)N}-  \frac{\delta}{I+I_0}{\bar y}\left(r,I\right)N=0$. 

{\bf (Step 2)} Next, we discuss the users' equilibrium strategies. According to (\ref{equ:userpayoff}), since $V {+ \theta {\bar y}\left(r,I\right)N}-  \frac{\delta}{I+I_0}{\bar y}\left(r,I\right)N=0$, a user's payoffs under decision $d=1$ and decision $d=2$ are the same, i.e., $U\omega-c$. Hence, when $c\ge U\omega$, the user chooses $d=0$; otherwise, the user chooses between $d=1$ and $2$. 
This implies that (i) all the users with $\omega=0$ choose $d=0$ at the equilibrium, (ii) the users with $\omega=1$ and $c\ge U$ choose $d=0$ at the equilibrium, (iii) the users with $\omega=1$ and $c<U$ choose between $d=1$ and $2$ at the equilibrium. Next, we discuss the detailed equilibrium strategies of the users with $\omega=1$ and $c<U$.

From $V {+ \theta {\bar y}\left(r,I\right)N}-  \frac{\delta}{I+I_0}{\bar y}\left(r,I\right)N=0$, we have
\begin{align}
{\bar y}\left(r,I\right)N=\frac{V}{\left(\frac{\delta}{I+I_0}-\theta\right)}.\label{SM:equ:proposition2:y}
\end{align}
That is to say, among the users with $\omega=1$ and $c<U$, the mass of users choosing $d=2$ is $\frac{V}{\left(\frac{\delta}{I+I_0}-\theta\right)}$, and the remaining users with $\omega=1$ and $c<U$ choose $d=1$. For a set $\hat {\cal C}$, it can represent the set of transportation costs of the users who have $\omega=1$ and $c<U$ and choose $d=2$ if and only if
\begin{align}
\eta N \int_{0}^{U} \frac{1}{c_{\max}} {\boldsymbol 1}_{\left\{c\in{\hat {\cal C}}\right\}} dc=\frac{V}{\left(\frac{\delta}{I+I_0}-\theta\right)}.
\end{align}
Therefore, we conclude that when $r=1$ and $I+I_0\le I_{\rm th}$, a type-$\left(\omega,c\right)$ user's optimal strategy is 
\begin{align}
d^*\left(\omega,c,r,I\right)= \left\{ {\begin{array}{*{20}{l}}
{2,}&{{\rm~if~}c\in{\hat{\cal C}}{\rm~and~}\omega=1,}\\
{1,}&{{\rm~if~}c\in[0,U\omega)\setminus {\hat{\cal C}},}\\
{0,}&{{\rm~if~}c\in\left[U\omega,c_{\max}\right],}\\
\end{array}} \right.\label{SM:equ:proposition2}
\end{align}
where $\omega\in\left\{0,1\right\}$, $c\in\left[0,c_{\max}\right]$, and ${\hat{\cal C}}\subseteq \left[0,U\right)$ can be any set that satisfies $\eta\int_{0}^{U}\frac{1}{{c_{\max}}}{{\boldsymbol 1}_{\left\{c\in{\hat{\cal C}}\right\}}} dc={\frac{V}{\left(\frac{\delta}{I+I_0}-\theta\right) N}}$. 
We can easily derive ${\bar x}\left(r,I\right)=\eta \frac{U}{c_{\max}}$ and ${\bar y}\left(r,I\right)={\frac{V}{\left(\frac{\delta}{I+I_0}-\theta\right) N}}$ from (\ref{SM:equ:proposition2}) and (\ref{SM:equ:proposition2:y}), respectively.
\endproof

\section{Proof of Proposition \ref{proposition:stageIII:c} in Section \ref{sec:stageIII}}\label{SM:proposition3}
\proof
{\bf(Step 1)} we prove that $V + \theta {\bar y}\left(r,I\right)N-  \frac{\delta}{I+I_0}{\bar y}\left(r,I\right)N>0$ holds at the equilibrium (a user's net payoff of interacting with the POI at the equilibrium is positive). We prove it by contradiction, i.e., we assume that $V {+ \theta {\bar y}\left(r,I\right)N}-  \frac{\delta}{I+I_0}{\bar y}\left(r,I\right)N\le0$.

The proof of the impossibility of $V {+ \theta {\bar y}\left(r,I\right)N}-  \frac{\delta}{I+I_0}{\bar y}\left(r,I\right)N<0$ is the same as that in the proof of Proposition \ref{proposition:stageIII:b} (Section \ref{SM:proposition2}), and is omitted here. 

Next we discuss the possibility of $V + \theta {\bar y}\left(r,I\right)N-\frac{\delta}{I+I_0}{\bar y}\left(r,I\right)N=0$. 
According to (\ref{equ:userpayoff}), when $V {+ \theta {\bar y}\left(r,I\right)N}-\frac{\delta}{I+I_0}{\bar y}\left(r,I\right)N=0$, a user's payoff under decision $d=2$ is $U\omega-c$. Hence, a user may choose $d=2$ only if $c<U\omega$. That is to say, the fraction of users choosing $d=2$ is no greater than the fraction of users with $c<U\omega$, i.e., ${\bar y}\left(r,I\right)\le \eta \frac{U}{c_{\max}}$. 
Recall that the conditions of Case C include $I+I_0> I_{\rm th}=\frac{\delta}{\theta+\frac{Vc_{\max}}{\eta UN}}$. We can easily derive the following relation
\begin{align}
\theta {\bar y}\left(r,I\right)N-\frac{\delta}{I+I_0}{\bar y}\left(r,I\right)N> - \frac{V c_{\max}}{\eta U} {\bar y}\left(r,I\right).
\end{align}
Considering ${\bar y}\left(r,I\right)\le \eta \frac{U}{c_{\max}}$, we have
\begin{align}
\theta {\bar y}\left(r,I\right)N-\frac{\delta}{I+I_0}{\bar y}\left(r,I\right)N> -V.
\end{align}
This contradicts with $V {+ \theta {\bar y}\left(r,I\right)N}-\frac{\delta}{I+I_0}{\bar y}\left(r,I\right)N=0$. Therefore, it is impossible that $V {+ \theta {\bar y}\left(r,I\right)N}-  \frac{\delta}{I+I_0}{\bar y}\left(r,I\right)N=0$.

Combing the above analysis, we conclude that $V {+ \theta {\bar y}\left(r,I\right)N}-  \frac{\delta}{I+I_0}{\bar y}\left(r,I\right)N>0$. 

{\bf(Step 2)} Next, we discuss the users' equilibrium strategies. From $V {+ \theta {\bar y}\left(r,I\right)N}-  \frac{\delta}{I+I_0}{\bar y}\left(r,I\right)N>0$ and (\ref{equ:userpayoff}), a user's payoff under $d=2$ is always strictly larger than its payoff under $d=1$. That is to say, no user chooses $d=1$ at the equilibrium. Comparing a user's payoffs under $d=0$ and $d=2$, we conclude that a user whose $c$ and $\omega$ satisfy $U\omega-c+V {+ \theta {\bar y}\left(r,I\right)N}-  \frac{\delta}{I+I_0}{\bar y}\left(r,I\right)N>0$ chooses $d=2$, and a user whose $c$ and $\omega$ satisfy $U\omega-c+V {+ \theta {\bar y}\left(r,I\right)N}-  \frac{\delta}{I+I_0}{\bar y}\left(r,I\right)N\le0$ chooses $d=0$. Since we assume $U+V+\theta N<c_{\max}$ in Section \ref{subsec:stackelberg}, we have $U+V {+ \theta {\bar y}\left(r,I\right)N}-  \frac{\delta}{I+I_0}{\bar y}\left(r,I\right)N<c_{\max}$. 
Therefore, we can compute ${\bar y}\left(r,I\right)$ as
\begin{align}
\nonumber
{\bar y}\left(r,I\right)=& \eta \frac{1}{c_{\max}}\left( U+V {+ \theta {\bar y}\left(r,I\right)N}-  \frac{\delta}{I+I_0}{\bar y}\left(r,I\right)N \right) \\
\nonumber
& \!\!\!\!\!+ \left(1-\eta\right) \frac{1}{c_{\max}}\left(V {+ \theta {\bar y}\left(r,I\right)N}-  \frac{\delta}{I+I_0}{\bar y}\left(r,I\right)N \right) \\
=& \frac{1}{c_{\max}} \left(\eta U+ V {+ \theta {\bar y}\left(r,I\right)N}-  \frac{\delta}{I+I_0}{\bar y}\left(r,I\right)N \right).
\end{align}
After arrangement, we have
\begin{align}
\nonumber
&{\bar y}\left(r,I\right)=\frac{\left(\eta U+V\right)\left(I+I_0\right)}{\left(c_{\max}-\theta N\right)\left(I+I_0\right)+\delta N} \\
\nonumber
&=\eta\frac{U}{c_{\max}}+{\frac{Vc_{\max}\left(I+I_0\right)-\eta UN\delta+\eta UN\theta \left(I+I_0\right)}{c_{\max}^2\left(I+I_0\right)+c_{\max}N\delta -c_{\max}N\theta\left(I+I_0\right)}}\\
&=\frac{\eta U+c_t}{c_{\max}}.\label{SM:equ:proposition3:y}
\end{align}
Recall that a user whose $c$ and $\omega$ satisfy $U\omega-c+V {+ \theta {\bar y}\left(r,I\right)N}-  \frac{\delta}{I+I_0}{\bar y}\left(r,I\right)N>0$ chooses $d=2$, and a user whose $c$ and $\omega$ satisfy $U\omega-c+V {+ \theta {\bar y}\left(r,I\right)N}-  \frac{\delta}{I+I_0}{\bar y}\left(r,I\right)N\le0$ chooses $d=0$. Based on (\ref{SM:equ:proposition3:y}), we conclude that when $r=1$ and $I+I_0>I_{\rm th}$, a type-$\left(\omega,c\right)$ user's optimal strategy is
\begin{align}
d^*\left(\omega,c,r,I\right)= \left\{ {\begin{array}{*{20}{l}}
{2,}&{{\rm~if~}c\in\left[0,U\omega+c_t\right),}\\
{0,}&{{\rm~if~}c\in\left[U\omega+c_t,c_{\max}\right],}\\
\end{array}} \right.\label{SM:equ:proposition3}
\end{align}
where $\omega\in\left\{0,1\right\}$ and $c\in\left[0,c_{\max}\right]$. From (\ref{SM:equ:proposition3}), we can easily compute ${\bar x}\left(r,I\right)=\eta\frac{U+c_t}{c_{\max}}$. Moreover, fraction ${\bar y}\left(r,I\right)$ is given in (\ref{SM:equ:proposition3:y}).
\endproof

\section{Proof of Proposition \ref{proposition:stageII:situationI} in Section \ref{sec:stageII}}\label{SM:proposition4}
\proof 
{\bf(Step 1)} We compute the venue's optimal payoffs in the following three decision regions: (i) $r=0$; (ii) $r=1$ and $I\in\left[0,I_{\rm th}-I_0\right]$; (iii) $r=1$ and $I\in\left(I_{\rm th}-I_0,\infty\right)$. 

First, in decision region (i) (i.e., $r=0$), the venue's payoff is
\begin{align}
\nonumber
\Pi^{\rm venue}\left(0,I,l,p\right)& = b N {\bar x}\left(0,I\right)-k I- 0\cdot \left(l+pN {\bar y}\left(0,I\right)\right)\\
&= b N \eta \frac{U}{c_{\max}}-kI,
\end{align}
where ${\bar x}\left(0,I\right)$ is computed by Proposition \ref{proposition:stageIII:a}. It is easy to see that the venue's optimal payoff in this decision region is $b N \eta \frac{U}{c_{\max}}$, and the optimal investment level is $0$. 

Second, in decision region (ii) (i.e., $r=1$ and $I\in\left[0,I_{\rm th}-I_0\right]$), the venue's payoff is 
\begin{align}
\nonumber
\Pi^{\rm venue}\left(1,I,l,p\right)& = b N {\bar x}\left(1,I\right)-k I-1\cdot \left(l+pN {\bar y}\left(1,I\right)\right)\\
& =  b N \eta\frac{U}{c_{\max}}-k I-\left(l+p {\frac{V}{\frac{\delta}{I+I_0}-\theta}}\right),
\end{align}
where ${\bar x}\left(1,I\right)$ and ${\bar y}\left(1,I\right)$ are computed by Proposition \ref{proposition:stageIII:b}. By examining $\frac{\partial \Pi^{\rm venue}\left(1,I,l,p\right)}{\partial I}$, we can find that $\Pi^{\rm venue}\left(1,I,l,p\right)$ either (a) monotonically changes or (b) first decreases and then increases in $I\in\left[0,I_{\rm th}-I_0\right]$. Hence, in order to obtain the venue's optimal payoff in this decision region, we only need to compare its payoffs under $I=0$ and $I=I_{\rm th}-I_0$. Specifically, we have
\begin{align}
& \Pi^{\rm venue}\left(1,0,l,p\right)=bN\eta\frac{U}{c_{\max}}-l-{\frac{V}{\frac{\delta}{I_0}-\theta}p}.\\
\nonumber
& \Pi^{\rm venue}\left(1,I_{\rm th}-I_0,l,p\right)=b N \eta\frac{U}{c_{\max}}-k \left(I_{\rm th}-I_0\right)\\
&{~~~~~~~~~~~~~~~~~~~~~~~~~~~~~~~~~~~~~~~~} -\left(l+p \frac{\eta UN}{c_{\max}}\right).\label{SM:equ:thresholdinvest}
\end{align}
Comparing $\Pi^{\rm venue}\left(1,0,l,p\right)$ and $\Pi^{\rm venue}\left(1,I_{\rm th}-I_0,l,p\right)$, we have $\Pi^{\rm venue}\left(1,0,l,p\right)>\Pi^{\rm venue}\left(1,I_{\rm th}-I_0,l,p\right)$ if and only if $p>p_0= -k\frac{\left(\delta-\theta I_0\right)c_{\max}}{Vc_{\max}+\theta\eta UN}$. 
Therefore, if $p>p_0$, the venue's optimal payoff in decision region (ii) is $\Pi^{\rm venue}\left(1,0,l,p\right)=bN\eta\frac{U}{c_{\max}}-l-{\frac{V}{\frac{\delta}{I_0}-\theta}p}$, and the optimal investment level is $0$; if $p\le p_0$, the venue's optimal payoff in this decision region is $\Pi^{\rm venue}\left(1,I_{\rm th}-I_0,l,p\right)=b N \eta\frac{U}{c_{\max}}-k \left(I_{\rm th}-I_0\right)-\left(l+p \frac{\eta UN}{c_{\max}}\right)$, and the optimal investment level is $I_{\rm th}-I_0$. 

Third, in decision region (iii) (i.e., $r=1$ and $I\in\left(I_{\rm th}-I_0,\infty\right)$), the venue's payoff is given as follows. 
\begin{align}
\nonumber
& \Pi^{\rm venue}\left(1,I,l,p\right)= -k I-l\\
\nonumber
& \!+\!b N {\left(\eta\frac{U}{c_{\max}}\!+\!\eta{\frac{Vc_{\max}\left(I+I_0\right)-\eta UN\delta+\eta UN\theta \left(I+I_0\right)}{c_{\max}^2\left(I+I_0\right)+c_{\max}N\delta -c_{\max}N\theta\left(I+I_0\right)}}\!\right)}\\
& \!-\!pN{\left(\eta\frac{U}{c_{\max}}\!+\!{\frac{Vc_{\max}\left(I+I_0\right)-\eta UN\delta+\eta UN\theta \left(I+I_0\right)}{c_{\max}^2\left(I+I_0\right)+c_{\max}N\delta -c_{\max}N\theta\left(I+I_0\right)}}\!\right)}.\label{SM:equ:longregion3}
\end{align}
We can compute
\begin{align}
& \frac{\partial \Pi^{\rm venue}\left(1,I,l,p\right)}{\partial I}=\frac{\delta N^2 \left(b\eta-p\right)\left(V+\eta U\right)}{\left(\delta N +{\left(I+I_0\right)} {\left(c_{\max}-N\theta\right)}\right)^2}-k. \label{SM:equ:derivative}
\end{align}

If $p\ge b\eta$, payoff $\Pi^{\rm venue}\left(1,I,l,p\right)$ decreases with $I$ for $I>I_{\rm th}-I_0$. The venue's optimal payoff in this decision region is smaller than $\left(b-p\right)N\eta\frac{U}{c_{\max}}-k\left(I_{\rm th}-I_0\right)-l$, which is exactly $\Pi^{\rm venue}\left(1,I_{\rm th}-I_0,l,p\right)$ given in (\ref{SM:equ:thresholdinvest}). 

If $p<b\eta $, we can derive the relation $\frac{\partial^2 \Pi^{\rm venue}\left(1,I,l,p\right)}{\partial I^2}<0$. 
Based on (\ref{SM:equ:derivative}), the solution to $\frac{\partial \Pi^{\rm venue}\left(1,I,l,p\right)}{\partial I}=0$ is $I=\frac{N}{{c_{\max}-N\theta}}\sqrt{\frac{\delta\left(V+\eta U\right)\left(b\eta -p\right)}{k}}-\delta \frac{N}{{c_{\max}-N\theta}} {-I_0}$. 
Hence, when $\frac{N}{{c_{\max}-N\theta}}\sqrt{\frac{\delta\left(V+\eta U\right)\left(b\eta -p\right)}{k}}-\delta \frac{N}{{c_{\max}-N\theta}} {-I_0}> I_{\rm th}-I_0$, $\Pi^{\rm venue}\left(1,I,l,p\right)$ increases with $I$ for $I_{\rm th}-I_0< I\le \frac{N}{{c_{\max}-N\theta}}\sqrt{\frac{\delta\left(V+\eta U\right)\left(b\eta -p\right)}{k}}-\delta \frac{N}{{c_{\max}-N\theta}} {-I_0}$, and decreases with $I$ for $I\ge \frac{N}{{c_{\max}-N\theta}}\sqrt{\frac{\delta\left(V+\eta U\right)\left(b\eta -p\right)}{k}}-\delta \frac{N}{{c_{\max}-N\theta}} {-I_0}$. Therefore, the venue's optimal investment level in this region is $\frac{N}{{c_{\max}-N\theta}}\sqrt{\frac{\delta\left(V+\eta U\right)\left(b\eta -p\right)}{k}}-\delta \frac{N}{{c_{\max}-N\theta}} {-I_0}$, and the corresponding optimal payoff is $\frac{N}{c_{\max}}b\eta\left(1-\eta\right)U+\frac{N}{c_{\max}-N\theta}\left(\sqrt{\left(V+\eta U\right)\left(b\eta -p\right)}-\sqrt{\delta k}\right)^2-l+kI_0$. Furthermore, this optimal payoff is greater than $\left(b-p\right)N\eta\frac{U}{c_{\max}}-k\left(I_{\rm th}-I_0\right)-l$, which is exactly $\Pi^{\rm venue}\left(1,I_{\rm th}-I_0,l,p\right)$ given in (\ref{SM:equ:thresholdinvest}). 
When $\frac{N}{{c_{\max}-N\theta}}\sqrt{\frac{\delta\left(V+\eta U\right)\left(b\eta -p\right)}{k}}-\delta \frac{N}{{c_{\max}-N\theta}} {-I_0}\le I_{\rm th}-I_0$, $\Pi^{\rm venue}\left(1,I,l,p\right)$ decreases with $I$ for $I> I_{\rm th}-I_0$. Therefore, the venue's optimal payoff in this decision region is smaller than $\left(b-p\right)N\eta\frac{U}{c_{\max}}-k\left(I_{\rm th}-I_0\right)-l$, which is exactly $\Pi^{\rm venue}\left(1,I_{\rm th}-I_0,l,p\right)$ given in (\ref{SM:equ:thresholdinvest}).

Note that $\frac{N}{{c_{\max}-N\theta}}\sqrt{\frac{\delta\left(V+\eta U\right)\left(b\eta -p\right)}{k}}-\delta \frac{N}{{c_{\max}-N\theta}} {-I_0}> I_{\rm th}-I_0= \frac{\delta}{\theta+\frac{Vc_{\max}}{\eta UN}}-I_0$ is equivalent to $p<p_1 = b\eta-\frac{\delta k \left(V+\eta U\right)c_{\max}^2}{\left(Vc_{\max}+\theta\eta UN\right)^2}$. Therefore, we can summarize the venue's optimal payoff in decision region (iii) as follows. If $p<p_1$, the venue's optimal payoff is $\frac{N}{c_{\max}}b\eta\left(1-\eta\right)U+\frac{N}{c_{\max}-N\theta}\left(\sqrt{\left(V+\eta U\right)\left(b\eta -p\right)}-\sqrt{\delta k}\right)^2-l+kI_0$, and the optimal investment level is $\frac{N}{{c_{\max}-N\theta}}\sqrt{\frac{\delta\left(V+\eta U\right)\left(b\eta -p\right)}{k}}-\delta \frac{N}{{c_{\max}-N\theta}} {-I_0}$. Furthermore, the venue's optimal payoff is greater than $\Pi^{\rm venue}\left(1,I_{\rm th}-I_0,l,p\right)$ given in (\ref{SM:equ:thresholdinvest}). If $p\ge p_1$, the venue's optimal payoff is smaller than $\Pi^{\rm venue}\left(1,I_{\rm th}-I_0,l,p\right)$.

So far, we have analyzed the venue's optimal payoffs in different decision regions. We will jointly consider and compare these optimal payoffs to determine the venue's equilibrium strategies in situation I. 

{\bf (Step 2)} We discuss the venue's equilibrium strategies under $p<p_1$. In Situation I, we have $\delta>\delta_{\rm th}$. Based on the definitions of $p_0$ and $p_1$, it is easy to show that $p_0>p_1$. Hence, we have $p<p_1<p_0$. Based on {\bf Step 1}, the venue's optimal payoff in decision region (i) is $\Pi^{\rm venue}\left(0,0,l,p\right)=b N \eta \frac{U}{c_{\max}}$, its optimal payoff in decision region (ii) is $\Pi^{\rm venue}\left(1,I_{\rm th}-I_0,l,p\right)$, and its optimal payoff in decision region (iii) is $\frac{N}{c_{\max}}b\eta\left(1-\eta\right)U+\frac{N}{c_{\max}-N\theta}\left(\sqrt{\left(V+\eta U\right)\left(b\eta -p\right)}-\sqrt{\delta k}\right)^2-l+kI_0$, which is greater than $\Pi^{\rm venue}\left(1,I_{\rm th}-I_0,l,p\right)$. Considering the three decision regions, we conclude that if $l\le -\frac{N}{c_{\max}}b\eta^2U+\frac{N}{c_{\max}-N\theta}\left(\sqrt{\left(V+\eta U\right)\left(b\eta -p\right)}-\sqrt{\delta k}\right)^2+kI_0$, the venue's optimal payoff is $\frac{N}{c_{\max}}b\eta\left(1-\eta\right)U+\frac{N}{c_{\max}-N\theta}\left(\sqrt{\left(V+\eta U\right)\left(b\eta -p\right)}-\sqrt{\delta k}\right)^2-l+kI_0$, and the corresponding optimal strategies are $r^*\left(l,p\right)=1$ and $I^*\left(l,p\right)=\frac{N}{{c_{\max}-N\theta}}\sqrt{\frac{\delta\left(V+\eta U\right)\left(b\eta -p\right)}{k}}-\delta \frac{N}{{c_{\max}-N\theta}} {-I_0}$; otherwise, the venue's optimal payoff is $b N \eta \frac{U}{c_{\max}}$, and the corresponding optimal strategies are $r^*\left(l,p\right)=0$ and $I^*\left(l,p\right)=0$. 

{\bf (Step 3)} We discuss the venue's equilibrium strategies under $p_1\le p\le p_0$. Based on {\bf Step 1}, the venue's optimal payoff in decision region (i) is $\Pi^{\rm venue}\left(0,0,l,p\right)=b N \eta \frac{U}{c_{\max}}$, its optimal payoff in decision region (ii) is $\Pi^{\rm venue}\left(1,I_{\rm th}-I_0,l,p\right)=b N \eta\frac{U}{c_{\max}}-k \left(I_{\rm th}-I_0\right)-\left(l+p \frac{\eta UN}{c_{\max}}\right)$, and its optimal payoff in decision region (iii) is smaller than $\Pi^{\rm venue}\left(1,I_{\rm th}-I_0,l,p\right)$. 
Considering the three decision regions, we conclude that if $l\le -pN\eta \frac{U}{c_{\max}}-k\left(I_{\rm th}-I_0\right)$, the venue's optimal payoff is $b N \eta\frac{U}{c_{\max}}-k \left(I_{\rm th}-I_0\right)-\left(l+p \frac{\eta UN}{c_{\max}}\right)$, and the corresponding optimal strategies are $r^*\left(l,p\right)=1$ and $I^*\left(l,p\right)=I_{\rm th}-I_0$; otherwise, the venue's optimal payoff is $b N \eta \frac{U}{c_{\max}}$, and the corresponding optimal strategies are $r^*\left(l,p\right)=0$ and $I^*\left(l,p\right)=0$. 

{\bf (Step 4)} We discuss the venue's equilibrium strategies under $p> p_0$. Since $p_0>p_1$, we have $p>p_0>p_1$. 
Based on {\bf Step 1}, the venue's optimal payoff in decision region (i) is $\Pi^{\rm venue}\left(0,0,l,p\right)=b N \eta \frac{U}{c_{\max}}$, its optimal payoff in decision region (ii) is $\Pi^{\rm venue}\left(1,0,l,p\right)=bN\eta\frac{U}{c_{\max}}-l-{\frac{V}{\frac{\delta}{I_0}-\theta}p}$, which is greater than $\Pi^{\rm venue}\left(1,I_{\rm th}-I_0,l,p\right)$, and its optimal payoff in decision region (iii) is smaller than $\Pi^{\rm venue}\left(1,I_{\rm th}-I_0,l,p\right)$. 
Considering the three decision regions, we conclude that if $l\le -{\frac{V}{\frac{\delta}{I_0}-\theta}p}$, the venue's optimal payoff is $bN\eta\frac{U}{c_{\max}}-l-{\frac{V}{\frac{\delta}{I_0}-\theta}p}$, and the corresponding optimal strategies are $r^*\left(l,p\right)=1$ and $I^*\left(l,p\right)=0$; otherwise, the venue's optimal payoff is $b N \eta \frac{U}{c_{\max}}$, and the corresponding optimal strategies are $r^*\left(l,p\right)=0$ and $I^*\left(l,p\right)=0$. 

Summarizing {\bf Step 2}, {\bf Step 3}, and {\bf Step 4}, we derive the venue's equilibrium strategies as
{\small{\begin{align}
\nonumber
& \!\left(r^*\left(l,p\right),I^*\left(l,p\right)\right)\!=\!\\
\nonumber
& \!\!\!\left\{ {\begin{array}{{l}{l}}
{\!\!\!\left(0,0\right),}&{\!\!\!\!\!{\rm if~}l\!>\!H_1\!\left(p \right)\!,}\\
{\!\!\!\!\!\!\left(\!1,\!\frac{N}{{c_{\max}\!-\!N\theta}}\!\sqrt{\!\frac{\delta\left(V+\eta U\right)\left(b\eta -p\right)}{k}}\!-\! \!\frac{\delta N}{{c_{\max}-N\theta}} {-\!I_0}\!\!\right)\!\!,}&{\!\!\!\!\!{\rm if~}l\!\le\! H_1\!\left(p \right)\!,p\!<\! p_1,}\\
{\!\!\!\left(1,I_{\rm th}-I_0\right),}&{\!\!\!\!\!{\rm if~}l\!\le\! H_1\!\left(p \right)\!,p_1\!\le \!p\!\le\! p_0,}\\
{\!\!\!\left(1,0\right),}&{\!\!\!\!\!{\rm if~}l\!\le\! H_1\!\left(p \right)\!,p\!>\!p_0,}
\end{array}} \right.
\end{align}}}
which is exactly (\ref{equ:venue:situationI}).
\endproof

\section{Proof of Lemma \ref{lemma:p2unique} in Section \ref{sec:stageII}}\label{SM:uniquep2}
\proof 
First of all, we can easily show that $\delta\le\delta_{\rm th}$ implies $p_0\le p_1$. Then, we define function $G\left(p\right)$ as 
\begin{align}
\nonumber
G\left(p\right)\triangleq & \frac{N}{c_{\max}-N\theta}\left(\sqrt{\left(V+\eta U\right)\left(b\eta -p\right)}-\sqrt{\delta k}\right)^2\\
& -\frac{N}{c_{\max}}b\eta^2U+kI_0+\frac{V}{\frac{\delta}{I_0}-\theta}p.\label{SM:equ:functionG}
\end{align}
We will prove that when $I_0\le I_{\rm th}$ and $\delta\le\delta_{\rm th}$, there is a unique $p$ satisfying $p\in\left[p_0,p_1\right]$ and $G\left(p\right)=0$. 

{\bf(Step 1)} We prove that $G\left(p\right)$ is strictly decreasing in $p\in\left[p_0,p_1\right)$ in Situation II ($I_0\le I_{\rm th}$ and $\delta\le\delta_{\rm th}$). From (\ref{SM:equ:functionG}), we can compute 
\begin{align}
\nonumber
\frac{d G\left(p\right)}{d p} =&-\frac{N\left(V+\eta U\right)}{c_{\max}-\theta N}+\frac{N\sqrt{\delta k}\sqrt{V+\eta U}}{\left(c_{\max}-N\theta \right)\sqrt{b\eta-p}}+\frac{V}{\frac{\delta}{I_0}-\theta}\\
\nonumber
\overset{(a)}{<}& -\frac{N\eta U}{c_{\max}-N\theta}+\frac{N\theta\eta UN}{\left(c_{\max}-\theta N\right)c_{\max}}+\frac{V}{\frac{\delta}{I_0}-\theta}\\
\nonumber
\overset{(b)}{\le}& -\frac{N\eta U}{c_{\max}-N\theta}+\frac{N\theta\eta UN}{\left(c_{\max}-\theta N\right)c_{\max}}+\frac{\eta UN}{c_{\max}}\\
=&0.
\end{align}
The inequality (a) is because of $p< p_1$, and inequality (b) is because of $I\le I_{\rm th}$. Since $\frac{d G\left(p\right)}{d p}<0$ for $p<p_1$, $G\left(p\right)$ is strictly decreasing in $p\in\left[p_0,p_1\right)$. 

{\bf(Step 2)} We prove that $G\left(p_1\right)\le0$ in Situation II ($I_0\le I_{\rm th}$ and $\delta\le\delta_{\rm th}$). By plugging $p_1 = b\eta-\frac{\delta k \left(V+\eta U\right)c_{\max}^2}{\left(Vc_{\max}+\theta\eta UN\right)^2}$ into (\ref{SM:equ:functionG}), we have
\begin{align}
\nonumber
& G\left(p_1\right)=\frac{N}{c_{\max}-N\theta}\left(\frac{\sqrt{\delta k}\left(V+\eta U\right)c_{\max}}{Vc_{\max}+\theta\eta U N}-\sqrt{\delta k}\right)^2 \\
& -\frac{N}{c_{\max}}b\eta^2U+kI_0 +\frac{V}{\frac{\delta}{I_0}-\theta}\left(b\eta-\frac{\delta k \left(V+\eta U\right)c_{\max}^2}{\left(Vc_{\max}+\theta\eta UN\right)^2}\right).\label{SM:equ:Gp1}
\end{align}
In order to prove that $G\left(p_1\right)\le0$ for $I_0\le I_{\rm th}$ and $\delta\le\delta_{\rm th}$, we define
\begin{align}
\nonumber
& Q\left(I_0\right)\triangleq \frac{N}{c_{\max}-N\theta}\left(\frac{\sqrt{\delta k}\left(V+\eta U\right)c_{\max}}{Vc_{\max}+\theta\eta U N}-\sqrt{\delta k}\right)^2 \\
& -\frac{N}{c_{\max}}b\eta^2U+kI_0 +\frac{V}{\frac{\delta}{I_0}-\theta}\left(b\eta-\frac{\delta k \left(V+\eta U\right)c_{\max}^2}{\left(Vc_{\max}+\theta\eta UN\right)^2}\right).\label{SM:equ:QI0}
\end{align}

First, we plug $I_0=I_{\rm th}= \frac{\delta}{\theta+\frac{Vc_{\max}}{\eta UN}}$ into the above equation, and obtain $Q\left(I_{\rm th}\right)=0$. 

Second, we prove that $Q\left(I_0\right)$ is increasing in $I_0\in\left[0,I_{\rm th}\right]$. From (\ref{SM:equ:QI0}), we can compute 
\begin{align}
\nonumber
& \left(\delta-\theta I_0\right)^2\left(Vc_{\max}+\theta\eta UN\right) \frac{d Q\left(I_0\right)}{d I_0}=-{V\delta^2 k\left(\eta U+V\right)c_{\max}^2}\\
& +{k\left(\delta-\theta I_0\right)^2 \left(Vc_{\max}+\theta\eta UN\right)^2}+V\delta b\eta \left(Vc_{\max}+\theta\eta UN\right)^2.
\end{align}

Since $\delta\le \delta_{\rm th}= \frac{\left(Vc_{\max}+\theta\eta UN\right)\left(b\eta \left(Vc_{\max}+\theta\eta U N\right)-\theta I_0 c_{\max}k \right)}{k c_{\max} \eta U \left(c_{\max}-\theta N\right)}$, we further have
\begin{align}
\nonumber
& \left(\delta-\theta I_0\right)^2\left(Vc_{\max}+\theta\eta UN\right) \frac{d Q\left(I_0\right)}{d I_0}\ge -{V\delta^2 k\left(\eta U+V\right)c_{\max}^2}\\
\nonumber
& +{k\left(\delta-\theta I_0\right)^2 \left(Vc_{\max}+\theta\eta UN\right)^2}+{V\delta^2 kc_{\max}\eta U \left(c_{\max}-\theta N\right) }\\
& +{\theta I_0 c_{\max} k \left(Vc_{\max}+\theta \eta UN\right)V\delta }.\label{SM:equ:longinequality}
\end{align}
We define function $S\left(I_0\right)$ as the right-hand side of (\ref{SM:equ:longinequality}):
\begin{align}
\nonumber
& S\!\left(I_0\right) \! \triangleq \! -{V\delta^2 k\left(\eta U\!+\!V\right)c_{\max}^2}\!+\!{k\left(\delta-\theta I_0\right)^2\! \left(Vc_{\max}\!+\!\theta\eta UN\right)^2}\\
& \!+\!{V\delta^2 kc_{\max}\eta U \left(c_{\max}-\theta N\right) }\!+\!{\theta I_0 c_{\max} k \left(Vc_{\max}+\theta \eta UN\right)V\delta }.
\end{align}
We can easily show that $\frac{d S\left(I_0\right)}{d I_0}\le 0$ for $I_0\in\left[0,I_{\rm th}\right]$, and $S\left(I_{\rm th}\right)=0$. Hence, $S\left(I_0\right)\ge0$ for $I_0\in\left[0,I_{\rm th}\right]$. Because $S\left(I_0\right)$ is the right-hand side of (\ref{SM:equ:longinequality}), we have $\frac{d Q\left(I_0\right)}{d I_0}\ge0$ for $I_0\in\left[0,I_{\rm th}\right]$. 

Combing $Q\left(I_{\rm th}\right)=0$ and $\frac{d Q\left(I_0\right)}{d I_0}\ge0$ for $I_0\in\left[0,I_{\rm th}\right]$, we can see that $Q\left(I_0\right)\le0$ for $I_0\in\left[0,I_{\rm th}\right]$. Since the expression of $Q\left(I_0\right)$ in (\ref{SM:equ:QI0}) is the same as the expression of $G\left(p_1\right)$ in (\ref{SM:equ:Gp1}), we conclude that $G\left(p_1\right)\le0$ in Situation II. 

{\bf(Step 3)} We prove that $G\left(p_0\right)\ge0$ in Situation II ($I_0\le I_{\rm th}$ and $\delta\le\delta_{\rm th}$). By plugging $p_0= -k\frac{\left(\delta-\theta I_0\right)c_{\max}}{Vc_{\max}+\theta\eta UN}$ into (\ref{SM:equ:functionG}), we have
\begin{align}
\nonumber
& G\left(p_0\right)=-\frac{N}{c_{\max}}b\eta^2U+ \frac{\theta\eta UN kI_0}{Vc_{\max}+\theta\eta UN} \\
& \!+\!\frac{N}{c_{\max}\!-\!N\theta}\left(\sqrt{\left(V\!+\!\eta U\right)\left(b\eta \!+\!k\frac{\left(\delta-\theta I_0\right)c_{\max}}{Vc_{\max}+\theta\eta UN}\right)}\!-\!\sqrt{\delta k}\!\right)^2.\label{SM:equ:Gp0}
\end{align}

In order to prove that $G\left(p_0\right)\ge0$ for $I_0\le I_{\rm th}$ and $\delta\le\delta_{\rm th}$, we define 
\begin{align}
\nonumber
& W\left(\delta\right)\triangleq -\frac{N}{c_{\max}}b\eta^2U+ \frac{\theta\eta UN kI_0}{Vc_{\max}+\theta\eta UN} \\
& \!+\!\frac{N}{c_{\max}\!-\!N\theta}\left(\sqrt{\left(V\!+\!\eta U\right)\left(b\eta \!+\!k\frac{\left(\delta-\theta I_0\right)c_{\max}}{Vc_{\max}+\theta\eta UN}\right)}\!-\!\sqrt{\delta k}\!\right)^2.\label{SM:equ:Wdelta}
\end{align} 

First, we prove $\sqrt{\left(V+\eta U\right)\left(b\eta +k\frac{\left(\delta-\theta I_0\right)c_{\max}}{Vc_{\max}+\theta\eta UN}\right)}>\sqrt{\delta k}$. Note that $\sqrt{{\left(V+\eta U\right)}\left({b\eta -p_1}\right)}=\sqrt{\delta k}\frac{Vc_{\max}+\eta U c_{\max}}{V c_{\max}+\eta U \theta N }>\sqrt{\delta k}$, where the inequality is because of the assumption $c_{\max}>U+V+\theta N$. In Situation II, $p_0\le p_1$. Hence, $\sqrt{{\left(V+\eta U\right)}\left({b\eta -p_0}\right)}\ge \sqrt{{\left(V+\eta U\right)}\left({b\eta -p_1}\right)}>\sqrt{\delta k}$. Because $p_0= -k\frac{\left(\delta-\theta I_0\right)c_{\max}}{Vc_{\max}+\theta\eta UN}$, we can see that $\sqrt{\left(V+\eta U\right)\left(b\eta +k\frac{\left(\delta-\theta I_0\right)c_{\max}}{Vc_{\max}+\theta\eta UN}\right)}>\sqrt{\delta k}$.

Second, we can prove $W\left(\delta_{\rm th}\right)=0$ by plugging $\delta= \delta_{\rm th}= \frac{\left(Vc_{\max}+\theta\eta UN\right)\left(b\eta \left(Vc_{\max}+\theta\eta U N\right)-\theta I_0 c_{\max}k \right)}{k c_{\max} \eta U \left(c_{\max}-\theta N\right)}$ into (\ref{SM:equ:Wdelta}). 

Third, we prove that function $W\left(\delta\right)$ is decreasing in $\delta\in\left(0,\delta_{\rm th}\right]$. Since we already have (\ref{SM:equ:Wdelta}) and the result $\sqrt{\left(V+\eta U\right)\left(b\eta +k\frac{\left(\delta-\theta I_0\right)c_{\max}}{Vc_{\max}+\theta\eta UN}\right)}>\sqrt{\delta k}$, we only need to prove that function ${\hat W}\left(\delta\right) \triangleq \sqrt{\left(V+\eta U\right)\left(b\eta +k\frac{\left(\delta-\theta I_0\right)c_{\max}}{Vc_{\max}+\theta\eta UN}\right)}-\sqrt{\delta k}$ is decreasing in $\delta\in\left(0,\delta_{\rm th}\right]$. 
We compute $\frac{d {\hat W}\left(\delta\right)}{d \delta}$ as
{\small
\begin{align}
\nonumber
& \frac{d {\hat W}\left(\delta\right)}{d \delta}=-\frac{\sqrt{k}}{2\sqrt{\delta}}\\
& + \frac{1}{2}\sqrt{\frac{V+\eta U}{Vc_{\max}+\theta \eta UN}}\frac{kc_{\max}}{\sqrt{b\eta V c_{\max}+b\eta^2\theta UN+k\left(\delta-\theta I_0\right)c_{\max}}}.\label{SM:equ:Wdelta:derivative}
\end{align}}
Note that $\frac{d {\hat W}\left(\delta\right)}{d \delta}$ has two terms, and we compare these two terms as follows. We compute the ratio between the two terms' squares as 
\begin{align}
\frac{\left(V c_{\max}+ \theta\eta UN\right)\left(\frac{b\eta V c_{\max}+b\eta^2\theta UN-k \theta I_0 c_{\max}}{\delta}+kc_{\max}\right)}{\left(V+\eta U\right)k c_{\max}^2}.\label{SM:equ:ratio}
\end{align}
When $\delta \le \delta_{\rm th}= \frac{\left(Vc_{\max}+\theta\eta UN\right)\left(b\eta \left(Vc_{\max}+\theta\eta U N\right)-\theta I_0 c_{\max}k \right)}{k c_{\max} \eta U \left(c_{\max}-\theta N\right)}$ and $\delta>0$, we can see that $b\eta V c_{\max}+b\eta^2\theta UN-k \theta I_0 c_{\max}>0$, and the ratio in (\ref{SM:equ:ratio}) is strictly decreasing in $\delta$. Moreover, when $\delta =\delta_{\rm th}= \frac{\left(Vc_{\max}+\theta\eta UN\right)\left(b\eta \left(Vc_{\max}+\theta\eta U N\right)-\theta I_0 c_{\max}k \right)}{k c_{\max} \eta U \left(c_{\max}-\theta N\right)}$, we can see that the ratio in (\ref{SM:equ:ratio}) is $1$. Therefore, when $\delta \le \delta_{\rm th}$, the ratio is no less than $1$. 
Because it is the ratio between the squares of the two terms in (\ref{SM:equ:Wdelta:derivative}), we can conclude that $ \frac{d {\hat W}\left(\delta\right)}{d \delta}\le 0$ for $\delta\in\left(0,\delta_{\rm th}\right]$. This implies that $ \frac{d W\left(\delta\right)}{d \delta}\le 0$ for $\delta\in\left(0,\delta_{\rm th}\right]$.

Combing $W\left(\delta_{\rm th}\right)=0$ and $\frac{d W\left(\delta\right)}{d \delta}\le0$ for $\delta\in\left(0,\delta_{\rm th}\right]$, we can see that $W\left(\delta\right)\ge0$ for $\delta\in\left(0,\delta_{\rm th}\right]$. Since the expression of $W\left(\delta\right)$ in (\ref{SM:equ:Wdelta}) is the same as the expression of $G\left(p_0\right)$ in (\ref{SM:equ:Gp0}), we have $G\left(p_0\right)\ge0$ in Situation II. 

{\bf(Step 4)} Now we have proved that in Situation II, $G\left(p\right)$ is strictly decreasing in $p\in\left[p_0,p_1\right)$, $G\left(p_1\right)\le0$, and $G\left(p_0\right)\ge0$. Note that $G\left(p\right)$ is continuous in $p\in\left[p_0,p_1\right]$. It is easy to see that there is a unique $p$ satisfying $p\in\left[p_0,p_1\right]$ and $G\left(p\right)=0$. 
\endproof

\section{Illustrations of Propositions \ref{proposition:stageII:situationII} and \ref{proposition:stageII:situationIII} in Section \ref{sec:stageII}}\label{SM:illustration:pro5}

\begin{figure}[h]
  \centering
  \includegraphics[scale=0.43]{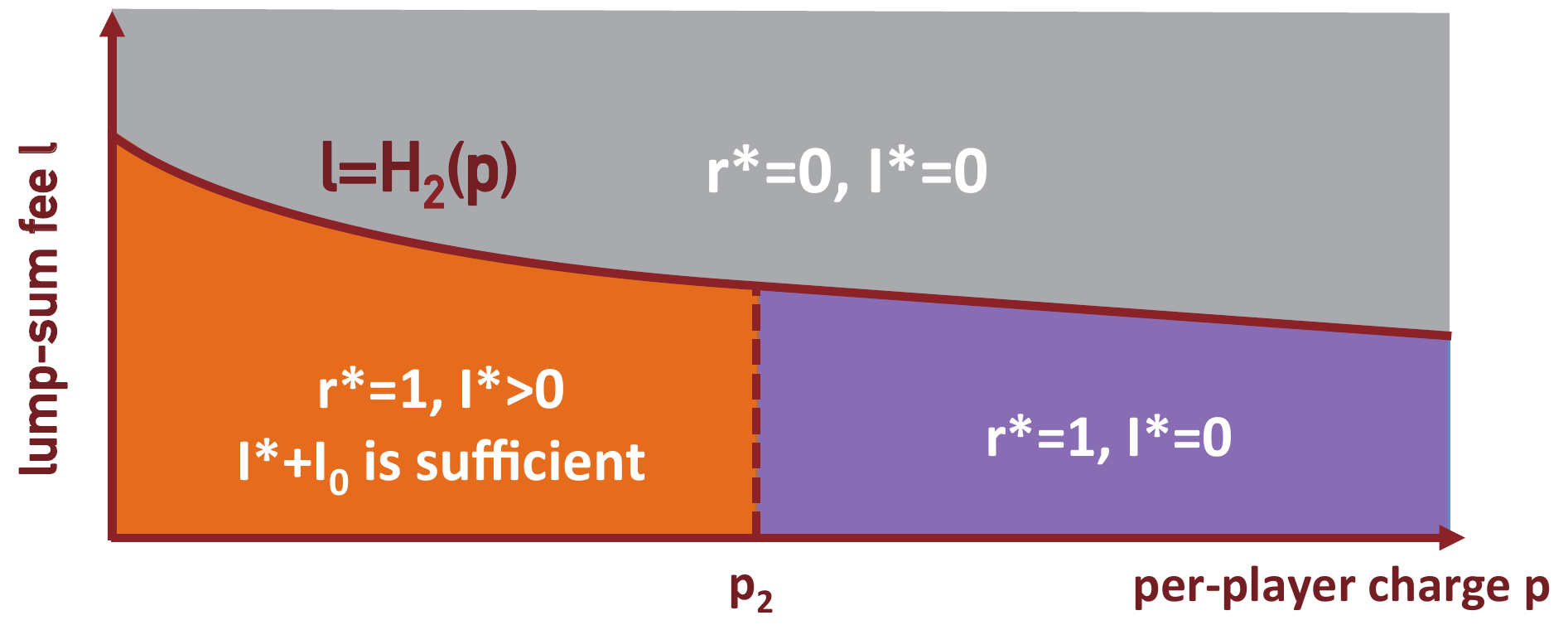}\\
  \caption{Situation II of Venue's POI Decision $r^*$ and Investment $I^*$ in Stage II.}
  \label{SM:fig:stageII:a}
  \vspace{-0.5cm}
\end{figure}

\begin{figure}[h]
  \centering
  \includegraphics[scale=0.43]{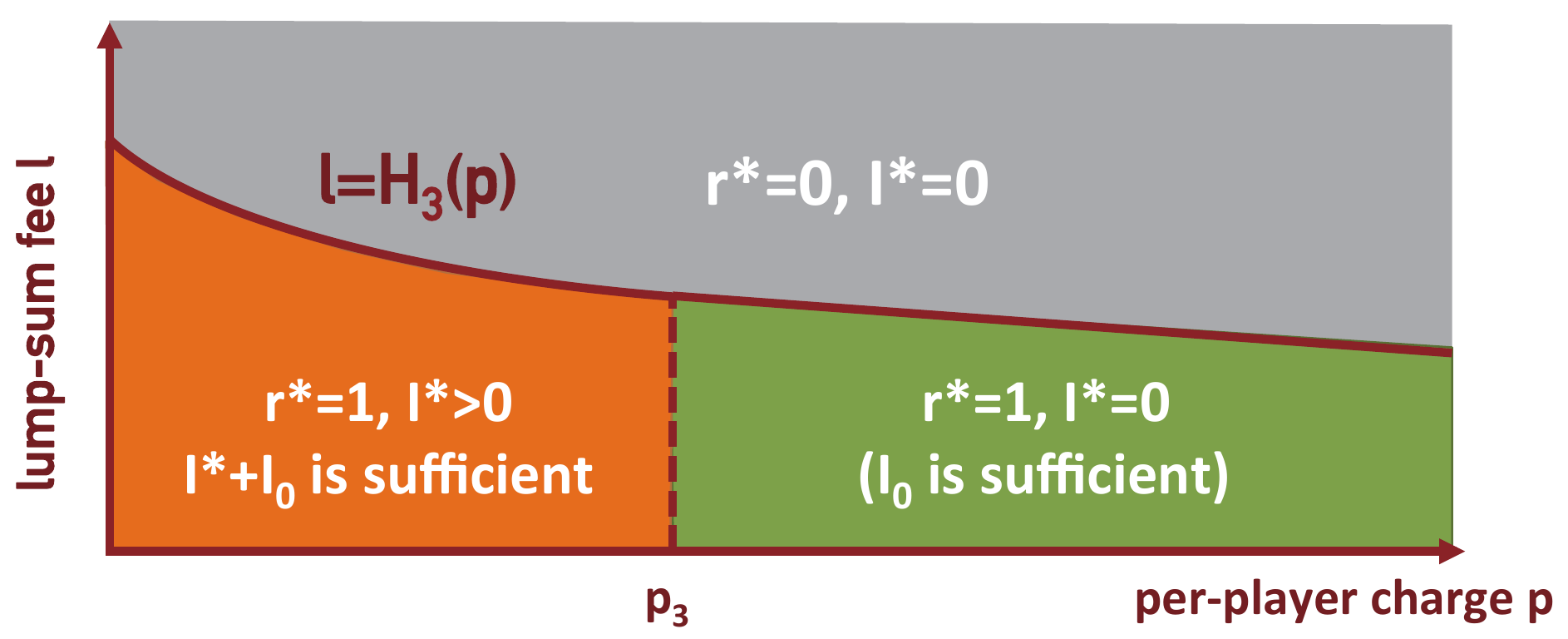}\\
  \caption{Situation III of Venue's POI Decision $r^*$ and Investment $I^*$ in Stage II.}
  \label{SM:fig:stageII:b}
  \vspace{-0.5cm}
\end{figure}

In Proposition \ref{proposition:stageII:situationII}, we summarize the venue's POI decision $r^*\left(l,p\right)$ and investment $I^*\left(l,p\right)$ in Stage II, under $I_0\le I_{\rm th}$ and $\delta\le \delta_{\rm th}$. In Proposition \ref{proposition:stageII:situationIII}, we summarize the venue's decisions under $I_0> I_{\rm th}$. We illustrate the results in Fig. \ref{SM:fig:stageII:a} and Fig. \ref{SM:fig:stageII:b}, respectively.

\section{Proof of Proposition \ref{proposition:stageII:situationII} in Section \ref{sec:stageII}}\label{SM:proposition5}
\proof 
{\bf(Step 1)} We can compute the venue's optimal payoffs in the following three decision regions: (i) $r=0$; (ii) $r=1$ and $I\in\left[0,I_{\rm th}-I_0\right]$; (iii) $r=1$ and $I\in\left(I_{\rm th}-I_0,\infty\right)$. The analysis is the same as that in the proof of Proposition \ref{proposition:stageII:situationI}. Hence, we omit the detailed analysis, and simply reemphasize the results here. 

In decision region (i), the venue's optimal payoff is $b N \eta \frac{U}{c_{\max}}$, and the optimal investment level is $0$. In decision region (ii), if $p>p_0$, the venue's optimal payoff is $\Pi^{\rm venue}\left(1,0,l,p\right)=bN\eta\frac{U}{c_{\max}}-l-{\frac{V}{\frac{\delta}{I_0}-\theta}p}$, and the optimal investment level is $0$; if $p\le p_0$, the venue's optimal payoff is $\Pi^{\rm venue}\left(1,I_{\rm th}-I_0,l,p\right)=b N \eta\frac{U}{c_{\max}}-k \left(I_{\rm th}-I_0\right)-\left(l+p \frac{\eta UN}{c_{\max}}\right)$, and the optimal investment level is $I_{\rm th}-I_0$. In decision region (iii), if $p<p_1$, the venue's optimal payoff is $\frac{N}{c_{\max}}b\eta\left(1-\eta\right)U+\frac{N}{c_{\max}-N\theta}\left(\sqrt{\left(V+\eta U\right)\left(b\eta -p\right)}-\sqrt{\delta k}\right)^2-l+kI_0$, and the optimal investment level is $\frac{N}{{c_{\max}-N\theta}}\sqrt{\frac{\delta\left(V+\eta U\right)\left(b\eta -p\right)}{k}}-\delta \frac{N}{{c_{\max}-N\theta}} {-I_0}$. Furthermore, the venue's optimal payoff is greater than $\Pi^{\rm venue}\left(1,I_{\rm th}-I_0,l,p\right)$. If $p\ge p_1$, the venue's optimal payoff is smaller than $\Pi^{\rm venue}\left(1,I_{\rm th}-I_0,l,p\right)$.

{\bf(Step 2)} We discuss the venue's equilibrium strategies under $p<p_0\le p_1$. The venue's optimal payoff in region (i) is $b N \eta \frac{U}{c_{\max}}$, its optimal payoff in region (ii) is $\Pi^{\rm venue}\left(1,I_{\rm th}-I_0,l,p\right)$, and its optimal payoff in region (iii) is $\frac{N}{c_{\max}}b\eta\left(1-\eta\right)U+\frac{N}{c_{\max}-N\theta}\left(\sqrt{\left(V+\eta U\right)\left(b\eta -p\right)}-\sqrt{\delta k}\right)^2-l+kI_0$, which is greater than $\Pi^{\rm venue}\left(1,I_{\rm th}-I_0,l,p\right)$. 
Considering the three decision regions, we conclude that if $l\le -\frac{N}{c_{\max}}b\eta^2U+\frac{N}{c_{\max}-N\theta}\left(\sqrt{\left(V+\eta U\right)\left(b\eta -p\right)}-\sqrt{\delta k}\right)^2+kI_0$, the venue's optimal strategies are $r^*\left(l,p\right)=1$ and $I^*\left(l,p\right)=\frac{N}{{c_{\max}-N\theta}}\sqrt{\frac{\delta\left(V+\eta U\right)\left(b\eta -p\right)}{k}}-\delta \frac{N}{{c_{\max}-N\theta}} {-I_0}$; otherwise, the venue's optimal strategies are $r^*\left(l,p\right)=0$ and $I^*\left(l,p\right)=0$. 

{\bf(Step 3)} We discuss the venue's equilibrium strategies under $p_0 \le p<p_2$. The venue's optimal payoff in region (i) is $b N \eta \frac{U}{c_{\max}}$, its optimal payoff in region (ii) is $\Pi^{\rm venue}\left(1,0,l,p\right)=bN\eta\frac{U}{c_{\max}}-l-{\frac{V}{\frac{\delta}{I_0}-\theta}p}$, and its optimal payoff in region (iii) is $\frac{N}{c_{\max}}b\eta\left(1-\eta\right)U+\frac{N}{c_{\max}-N\theta}\left(\sqrt{\left(V+\eta U\right)\left(b\eta -p\right)}-\sqrt{\delta k}\right)^2-l+kI_0$. According to our proof of the uniqueness of $p_2$, the function
\begin{align}
\nonumber
G\left(p\right)= &\frac{N}{c_{\max}-N\theta}\left(\sqrt{\left(V+\eta U\right)\left(b\eta -p\right)}-\sqrt{\delta k}\right)^2\\
& -\frac{N}{c_{\max}}b\eta^2U+kI_0+\frac{V}{\frac{\delta}{I_0}-\theta}p
\end{align} 
is $0$ when $p=p_2$, and is strictly decreasing in $p\in\left[p_0,p_2\right]$. Therefore, we have the following relation 
\begin{multline}
\frac{N}{c_{\max}-N\theta}\left(\sqrt{\left(V+\eta U\right)\left(b\eta -p\right)}-\sqrt{\delta k}\right)^2\\-\frac{N}{c_{\max}}b\eta^2U
+kI_0+\frac{V}{\frac{\delta}{I_0}-\theta}p\ge0
\end{multline} 
for $p\in\left[p_0,p_2\right]$. This implies that the venue's optimal payoff in decision region (iii) is no less than its optimal payoff in decision region (ii), and we only need to compare the venue's optimal payoffs in decision regions (i) and (iii). We conclude that if $l\le -\frac{N}{c_{\max}}b\eta^2U+\frac{N}{c_{\max}-N\theta}\left(\sqrt{\left(V+\eta U\right)\left(b\eta -p\right)}-\sqrt{\delta k}\right)^2+kI_0$, the venue's optimal strategies are $r^*\left(l,p\right)=1$ and $I^*\left(l,p\right)=\frac{N}{{c_{\max}-N\theta}}\sqrt{\frac{\delta\left(V+\eta U\right)\left(b\eta -p\right)}{k}}-\delta \frac{N}{{c_{\max}-N\theta}} {-I_0}$; otherwise, the venue's optimal strategies are $r^*\left(l,p\right)=0$ and $I^*\left(l,p\right)=0$.  

{\bf(Step 4)} We discuss the venue's equilibrium strategies under $p_2 \le p\le p_1$. The venue's optimal payoff in region (i) is $b N \eta \frac{U}{c_{\max}}$, its optimal payoff in region (ii) is $\Pi^{\rm venue}\left(1,0,l,p\right)=bN\eta\frac{U}{c_{\max}}-l-{\frac{V}{\frac{\delta}{I_0}-\theta}p}$, and its optimal payoff in region (iii) is $\frac{N}{c_{\max}}b\eta\left(1-\eta\right)U+\frac{N}{c_{\max}-N\theta}\left(\sqrt{\left(V+\eta U\right)\left(b\eta -p\right)}-\sqrt{\delta k}\right)^2-l+kI_0$. According to our proof of the uniqueness of $p_2$, the function
\begin{align}
\nonumber
G\left(p\right)= & \frac{N}{c_{\max}-N\theta}\left(\sqrt{\left(V+\eta U\right)\left(b\eta -p\right)}-\sqrt{\delta k}\right)^2\\
& -\frac{N}{c_{\max}}b\eta^2U+kI_0+\frac{V}{\frac{\delta}{I_0}-\theta}p
\end{align} 
is $0$ when $p=p_2$, is strictly decreasing in $p\in\left[p_2,p_1\right)$, and is continuous at $p=p_1$. Therefore, we have the following relation 
\begin{multline}
\frac{N}{c_{\max}-N\theta}\left(\sqrt{\left(V+\eta U\right)\left(b\eta -p\right)}-\sqrt{\delta k}\right)^2\\-\frac{N}{c_{\max}}b\eta^2U+kI_0+\frac{V}{\frac{\delta}{I_0}-\theta}p\le0
\end{multline} 
for $p\in\left[p_2,p_1\right]$. This implies that the venue's optimal payoff in decision region (iii) is no greater than its optimal payoff in decision region (ii), and we only need to compare the venue's optimal payoffs in decision regions (i) and (ii). We conclude that if $l\le -{\frac{V}{\frac{\delta}{I_0}-\theta}p}$, the venue's optimal strategies are $r^*\left(l,p\right)=1$ and $I^*\left(l,p\right)=0$; otherwise, the venue's optimal strategies are $r^*\left(l,p\right)=0$ and $I^*\left(l,p\right)=0$. 

{\bf(Step 5)} We discuss the venue's equilibrium strategies under $p> p_1\ge p_0$. 
The venue's optimal payoff in decision region (i) is $b N \eta \frac{U}{c_{\max}}$, its optimal payoff in decision region (ii) is $bN\eta\frac{U}{c_{\max}}-l-{\frac{V}{\frac{\delta}{I_0}-\theta}p}$, which is greater than $\Pi^{\rm venue}\left(1,I_{\rm th}-I_0,l,p\right)$, and its optimal payoff in decision region (iii) is smaller than $\Pi^{\rm venue}\left(1,I_{\rm th}-I_0,l,p\right)$. We conclude that if $l\le -{\frac{V}{\frac{\delta}{I_0}-\theta}p}$, the venue's optimal strategies are $r^*\left(l,p\right)=1$ and $I^*\left(l,p\right)=0$; otherwise, the venue's optimal strategies are $r^*\left(l,p\right)=0$ and $I^*\left(l,p\right)=0$. 

Summarizing {\bf Step 2}, {\bf Step 3}, {\bf Step 4}, and {\bf Step 5}, we derive the venue's equilibrium strategies as follows
{\small
\begin{align}
\nonumber
& \left(r^*\left(l,p\right),I^*\left(l,p\right)\right)= \\
\nonumber
& \!\!\!\left\{ {\begin{array}{{l}{l}}
{\!\!\!\!\left(0,0\right),}&{\!\!\!\!\!{\rm if~}\!l\!>\!H_2\left(p \right)\!,}\\
{\!\!\!\!\!\left(1,\!\frac{N}{{c_{\max}-N\theta}}\!\sqrt{\frac{\delta\left(V+\eta U\right)\left(b\eta -p\right)}{k}}\!-\! \frac{\delta N}{{c_{\max}-N\theta}} {-I_0}\!\!\right)\!\!,}&{\!\!\!\!\!{\rm if~}\!l\!\le\! H_2\!\left(p \right)\!,p\!<\!p_2,}\\
{\!\!\!\!\left(1,0\right),}&{\!\!\!\!\!{\rm if~}\!l\!\le \!H_2\!\left(p \right)\!,p\!\ge\! p_2,}
\end{array}} \right.
\end{align}}
which is exactly (\ref{equ:venue:situationII}).
\endproof



\section{Proof of Proposition \ref{proposition:stageII:situationIII} in Section \ref{sec:stageII}}\label{SM:proposition6}
\proof
{\bf(Step 1)} We compute the venue's optimal payoffs in the following two decision regions: (i) $r=0$; (ii) $r=1$ and $I\in\left[0,\infty\right)$. In decision region (i), the venue's optimal payoff is $b N \eta \frac{U}{c_{\max}}$, and the optimal investment level is $0$. 

Next, we analyze the venue's optimal payoff in decision region (ii) (i.e., $r=1$ and $I\in\left[0,\infty\right)$). In this decision region, the venue's payoff is given as
\begin{align}
\nonumber
& \Pi^{\rm venue}\left(1,I,l,p\right)=-k I-l\\
\nonumber
& \!+\!b N {\left(\eta\frac{U}{c_{\max}}\!+\!\eta{\frac{Vc_{\max}\left(I+I_0\right)\!-\!\eta UN\delta\!+\!\eta UN\theta \left(I+I_0\right)}{c_{\max}^2\left(I+I_0\right)+c_{\max}N\delta -c_{\max}N\theta\left(I+I_0\right)}}\!\right)}\\
& \!-\!pN{\left(\eta\frac{U}{c_{\max}}\!+\!{\frac{Vc_{\max}\left(I+I_0\right)\!-\!\eta UN\delta\!+\!\eta UN\theta \left(I+I_0\right)}{c_{\max}^2\left(I+I_0\right)+c_{\max}N\delta -c_{\max}N\theta\left(I+I_0\right)}}\!\right)}.\label{SM:equ:longregion6}
\end{align}
We can compute
\begin{align}
& \frac{\partial \Pi^{\rm venue}\left(1,I,l,p\right)}{\partial I}=\frac{\delta N^2 \left(b\eta-p\right)\left(V+\eta U\right)}{\left(\delta N +{\left(I+I_0\right)} {\left(c_{\max}-N\theta\right)}\right)^2}-k. \label{SM:equ:derivative:2}
\end{align}

If $p\ge b\eta$, $\Pi^{\rm venue}\left(1,I,l,p\right)$ decreases with $I$ for $I\ge0$. In this case, the venue's optimal payoff in decision region (ii) is $\Pi^{\rm venue}\left(1,0,l,p\right)=b N {\left(\eta\frac{U}{c_{\max}}+\eta{\frac{Vc_{\max}I_0-\eta UN\delta+\eta UN\theta I_0}{c_{\max}^2I_0+c_{\max}N\delta -c_{\max}N\theta I_0}}\right)}-l -pN{\left(\eta\frac{U}{c_{\max}}+{\frac{Vc_{\max}I_0-\eta UN\delta+\eta UN\theta I_0}{c_{\max}^2I_0+c_{\max}N\delta -c_{\max}N\theta I_0}}\right)}$.

If $p<b\eta $, we can derive the relation $\frac{\partial^2 \Pi^{\rm venue}\left(1,I,l,p\right)}{\partial I^2}<0$. 
Based on (\ref{SM:equ:derivative:2}), the solution to $\frac{\partial \Pi^{\rm venue}\left(1,I,l,p\right)}{\partial I}=0$ is $I=\frac{N}{{c_{\max}-N\theta}}\sqrt{\frac{\delta\left(V+\eta U\right)\left(b\eta -p\right)}{k}}-\delta \frac{N}{{c_{\max}-N\theta}} {-I_0}$. 
Hence, when $\frac{N}{{c_{\max}-N\theta}}\sqrt{\frac{\delta\left(V+\eta U\right)\left(b\eta -p\right)}{k}}-\delta \frac{N}{{c_{\max}-N\theta}} {-I_0}>0$, $\Pi^{\rm venue}\left(1,I,l,p\right)$ increases with $I$ for $0\le I\le \frac{N}{{c_{\max}-N\theta}}\sqrt{\frac{\delta\left(V+\eta U\right)\left(b\eta -p\right)}{k}}-\delta \frac{N}{{c_{\max}-N\theta}} {-I_0}$, and decreases with $I$ for $I\ge \frac{N}{{c_{\max}-N\theta}}\sqrt{\frac{\delta\left(V+\eta U\right)\left(b\eta -p\right)}{k}}-\delta \frac{N}{{c_{\max}-N\theta}} {-I_0}$. Therefore, the venue's optimal investment level in region (ii) is $\frac{N}{{c_{\max}-N\theta}}\sqrt{\frac{\delta\left(V+\eta U\right)\left(b\eta -p\right)}{k}}-\delta \frac{N}{{c_{\max}-N\theta}} {-I_0}$, and the corresponding optimal payoff is $\frac{N}{c_{\max}}b\eta\left(1-\eta\right)U+\frac{N}{c_{\max}-N\theta}\left(\sqrt{\left(V+\eta U\right)\left(b\eta -p\right)}-\sqrt{\delta k}\right)^2-l+kI_0$. When $\frac{N}{{c_{\max}-N\theta}}\sqrt{\frac{\delta\left(V+\eta U\right)\left(b\eta -p\right)}{k}}-\delta \frac{N}{{c_{\max}-N\theta}} {-I_0}\le 0$, $\Pi^{\rm venue}\left(1,I,l,p\right)$ decreases with $I$ for $I\ge0$. In this case, the venue's optimal payoff in decision region (ii) is $\Pi^{\rm venue}\left(1,0,l,p\right)=b N {\left(\eta\frac{U}{c_{\max}}+\eta{\frac{Vc_{\max}I_0-\eta UN\delta+\eta UN\theta I_0}{c_{\max}^2I_0+c_{\max}N\delta -c_{\max}N\theta I_0}}\right)}-l -pN{\left(\eta\frac{U}{c_{\max}}+{\frac{Vc_{\max}I_0-\eta UN\delta+\eta UN\theta I_0}{c_{\max}^2I_0+c_{\max}N\delta -c_{\max}N\theta I_0}}\right)}$.

Note that $\frac{N}{{c_{\max}-N\theta}}\sqrt{\frac{\delta\left(V+\eta U\right)\left(b\eta -p\right)}{k}}-\delta \frac{N}{{c_{\max}-N\theta}} {-I_0}> 0$ is equivalent to $p<p_3 = b\eta-\frac{k\left(\left(c_{\max}-\theta N\right)I_0+\delta N\right)^2}{\delta\left(V+\eta U\right)N^2}$. Therefore, we can summarize the venue's optimal payoff in decision region (ii) as follows. If $p<p_3$, the venue's optimal payoff is $\frac{N}{c_{\max}}b\eta\left(1-\eta\right)U+\frac{N}{c_{\max}-N\theta}\left(\sqrt{\left(V+\eta U\right)\left(b\eta -p\right)}-\sqrt{\delta k}\right)^2-l+kI_0$, and the optimal investment level is $\frac{N}{{c_{\max}-N\theta}}\sqrt{\frac{\delta\left(V+\eta U\right)\left(b\eta -p\right)}{k}}-\delta \frac{N}{{c_{\max}-N\theta}} {-I_0}$. If $p\ge p_3$, the venue's optimal payoff is $b N {\left(\eta\frac{U}{c_{\max}}+\eta{\frac{Vc_{\max}I_0-\eta UN\delta+\eta UN\theta I_0}{c_{\max}^2I_0+c_{\max}N\delta -c_{\max}N\theta I_0}}\right)}-l -pN{\left(\eta\frac{U}{c_{\max}}+{\frac{Vc_{\max}I_0-\eta UN\delta+\eta UN\theta I_0}{c_{\max}^2I_0+c_{\max}N\delta -c_{\max}N\theta I_0}}\right)}$, and the optimal investment level is $0$. 

{\bf(Step 2)} We discuss the venue's equilibrium strategies under $p<p_3$. The venue's optimal payoff in region (i) is $b N \eta \frac{U}{c_{\max}}$, and its optimal payoff in region (ii) is $\frac{N}{c_{\max}}b\eta\left(1-\eta\right)U+\frac{N}{c_{\max}-N\theta}\left(\sqrt{\left(V+\eta U\right)\left(b\eta -p\right)}-\sqrt{\delta k}\right)^2-l+kI_0$. We conclude that if $l\le -\frac{N}{c_{\max}}b\eta^2U+\frac{N}{c_{\max}-N\theta}\left(\sqrt{\left(V+\eta U\right)\left(b\eta -p\right)}-\sqrt{\delta k}\right)^2+kI_0$, the venue's optimal strategies are $r^*\left(l,p\right)=1$ and $I^*\left(l,p\right)=\frac{N}{{c_{\max}-N\theta}}\sqrt{\frac{\delta\left(V+\eta U\right)\left(b\eta -p\right)}{k}}-\delta \frac{N}{{c_{\max}-N\theta}} {-I_0}$; otherwise, the venue's optimal strategies are $r^*\left(l,p\right)=0$ and $I^*\left(l,p\right)=0$. 

{\bf(Step 3)} We discuss the venue's equilibrium strategies under $p\ge p_3$. 
The venue's optimal payoff in decision region (i) is $b N \eta \frac{U}{c_{\max}}$, and its optimal payoff in decision region (ii) is $b N {\left(\eta\frac{U}{c_{\max}}+\eta{\frac{Vc_{\max}I_0-\eta UN\delta+\eta UN\theta I_0}{c_{\max}^2I_0+c_{\max}N\delta -c_{\max}N\theta I_0}}\right)}-l -pN{\left(\eta\frac{U}{c_{\max}}+{\frac{Vc_{\max}I_0-\eta UN\delta+\eta UN\theta I_0}{c_{\max}^2I_0+c_{\max}N\delta -c_{\max}N\theta I_0}}\right)}$. 
We conclude that if $l\le \left(b\eta-p\right) N {\left({\frac{Vc_{\max}I_0-\eta UN\delta+\eta UN\theta I_0}{c_{\max}^2I_0+c_{\max}N\delta -c_{\max}N\theta I_0}}\right)} -pN \eta\frac{U}{c_{\max}}$, the venue's optimal strategies are $r^*\left(l,p\right)=1$ and $I^*\left(l,p\right)=0$; otherwise, the venue's optimal strategies are $r^*\left(l,p\right)=0$ and $I^*\left(l,p\right)=0$. 

Summarizing {\bf Step 2} and {\bf Step 3}, we derive the venue's equilibrium strategies as follows
{\small{
\begin{align}
\nonumber 
&\left(r^*\left(l,p\right),I^*\left(l,p\right)\right)=\\
\nonumber
& \!\!\!\left\{ {\begin{array}{{l}{l}}
{\!\!\!\!\left(0,0\right),}&{\!\!\!\!\!{\rm if~}\!l\!>\!H_3\!\left(p \right)\!,}\\
{\!\!\!\!\!\left(1,\!\frac{N}{{c_{\max}\!-\!N\theta}}\!\sqrt{\frac{\delta\left(V+\eta U\right)\left(b\eta -p\right)}{k}}\!-\! \frac{\delta N}{{c_{\max}-N\theta}} {-I_0}\!\right)\!,}&{\!\!\!\!\!{\rm if~}\!l\!\le\! H_3\!\left(p \right)\!,p\!<\!p_3,}\\
{\!\!\!\!\left(1,0\right),}&{\!\!\!\!\!{\rm if~}\!l\!\le \!H_3\!\left(p \right)\!,p\!\ge\! p_3,}
\end{array}} \right.
\end{align}}}
which is exactly (\ref{equ:venue:situationIII}).
\endproof

\section{Proof of Theorem \ref{theorem:optimaltariff} in Section \ref{sec:stageI}}\label{SM:theorem:O2}
\proof
{\bf(Step 1)} We prove that the venue's payoff under the optimal two-part tariff $\left(l^*,p^*\right)$ is $bN\eta\frac{U}{c_{\max}}$, i.e., $\Pi^{\rm venue}\left(r^*\left(l^*,p^*\right),I^*\left(l^*,p^*\right),l^*,p^*\right)=bN\eta\frac{U}{c_{\max}}$. 

First, when $r^*\left(l^*,p^*\right)=0$, the venue does not become a POI under the optimal two-part tariff. In this case, we have ${\bar x}\left(r^*\left(l^*,p^*\right),I^*\left(l^*,p^*\right)\right)=\eta \frac{U}{c_{\max}}$ based on Proposition \ref{proposition:stageIII:a}. The venue's payoff is $bN\eta\frac{U}{c_{\max}}$.

Second, when $r^*\left(l^*,p^*\right)=1$, the venue becomes a POI under the optimal two-part tariff. We prove $\Pi^{\rm venue}\left(r^*\left(l^*,p^*\right),I^*\left(l^*,p^*\right),l^*,p^*\right)=bN\eta\frac{U}{c_{\max}}$ by contradiction. 
If $\Pi^{\rm venue}\left(r^*\left(l^*,p^*\right),I^*\left(l^*,p^*\right),l^*,p^*\right)<bN\eta\frac{U}{c_{\max}}$, the venue can switch its POI choice from $1$ to $0$, and achieve a larger payoff (i.e., $bN\eta\frac{U}{c_{\max}}$). This violates the condition $r^*\left(l^*,p^*\right)=1$. 
If $\Pi^{\rm venue}\left(r^*\left(l^*,p^*\right),I^*\left(l^*,p^*\right),l^*,p^*\right)>bN\eta\frac{U}{c_{\max}}$, the app can always increase its lump-sum fee to extract more revenue from the venue, without changing the venue's POI choice and investment choice. For example, the app can increase its lump-sum fee from $l^*$ to $l^*+\zeta$, where $\zeta\in\left(0,\Pi^{\rm venue}\left(r^*\left(l^*,p^*\right),I^*\left(l^*,p^*\right),l^*,p^*\right)-bN\eta\frac{U}{c_{\max}}\right]$. This violates the fact that $l^*$ is the app's optimal lump-sum fee. 

Combing the case $r^*\left(l^*,p^*\right)=0$ and case $r^*\left(l^*,p^*\right)=1$, we can easily see that $\Pi^{\rm venue}\left(r^*\left(l^*,p^*\right),I^*\left(l^*,p^*\right),l^*,p^*\right)=bN\eta\frac{U}{c_{\max}}$. 

{\bf(Step 2)} We compute the upper bound of $R^{\rm app}\left(l^*,p^*\right)$. Based on the definition of $\Pi^{\rm venue}\left(r,I,l,p\right)$, we have
\begin{align}
\nonumber
&\Pi^{\rm venue}\!\left(r^*\!\left(l^*,p^*\right),I^*\!\left(l^*,p^*\right),l,p\right)\!=\! b N {\bar x}\left(r^*\!\left(l^*,p^*\right),I^*\!\left(l^*,p^*\right)\right)\\
& \!-\!k I^*\left(l^*,p^*\right) \!-\! r^*\left(l^*,p^*\right) \left(l^*\!+\!p^*N {\bar y}\left(r^*\left(l^*,p^*\right),I^*\left(l^*,p^*\right)\right)\right).\label{SM:add:a}
\end{align}
As shown in {\bf Step 1}, $\Pi^{\rm venue}\left(r^*\left(l^*,p^*\right),I^*\left(l^*,p^*\right),l^*,p^*\right)=bN\eta\frac{U}{c_{\max}}$. Hence, we have
\begin{align}
\nonumber
& r^*\left(l^*,p^*\right) \left(l^*+p^*N {\bar y}\left(r^*\left(l^*,p^*\right),I^*\left(l^*,p^*\right)\right)\right)=\\
& b N {\bar x}\left(r^*\left(l^*,p^*\right),I^*\left(l^*,p^*\right)\right)-k I^*\left(l^*,p^*\right)-bN\eta\frac{U}{c_{\max}}.\label{SM:theorem1:long}
\end{align}
Recall that the app's optimal revenue is
\begin{align}
\nonumber
& R^{\rm app}\left(l^*,p^*\right)=\phi N{\bar y}\left(r^*\left(l^*,p^*\right),I^*\left(l^*,p^*\right)\right)\\ 
&{~~~~} +{r^*}\left(l^*,p^*\right)\Bigl(l^*+p^* N{\bar y}\left(r^*\left(l^*,p^*\right),I^*\left(l^*,p^*\right)\right)\Bigr).
\end{align}
Based on (\ref{SM:theorem1:long}), we can rewrite $R^{\rm app}\left(l^*,p^*\right)$ as
\begin{align}
\nonumber
& R^{\rm app}\left(l^*,p^*\right)=\phi N{\bar y}\left(r^*\left(l^*,p^*\right),I^*\left(l^*,p^*\right)\right)-bN\eta\frac{U}{c_{\max}}\\
& +b N {\bar x}\left(r^*\left(l^*,p^*\right),I^*\left(l^*,p^*\right)\right)-k I^*\left(l^*,p^*\right).\label{SM:theorem1:revenue}
\end{align}
The right-hand side of the above equation can be determined by $r^*\left(l^*,p^*\right)$ and $I^*\left(l^*,p^*\right)$. Therefore, we can compute the upper bound of $R^{\rm app}\left(l^*,p^*\right)$ as
\begin{align}
\nonumber
R^{\rm app}\left(l^*,p^*\right)\le \mathop{\max}_{r\in\left\{0,1\right\},I\ge0} & \phi N{\bar y}\left(r,I\right)+b N {\bar x}\left(r,I\right)-k I \\
& -bN\eta\frac{U}{c_{\max}}.
\end{align}
Because ${\bar y}\left(1,I\right)\ge {\bar y}\left(0,I\right)$ and ${\bar x}\left(1,I\right)\ge{\bar x}\left(0,I\right)$, we have $\phi N{\bar y}\left(1,I\right)+b N {\bar x}\left(1,I\right)\ge\phi N{\bar y}\left(0,I\right)+b N {\bar x}\left(0,I\right)$. Hence, we have the following relation:
\begin{align}
R^{\rm app}\!\left(l^*,p^*\right)\!\le\! \mathop{\max}_{I\ge0} \phi N{\bar y}\left(1,I\right)\!+\!b N {\bar x}\left(1,I\right)\!-\!k I\!-\!bN\eta\frac{U}{c_{\max}}.\label{SM:theorem1:upperbound}
\end{align}

{\bf (Step 3)} We show that if the app chooses $\left(\hat l,\hat p\right)=\left({\tilde H}\left(-\phi\right),-\phi\right)$, its revenue achieves the upper bound shown in (\ref{SM:theorem1:upperbound}). According to the definition of ${\tilde H}\left(p\right)$ and Propositions \ref{proposition:stageII:situationI}, \ref{proposition:stageII:situationII}, and \ref{proposition:stageII:situationIII}, the venue chooses $r^*\left(\hat l,\hat p\right)=1$ when $\hat l={\tilde H}\left(-\phi\right)$ and $\hat p=-\phi$. Furthermore, according to Problem \ref{problem:venue}, the venue chooses $I^*\left(\hat l,\hat p\right)$ that satisfies 
\begin{align}
I^*\left(\hat l,\hat p\right)
= \mathop{\arg\max}_{I\ge0} \phi N {\bar y}\left(1,I\right)+b N {\bar x}\left(1,I\right)-k I.
\end{align}
It is easy to compute that the venue's payoff under the tariff $\left(\hat l,\hat p\right)=\left({\tilde H}\left(-\phi\right),-\phi\right)$ is $bN\eta\frac{U}{c_{\max}}$, i.e., $\Pi^{\rm venue}\left(r^*\left(\hat l,\hat p\right),I^*\left(\hat l,\hat p\right),\hat l,\hat p\right)=bN\eta\frac{U}{c_{\max}}$. Based on the approach we derive (\ref{SM:add:a})-(\ref{SM:theorem1:revenue}), the app's revenue under the tariff $\left(\hat l,\hat p\right)$ can be written as
\begin{align}
\nonumber
R^{\rm app}\left(\hat l,\hat p\right)= & \phi N{\bar y}\left(1,I^*\left(\hat l,\hat p\right)\right)+b N {\bar x}\left(1,I^*\left(\hat l,\hat p\right)\right) \\
& -k I^*\left(\hat l,\hat p\right)-bN\eta\frac{U}{c_{\max}},\label{SM:theorem1:obtain}
\end{align}
where $I^*\left(\hat l,\hat p\right) = \mathop{\arg\max}_{I\ge0} \phi N {\bar y}\left(1,I\right)+b N {\bar x}\left(1,I\right)-k I$. Comparing (\ref{SM:theorem1:obtain}) with (\ref{SM:theorem1:upperbound}), we can see that $R^{\rm app}\left(\hat l,\hat p\right)$ reaches the upper bound of the app's revenue. This means that $\left(\hat l,\hat p\right)=\left({\tilde H}\left(-\phi\right),-\phi\right)$ is the app's optimal two-part tariff. 
\endproof

\section{Proof of Theorem \ref{theorem:optimalforall} in Section \ref{sec:stageI}}\label{SM:theorem2}
\proof
Since we will analyze the equilibrium outcomes (e.g., the venue's payoff and the app's revenue) under a general $\left(r,I\right)${-dependent} tariff scheme $T$ that might not be a two-part tariff scheme, we need to first introduce some new notations. Specifically, we use $\Pi_g^{\rm venue}\left(r,I,T\right)$ to denote the venue's payoff under a general $\left(r,I\right)$-dependent tariff scheme $T$ (the subscript $g$ stands for ``general''). Similar as (\ref{equ:venueopt:obj}), $\Pi_g^{\rm venue}\left(r,I,T\right)$ is given by
\begin{align}
\Pi_g^{\rm venue}\left(r,I,T\right)=b N {\bar x}\left(r,I\right)-k I- T.\label{SM:equ:g:venuepayoff}
\end{align}
Note that the $\left(r,I\right)$-dependent tariff scheme $T$ is a function of $r$ and $I$. As defined in Section \ref{sec:stageI}, a feasible tariff scheme $T$ satisfies $T=0$ when $r=0$ (the venue does not become a POI). Here, we use $T$ rather than the function form $T\left(r,I\right)$ to simplify the expression. 

We further use $r_g^*\left(T\right)$ and $I_g^*\left(T\right)$ to denote the venue's optimal POI and investment decisions under a general $\left(r,I\right)$-dependent tariff scheme $T$. Then, we use $R_g^{\rm app}\left(T\right)$ to denote the app's revenue under $T$, which is given by 
\begin{align}
R_g^{\rm app}\left(T\right)=T+\phi N {\bar y}\left(r_g^*\left(T\right),I_g^*\left(T\right)\right).\label{SM:equ:g:appvenue}
\end{align}
With these new notations, we are ready to introduce the proof, which is actually similar to the proof of Theorem \ref{theorem:optimaltariff}.

{\bf (Step 1)} We characterize the upper bound of $R_g^{\rm app}\left(T\right)$ under a general $\left(r,I\right)$-dependent tariff scheme $T$. From (\ref{SM:equ:g:venuepayoff}) and (\ref{SM:equ:g:appvenue}), we can see that the summation of the app's revenue and venue's payoff under $T$ satisfies the following relation:
\begin{align}
\nonumber
& R_g^{\rm app}\left(T\right)+\Pi_g^{\rm venue}\left(r_g^*\left(T\right),I_g^*\left(T\right),T\right)=\\
& b N {\bar x}\left(r_g^*\left(T\right),I_g^*\left(T\right)\right)-k I_g^*\left(T\right)+\phi N {\bar y}\left(r_g^*\left(T\right),I_g^*\left(T\right)\right).\label{SM:equ:g:sum}
\end{align}
When the venue does not become a POI and does not invest, its payoff is always $bN\eta \frac{U}{c_{\max}}$. Hence, the venue's optimal payoff under $T$ is always no smaller than $bN\eta \frac{U}{c_{\max}}$, i.e., $\Pi_g^{\rm venue}\left(r_g^*\left(T\right),I_g^*\left(T\right),T\right)\ge bN\eta \frac{U}{c_{\max}}$. Considering this inequality and (\ref{SM:equ:g:sum}), we have the following inequality:
\begin{align}
\nonumber
R_g^{\rm app}\left(T\right)\le & b N {\bar x}\left(r_g^*\left(T\right),I_g^*\left(T\right)\right)-k I_g^*\left(T\right)\\
& +\phi N {\bar y}\left(r_g^*\left(T\right),I_g^*\left(T\right)\right)-bN\eta \frac{U}{c_{\max}}.
\end{align}
The right-hand side of the above inequality can be determined by $r_g^*\left(T\right)$ and $I_g^*\left(T\right)$. Therefore, we can compute the upper bound of $R_g^{\rm app}\left(T\right)$ as
\begin{align}
R_g^{\rm app}\!\left(T\right)\le \!\max_{r\in\left\{0,1\right\},I\ge0} \!bN {\bar x}\left(r,I\right)\!-\!kI\!+\!\phi N {\bar y}\left(r,I\right)\!-\!bN\eta \frac{U}{c_{\max}}.
\end{align}
Because ${\bar x}\left(1,I\right)\ge{\bar x}\left(0,I\right)$ and ${\bar y}\left(1,I\right)\ge {\bar y}\left(0,I\right)$, we have $b N {\bar x}\left(1,I\right)+\phi N{\bar y}\left(1,I\right) \ge b N {\bar x}\left(0,I\right)+\phi N{\bar y}\left(0,I\right)$. Hence, we have the following relation:
\begin{align}
R_g^{\rm app}\left(T\right)\le \max_{I\ge0} bN {\bar x}\left(1,I\right)-kI+\phi N {\bar y}\left(1,I\right)-bN\eta \frac{U}{c_{\max}}.\label{SM:equ:g:upper}
\end{align}

{\bf (Step 2)} We show that if the app chooses ${\hat T}=r\left({\tilde H}\left(-\phi\right)-\phi N {\bar y}\left(r,I\right) \right)$, i.e., our optimal two-part tariff scheme introduced in Theorem \ref{theorem:optimaltariff}, its revenue achieves the upper bound in (\ref{SM:equ:g:upper}). Based on the proof of Theorem \ref{theorem:optimaltariff}, under the optimal two-part tariff scheme, (i) the venue becomes a POI, i.e., $r_g^*\left({\hat T}\right)=1$; (ii) the venue's investment level $I_g^*\left({\hat T}\right)$ satisfies $I_g^*\left(\hat T\right)= {\arg\max}_{I\ge0} b N {\bar x}\left(1,I\right)+\phi N {\bar y}\left(1,I\right)-k I$; (iii) the venue's optimal payoff $\Pi_g^{\rm venue}\left(r_g^*\left({\hat T}\right),I_g^*\left({\hat T}\right),{\hat T}\right)=bN\eta \frac{U}{c_{\max}}$. 
Considering these results and (\ref{SM:equ:g:sum}), the app's revenue under ${\hat T}$ satisfies the following relation:
\begin{align}
\nonumber
R_g^{\rm app}\left({\hat T}\right)=& b N {\bar x}\left(1,I_g^*\left({\hat T}\right)\right)-k I_g^*\left({\hat T}\right)\\
& +\phi N {\bar y}\left(1,I_g^*\left({\hat T}\right)\right)-bN\eta \frac{U}{c_{\max}},\label{SM:equ:g:bound}
\end{align}
where $I_g^*\left(\hat T\right)= {\arg\max}_{I\ge0} b N {\bar x}\left(1,I\right)+\phi N {\bar y}\left(1,I\right)-k I$. Comparing (\ref{SM:equ:g:bound}) with (\ref{SM:equ:g:upper}), we can see that $R_g^{\rm app}\left({\hat T}\right)$ achieves the upper bound of the app's revenue under any feasible $\left(r,I\right)$-dependent tariff scheme $T$. This means that our optimal two-part tariff scheme ${\hat T}=r\left({\tilde H}\left(-\phi\right)-\phi N {\bar y}\left(r,I\right) \right)$ achieves the highest app's revenue among all feasible $\left(r,I\right)${-dependent tariff schemes}.
\endproof

\section{Proof of Corollary \ref{corollary:optimalrevenue} in Section \ref{sec:stageI}}\label{SM:corollary1}
\proof 
First, we prove that $R^{\rm app}\left(l^*,p^*\right)={\tilde H}\left(-\phi\right)$. Recall that the app's optimal revenue is
\begin{align}
\nonumber
& R^{\rm app}\left(l^*,p^*\right)=\phi N{\bar y}\left(r^*\left(l^*,p^*\right),I^*\left(l^*,p^*\right)\right) \\
& {~~~~}+{r^*}\left(l^*,p^*\right)\Bigl(l^*+p^* N{\bar y}\left(r^*\left(l^*,p^*\right),I^*\left(l^*,p^*\right)\right)\Bigr).
\end{align}
Since $\left(l^*,p^*\right)=\left({\tilde H}\left(-\phi\right),-\phi\right)$, we have $l^*={\tilde H}\left(p^*\right)$. According to the definition of ${\tilde H}\left(p\right)$ and Propositions \ref{proposition:stageII:situationI}, \ref{proposition:stageII:situationII}, and \ref{proposition:stageII:situationIII}, we have $r^*\left(l^*, p^*\right)=1$. Hence, the app's optimal revenue is simplified as
\begin{align}
R^{\rm app}\left(l^*,p^*\right)=& \left(\phi+p^*\right) N{\bar y}\left(1,I^*\left(l^*,p^*\right)\right)+l^*.
\end{align}
Because $p^*=-\phi$ and $l^*={\tilde H}\left(-\phi\right)$, we can see that $R^{\rm app}\left(l^*,p^*\right)={\tilde H}\left(-\phi\right)$. 

Second, we prove that $R^{\rm app}\left(l^*,p^*\right)\ge0$. Since $\left(l^*,p^*\right)=\left({\tilde H}\left(-\phi\right),-\phi\right)$ is the app's optimal tariff, the relation
\begin{align}
R^{\rm app}\!\left(l^*,p^*\right) \!=\! R^{\rm app}\!\left({\tilde H}\left(-\phi\right),-\phi\right) \!\ge\! R^{\rm app}\left({\tilde H}\left(-\phi\right)+\epsilon ,-\phi\right)\label{SM:equ:corollary}
\end{align}
holds for any positive constant $\epsilon$. Based on the definition of ${\tilde H}\left(p\right)$ and Propositions \ref{proposition:stageII:situationI}, \ref{proposition:stageII:situationII}, and \ref{proposition:stageII:situationIII}, we have $R^{\rm app}\left({\tilde H}\left(-\phi\right)+\epsilon ,-\phi\right)=0$. Considering (\ref{SM:equ:corollary}), we have $R^{\rm app}\left(l^*,p^*\right)\ge0$.

Third, we have already proved the result $\Pi^{\rm venue}\left(r^*\left(l^*,p^*\right),I^*\left(l^*,p^*\right),l^*,p^*\right)=b N \eta \frac{U}{c_{\max}}$ in {\bf Step 1} of the proof of Theorem \ref{theorem:optimaltariff} (in Section \ref{SM:theorem:O2}).
\endproof

\section{Proofs of Proposition \ref{proposition:impactU:smalldelta}, Proposition \ref{proposition:impactU:largedelta}, and Corollary \ref{corollary:impactU}}

\subsection{Proofs of Propositions \ref{proposition:impactU:smalldelta} and \ref{proposition:impactU:largedelta} in Section \ref{subsec:influence:quality}}\label{SM:proposition78}
\proof 
From Theorem \ref{theorem:optimaltariff} and Corollary \ref{corollary:optimalrevenue}, when ${I_0\le I_{\rm th}}{~\rm and~}{\delta>\delta_{\rm th}}$, $R^{\rm app}\left(l^*,p^*\right)$ is given in (\ref{SM:equ:H1}); when ${I_0\le I_{\rm th}}{~\rm and~}{\delta\le\delta_{\rm th}}$, $R^{\rm app}\left(l^*,p^*\right)$ is given in (\ref{SM:equ:H2}); when ${I_0>I_{\rm th}}$, $R^{\rm app}\left(l^*,p^*\right)$ is given in (\ref{SM:equ:H3}). We can also easily verify that $R^{\rm app}\left(l^*,p^*\right)$ is continuous in $U$.  

{\bf(Step 1)} We analyze the monotonicity of the following four functions of $U$. 
\begin{align}
\nonumber
A_1\left(U\right)\triangleq &\frac{N}{c_{\max}-N\theta}\left(\sqrt{\left(V+\eta U\right)\left(b\eta +\phi\right)}-\sqrt{\delta k}\right)^2 \\
& -\frac{N}{c_{\max}}b\eta^2U+kI_0,\\
A_2\left(U\right)\triangleq &\phi N\eta \frac{U}{c_{\max}}-k I_{\rm th}+k I_0,\\
A_3\left(U\right)\triangleq &\frac{V}{\frac{\delta}{I_0}-\theta}\phi,\\
\nonumber
A_4\left(U\right)\triangleq &\left(b\eta+ \phi\right)N {\frac{Vc_{\max}I_0-\eta UN\delta+\eta UN\theta I_0}{c_{\max}^2I_0+c_{\max}N\delta -c_{\max}N\theta I_0}}\\
& +\phi N\eta\frac{U}{c_{\max}}.
\end{align}
Through simple computation, we can easily show that (i) $A_1\left(U\right)$ decreases with $U$ for $U\le\frac{\delta k\left(b\eta+\phi\right)}{\eta \left(\frac{N\theta b \eta }{c_{\max}}+\phi\right)^2}-\frac{V}{\eta}$, and increases with $U$ for $U\ge\frac{\delta k\left(b\eta+\phi\right)}{\eta \left(\frac{N\theta b \eta }{c_{\max}}+\phi\right)^2}-\frac{V}{\eta}$; (ii) $A_2\left(U\right)$ increases with $U$ when $I_0\le I_{\rm th}$ and $-\phi\le p_0= -k\frac{\left(\delta-\theta I_0\right)c_{\max}}{Vc_{\max}+\theta\eta UN}$; (iii) $A_3\left(U\right)$ does not change with $U$; (iv) $A_4\left(U\right)$ increases with $U$ when $b\eta N\theta I_0+\phi c_{\max} I_0-b\eta N\delta\ge0$ and decreases with $U$ otherwise.

{\bf(Step 2)} We prove Proposition \ref{proposition:impactU:smalldelta}, i.e., when $\delta\le\frac{\left(b\eta N\theta +\phi c_{\max}\right)I_0}{b\eta N}$, $R^{\rm app}\left(l^*,p^*\right)$ increases with $U\in\left(0,\infty\right)$. 

First, we consider $R^{\rm app}\left(l^*,p^*\right)$ in the following regions: (a) ${I_0\le I_{\rm th}}$, ${\delta>\delta_{\rm th}}$, and $-\phi<p_1$; (b) ${I_0\le I_{\rm th}}$, ${\delta\le\delta_{\rm th}}$, and $-\phi<p_2$; (c) ${I_0>I_{\rm th}}$ and $-\phi<p_3$. In these regions, $R^{\rm app}\left(l^*,p^*\right)$'s expression is the same as $A_1\left(U\right)$, which decreases with $U$ for $U\le\frac{\delta k\left(b\eta+\phi\right)}{\eta \left(\frac{N\theta b \eta }{c_{\max}}+\phi\right)^2}-\frac{V}{\eta}$, and increases with $U$ for $U\ge\frac{\delta k\left(b\eta+\phi\right)}{\eta \left(\frac{N\theta b \eta }{c_{\max}}+\phi\right)^2}-\frac{V}{\eta}$. Next, we show that when the parameters' values fall into region (a), region (b), or region (c), we will have $U\ge\frac{\delta k\left(b\eta+\phi\right)}{\eta \left(\frac{N\theta b \eta }{c_{\max}}+\phi\right)^2}-\frac{V}{\eta}$, which implies that $R^{\rm app}\left(l^*,p^*\right)$ increases with $U$ in these regions. 

In region (a), ${I_0\le I_{\rm th}}$, ${\delta>\delta_{\rm th}}$, and $-\phi<p_1$. In region (b), ${I_0\le I_{\rm th}}$, ${\delta\le\delta_{\rm th}}$, and $-\phi<p_2$. Note that in region (b), $p_2\le p_1$. Hence, the relations ${I_0\le I_{\rm th}}$ and $-\phi<p_1$ hold for both region (a) and region (b). Since $p_1=b\eta-\frac{\delta k \left(V+\eta U\right)c_{\max}^2}{\left(Vc_{\max}+\theta\eta UN\right)^2}$, we can rewrite the relation $-\phi<p_1$ as
\begin{align}
-\phi<b\eta-\frac{\delta k \left(V+\eta U\right)c_{\max}^2}{\left(Vc_{\max}+\theta\eta UN\right)^2}.
\end{align}
After arrangement, we can see that the following relation holds for both region (a) and region (b):
\begin{align}
\delta<\frac{\left(\phi+b\eta\right)\left(Vc_{\max}+\theta\eta UN\right)^2}{k\left(V+\eta U\right)c_{\max}^2}.\label{SM:deltainequality}
\end{align}
Next, we compare term $\frac{\left(\phi+b\eta\right)\left(Vc_{\max}+\theta\eta UN\right)^2}{k\left(V+\eta U\right)c_{\max}^2}$ (i.e., the right-hand side of (\ref{SM:deltainequality})) with term $\frac{\left(\eta U+V\right)\left(\frac{N\theta b\eta}{c_{\max}}+\phi\right)^2}{k\left(b\eta+\phi\right)}$. We compute the ratio between them as
\begin{align}
\nonumber
& \frac{\left(\phi+b\eta\right)\left(Vc_{\max}+\theta\eta UN\right)^2}{k\left(V+\eta U\right)c_{\max}^2} \frac{k\left(b\eta+\phi\right)}{\left(\eta U+V\right)\left(\frac{N\theta b\eta}{c_{\max}}+\phi\right)^2} \\
\nonumber
& = \frac{\left(\phi+b\eta\right)^2}{\left(V+ U \eta \right)^2 } \frac{\left(Vc_{\max}+U \theta\eta N\right)^2}{\left(\phi c_{\max}+{b\theta \eta N}\right)^2}\\
& =\left(\frac{\phi V c_{\max}+b \eta^2 U\theta N+V b\eta c_{\max} +\phi U\eta \theta N}{\phi Vc_{\max}+b\eta^2 U\theta N +V b\eta \theta N+\phi U \eta c_{\max}}\right)^2.\label{SM:proposition7:ratio}
\end{align}
Based on ${I_0\le I_{\rm th}}=\frac{\delta}{\theta+\frac{Vc_{\max}}{\eta UN}}$ (i.e., a relation that holds for region (a) and region (b)) and $\delta\le\frac{\left(b\eta N\theta +\phi c_{\max}\right)I_0}{b\eta N}$ (i.e., the condition of Proposition \ref{proposition:impactU:smalldelta}), we can derive the relation $\frac{\delta}{\theta+\frac{\phi c_{\max}}{b\eta N}}\le \frac{\delta}{\theta+\frac{Vc_{\max}}{\eta UN}}$, which implies $\phi U \ge b V$. Considering $c_{\max}-\theta N>0$ (according to our assumption in Section \ref{sec:model}), we further have $\phi U \left(c_{\max}-\theta N\right)\eta \ge bV \left(c_{\max}-\theta N\right)\eta$. After arrangement, we obtain the following relation:
\begin{align}
\phi U\eta c_{\max} + Vb \eta\theta N  \ge Vb\eta c_{\max}+\phi U \eta\theta N .
\end{align}
Based on the above relation, we can see that the ratio in (\ref{SM:proposition7:ratio}) is no larger than $1$. Considering (\ref{SM:deltainequality}), we have 
\begin{align}
\delta<\frac{\left(\phi+b\eta\right)\left(Vc_{\max}+\theta\eta UN\right)^2}{k\left(V+\eta U\right)c_{\max}^2}\le\frac{\left(\eta U+V\right)\left(\frac{N\theta b\eta}{c_{\max}}+\phi\right)^2}{k\left(b\eta+\phi\right)}.
\end{align}
After arrangement, we can see the following relation holds for region (a) and region (b):
\begin{align}
U\ge\frac{\delta k\left(b\eta+\phi\right)}{\eta \left(\frac{N\theta b \eta }{c_{\max}}+\phi\right)^2}-\frac{V}{\eta}.
\end{align}
Recall that $A_1\left(U\right)$ increases with $U$ for $U\ge\frac{\delta k\left(b\eta+\phi\right)}{\eta \left(\frac{N\theta b \eta }{c_{\max}}+\phi\right)^2}-\frac{V}{\eta}$. Hence, revenue $R^{\rm app}\left(l^*,p^*\right)$ increases with $U$ in region (a) and region (b). 

In region (c), ${I_0>I_{\rm th}}$ and $-\phi<p_3$. We can rewrite the relation $-\phi<p_3$ as 
\begin{align}
-\phi< b\eta-\frac{k\left(\left(c_{\max}-\theta N\right)I_0+\delta N\right)^2}{\delta\left(V+\eta U\right)N^2}.
\end{align}
After arrangement, we have the following relation
\begin{align}
U> \frac{k\left(\left(c_{\max}-\theta N\right)I_0+\delta N\right)^2}{\eta\left(\phi+b\eta \right)\delta N^2}-\frac{V}{\eta}.
\end{align}
Since $\delta\le \frac{\left(b\eta N\theta +\phi c_{\max}\right)I_0}{b\eta N}$ (the condition of Proposition \ref{proposition:impactU:smalldelta}), we can see that the following relation holds for region (c)
\begin{align}
U>\frac{\delta k\left(b\eta+\phi\right)}{\eta \left(\frac{N\theta b \eta }{c_{\max}}+\phi\right)^2}-\frac{V}{\eta}.
\end{align}
Recall that $A_1\left(U\right)$ increases with $U$ for $U\ge\frac{\delta k\left(b\eta+\phi\right)}{\eta \left(\frac{N\theta b \eta }{c_{\max}}+\phi\right)^2}-\frac{V}{\eta}$. Hence, revenue $R^{\rm app}\left(l^*,p^*\right)$ increases with $U$ in region (c). 

Second, we consider $R^{\rm app}\left(l^*,p^*\right)$ in the following region: ${I_0\le I_{\rm th}}$, ${\delta>\delta_{\rm th}}$, and $p_1\le -\phi \le p_0$. In this region, $R^{\rm app}\left(l^*,p^*\right)$'s expression is the same as $A_2\left(U\right)$. Recall that $A_2\left(U\right)$ increases with $U$ when $I_0\le I_{\rm th}$ and $-\phi\le p_0$. We can see that $R^{\rm app}\left(l^*,p^*\right)$ increases with $U$ in this region. 

Third, we consider $R^{\rm app}\left(l^*,p^*\right)$ in the following regions: (a) ${I_0\le I_{\rm th}}$, ${\delta>\delta_{\rm th}}$, and $-\phi>p_0$, and (b) ${I_0\le I_{\rm th}}$, ${\delta\le\delta_{\rm th}}$, and $-\phi\ge p_2$. In these regions, $R^{\rm app}\left(l^*,p^*\right)$'s expression is the same as $A_3\left(U\right)$, which does not change with $U$.

Fourth, we consider $R^{\rm app}\left(l^*,p^*\right)$ in the following region: ${I_0> I_{\rm th}}$ and $-\phi\ge p_3$. In this region, $R^{\rm app}\left(l^*,p^*\right)$'s expression is the same as $A_4\left(U\right)$, which increases with $U$ when $b\eta N\theta I_0+\phi c_{\max} I_0-b\eta N\delta\ge0$. Recall that the condition of Proposition  \ref{proposition:impactU:smalldelta} is $\delta\le \frac{\left(b\eta N\theta +\phi c_{\max}\right)I_0}{b\eta N}$. We can see that $R^{\rm app}\left(l^*,p^*\right)$ increases with $U$ in the region ${I_0> I_{\rm th}}$ and $-\phi\ge p_3$.

Considering the analysis for all the above regions and the fact that $R^{\rm app}\left(l^*,p^*\right)$ is continuous in $U$, we complete the proof of Proposition  \ref{proposition:impactU:smalldelta}.

{\bf(Step 3)} We prove Proposition \ref{proposition:impactU:largedelta}, i.e., when $\delta>\frac{\left(b\eta N\theta +\phi c_{\max}\right)I_0}{b\eta N}$, $R^{\rm app}\left(l^*,p^*\right)$ decreases with $U$ for $U\le\frac{\delta k\left(b\eta+\phi\right)}{\eta \left(\frac{N\theta b \eta }{c_{\max}}+\phi\right)^2}-\frac{V}{\eta}$, and increases with $U$ for $U\ge\frac{\delta k\left(b\eta+\phi\right)}{\eta \left(\frac{N\theta b \eta }{c_{\max}}+\phi\right)^2}-\frac{V}{\eta}$. 

First, we consider $R^{\rm app}\left(l^*,p^*\right)$ in the following regions: (a) ${I_0\le I_{\rm th}}$, ${\delta>\delta_{\rm th}}$, and $-\phi<p_1$; (b) ${I_0\le I_{\rm th}}$, ${\delta\le\delta_{\rm th}}$, and $-\phi<p_2$; (c) ${I_0>I_{\rm th}}$ and $-\phi<p_3$. In these regions, $R^{\rm app}\left(l^*,p^*\right)$'s expression is the same as $A_1\left(U\right)$, which decreases with $U$ for $U\le\frac{\delta k\left(b\eta+\phi\right)}{\eta \left(\frac{N\theta b \eta }{c_{\max}}+\phi\right)^2}-\frac{V}{\eta}$, and increases with $U$ for $U\ge\frac{\delta k\left(b\eta+\phi\right)}{\eta \left(\frac{N\theta b \eta }{c_{\max}}+\phi\right)^2}-\frac{V}{\eta}$.    

Second, we consider $R^{\rm app}\left(l^*,p^*\right)$ in the following region: ${I_0\le I_{\rm th}}$, ${\delta>\delta_{\rm th}}$, and $p_1\le -\phi \le p_0$. In this region, $R^{\rm app}\left(l^*,p^*\right)$'s expression is the same as $A_2\left(U\right)$. Recall that $A_2\left(U\right)$ increases with $U$ when $I_0\le I_{\rm th}$ and $-\phi\le p_0$. We can see that $R^{\rm app}\left(l^*,p^*\right)$ increases with $U$ in this region. Next, we prove that the relation $U\ge\frac{\delta k\left(b\eta+\phi\right)}{\eta \left(\frac{N\theta b \eta }{c_{\max}}+\phi\right)^2}-\frac{V}{\eta}$ holds in this region. Note that $p_1\le -\phi$ and $ -\phi \le p_0$ imply that 
\begin{align}
b\eta-\frac{\delta k \left(V+\eta U\right)c_{\max}^2}{\left(Vc_{\max}+\theta\eta UN\right)^2} \le -\phi \le -k\frac{\left(\delta-\theta I_0\right)c_{\max}}{Vc_{\max}+\theta\eta UN}.
\end{align}
After arrangement, we have the following two inequalities:
\begin{align}
U\ge\frac{\left(\phi+b\eta\right)\left(Vc_{\max}+\theta\eta UN\right)^2}{\delta k c_{\max}^2 \eta}-\frac{V}{\eta},\\
\left(Vc_{\max}+\theta\eta UN\right)^2\ge \frac{k^2c_{\max}^2\left(\delta-\theta I_0\right)^2}{\phi^2}.
\end{align}
Then we can easily derive the following inequality:
\begin{align}
U\ge\frac{\left(\phi+b\eta\right) k \left(\delta-\theta I_0\right)^2}{{\delta  \eta}  \phi^2} -\frac{V}{\eta}.
\end{align}
Based on $\delta>\frac{\left(b\eta N\theta +\phi c_{\max}\right)I_0}{b\eta N}$ (the condition of Proposition \ref{proposition:impactU:largedelta}), we can obtain
\begin{align}
U\ge\frac{\delta k\left(b\eta+\phi\right)}{\eta \left(\frac{N\theta b \eta }{c_{\max}}+\phi\right)^2}-\frac{V}{\eta}.
\end{align}
Hence, we have shown that in the region ${I_0\le I_{\rm th}}$, ${\delta>\delta_{\rm th}}$, and $p_1\le -\phi \le p_0$, $R^{\rm app}\left(l^*,p^*\right)$ increases with $U$, and $U$ satisfies $U\ge\frac{\delta k\left(b\eta+\phi\right)}{\eta \left(\frac{N\theta b \eta }{c_{\max}}+\phi\right)^2}-\frac{V}{\eta}$.

Third, we consider $R^{\rm app}\left(l^*,p^*\right)$ in the following regions: (a) ${I_0\le I_{\rm th}}$, ${\delta>\delta_{\rm th}}$, and $-\phi>p_0$, and (b) ${I_0\le I_{\rm th}}$, ${\delta\le\delta_{\rm th}}$, and $-\phi\ge p_2$. In these regions, $R^{\rm app}\left(l^*,p^*\right)$'s expression is the same as $A_3\left(U\right)$, which does not change with $U$.

Fourth, we consider $R^{\rm app}\left(l^*,p^*\right)$ in the following region: ${I_0> I_{\rm th}}$ and $-\phi\ge p_3$. In this region, $R^{\rm app}\left(l^*,p^*\right)$'s expression is the same as $A_4\left(U\right)$, which decreases with $U$ when $b\eta N\theta I_0+\phi c_{\max} I_0-b\eta N\delta<0$. 
Recall that the condition of Proposition \ref{proposition:impactU:largedelta} is $\delta>\frac{\left(b\eta N\theta +\phi c_{\max}\right)I_0}{b\eta N}$. We can see that $R^{\rm app}\left(l^*,p^*\right)$ decreases with $U$ in this region. Next, we show that the relation $U<\frac{\delta k\left(b\eta+\phi\right)}{\eta \left(\frac{N\theta b \eta }{c_{\max}}+\phi\right)^2}-\frac{V}{\eta}$ holds in this region. We can rewrite $-\phi\ge p_3$ as 
\begin{align}
-\phi\ge b\eta-\frac{k\left(\left(c_{\max}-\theta N\right)I_0+\delta N\right)^2}{\delta\left(V+\eta U\right)N^2}.
\end{align}
After arrangement, we obtain the following relation:
\begin{align}
U\le \frac{k\left(\left(c_{\max}-\theta N\right)I_0+\delta N\right)^2}{\eta\left(\phi+b\eta \right)\delta N^2}-\frac{V}{\eta}.
\end{align}
Since $\delta>\frac{\left(b\eta N\theta +\phi c_{\max}\right)I_0}{b\eta N}$ (the condition of Proposition \ref{proposition:impactU:largedelta}), we can see that the following relation holds in this region:
\begin{align}
U<\frac{\delta k\left(b\eta+\phi\right)}{\eta \left(\frac{N\theta b \eta }{c_{\max}}+\phi\right)^2}-\frac{V}{\eta}.
\end{align}
Hence, we have shown that in the region ${I_0> I_{\rm th}}$ and $-\phi\ge p_3$, $R^{\rm app}\left(l^*,p^*\right)$ decreases with $U$, and $U$ satisfies $U<\frac{\delta k\left(b\eta+\phi\right)}{\eta \left(\frac{N\theta b \eta }{c_{\max}}+\phi\right)^2}-\frac{V}{\eta}$.

Considering the analysis for all the above regions and the fact that $R^{\rm app}\left(l^*,p^*\right)$ is continuous in $U$, we complete the proof of Proposition  \ref{proposition:impactU:largedelta}.
\endproof

\subsection{Proof of Corollary \ref{corollary:impactU} in Section \ref{subsec:influence:quality}}\label{SM:corollary:impactU}
\proof 
The proof is very straightforward. Based on Proposition \ref{proposition:impactU:largedelta}, when $\delta > \frac{\phi c_{\max} I_0}{b\eta N}+\theta I_0$, $R^{\rm app}\left(l^*,p^*\right)$ decreases with $U$ for $U\in\left(0,\frac{\delta k\left(b\eta+\phi\right)}{\eta \left(\frac{N\theta b \eta }{c_{\max}}+\phi\right)^2}-\frac{V}{\eta}\right]$. Moreover, when $\delta>\frac{\left(\frac{N\theta b \eta }{c_{\max}}+\phi\right)^2\left(\eta c_{\max}+\left(1-\eta\right)V-\eta \theta N\right)}{k\left(b\eta+\phi\right)}$, we can easily see that $\frac{\delta k\left(b\eta+\phi\right)}{\eta \left(\frac{N\theta b \eta }{c_{\max}}+\phi\right)^2}-\frac{V}{\eta} > c_{\max}-V-\theta N$. 
Therefore, when $\delta> \max\left\{\frac{\phi c_{\max} I_0}{b\eta N}+\theta I_0,\frac{\left(\frac{N\theta b \eta }{c_{\max}}+\phi\right)^2\left(\eta c_{\max}+\left(1-\eta\right)V-\eta \theta N\right)}{k\left(b\eta+\phi\right)}\right\}$, we can conclude that $R^{\rm app}\left(l^*,p^*\right)$ decreases with $U$ for $U\in\left(0,c_{\max}-V-\theta N\right)$.
\endproof

\section{Proof of Proposition \ref{proposition:impacteta:largephi} in Section \ref{subsec:impacteta}}\label{SM:proposition9}
\proof

\begin{figure*}
{\small{
\begin{align}
& \left\{ {\begin{array}{*{20}{l}}
{-\frac{N}{c_{\max}}b\eta^2U+\frac{N}{c_{\max}-N\theta}\left(\sqrt{\left(V+\eta U\right)\left(b\eta +\phi\right)}-\sqrt{\delta k}\right)^2+kI_0,}&{{\rm if~}-\phi<p_1 ,}\\
{\phi N\eta \frac{U}{c_{\max}}-k I_{\rm th}+k I_0,}&{{\rm if~} p_1 \le -\phi \le p_0,}\\
{\frac{V}{\frac{\delta}{I_0}-\theta}\phi,}&{{\rm if~}-\phi> p_0.}\\
\end{array}} \right.\label{SM:equ:H1}\\
& \left\{ {\begin{array}{*{20}{l}}
{-\frac{N}{c_{\max}}b\eta^2U+\frac{N}{c_{\max}-N\theta}\left(\sqrt{\left(V+\eta U\right)\left(b\eta +\phi\right)}-\sqrt{\delta k}\right)^2+kI_0,}&{{\rm if~}-\phi< p_2,}\\
{\frac{V}{\frac{\delta}{I_0}-\theta}\phi,}&{{\rm if~}-\phi\ge p_2.}\\
\end{array}} \right.\label{SM:equ:H2}\\
& \left\{ {\begin{array}{*{20}{l}}
{-\frac{N}{c_{\max}}b\eta^2U+\frac{N}{c_{\max}-N\theta}\left(\sqrt{\left(V+\eta U\right)\left(b\eta +\phi\right)}-\sqrt{\delta k}\right)^2+kI_0,}&{{\rm if~}-\phi< p_3,}\\
{{\left(b\eta+\phi\right)N {\frac{Vc_{\max}I_0-\eta UN\delta+\eta UN\theta I_0}{c_{\max}^2I_0+c_{\max}N\delta -c_{\max}N\theta I_0}}+\phi N\eta\frac{U}{c_{\max}}},}&{{\rm if~}-\phi\ge p_3.}
\end{array}} \right.\label{SM:equ:H3}
\end{align}}}
\vspace{-0.5cm}
\hrulefill
\end{figure*}
\vspace{-0.12cm}

From Theorem \ref{theorem:optimaltariff} and Corollary \ref{corollary:optimalrevenue}, when ${I_0\le I_{\rm th}}{~\rm and~}{\delta>\delta_{\rm th}}$, $R^{\rm app}\left(l^*,p^*\right)$ is given in (\ref{SM:equ:H1}); when ${I_0\le I_{\rm th}}{~\rm and~}{\delta\le\delta_{\rm th}}$, $R^{\rm app}\left(l^*,p^*\right)$ is given in (\ref{SM:equ:H2}); when ${I_0>I_{\rm th}}$, $R^{\rm app}\left(l^*,p^*\right)$ is given in (\ref{SM:equ:H3}). We can also easily verify that $R^{\rm app}\left(l^*,p^*\right)$ is continuous in $\eta$. 

{\bf(Step 1)} We study the impact of $\eta$ on $R^{\rm app}\left(l^*,p^*\right)$ in the following situations: (a) ${I_0\le I_{\rm th}}$, ${\delta>\delta_{\rm th}}$, and $-\phi>p_0$; (b) ${I_0\le I_{\rm th}}$, ${\delta\le\delta_{\rm th}}$, and $-\phi\ge p_2$. In these situations, $R^{\rm app}\left(l^*,p^*\right)$'s expression is $\frac{V}{\frac{\delta}{I_0}-\theta}\phi$, which is independent of $\eta$.  

{\bf(Step 2)} We study the impact of $\eta$ on $R^{\rm app}\left(l^*,p^*\right)$ in the following situation: ${I_0\le I_{\rm th}}$, ${\delta>\delta_{\rm th}}$, and $p_1\le -\phi \le p_0$. In this situation, $R^{\rm app}\left(l^*,p^*\right)$'s expression is the same as the following function:
\begin{align}
B_1\left(\eta\right)\triangleq &\phi N\eta \frac{U}{c_{\max}}-k I_{\rm th}+k I_0.
\end{align}
From the definition of $I_{\rm th}$, we have $I_{\rm th}=\frac{\delta}{\theta+\frac{Vc_{\max}}{\eta UN}}$. By taking the derivative of $B_1\left(\eta\right)$ with respect to $\eta$ and utilizing the conditions $-\phi\le p_0= -k\frac{\left(\delta-\theta I_0\right)c_{\max}}{Vc_{\max}+\theta\eta UN}$ and ${I_0\le I_{\rm th}}$, we can easily prove that $\frac{d B_1\left(\eta\right)}{d\eta}\ge0$. Hence, in the considered situation, $R^{\rm app}\left(l^*,p^*\right)$ increases with $\eta$. 

{\bf(Step 3)} We assume that $\phi \ge \frac{bV}{U}$, and study the impact of $\eta$ on $R^{\rm app}\left(l^*,p^*\right)$ in the following situations: (a) ${I_0\le I_{\rm th}}$, ${\delta>\delta_{\rm th}}$, and $-\phi<p_1$; (b) ${I_0\le I_{\rm th}}$, ${\delta\le\delta_{\rm th}}$, and $-\phi<p_2$; (c) ${I_0>I_{\rm th}}$ and $-\phi<p_3$. In these situations, $R^{\rm app}\left(l^*,p^*\right)$'s expression is the same as the following function:
\begin{align}
\nonumber
B_2\left(\eta\right)\triangleq &\frac{N}{c_{\max}-N\theta}\left(\sqrt{\left(V+\eta U\right)\left(b\eta +\phi\right)}-\sqrt{\delta k}\right)^2 \\
& -\frac{N}{c_{\max}}b\eta^2U+kI_0.
\end{align}
We compute $\frac{d B_2\left(\eta\right)}{d \eta}$ as
\begin{align}
\nonumber
& \frac{d B_2\left(\eta\right)}{d \eta}=-2bU\eta\frac{N}{c_{\max}}\\
& +\frac{N}{c_{\max}\!-\!N\theta} \left(bV\!+\!2 b\eta U\!+\!U\phi \right)\left(1\!-\!\sqrt{\frac{\delta k}{\left(b\eta+\phi\right)\left(V+\eta U\right)}}\right).\label{SM:equ:impacteta:1}
\end{align}
In situations (a), (b), and (c), we can prove that the condition $-\phi < b\eta-\frac{\delta k \left(V+\eta U\right)c_{\max}^2}{\left(Vc_{\max}+\theta\eta UN\right)^2}$ always holds. Hence, in these situations, we have $1-\sqrt{\frac{\delta k}{\left(b\eta+\phi\right)\left(V+\eta U\right)}}>\frac{c_{\max} \eta U-\theta N\eta U}{Vc_{\max}+c_{\max} \eta U}>0$. Then, we can obtain the following relation:
\begin{align}
\nonumber
& \frac{d B_2\left(\eta\right)}{d \eta}> -2bU\eta\frac{N}{c_{\max}}\\
& +\frac{N}{c_{\max}-N\theta} \left(bV+2 b\eta U+U\phi \right) \frac{c_{\max} \eta U-\theta N\eta U}{Vc_{\max}+c_{\max} \eta U}.
\end{align}
Based on $\phi \ge \frac{bV}{U}$ (the assumption of {\bf Step 3}), we further have the following relation:
\begin{align}
\nonumber
\frac{d B_2\left(\eta\right)}{d \eta}&\!>\! -2bU\eta\frac{N}{c_{\max}}\!+\!\frac{2bN\left(V\!+\!\eta U \right)}{c_{\max}\!-\!N\theta} \frac{c_{\max} \eta U\!-\!\theta N\eta U}{Vc_{\max}\!+\!c_{\max} \eta U}\\
& \!=\!0.
\end{align}
Therefore, when $\phi \ge \frac{bV}{U}$, $R^{\rm app}\left(l^*,p^*\right)$ increases with $\eta$ in the considered situations. 

{\bf(Step 4)} We assume that $\phi \ge \frac{bV}{U}$, and study the impact of $\eta$ on $R^{\rm app}\left(l^*,p^*\right)$ in the situation ${I_0> I_{\rm th}}$ and $-\phi\ge p_3$. In this situation, $R^{\rm app}\left(l^*,p^*\right)$'s expression is the same as the following function:
\begin{align}
\nonumber
B_3\left(\eta\right)\triangleq &\left(b\eta+ \phi\right)N {\frac{Vc_{\max}I_0-\eta UN\delta+\eta UN\theta I_0}{c_{\max}^2I_0+c_{\max}N\delta -c_{\max}N\theta I_0}}\\
& +\phi N\eta\frac{U}{c_{\max}}.
\end{align}
When $\delta\le \theta I_0$, we can easily see that $B_3\left(\eta\right)$ increases with $\eta$. Next, we focus on the case where $\delta>\theta I_0$. We can compute $\frac{d B_3\left(\eta\right)}{d \eta}$ as follows:
\begin{align}
\frac{d B_3\left(\eta\right)}{d \eta}=\frac{\left(bV+\phi U\right)c_{\max} I_0-2b\eta UN\left(\delta-\theta I_0\right)}{c_{\max}^2I_0+c_{\max}N\delta -c_{\max}N\theta I_0}N.
\end{align}
From $I_0>I_{\rm th} =\frac{\delta}{\theta+\frac{Vc_{\max}}{\eta UN}}$, we have the relation $\eta<\frac{Vc_{\max}I_0}{UN\left(\delta-\theta I_0\right)}$. Hence, we can see that
\begin{align}
\frac{d B_3\left(\eta\right)}{d \eta}>\frac{\phi U c_{\max} I_0-b Vc_{\max}I_0}{c_{\max}^2I_0+c_{\max}N\delta -c_{\max}N\theta I_0}N.
\end{align}
From the assumption $\phi \ge \frac{bV}{U}$, we can see that $\frac{d B_3\left(\eta\right)}{d \eta}>0$. Combing the results of the case $\delta\le \theta I_0$ and the case $\delta>\theta I_0$, we can see that when $\phi \ge \frac{bV}{U}$, $R^{\rm app}\left(l^*,p^*\right)$ increases with $\eta$ in the considered situation.

According to the fact that $R^{\rm app}\left(l^*,p^*\right)$ is continuous in $\eta$ and the results in {\bf Step 1}, {\bf Step 2}, {\bf Step 3}, and {\bf Step 4}, we conclude that when $\phi \ge \frac{bV}{U}$, $R^{\rm app}\left(l^*,p^*\right)$ increases with $\eta$.
\endproof

\section{Proof of Lemma \ref{lemma:delta1} in Section \ref{subsec:impacteta}}\label{SM:lemma2}
\proof 
We define function $E\left(\delta\right)\triangleq \delta^2-2\theta I_0\delta -\frac{bV^2}{kU} \delta+\theta^2 I_0^2$. It is easy to show that (a) $E\left(\delta\right)$ is continuous in $\delta\in\left(-\infty,\infty\right)$; (b) $E\left(\delta\right)$ strictly decreases in $\delta\in\left(-\infty,\theta I_0+\frac{bV^2}{2kU}\right)$; (c) $E\left(\delta\right)$ strictly convexly increases in $\delta\in\left(\theta I_0+\frac{bV^2}{2kU},\infty\right)$ with $\frac{d^2 E\left(\delta\right)}{d \delta^2}=2$. Moreover, we can verify that $E\left(\delta\right)|_{\delta=\theta I_0}<0$. Hence, we can see that there is a unique $\delta\in\left(\theta I_0,\infty\right)$ that satisfies $E\left(\delta\right)=0$. 
\endproof

\section{Proof of Proposition \ref{proposition:eta:no} in Section \ref{subsec:impacteta}}\label{SM:proposition10}
\proof 
{\bf(Step 1)} We show that when $\delta>\delta_2$, $\phi<\frac{bV}{U}$, and $\eta<\frac{Vc_{\max}I_0}{UN\left(\delta-\theta I_0\right)}$, we have $-\phi \ge b\eta-\frac{k\left(\left(c_{\max}-\theta N\right)I_0+\delta N\right)^2}{\delta\left(V+\eta U\right)N^2}$. We prove it by contradiction, i.e., we assume that when $\delta>\delta_2$, $\phi<\frac{bV}{U}$, and $\eta<\frac{Vc_{\max}I_0}{UN\left(\delta-\theta I_0\right)}$, we have $-\phi < b\eta-\frac{k\left(\left(c_{\max}-\theta N\right)I_0+\delta N\right)^2}{\delta\left(V+\eta U\right)N^2}$. 

Rearranging $-\phi < b\eta-\frac{k\left(\left(c_{\max}-\theta N\right)I_0+\delta N\right)^2}{\delta\left(V+\eta U\right)N^2}$, we obtain the following relation:
\begin{align}
k\left(\left(c_{\max}\!-\!\theta N\right)I_0+\delta N\right)^2\!<\!\left(\phi+b\eta\right) \delta\left(V+\eta U\right)N^2.
\end{align}
From the condition $\phi<\frac{bV}{U}$, we further have the following relation:
\begin{align}
k\left(\left(c_{\max}-\theta N\right)I_0+\delta N\right)^2<\frac{b}{U} \delta\left(V+\eta U\right)^2 N^2.
\end{align}
Considering the condition $\eta<\frac{Vc_{\max}I_0}{UN\left(\delta-\theta I_0\right)}$, we further have the following relation:
\begin{align}
k\!\left(\left(c_{\max}\!-\!\theta N\right)\!I_0\!+\!\delta N\right)^2\!<\!\frac{b}{U} \delta\!\left(\frac{N\left(\delta\!-\!\theta I_0\right)\!+\!c_{\max}I_0}{N\left(\delta\!-\!\theta I_0\right)}\right)^2 V^2N^2.
\end{align}
After the rearrangement of the above inequality, we obtain the following result:
\begin{align}
\delta^2-2\theta I_0\delta -\frac{bV^2}{kU} \delta+\theta^2 I_0^2<0.\label{SM:equ:impacteta:2}
\end{align}

Based on Lemma \ref{lemma:delta1}, we have $\delta_2^2-2\theta I_0\delta_2 -\frac{bV^2}{kU} \delta_2+\theta^2 I_0^2=0$. Moreover, it is easy to show that $\delta^2-2\theta I_0\delta -\frac{bV^2}{kU} \delta+\theta^2 I_0^2$ strictly increases with $\delta$ in $\delta\in\left[\delta_2,\infty\right)$. Hence, when $\delta>\delta_2$, we have the relation $\delta^2-2\theta I_0\delta -\frac{bV^2}{kU} \delta+\theta^2 I_0^2>0$. This contradicts with (\ref{SM:equ:impacteta:2}). Therefore, we can conclude that when $\delta>\delta_2$, $\phi<\frac{bV}{U}$, and $\eta<\frac{Vc_{\max}I_0}{UN\left(\delta-\theta I_0\right)}$, we have $-\phi \ge b\eta-\frac{k\left(\left(c_{\max}-\theta N\right)I_0+\delta N\right)^2}{\delta\left(V+\eta U\right)N^2}$.

{\bf(Step 2)} We show that when $\phi<\frac{bV}{U}$ and $\delta>\delta_2$, we have the relation $\eta_A<\eta_B$. When $\delta>\delta_2>\theta I_0$, we can see that $\eta_A= \frac{\left(bV+\phi U\right)c_{\max}I_0}{2bUN\left(\delta-\theta I_0\right)}>0$ and $\eta_B = \frac{Vc_{\max}I_0}{UN\left(\delta-\theta I_0\right)}>0$. Moreover, we can compute the ratio between $\eta_A$ and $\eta_B$ as $\frac{\eta_A}{\eta_B}=\frac{bV+\phi U}{2bV}$, which is smaller than $1$ under the condition $\phi<\frac{bV}{U}$. Hence, when $\phi<\frac{bV}{U}$ and $\delta>\delta_2$, we have the relation $\eta_A<\eta_B$. 

{\bf(Step 3)} We show that when $\phi<\frac{bV}{U}$ and $\delta>\delta_2$, $R^{\rm app}\left(l^*,p^*\right)$ decreases with $\eta$ for $\eta\in\left(\eta_A,\eta_B\right)$. 

Based on our result in {\bf Step 1}, when $\delta>\delta_2$, $\phi<\frac{bV}{U}$, and $\eta<\eta_B=\frac{Vc_{\max}I_0}{UN\left(\delta-\theta I_0\right)}$, we have $-\phi \ge b\eta-\frac{k\left(\left(c_{\max}-\theta N\right)I_0+\delta N\right)^2}{\delta\left(V+\eta U\right)N^2}=p_3$. Moreover, when $\delta>\delta_2$, we can rearrange the relation $\eta<\frac{Vc_{\max}I_0}{UN\left(\delta-\theta I_0\right)}$ to obtain the relation $I_0>\frac{\delta}{\theta+\frac{Vc_{\max}}{\eta UN}}=I_{\rm th} $. According to Theorem \ref{theorem:optimaltariff} and Corollary \ref{corollary:optimalrevenue}, under $-\phi \ge p_3$ and $I_0>I_{\rm th} $, $R^{\rm app}\left(l^*,p^*\right)$'s expression is given as follows:
\begin{align}
\nonumber
R^{\rm app}\left(l^*,p^*\right)=&\left(b\eta+ \phi\right)N {\frac{Vc_{\max}I_0-\eta UN\delta+\eta UN\theta I_0}{c_{\max}^2I_0+c_{\max}N\delta -c_{\max}N\theta I_0}}\\
& +\phi N\eta\frac{U}{c_{\max}}.
\end{align}
Based on our computation in {\bf Step 4} in Section \ref{SM:proposition9}, the above $R^{\rm app}\left(l^*,p^*\right)$'s derivative with respective to $\eta$ is $\frac{\left(bV+\phi U\right)c_{\max} I_0-2b\eta UN\left(\delta-\theta I_0\right)}{c_{\max}^2I_0+c_{\max}N\delta -c_{\max}N\theta I_0}N$. When $\eta>\eta_A=\frac{\left(bV+\phi U\right)c_{\max}I_0}{2bUN\left(\delta-\theta I_0\right)}$, we can easily see that such a derivative is negative. That is to say, when $\phi<\frac{bV}{U}$, $\delta>\delta_2$, and $\eta_A<\eta<\eta_B$, $R^{\rm app}\left(l^*,p^*\right)$ decreases with $\eta$.
\endproof

\section{Proof of Proposition \ref{proposition:N:largephi} in Section \ref{subsec:impactN}}\label{SM:proposition11}
\proof 
From Theorem \ref{theorem:optimaltariff} and Corollary \ref{corollary:optimalrevenue}, when ${I_0\le I_{\rm th}}{~\rm and~}{\delta>\delta_{\rm th}}$, $R^{\rm app}\left(l^*,p^*\right)$ is given in (\ref{SM:equ:H1}); when ${I_0\le I_{\rm th}}{~\rm and~}{\delta\le\delta_{\rm th}}$, $R^{\rm app}\left(l^*,p^*\right)$ is given in (\ref{SM:equ:H2}); when ${I_0>I_{\rm th}}$, $R^{\rm app}\left(l^*,p^*\right)$ is given in (\ref{SM:equ:H3}). We can also easily verify that $R^{\rm app}\left(l^*,p^*\right)$ is continuous in $N$. 

{\bf(Step 1)} We study the impact of $N$ on $R^{\rm app}\left(l^*,p^*\right)$ in the following situations: (a) ${I_0\le I_{\rm th}}$, ${\delta>\delta_{\rm th}}$, and $-\phi>p_0$; (b) ${I_0\le I_{\rm th}}$, ${\delta\le\delta_{\rm th}}$, and $-\phi\ge p_2$. In these situations, $R^{\rm app}\left(l^*,p^*\right)$'s expression is $\frac{V}{\frac{\delta}{I_0}-\theta}\phi$, which is independent of $N$.  

{\bf(Step 2)} We study the impact of $N$ on $R^{\rm app}\left(l^*,p^*\right)$ in the following situation: ${I_0\le I_{\rm th}}$, ${\delta>\delta_{\rm th}}$, and $p_1\le -\phi \le p_0$. In this situation, $R^{\rm app}\left(l^*,p^*\right)$'s expression is the same as the following function:
\begin{align}
C_1\left(N\right)\triangleq &\phi N\eta \frac{U}{c_{\max}}-k I_{\rm th}+k I_0.
\end{align}
From the definition of $I_{\rm th}$, we have $I_{\rm th}=\frac{\delta}{\theta+\frac{Vc_{\max}}{\eta UN}}$. By taking the derivative of $C_1\left(N\right)$ with respect to $N$ and utilizing the conditions $-\phi\le p_0= -k\frac{\left(\delta-\theta I_0\right)c_{\max}}{Vc_{\max}+\theta\eta UN}$ and ${I_0\le I_{\rm th}}$, we can easily prove that $\frac{d C_1\left(N\right)}{d N}\ge0$. Hence, in the considered situation, $R^{\rm app}\left(l^*,p^*\right)$ increases with $N$. 

{\bf(Step 3)} We assume that $\phi \ge \frac{bV}{U}$, and study the impact of $N$ on $R^{\rm app}\left(l^*,p^*\right)$ in the following situations: (a) ${I_0\le I_{\rm th}}$, ${\delta>\delta_{\rm th}}$, and $-\phi<p_1$; (b) ${I_0\le I_{\rm th}}$, ${\delta\le\delta_{\rm th}}$, and $-\phi<p_2$; (c) ${I_0>I_{\rm th}}$ and $-\phi<p_3$. In these situations, $R^{\rm app}\left(l^*,p^*\right)$'s expression is the same as the following function:
\begin{align}
\nonumber
C_2\left(N\right)\triangleq &\frac{N}{c_{\max}-N\theta}\left(\sqrt{\left(V+\eta U\right)\left(b\eta +\phi\right)}-\sqrt{\delta k}\right)^2 \\
& -\frac{N}{c_{\max}}b\eta^2U+kI_0.
\end{align}
We compute $\frac{d C_2\left(N\right)}{d N}$ as
{\small
\begin{align}
\frac{d C_2\!\left(N\right)}{d N}\!=\!\frac{c_{\max}^2 \!\left(\!\sqrt{\!\left(V+\eta U\right)\!\left(b\eta+\phi\right)}\!-\!\sqrt{\delta k}\!\right)^2\!-\!b \eta^2 U \left(c_{\max}\!-\!N\theta\right)^2}{\left(c_{\max}-N\theta\right)^2 c_{\max}}.\label{SM:equ:impactN:1}
\end{align}}
In situations (a), (b), and (c), we can prove that the condition $-\phi < b\eta-\frac{\delta k \left(V+\eta U\right)c_{\max}^2}{\left(Vc_{\max}+\theta\eta UN\right)^2}$ always holds. Hence, in these situations, we have $\delta<\frac{\left(b\eta+\phi\right)\left(Vc_{\max}+\theta \eta UN\right)^2}{k \left(V+\eta U\right)c_{\max}^2 }$. Furthermore, we have the following relation: 
\begin{align}
\sqrt{\!\left(V\!+\!\eta U\right)\!\left(b\eta\!+\!\phi\right)}\!-\!\sqrt{\delta k}\!>\!\sqrt{\frac{b\eta\!+\!\phi}{V\!+\!\eta U}}\eta U\!\left(1\!-\!\frac{\theta N}{c_{\max}}\!\right)\!>\!0.\label{SM:equ:impactN:2}
\end{align}
The first inequality is based on $\delta<\frac{\left(b\eta+\phi\right)\left(Vc_{\max}+\theta \eta UN\right)^2}{k \left(V+\eta U\right)c_{\max}^2 }$, and the second inequality is based on the assumption $c_{\max}>\theta N+U+V$ (made in Section \ref{subsec:stackelberg}). According to (\ref{SM:equ:impactN:2}), we can see that $\frac{d C_2\left(N\right)}{d N}$ in (\ref{SM:equ:impactN:1}) decreases with $\delta$. By utilizing the relation $\delta<\frac{\left(b\eta+\phi\right)\left(Vc_{\max}+\theta \eta UN\right)^2}{k \left(V+\eta U\right)c_{\max}^2 }$, we can obtain the following relation:
\begin{align}
\frac{d C_2\left(N\right)}{d N}> \frac{\phi U -bV}{\left(V+\eta U\right) c_{\max}}\eta^2 U.
\end{align}
Recall that we assume $\phi \ge \frac{bV}{U}$ in {\bf Step 3}. Hence, we have $\frac{d C_2\left(N\right)}{d N}>0$. This implies that when $\phi \ge \frac{bV}{U}$, $R^{\rm app}\left(l^*,p^*\right)$ increases with $N$ in the considered situations. 

{\bf(Step 4)} We assume that $\phi \ge \frac{bV}{U}$, and study the impact of $N$ on $R^{\rm app}\left(l^*,p^*\right)$ in the situation ${I_0> I_{\rm th}}$ and $-\phi\ge p_3$. In this situation, $R^{\rm app}\left(l^*,p^*\right)$'s expression is the same as the following function:
\begin{align}
\nonumber
C_3\left(N\right)\triangleq & \left(b\eta+ \phi\right)N {\frac{Vc_{\max}I_0-\eta UN\delta+\eta UN\theta I_0}{c_{\max}^2I_0+c_{\max}N\delta -c_{\max}N\theta I_0}}\\
& +\phi N\eta\frac{U}{c_{\max}}.
\end{align}
We can compute $\frac{d C_3\left(N\right)}{d N}$ as (\ref{SM:equ:TMC1}).

\begin{figure*}
\begin{align}
\frac{d C_3\left(N\right)}{d N} = \frac{b\eta c_{\max}\left(Vc_{\max}^2 I_0^2 +2 c_{\max} I_0 \eta U N \left(\theta I_0-\delta\right)-\eta U N^2 \left(\delta-\theta I_0\right)^2 \right)+\phi\left(\eta U+V\right)c_{\max}^3 I_0^2}{\left({c_{\max}^2I_0+c_{\max}N\delta -c_{\max}N\theta I_0}\right)^2}.\label{SM:equ:TMC1}
\end{align}
\hrulefill

\begin{align}
{\hat C}_3\left(N\right)\triangleq b\eta c_{\max}\left(Vc_{\max}^2 I_0^2 +2 c_{\max} I_0 \eta U N \left(\theta I_0-\delta\right)-\eta U N^2 \left(\delta-\theta I_0\right)^2 \right)+\phi\left(\eta U+V\right)c_{\max}^3 I_0^2.\label{SM:equ:TMC2}
\end{align}
\hrulefill
\end{figure*}

When $\delta \le \theta I_0$, we can utilize the relation $c_{\max}>\theta N+U+V$ (our assumption made in Section \ref{subsec:stackelberg}) to verify that $\frac{d C_3\left(N\right)}{d N}>0$. Next, we focus on the case where $\delta > \theta I_0$. We define function ${\hat C}_3\left(N\right)$ as (\ref{SM:equ:TMC2}).

Note that ${\hat C}_3\left(N\right)$ is actually the numerator of $\frac{d C_3\left(N\right)}{d N}$. Rearranging ${I_0> I_{\rm th}}=\frac{\delta}{\theta+\frac{Vc_{\max}}{\eta UN}}$ and utilizing the condition $\delta > \theta I_0$, we obtain the relation $N<\frac{Vc_{\max}I_0}{\eta U \left(\delta-\theta I_0\right)}$. We can easily see that ${\hat C}_3\left(N\right)$ is a quadratic function of $N$, and decreases in $N\in\left(0,\frac{Vc_{\max}I_0}{\eta U \left(\delta-\theta I_0\right)}\right]$. Therefore, under $0<N<\frac{Vc_{\max}I_0}{\eta U \left(\delta-\theta I_0\right)}$, the following result holds: 
\begin{align}
{\hat C}_3\left(N\right)\!>\! {\hat C}_3\left(N\right)|_{N=\frac{Vc_{\max}I_0}{\eta U \left(\delta-\theta I_0\right)}}\!=\!\left(\eta U\!+\!V\right) c_{\max}^3 I_0^2 \left(\phi-b\frac{V}{U}\right).
\end{align}
Recall that we assume that $\phi \ge \frac{bV}{U}$ holds in {\bf Step 4}. Hence, we have ${\hat C}_3\left(N\right)>0$, which implies that $\frac{d C_3\left(N\right)}{d N}>0$. 
Now we have proved that $\frac{d C_3\left(N\right)}{d N}>0$ holds for both the case $\delta \le \theta I_0$ and the case $\delta > \theta I_0$. We can see that when $\phi \ge \frac{bV}{U}$, $R^{\rm app}\left(l^*,p^*\right)$ increases with $N$ in the considered situation. 

According to the fact that $R^{\rm app}\left(l^*,p^*\right)$ is continuous in $N$ and the results in {\bf Step 1}, {\bf Step 2}, {\bf Step 3}, and {\bf Step 4}, we conclude that when $\phi \ge \frac{bV}{U}$, $R^{\rm app}\left(l^*,p^*\right)$ increases with $N$.
\endproof

\section{Proofs of Lemmas \ref{lemma:delta:2} and \ref{lemma:N:1}}

\subsection{Proof of Lemma \ref{lemma:delta:2} in Section \ref{subsec:impactN}}\label{SM:lemma3}
\proof 
We define function $L\left(\delta\right)\triangleq \delta^2-2\theta I_0 \delta -\frac{V^2\left(b\eta+\phi\right)}{k\left(V+\eta U\right)}\delta + \theta^2 I_0^2$. 
It is easy to show that (a) $L\left(\delta\right)$ is continuous in $\delta\in\left(-\infty,\infty\right)$; (b) $L\left(\delta\right)$ strictly decreases in $\delta\in\left(-\infty,\theta I_0+\frac{V^2\left(b\eta+\phi\right)}{2k\left(V+\eta U\right)}\right)$; (c) $L\left(\delta\right)$ strictly convexly increases in $\delta\in\left(\theta I_0+\frac{V^2\left(b\eta+\phi\right)}{2k\left(V+\eta U\right)},\infty\right)$ with $\frac{d^2 L\left(\delta\right)}{d \delta^2}=2$. Moreover, we can verify that $L\left(\delta\right)|_{\delta=\theta I_0}<0$. Hence, we can see that there is a unique $\delta\in\left(\theta I_0,\infty\right)$ that satisfies $L\left(\delta\right)=0$. 
\endproof

\subsection{Proof of Lemma \ref{lemma:N:1} in Section \ref{subsec:impactN}}\label{SM:lemma4}
\proof
Recall that in the proof of Proposition \ref{proposition:N:largephi} in Section \ref{SM:proposition11}, we have defined a function ${\hat C}_3\left(N\right)$, which has the same expression as the left-hand side of the following equation: 
\begin{multline}
-b \eta^2 c_{\max} U \left(\delta-\theta I_0\right)^2 N^2 +2b \eta^2 c_{\max}^2 I_0 U\left(\theta I_0-\delta\right) N \\
+b\eta c_{\max}^3 I_0^2 V+\phi \left(\eta U+V\right) c_{\max}^3 I_0^2=0.
\end{multline}
We can easily see that ${\hat C}_3\left(N\right)$ is a continuous function, and strictly decreases in $N\in\left(-\frac{c_{\max} I_0}{\delta-\theta I_0},\infty\right)$. When $\delta>\delta_3$, we can utilize $\delta_3>\theta I_0$ to prove that $\delta>\theta I_0$. This implies the relation $-\frac{c_{\max} I_0}{\delta-\theta I_0}\le 0 \le \frac{Vc_{\max}I_0}{\eta U \left(\delta-\theta I_0\right)}$. Hence, ${\hat C}_3\left(N\right)$ strictly decreases in $N\in \left(0,\frac{Vc_{\max}I_0}{\eta U \left(\delta-\theta I_0\right)}\right)$. 
Next, we investigate ${\hat C}_3\left(N\right)$'s values at points $N=0$ and $N=\frac{Vc_{\max}I_0}{\eta U \left(\delta-\theta I_0\right)}$.

We can see that ${\hat C}_3\left(N\right)|_{N=0}=b\eta c_{\max}^3 I_0^2 V+\phi \left(\eta U+V\right) c_{\max}^3 I_0^2>0$. Moreover, when $\phi<\frac{bV}{U}$, we can verify that
\begin{align}
{\hat C}_3\left(N\right)|_{N=\frac{Vc_{\max}I_0}{\eta U \left(\delta-\theta I_0\right)}}=\left(\eta U+V\right) c_{\max}^3 I_0^2 \left(\phi-b\frac{V}{U}\right)<0.
\end{align} 
Recall that ${\hat C}_3\left(N\right)$ is continuous, and strictly decreases in $N\in \left(0,\frac{Vc_{\max}I_0}{\eta U \left(\delta-\theta I_0\right)}\right)$. We can conclude that when $\phi<\frac{bV}{U}$ and $\delta>\delta_3$, there is a unique $N\in\left(0,\frac{Vc_{\max}I_0}{\eta U\left(\delta-\theta I_0\right)}\right)$ that satisfies ${\hat C}_3\left(N\right)=0$.
\endproof

\section{Proof of Proposition \ref{proposition:N:no} in Section \ref{subsec:impactN}}\label{SM:proposition12}
\proof
{\bf(Step 1)} We show that when $\delta>\delta_3$, $\phi<\frac{bV}{U}$, and $N<N_B
$, we have $-\phi \ge b\eta-\frac{k\left(\left(c_{\max}-\theta N\right)I_0+\delta N\right)^2}{\delta\left(V+\eta U\right)N^2}$. We prove it by contradiction, i.e., we assume that when $\delta>\delta_3$, $\phi<\frac{bV}{U}$, and $N<N_B$, we have $-\phi < b\eta-\frac{k\left(\left(c_{\max}-\theta N\right)I_0+\delta N\right)^2}{\delta\left(V+\eta U\right)N^2}$. 

Rearranging $-\phi < b\eta-\frac{k\left(\left(c_{\max}-\theta N\right)I_0+\delta N\right)^2}{\delta\left(V+\eta U\right)N^2}$, we obtain the following relation:
\begin{multline}
\left(k\left(\delta-\theta I_0\right)^2-\delta\left(b\eta+\phi\right)\left(V+\eta U\right)\right) N^2 \\
+ 2k\left(\delta-\theta I_0\right) c_{\max} I_0 N +kc_{\max}^2 I_0^2 <0.\label{SM:equ:impactN:3}
\end{multline}
We define function $M\left(N\right)$ as the left-hand side of the above inequality:
\begin{align}
\nonumber
M\left(N\right)\triangleq & \left(k\left(\delta-\theta I_0\right)^2-\delta\left(b\eta+\phi\right)\left(V+\eta U\right)\right) N^2 \\
& + 2k\left(\delta-\theta I_0\right) c_{\max} I_0 N +kc_{\max}^2 I_0^2.
\end{align}
Next, we discuss the case $k\left(\delta-\theta I_0\right)^2-\delta\left(b\eta+\phi\right)\left(V+\eta U\right)\ge0$ and the case $k\left(\delta-\theta I_0\right)^2-\delta\left(b\eta+\phi\right)\left(V+\eta U\right)<0$, separately.   

First, if $k\left(\delta-\theta I_0\right)^2-\delta\left(b\eta+\phi\right)\left(V+\eta U\right)\ge0$, we can utilize the relations $\delta>\delta_3$ and $\delta_3>\theta I_0$ to verify that $M\left(N\right)$ increases in $N\in\left[0,\infty\right)$. We can also see that $M\left(N\right)|_{N=0}=kc_{\max}^2 I_0^2\ge0$. Therefore, when $N<N_B= \frac{Vc_{\max}I_0}{\eta U\left(\delta-\theta I_0\right)}$ and $N>0$ (based on $N$'s definition, $N$ is positive), we have the relation $M\left(N\right)\ge0$, which contradicts with (\ref{SM:equ:impactN:3}). 

Second, if $k\left(\delta-\theta I_0\right)^2-\delta\left(b\eta+\phi\right)\left(V+\eta U\right)<0$, we can utilize the relations $\delta>\delta_3$ and $\delta_3>\theta I_0$ to verify that $M\left(N\right)$ is a unimodal function, and $M\left(N\right)\ge \min\left\{M\left(N\right)|_{N=0}, M\left(N\right)|_{N=N_B}\right\}$ for any $N\in\left(0,N_B\right)$. It is easy to see that $M\left(N\right)|_{N=0}=kc_{\max}^2 I_0^2\ge0$. Next, we investigate the relation between $M\left(N\right)|_{N=N_B}$ and zero. From $N_B= \frac{Vc_{\max}I_0}{\eta U\left(\delta-\theta I_0\right)}$, we can compute $M\left(N\right)|_{N=N_B}$ as
{\small
\begin{align}
\nonumber
& M\left(N\right)|_{N=N_B}=\\
& \frac{c_{\max}^2 I_0^2 }{\eta^2 U^2 \left(\delta\!-\!\theta I_0\right)^2} \left(V\!+\!\eta U\right) \left(\left(\delta\!-\!\theta I_0\right)^2 k \left(V\!+\!\eta U\right)\!-\!\delta \left(b\eta\!+\!\phi\right)V^2\right).\label{SM:equ:impactN:4}
\end{align}}
Based on Lemma \ref{lemma:delta:2}, we have $\delta_3^2-2\theta I_0 \delta_3 -\frac{V^2\left(b\eta+\phi\right)}{k\left(V+\eta U\right)}\delta_3 + \theta^2 I_0^2=0$. Moreover, it is easy to show that $\delta^2-2\theta I_0 \delta -\frac{V^2\left(b\eta+\phi\right)}{k\left(V+\eta U\right)}\delta + \theta^2 I_0^2$ strictly increases with $\delta$ in $\delta\in\left[\delta_3,\infty\right)$. Hence, when $\delta>\delta_3$, we have the relation $\delta^2-2\theta I_0 \delta -\frac{V^2\left(b\eta+\phi\right)}{k\left(V+\eta U\right)}\delta + \theta^2 I_0^2>0$. After rearrangement, we have the relation $k\left(V+\eta U\right) \left(\delta - \theta I_0\right)^2 -{V^2\left(b\eta+\phi\right)}\delta >0$. Then, we can easily see that $M\left(N\right)|_{N=N_B}$ in (\ref{SM:equ:impactN:4}) is non-negative. Based on the results $M\left(N\right)\ge \min\left\{M\left(N\right)|_{N=0}, M\left(N\right)|_{N=N_B}\right\}$ for any $N\in\left(0,N_B\right)$, $M\left(N\right)|_{N=0}\ge0$, and $M\left(N\right)|_{N=N_B}\ge0$, we can obtain the relation $M\left(N\right)\ge0$, which contradicts with (\ref{SM:equ:impactN:3}). 

Combining the discussions for the case $k\left(\delta-\theta I_0\right)^2-\delta\left(b\eta+\phi\right)\left(V+\eta U\right)\ge0$ and the case $k\left(\delta-\theta I_0\right)^2-\delta\left(b\eta+\phi\right)\left(V+\eta U\right)<0$, we can see that when $\delta>\delta_3$, $\phi<\frac{bV}{U}$, and $N<N_B
$, we have $-\phi \ge b\eta-\frac{k\left(\left(c_{\max}-\theta N\right)I_0+\delta N\right)^2}{\delta\left(V+\eta U\right)N^2}$.

{\bf(Step 2)} We show that when $\phi<\frac{bV}{U}$ and $\delta>\delta_3$, $R^{\rm app}\left(l^*,p^*\right)$ decreases with $N$ for $N\in\left(N_A,N_B\right)$. 

Based on our result in {\bf Step 1}, when $\delta>\delta_3$, $\phi<\frac{bV}{U}$, and $N<N_B=\frac{Vc_{\max}I_0}{\eta U\left(\delta-\theta I_0\right)}$, we have $-\phi \ge b\eta-\frac{k\left(\left(c_{\max}-\theta N\right)I_0+\delta N\right)^2}{\delta\left(V+\eta U\right)N^2}=p_3$. Moreover, when $\delta>\delta_3$, we can rearrange the relation $N<\frac{Vc_{\max}I_0}{\eta U\left(\delta-\theta I_0\right)}$ to obtain the relation $I_0>\frac{\delta}{\theta+\frac{Vc_{\max}}{\eta UN}}=I_{\rm th} $. According to Theorem \ref{theorem:optimaltariff} and Corollary \ref{corollary:optimalrevenue}, under $-\phi \ge p_3$ and $I_0>I_{\rm th} $, $R^{\rm app}\left(l^*,p^*\right)$'s expression is given as follows:
\begin{align}
\nonumber
R^{\rm app}\left(l^*,p^*\right)=& \left(b\eta+ \phi\right)N {\frac{Vc_{\max}I_0-\eta UN\delta+\eta UN\theta I_0}{c_{\max}^2I_0+c_{\max}N\delta -c_{\max}N\theta I_0}}\\
& +\phi N\eta\frac{U}{c_{\max}}.
\end{align}
Based on our computation in {\bf Step 4} in Section \ref{SM:proposition11}, the above $R^{\rm app}\left(l^*,p^*\right)$'s derivative with respective to $N$ is $\frac{b\eta c_{\max}\left(\!Vc_{\max}^2 I_0^2 \!+\!2 c_{\max} I_0 \eta U N \left(\theta I_0\!-\!\delta\right)\!-\!\eta U N^2 \left(\delta\!-\!\theta I_0\right)^2 \right)+\phi\left(\eta U+V\right)c_{\max}^3 I_0^2}{\left({c_{\max}^2I_0+c_{\max}N\delta -c_{\max}N\theta I_0}\right)^2}$. Note that the numerator's expression is the same as ${\hat C}_3\left(N\right)$, where ${\hat C}_3\left(N\right)$ is studied in our proof of Lemma \ref{lemma:N:1} in Section \ref{SM:lemma4}. Based on Section \ref{SM:lemma4}, we can show that when $\delta>\delta_3$ and $\phi<\frac{bV}{U}$, function ${\hat C}_3\left(N\right)$ strictly decreases in $N\in \left[N_A,\frac{Vc_{\max}I_0}{\eta U \left(\delta-\theta I_0\right)}\right)$. Moreover, from the definition of $N_A$, we have the relation ${\hat C}_3\left(N\right)|_{N=N_A}=0$. Therefore, when $N>N_A$, we have ${\hat C}_3\left(N\right)<0$. This means $R^{\rm app}\left(l^*,p^*\right)$'s derivative with respective to $N$ is negative. That is to say, when $\phi<\frac{bV}{U}$, $\delta>\delta_3$, and $N_A<N<N_B$, $R^{\rm app}\left(l^*,p^*\right)$ decreases with $N$.
\endproof

\section{Proof of Theorem \ref{theorem:uncertainty} in Section \ref{sec:uncertainty}}\label{SM:theoremuncertainty}
\proof 
The app's expected utility ${{\mathbb E}_{\phi}\left\{J\big(R^{\rm app}\left(l_u,p_u\right) \big)\right\}}$ has different expressions in situation $I_0>I_{\rm th}$, situation $I_0\le I_{\rm th},\delta>\delta_{\rm th}$, and situation $I_0\le I_{\rm th},\delta\le\delta_{\rm th}$. 
To simplify the presentation, we focus on the proof of Theorem \ref{theorem:uncertainty} in situation $I_0>I_{\rm th}$. The proofs of Theorem \ref{theorem:uncertainty} in the remaining two situations are similar and hence are omitted here. 

{\bf(Step 1)} We prove that we only need to focus on $\left(l_u,p_u\right)$ with $l_u={\tilde H}\left(p_u\right)$ to determine the optimal tariff. First, for any $\left(l_u,p_u\right)$ with $l_u<{\tilde H}\left(p_u\right)$, the app can always increase the lump-sum fee $l_u$ to ${\tilde H}\left(p_u\right)$ to improve its expected utility without changing the venue's POI and investment choices. Second, under any $\left(l_u,p_u\right)$ with $l_u>{\tilde H}\left(p_u\right)$, the venue does not become a POI, and the app's expected utility is $J\left(0\right)$. Under $p_u'=0$ and $l_u'={\tilde H}\left(p_u'\right)={\tilde H}\left(0\right)$, the app's revenue is
\begin{align}
R^{\rm app}\left(l_u',p_u'\right) ={\tilde H}\left(0\right)+\phi N{\bar y}\left(r^*\left(l_u',p_u'\right),I^*\left(l_u',p_u'\right)\right).
\end{align}
In Corollary \ref{corollary:optimalrevenue}, we have shown that ${\tilde H}\left(-\phi\right)\ge0$ for any $\phi\ge0$. Hence, it is easy to see that ${\tilde H}\left(0\right)\ge0$. Since function ${\bar y}\left(\cdot,\cdot\right)$ is the fraction of users interacting with the POI, we have ${\bar y}\left(r^*\left(l_u',p_u'\right),I^*\left(l_u',p_u'\right)\right)\ge0$. Therefore, we can see that $R^{\rm app}\left(l_u',p_u'\right)\ge0$. 

Since  $J'\left(z\right)\ge0$ for all $z\in{\mathbb R}$, we have the relation ${{\mathbb E}_{\phi}\left\{J\big(R^{\rm app}\left(l_u',p_u'\right)  \big)\right\}}\ge J\left(0\right)$. This implies that $\left(l_u',p_u'\right)$, which satisfies $l_u'={\tilde H}\left(p_u'\right)$, achieves at least the same app's expected utility as all $\left(l_u,p_u\right)$ with $l_u>{\tilde H}\left(p_u\right)$. 
Therefore, we can see that we only need to focus on $\left(l_u,p_u\right)$ with $l_u={\tilde H}\left(p_u\right)$ to determine the optimal tariff.

{\bf(Step 2)} We show $R^{\rm app}\left(l_u,p_u\right)$'s concrete expression under $l_u={\tilde H}\left(p_u\right)$ and $I_0>I_{\rm th}$, and compute $\frac{d {{\mathbb E}_{\phi}\left\{J\big(R^{\rm app}\left({\tilde H}\left(p_u\right),p_u\right) \big)\right\}}}{d p_u}$. Recall that when $l_u={\tilde H}\left(p_u\right)$, we can write $R^{\rm app}\left(l_u,p_u\right)$ as
\begin{align}
R^{\rm app}\left(l_u,p_u\right)={\tilde H}\left(p_u\right)+\left(\phi+p_u\right) N{\bar y}\left(1,I^*\left({\tilde H}\left(p_u\right),p_u\right)\right).
\end{align}
According to the concrete expressions of ${\tilde H}\left(p_u\right)$ and ${\bar y}\left(1,I^*\left({\tilde H}\left(p_u\right),p_u\right)\right)$ under $I_0>I_{\rm th}$, we show $R^{\rm app}\left({\tilde H}\left(p_u\right),p_u\right)$ as (\ref{SM:theorem2:long}).

\begin{figure*}
\begin{align}
\nonumber
& R^{\rm app}\left({\tilde H}\left(p_u\right),p_u\right)=\\
& \left\{ {\begin{array}{*{20}{l}}
{\!\!\!-\frac{N}{c_{\max}}b\eta^2U+\frac{N}{c_{\max}-N\theta}\left(\sqrt{\left(V+\eta U\right)\left(b\eta -p_u\right)}-\sqrt{\delta k}\right)^2+kI_0+\frac{\left(\phi+p_u\right) N}{c_{\max}-N\theta} \left({\eta U+V}-\sqrt{\frac{k\delta\left(V+\eta U\right)}{b\eta-p_u}}\right) ,}&{{\rm if~}p_u< p_3,}\\
{\!\!\!{\left(b\eta+\phi\right)N {\frac{Vc_{\max}I_0-\eta UN\delta+\eta UN\theta I_0}{c_{\max}^2I_0+c_{\max}N\delta -c_{\max}N\theta I_0}}+\phi N\eta\frac{U}{c_{\max}}},}&{{\rm if~}p_u\ge p_3.}
\end{array}} \right.\label{SM:theorem2:long}
\end{align}
\hrulefill

\begin{align}
\nonumber
& O\left(p_u,\phi\right)\triangleq \\
& \left\{ {\begin{array}{*{20}{l}}
{\!\!\!-\frac{N}{c_{\max}}b\eta^2U+\frac{N}{c_{\max}-N\theta}\left(\sqrt{\left(V+\eta U\right)\left(b\eta -p_u\right)}-\sqrt{\delta k}\right)^2+kI_0+\frac{\left(\phi+p_u\right) N}{c_{\max}-N\theta} \left({\eta U+V}-\sqrt{\frac{k\delta\left(V+\eta U\right)}{b\eta-p_u}}\right) ,}&{{\rm if~}p_u< p_3,}\\
{\!\!\!{\left(b\eta+\phi\right)N {\frac{Vc_{\max}I_0-\eta UN\delta+\eta UN\theta I_0}{c_{\max}^2I_0+c_{\max}N\delta -c_{\max}N\theta I_0}}+\phi N\eta\frac{U}{c_{\max}}},}&{{\rm if~}p_u\ge p_3.}
\end{array}} \right.\label{SM:theorem2:longsame}
\end{align}
\hrulefill
\end{figure*}

We can verify that $R^{\rm app}\left({\tilde H}\left(p_u\right),p_u\right)$ is continuous for $p_u\in\left(-\infty,\infty\right)$. Furthermore, when $p_u<p_3$, we compute 
\begin{align}
\frac{d \!R^{\rm app}\!\left(\!{\tilde H}\left(p_u\right)\!,p_u\!\right)}{d p_u}\!=\! -\frac{1}{2}\frac{N \sqrt{k\delta \left(V\!+\!\eta U\right)}}{c_{\max}-N\theta} \!\left(\phi\!+\!p_u\right) \left(b\eta\!-\!p_u\right)^{-\frac{3}{2}}\!;
\end{align}
when $p_u\ge p_3$, $R^{\rm app}\left({\tilde H}\left(p_u\right),p_u\right)$ does not change with $p_u$. 

We define $O\left(p_u,\phi\right)$ as (\ref{SM:theorem2:longsame}). Function $O\left(p_u,\phi\right)$ has the same expression as $R^{\rm app}\left({\tilde H}\left(p_u\right),p_u\right)$ (under the same $p_u$ and $\phi$). We use function $O\left(p_u,\phi\right)$ to capture the dependence of the expression's value on $\phi$. 

When $p_u<p_3$, we can compute $\frac{d {{\mathbb E}_{\phi}\left\{J\big(R^{\rm app}\left({\tilde H}\left(p_u\right),p_u\right) \big)\right\}}}{d p_u}$ as
{\small\begin{align}
\nonumber
& \frac{d {{\mathbb E}_{\phi}\left\{J\big(R^{\rm app}\left({\tilde H}\left(p_u\right),p_u\right) \big)\right\}}}{d p_u}= \\
& \!-\frac{1}{2}\!\frac{N \sqrt{k\delta \!\left(V\!+\!\eta U\right)}}{c_{\max}-N\theta} \! \left(b\eta\!-\!p_u\right)^{-\frac{3}{2}}\! \cdot \!\int_{\phi_{\min}}^{\phi_{\max}} \!\!J'\big( \!O\!\left(p_u,\phi\right)\! \big)\!   \left(\phi\!+\!p_u\right)\! d\!F\left(\phi\right),\label{SM:derivative:a}
\end{align}}
where $F\left(\phi\right)$ is the cumulative distribution function of $\phi$. When $p_u\ge p_3$, $\frac{d {{\mathbb E}_{\phi}\left\{J\big(R^{\rm app}\left({\tilde H}\left(p_u\right),p_u\right) \big)\right\}}}{d p_u}=0$. 

{\bf(Step 3)} We prove that for any utility function $J\left(z\right)$, we only need to consider $p_u\in\left[-\phi_{\max},-\phi_{\min}\right]$ to determine the optimal $p_u^*$. 

First, when $p_u<-\phi_{\max}$ and $p_u<p_3$, we have $\phi+p_u<0$ for all $\phi\in\left[\phi_{\min},\phi_{\max}\right]$. Recall that $J'\left(z\right)\ge0$ for all $z\in\left(-\infty,\infty\right)$. Based on (\ref{SM:derivative:a}), we have $\frac{d {{\mathbb E}_{\phi}\left\{J\big(R^{\rm app}\left({\tilde H}\left(p_u\right),p_u\right) \big)\right\}}}{d p_u}\ge 0$. When $p_u<-\phi_{\max}$ and $p_u\ge p_3$, $\frac{d {{\mathbb E}_{\phi}\left\{J\big(R^{\rm app}\left({\tilde H}\left(p_u\right),p_u\right) \big)\right\}}}{d p_u}=0$. We can see that ${{\mathbb E}_{\phi}\left\{J\big(R^{\rm app}\left({\tilde H}\left(p_u\right),p_u\right) \big)\right\}}$ is increasing in $p_u\in\left(-\infty,-\phi_{\max}\right)$. 

Second, when $p_u>-\phi_{\min}$ and $p_u<p_3$, we have $\phi+p_u>0$ for all $\phi\in\left[\phi_{\min},\phi_{\max}\right]$. Recall that $J'\left(z\right)\ge0$ for all $z\in\left(-\infty,\infty\right)$. Based on (\ref{SM:derivative:a}), we have $\frac{d {{\mathbb E}_{\phi}\left\{J\big(R^{\rm app}\left({\tilde H}\left(p_u\right),p_u\right) \big)\right\}}}{d p_u}\le 0$. When $p_u>-\phi_{\min}$ and $p_u\ge p_3$, $\frac{d {{\mathbb E}_{\phi}\left\{J\big(R^{\rm app}\left({\tilde H}\left(p_u\right),p_u\right) \big)\right\}}}{d p_u}=0$. We can see that ${{\mathbb E}_{\phi}\left\{J\big(R^{\rm app}\left({\tilde H}\left(p_u\right),p_u\right) \big)\right\}}$ is decreasing in $p_u\in\left(-\phi_{\min},\infty\right)$. 

Combining the analysis above, we show that we only need to consider $p_u\in\left[-\phi_{\max},-\phi_{\min}\right]$ to determine the optimal $p_u^*$. 

{\bf(Step 4)} We prove that when $J''\left(z\right)\le 0$ for $z\in\left(-\infty,\infty\right)$, ${{\mathbb E}_{\phi}\left\{J\big(R^{\rm app}\left({\tilde H}\left(p_u\right),p_u\right) \big)\right\}}$ is increasing in $p_u\in\left[-\phi_{\max},-{\mathbb{E}}\left\{\phi\right\}\right]$. 

As we can see in (\ref{SM:derivative:a}), the sign of $\frac{d {{\mathbb E}_{\phi}\left\{J\big(R^{\rm app}\left({\tilde H}\left(p_u\right),p_u\right) \big)\right\}}}{d p_u}$ under $p_u<p_3$ is mainly affected by the sign of $\int_{\phi_{\min}}^{\phi_{\max}} J'\big( O\left(p_u,\phi\right) \big)   \left(\phi+p_u\right) dF\left(\phi\right)$. Next, we analyze $\int_{\phi_{\min}}^{\phi_{\max}} J'\big( O\left(p_u,\phi\right) \big)   \left(\phi+p_u\right) dF\left(\phi\right)$. When $p_u\in\left[-\phi_{\max},-\phi_{\min}\right]$, we have
\begin{align}
\nonumber
& \int_{\phi_{\min}}^{\phi_{\max}} J'\big( O\left(p_u,\phi\right) \big)   \left(\phi+p_u\right) dF\left(\phi\right)=\\
\nonumber
&{~~~~~~}\int_{\phi_{\min}}^{-p_u} J'\big( O\left(p_u,\phi\right) \big)   \left(\phi+p_u\right) dF\left(\phi\right)\\
&{~~~~~~}+\int_{-p_u}^{\phi_{\max}} J'\big( O\left(p_u,\phi\right) \big)   \left(\phi+p_u\right) dF\left(\phi\right). \label{SM:theorem2:indexcombine}
\end{align}
When $\phi\in\left[\phi_{\min},-p_u\right]$, we have $\phi+p_u\le0$. Moreover, it is easy to verify that $O\left(p_u,\phi\right)$ is increasing in $\phi$. Hence, we have $O\left(p_u,\phi\right)\le O\left(p_u,-p_u\right)$. Because $J''\left(z\right)\le 0$ for $z\in\left(-\infty,\infty\right)$, we have $J'\big( O\left(p_u,\phi\right) \big)\ge J'\big( O\left(p_u,-p_u\right) \big)$. 
Then, we have the relation
\begin{align}
\nonumber
& \int_{\phi_{\min}}^{-p_u} J'\big( O\left(p_u,\phi\right) \big)   \left(\phi+p_u\right) dF\left(\phi\right)\\
& \le \int_{\phi_{\min}}^{-p_u} J'\big( O\left(p_u,-p_u\right) \big)   \left(\phi+p_u\right) dF\left(\phi\right).\label{SM:theorem2:index1}
\end{align}
When $\phi\in\left[-p_u,\phi_{\max}\right]$, we have $\phi+p_u\ge0$. Moreover, we have $O\left(p_u,\phi\right)\ge O\left(p_u,-p_u\right)$. Because $J''\left(z\right)\le 0$ for $z\in\left(-\infty,\infty\right)$, we have $J'\big( O\left(p_u,\phi\right) \big)\le J'\big( O\left(p_u,-p_u\right) \big)$. 
Then, we have the relation
\begin{align}
\nonumber
& \int_{-p_u}^{\phi_{\max}} J'\big( O\left(p_u,\phi\right) \big)   \left(\phi+p_u\right) dF\left(\phi\right)\\
& \le\int_{-p_u}^{\phi_{\max}} J'\big( O\left(p_u,-p_u\right) \big)   \left(\phi+p_u\right) dF\left(\phi\right).\label{SM:theorem2:index2}
\end{align}
Based on (\ref{SM:theorem2:indexcombine}), (\ref{SM:theorem2:index1}), and (\ref{SM:theorem2:index2}), we can derive the following result:
\begin{align}
\nonumber
&\int_{\phi_{\min}}^{\phi_{\max}} J'\big( O\left(p_u,\phi\right) \big)   \left(\phi+p_u\right) dF\left(\phi\right)\\
\nonumber
\le &\int_{\phi_{\min}}^{-p_u} J'\big( O\left(p_u,-p_u\right) \big)   \left(\phi+p_u\right) dF\left(\phi\right) \\
\nonumber
& +\int_{-p_u}^{\phi_{\max}} J'\big( O\left(p_u,-p_u\right) \big)   \left(\phi+p_u\right) dF\left(\phi\right)\\
\nonumber
=& J'\big( O\left(p_u,-p_u\right) \big) \int_{\phi_{\min}}^{\phi_{\max}}    \left(\phi+p_u\right) dF\left(\phi\right)\\
=& J'\big( O\left(p_u,-p_u\right) \big) \left(p_u+{\mathbb E}\left\{\phi\right\}\right).
\end{align}
We can see that if $p_u\le -{\mathbb E}\left\{\phi\right\}$, $\int_{\phi_{\min}}^{\phi_{\max}} J'\big( O\left(p_u,\phi\right) \big)   \left(\phi+p_u\right) dF\left(\phi\right)\le0$. Considering the analysis of $\frac{d {{\mathbb E}_{\phi}\left\{J\big(R^{\rm app}\left({\tilde H}\left(p_u\right),p_u\right) \big)\right\}}}{d p_u}$ in {\bf Step 2}, we conclude that if $p_u\le -{\mathbb E}\left\{\phi\right\}$, $\frac{d {{\mathbb E}_{\phi}\left\{J\big(R^{\rm app}\left({\tilde H}\left(p_u\right),p_u\right) \big)\right\}}}{d p_u}\ge0$. Therefore, we have proved that when $J''\left(z\right)\le 0$ for $z\in\left(-\infty,\infty\right)$, ${{\mathbb E}_{\phi}\left\{J\big(R^{\rm app}\left({\tilde H}\left(p_u\right),p_u\right) \big)\right\}}$ is increasing in $p_u\in\left[-\phi_{\max},-{\mathbb{E}}\left\{\phi\right\}\right]$. 

{\bf (Step 5)} Based on similar approaches as {\bf Step 4}, we can prove that when $J''\left(z\right)\ge 0$ for $z\in\left(-\infty,\infty\right)$, ${{\mathbb E}_{\phi}\left\{J\big(R^{\rm app}\left({\tilde H}\left(p_u\right),p_u\right) \big)\right\}}$ is decreasing in $p_u\in\left[-{\mathbb{E}}\left\{\phi\right\},-\phi_{\min}\right]$. We omit the details here. 

{\bf (Step 6)} Now we are ready to conclude the proof. 

When $J''\left(z\right)=0,z\in{\mathbb R}$, based on {\bf Step 4} and {\bf Step 5}, ${{\mathbb E}_{\phi}\left\{J\big(R^{\rm app}\left({\tilde H}\left(p_u\right),p_u\right) \big)\right\}}$ is increasing in $p_u\in\left[-\phi_{\max},-{\mathbb{E}}\left\{\phi\right\}\right]$ and decreasing in $p_u\in\left[-{\mathbb{E}}\left\{\phi\right\},-\phi_{\min}\right]$. Based on {\bf Step 3}, $p_u^*$ lies in $\left[-\phi_{\max},-\phi_{\min}\right]$. Hence, we have $p_u^*=-\mathbb{E}\left\{\phi\right\}$.

Based on {\bf Step 4}, when $J''\left(z\right)\le0,z\in{\mathbb R}$, ${{\mathbb E}_{\phi}\left\{J\big(R^{\rm app}\left({\tilde H}\left(p_u\right),p_u\right) \big)\right\}}$ is increasing in $p_u\in\left[-\phi_{\max},-{\mathbb{E}}\left\{\phi\right\}\right]$. Based on {\bf Step 3}, $p_u^*$ lies in $\left[-\phi_{\max},-\phi_{\min}\right]$. Hence, we have $-\mathbb{E}\left\{\phi\right\} \le p_u^* \le -\phi_{\min}$.

Based on {\bf Step 5}, when $J''\left(z\right)\ge0,z\in{\mathbb R}$, ${{\mathbb E}_{\phi}\left\{J\big(R^{\rm app}\left({\tilde H}\left(p_u\right),p_u\right) \big)\right\}}$ is decreasing in $p_u\in\left[-{\mathbb{E}}\left\{\phi\right\},-\phi_{\min}\right]$. Based on {\bf Step 3}, $p_u^*$ lies in $\left[-\phi_{\max},-\phi_{\min}\right]$. Hence, we have $-\phi_{\max} \le p_u^*\le-\mathbb{E}\left\{\phi\right\}$. 

Based on {\bf Step 1}, for any $J\left(\cdot\right)$, we have $l_u^*={\tilde H}\left(p_u^*\right)$. Note that function ${\tilde H}\left(p_u\right)$ is decreasing in $p_u$. According to the relation between $p_u^*$ and $-\mathbb{E}\left\{\phi\right\}$ for $J''\left(z\right)=0$, $J''\left(z\right)<0$, and $J''\left(z\right)>0$, we can easily derive the corresponding relation between $l_u^*$ and ${\tilde H}\left(-\mathbb{E}\left\{\phi\right\}\right)$.
\endproof

\vspace{0.5cm}



\section{Model Discussions}

\subsection{Modeling Congestion in General Infrastructure}\label{SM:model:congestion}

{{In this work, we focus on the venue's investment in app-related infrastructure, and assume that the infrastructure (such as Wi-Fi networks and smartphone chargers) is mainly used by the users interacting with the POI. It is interesting to study the scenario where the infrastructure is used by all users visiting the venue and the users who do not interact with the POI also cause congestion. Next, we briefly discuss the potential changes in the model and results when we study this scenario.
      
First, we need to modify the users' payoff function in Eq. (\ref{equ:userpayoff}). The original congestion term is $-\frac{\delta}{I+I_0}{\bar y}\left(r,I\right)N$, where ${\bar y}\left(r,I\right)$ is the fraction of users interacting with the POI. {{Only the payoffs of the users choosing $d=2$ (i.e., visiting the venue and interacting with the POI) include this congestion term.}} In the modified model, we need to define a new function, ${\bar z}\left(r,I\right)$, to capture the fraction of users who visit the venue, and modify the congestion term as $-\frac{\delta}{I+I_0}{\bar z}\left(r,I\right)N$. Then, {{the payoffs of both the users choosing $d=1$ (i.e., visiting the venue without interacting with the POI) and the users choosing $d=2$ will include this new congestion term.}}      
      
      
Second, the results of different decision makers' strategies will change. Because the congestion term changes from $-\frac{\delta}{I+I_0}{\bar y}\left(r,I\right)N$ to $-\frac{\delta}{I+I_0}{\bar z}\left(r,I\right)N$, we need to apply backward induction to analyze the new three-stage game and derive new results about the app's, venue's, and users' strategies at the equilibrium. Intuitively, since the users who do not interact with the POI also suffer from the congestion of using infrastructure, the venue may invest in the infrastructure even without any payment from the app.}}

\subsection{Modeling Non-Monetary Rewards}\label{SM:model:nonmonetary}

{{In this section, we use an example to briefly discuss the extension of our framework to a case where the app-venue collaboration is based on non-monetary rewards. 


      Suppose that the app does not charge the venue based on a two-part tariff. Instead, the app displays information about the venue to the players who use the app at the venue. For example, when the venue is a restaurant, the app can display information about the venue's food recommendations.
      
      {{To analyze this app-venue collaboration, we can modify the venue's payoff function and the app's revenue function as follows:}} 
      \begin{itemize}
      \item We can modify the venue's payoff function by replacing the payment term (i.e., the payment from the venue to the app) by a new term, which captures the increase in the venue's profit due to the app's display of the venue's information. Intuitively, if the app increases the degree of displaying the venue's information, the venue will increase its investment level on the app-related infrastructure, which enables more users to use the app. 
      \item When the app increases the degree of displaying the venue's information, the app will have less space to advertise for other advertisers. Hence, we also need to modify the app's revenue function by replacing the venue's payment term by an \emph{ad revenue loss term}, which describes the loss in the ad revenue due to the display of the venue's information. In this modified model, the app's problem is to determine the degree of displaying the venue's information rather than the lump-sum fee and per-player charge.
      \end{itemize}

We can still apply backward induction to analyze the multi-stage game and derive the app's and venue's optimal strategies.
}}


\begin{figure*}
\begin{align}
& \max \Bigl(R^{\rm app}\left(l_b,p_b\right) \Bigr)^{\gamma} \Bigl( \Pi^{\rm venue}\left({r^*}\left(l_b,p_b\right),I^*\left(l_b,p_b\right),l_b,p_b\right)-b\eta N\frac{U}{c_{\max}} \Bigr)^{1-\gamma} \label{SM:bargain:obj}\\
& {\rm s.t.~~~} R^{\rm app}\left(l_b,p_b\right)\ge0, \Pi^{\rm venue}\left({r^*}\left(l_b,p_b\right),I^*\left(l_b,p_b\right),l_b,p_b\right)\ge b\eta N\frac{U}{c_{\max}},\label{SM:bargain:constraint}\\
& {\rm var.~~~} l_b,p_b\in{\mathbb R}.\label{SM:bargain:var}
\end{align}
\hrulefill
\end{figure*}

\section{Bargaining Between App and Venue}\label{SM:bargaining}
In Section \ref{sec:stageI}, we study the case where the app has the market power and directly determines the tariff. In this section, we consider a bargaining-based negotiation between the app and venue in Stage I, where the app bargains with the venue to decide the tariff. 
We formulate the bargaining between the app and venue as Problem \ref{SM:problem:bargain}. 
Note that the bargaining formulation in Stage I does not change the analysis and results in Stages II and III. 

\begin{problem}\label{SM:problem:bargain}
The app and venue determine $\left(l_b^*,p_b^*\right)$ by solving (\ref{SM:bargain:obj})-(\ref{SM:bargain:var}).
\end{problem}

When the app and venue cannot reach an agreement on the tariff, the app will not tag the venue as a POI. In this case, the app's revenue is $0$, and the venue's payoff is $b\eta N\frac{U}{c_{\max}}$. As shown in (\ref{SM:bargain:constraint}), the bargaining solution $\left(l_b,p_b\right)$ ensures that the app's revenue and venue's payoff will be no smaller than $0$ and $b\eta N\frac{U}{c_{\max}}$, respectively. Furthermore, the optimal bargaining solution $\left(l_b^*,p_b^*\right)$ maximizes the product between $\Bigl(R^{\rm app}\left(l_b,p_b\right) \Bigr)^{\gamma}$ and $\Bigl( \Pi^{\rm venue}\left({r^*}\left(l_b,p_b\right),I^*\left(l_b,p_b\right),l_b,p_b\right)-b\eta N\frac{U}{c_{\max}} \Bigr)^{1-\gamma} $. Here, $R^{\rm app}\left(l_b,p_b\right)=R^{\rm app}\left(l_b,p_b\right)-0$ captures the difference between the app's revenue under $\left(l_b,p_b\right)$ and its revenue when it cannot reach an agreement with the venue (i.e., $0$). $\Pi^{\rm venue}\left({r^*}\left(l_b,p_b\right),I^*\left(l_b,p_b\right),l_b,p_b\right)-b\eta N\frac{U}{c_{\max}}$ is the difference between the venue's payoff under $\left(l_b,p_b\right)$ and its payoff when it cannot reach an agreement with the app. $\gamma\in\left[0,1\right]$ and $1-\gamma\in\left[0,1\right]$ represent the app's and venue's bargaining powers, respectively. When $\gamma=1$, Problem \ref{SM:problem:bargain} degenerates to Problem \ref{problem:app}, where the app has the market power and directly determines the tariff.  

We show the optimal bargaining solution in Theorem \ref{SM:theorem:bargaining}.
\begin{theorem}\label{SM:theorem:bargaining}
The two-part tariff under the bargaining is
\begin{align}
p_b^*=-\phi, l_b^*=\gamma {\tilde H}\left(-\phi\right).
\end{align}
\end{theorem}
\proof
{\bf(Step 1)} We show that we only need to focus on the $\left(l_b,p_b\right)$ satisfying $r^*\left(l_b,p_b\right)=1$ to determine the optimal bargaining solution. 
For $\left(l_b,p_b\right)$ with $r^*\left(l_b,p_b\right)=0$, we can see that the corresponding value of the objective function (\ref{SM:bargain:obj}) is zero. When $l_b=\gamma {\tilde H}\left(-\phi\right)$ and $p_b=-\phi$, we can see that $r^*\left(l_b,p_b\right)=1$, and compute the value of the objective function (\ref{SM:bargain:obj}) as $\gamma \left(1-\gamma\right)  {\tilde H}^2\left(-\phi\right)$, which is no less than zero. Because it is easy to check that $l_b=\gamma {\tilde H}\left(-\phi\right)$ and $p_b=-\phi$ satisfy $r^*\left(l_b,p_b\right)=1$, we only need to focus on the $\left(l_b,p_b\right)$ satisfying $r^*\left(l_b,p_b\right)=1$ to determine the optimal bargaining solution. 

{\bf(Step 2)} We prove that for any feasible bargaining solution $\left(l_b,p_b\right)$ with $p_b\ne -\phi$ and $r^*\left(l_b,p_b\right)=1$, we can always ensure that $p_b'=-\phi$ and 
\begin{align}
\nonumber
l_b'=& b N {\bar x}\left(1,I^*\left({\tilde H}\left(-\phi\right),-\phi\right)\right)-k I^*\left({\tilde H}\left(-\phi\right),-\phi\right)\\
\nonumber
& +  \phi N {\bar y}\left(1,I^*\left({\tilde H}\left(-\phi\right),-\phi\right)\right)-b N {\bar x}\left(1,I^*\left(l_b,p_b\right)\right)\\
& +k I^*\left(l_b,p_b\right)+  p_bN {\bar y}\left(1,I^*\left(l_b,p_b\right)\right)+l_b\label{SM:equ:lbprime}
\end{align}
constitute a feasible bargaining solution and achieve at least the same objective function's value as $p_b$ and $l_b$.

First, we prove $r^*\left(l_b',p_b'\right)=1$. When the venue does not become a POI under $\left(l_b',p_b'\right)$, it does not invest and has a payoff of $b\eta N\frac{U}{c_{\max}}$. When the venue becomes a POI and chooses $I^*\left({\tilde H}\left(-\phi\right),p_b'\right)$ as its investment level, its payoff becomes $b N {\bar x}\left(1,I^*\left({\tilde H}\left(-\phi\right),p_b'\right)\right)-k I^*\left({\tilde H}\left(-\phi\right),p_b'\right)-  \left(l_b'+p_b'N {\bar y}\left(1,I^*\left({\tilde H}\left(-\phi\right),p_b'\right)\right)\right)$. Plugging the expression of $l_b'$, we can rewrite the venue's payoff as $b N {\bar x}\left(1,I^*\left(l_b,p_b\right)\right)-k I^*\left(l_b,p_b\right)-  p_bN {\bar y}\left(1,I^*\left(l_b,p_b\right)\right)-l_b$, which is the same as the venue's optimal payoff under $\left(l_b,p_b\right)$. Note that the venue's optimal payoff under $\left(l_b,p_b\right)$ is no less than $b\eta N\frac{U}{c_{\max}}$. This implies that under $\left(l_b',p_b'\right)$, when the venue becomes a POI and chooses $I^*\left({\tilde H}\left(-\phi\right),p_b'\right)$, its payoff is no less that $b\eta N\frac{U}{c_{\max}}$. 
Recall that $b\eta N\frac{U}{c_{\max}}$ is the venue's payoff when it does not become a POI under $\left(l_b',p_b'\right)$. We can see that the venue should always become a POI under $\left(l_b',p_b'\right)$, i.e., $r^*\left(l_b',p_b'\right)=1$.

Second, we prove that $R^{\rm app}\left(l_b',p_b'\right)\ge R^{\rm app}\left(l_b,p_b\right)\ge0$. Note that $R^{\rm app}\left(l_b,p_b\right)$ can be written as
\begin{align}
R^{\rm app}\left(l_b,p_b\right)=l_b+\left(p_b+\phi\right) N{\bar y}\left(1,I^*\left(l_b,p_b\right)\right).
\end{align}
Since $\left(l_b,p_b\right)$ is feasible, $R^{\rm app}\left(l_b,p_b\right)$ is no less than $0$. Then, we compute $R^{\rm app}\left(l_b',p_b'\right)$ as
\begin{align}
R^{\rm app}\left(l_b',p_b'\right) =l_b'+\left(p_b'+\phi\right) N{\bar y}\left(1,I^*\left(l_b',p_b'\right)\right)=l_b'. \label{SM:equ:tired:end}
\end{align}
According to Propositions \ref{proposition:stageII:situationI}, \ref{proposition:stageII:situationII}, and \ref{proposition:stageII:situationIII}, when the venue becomes a POI, its investment level is not affected by the lump-sum fee. 
Since $r^*\left({\tilde H}\left(-\phi\right),p_b'\right)=r^*\left(l_b',p_b'\right)=1$, we have $I^*\left({\tilde H}\left(-\phi\right),-\phi\right)=I^*\left(l_b',-\phi\right)$. Therefore, we can rewrite $l_b'$ in (\ref{SM:equ:lbprime}) as
\begin{align}
\nonumber
 l_b'=& b N {\bar x}\left(1,I^*\left(l_b',-\phi\right)\right)-k I^*\left(l_b',-\phi\right)\\
\nonumber
& +  \phi N {\bar y}\left(1,I^*\left(l_b',-\phi\right)\right)-b N {\bar x}\left(1,I^*\left(l_b,p_b\right)\right)\\
& +k I^*\left(l_b,p_b\right)+  p_bN {\bar y}\left(1,I^*\left(l_b,p_b\right)\right)+l_b.\label{SM:equ:lbprimenew}
\end{align}
By plugging the above equation to (\ref{SM:equ:tired:end}), we have 
\begin{align}
\nonumber
R^{\rm app}\left(l_b',p_b'\right) =& l_b+b N {\bar x}\left(1,I^*\left(l_b',-\phi\right)\right)-k I^*\left(l_b',-\phi\right)\\
\nonumber
& +  \phi N {\bar y}\left(1,I^*\left(l_b',-\phi\right)\right)-b N {\bar x}\left(1,I^*\left(l_b,p_b\right)\right)\\
& +k I^*\left(l_b,p_b\right)+  p_bN {\bar y}\left(1,I^*\left(l_b,p_b\right)\right).
\end{align}
Next, we compare $R^{\rm app}\left(l_b',p_b'\right)$ with $R^{\rm app}\left(l_b,p_b\right)$. We compute $R^{\rm app}\left(l_b',p_b'\right)-R^{\rm app}\left(l_b,p_b\right)$ as 
\begin{align}
\nonumber
& b N {\bar x}\left(1,I^*\left(l_b',-\phi\right)\right)-k I^*\left(l_b',-\phi\right)+  \phi N {\bar y}\left(1,I^*\left(l_b',-\phi\right)\right)\\
& -b N {\bar x}\left(1,I^*\left(l_b,p_b\right)\right)+k I^*\left(l_b,p_b\right)- \phi N {\bar y}\left(1,I^*\left(l_b,p_b\right)\right). \label{SM:equ:gap}
\end{align}
Based on $r^*\left(l_b',p_b'\right)=r^*\left(l_b',-\phi\right)=1$ and the definition of $I^*\left(l_b',-\phi\right)$, we have
\begin{align}
I^*\left(l_b',-\phi\right)=\mathop{\arg\max}_{I\ge0} b N {\bar x}\left(1,I\right)-k I+\phi N {\bar y}\left(r,I\right).
\end{align}
Hence, (\ref{SM:equ:gap}) is always non-negative. This implies that $R^{\rm app}\left(l_b',p_b'\right)\ge R^{\rm app}\left(l_b,p_b\right)$. Moreover, since $R^{\rm app}\left(l_b,p_b\right)\ge0$ (according to the feasibility of $\left(l_b,p_b\right)$), we can see that $R^{\rm app}\left(l_b',p_b'\right)\ge R^{\rm app}\left(l_b,p_b\right)\ge0$. 

Third, we prove $\Pi^{\rm venue}\left({r^*}\left(l_b',p_b'\right),I^*\left(l_b',p_b'\right),l_b',p_b'\right)=\Pi^{\rm venue}\left({r^*}\left(l_b,p_b\right),I^*\left(l_b,p_b\right),l_b,p_b\right) \ge b\eta N\frac{U}{c_{\max}}$. The expression of $\Pi^{\rm venue}\left({r^*}\left(l_b,p_b\right),I^*\left(l_b,p_b\right),l_b,p_b\right)$ is \begin{align}
\nonumber
& \Pi^{\rm venue}\left({r^*}\left(l_b,p_b\right),I^*\left(l_b,p_b\right),l_b,p_b\right)=b N {\bar x}\left(1,I^*\left(l_b,p_b\right)\right)\\
&{~~~~~~~~~~} -k I^*\left(l_b,p_b\right)-  \left(l_b+p_bN {\bar y}\left(1,I^*\left(l_b,p_b\right)\right)\right).\label{SM:equ:bargain:first}
\end{align}
Since $\left(l_b,p_b\right)$ is feasible, the above expression is no less than $b\eta N\frac{U}{c_{\max}}$. Then, the expression of $\Pi^{\rm venue}\left({r^*}\left(l_b',p_b'\right),I^*\left(l_b',p_b'\right),l_b',p_b'\right)$ is
\begin{align}
\nonumber
& \Pi^{\rm venue}\left({r^*}\left(l_b',p_b'\right),I^*\left(l_b',p_b'\right),l_b',p_b'\right)=b N {\bar x}\left(1,I^*\left(l_b',p_b'\right)\right)\\
&{~~~~~~~~~~} -k I^*\left(l_b',p_b'\right)-  \left(l_b'+p_b'N {\bar y}\left(1,I^*\left(l_b',p_b'\right)\right)\right).
\end{align}
By plugging the expression of $l_b'$ in (\ref{SM:equ:lbprimenew}) to the above equation, we have
\begin{align}
\nonumber
& \Pi^{\rm venue}\left({r^*}\left(l_b',p_b'\right),I^*\left(l_b',p_b'\right),l_b',p_b'\right)=b N {\bar x}\left(1,I^*\left(l_b,p_b\right)\right)\\
&{~~~~~~~~~~} -k I^*\left(l_b,p_b\right) -  \left(l_b+p_bN {\bar y}\left(1,I^*\left(l_b,p_b\right)\right)\right).\label{SM:equ:bargain:second}
\end{align}
Comparing (\ref{SM:equ:bargain:second}) with (\ref{SM:equ:bargain:first}), we find that
\begin{align}
\nonumber
& \Pi^{\rm venue}\left({r^*}\left(l_b',p_b'\right),I^*\left(l_b',p_b'\right),l_b',p_b'\right)=\\
& \Pi^{\rm venue}\left({r^*}\left(l_b,p_b\right),I^*\left(l_b,p_b\right),l_b,p_b\right).
\end{align}
Because $\Pi^{\rm venue}\left({r^*}\left(l_b,p_b\right),I^*\left(l_b,p_b\right),l_b,p_b\right)\ge b\eta N\frac{U}{c_{\max}}$, we have $\Pi^{\rm venue}\left({r^*}\left(l_b',p_b'\right),I^*\left(l_b',p_b'\right),l_b',p_b'\right)=\Pi^{\rm venue}\left({r^*}\left(l_b,p_b\right),I^*\left(l_b,p_b\right),l_b,p_b\right) \ge b\eta N\frac{U}{c_{\max}}$. 

Fourth, since we have proved the relation $R^{\rm app}\left(l_b',p_b'\right)\ge R^{\rm app}\left(l_b,p_b\right)$ and relation $\Pi^{\rm venue}\left({r^*}\left(l_b',p_b'\right),I^*\left(l_b',p_b'\right),l_b',p_b'\right)=\Pi^{\rm venue}\left({r^*}\left(l_b,p_b\right),I^*\left(l_b,p_b\right),l_b,p_b\right)$, we can easily see that the value of the objective function (\ref{SM:bargain:obj}) under $\left(l_b',p_b'\right)$ is no less than that under $\left(l_b,p_b\right)$. 

{\bf(Step 3)} Based on the results of {\bf Step 1} and {\bf Step 2}, we only need to focus on the $\left(l_b,p_b\right)$ satisfying $p_b=-\phi$ and $r^*\left(l_b,p_b\right)=1$ to determine the optimal bargaining solution. 
When $p_b=-\phi$ and $r^*\left(l_b,p_b\right)=1$, we can derive the following relation 
\begin{align}
\nonumber
& R^{\rm app}\left(l_b,p_b\right)+ \Pi^{\rm venue}\left({r^*}\left(l_b,p_b\right),I^*\left(l_b,p_b\right),l_b,p_b\right)=\\
& b N {\bar x}\left(1,I^*\left(l_b,-\phi\right)\right)-k I^*\left(l_b,-\phi\right)+\phi N{\bar y}\left(1,I^*\left(l_b,-\phi\right)\right).\label{SM:equ:summation}
\end{align}
According to Propositions \ref{proposition:stageII:situationI}, \ref{proposition:stageII:situationII}, and \ref{proposition:stageII:situationIII}, when the venue becomes a POI, its investment level is not affected by the lump-sum fee. Hence, we have the relation $I^*\left(l_b,-\phi\right)=I^*\left({\tilde H}\left(-\phi\right),-\phi\right)$. Then, we can derive the following equation from (\ref{SM:equ:summation}):
\begin{align}
\nonumber
& R^{\rm app}\left(l_b,p_b\right)+ \Pi^{\rm venue}\left({r^*}\left(l_b,p_b\right),I^*\left(l_b,p_b\right),l_b,p_b\right)=\\
\nonumber
& b N {\bar x}\left(1,I^*\left({\tilde H}\left(-\phi\right),-\phi\right)\right)-k I^*\left({\tilde H}\left(-\phi\right),-\phi\right)\\
& +\phi N{\bar y}\left(1,I^*\left({\tilde H}\left(-\phi\right),-\phi\right)\right).
\end{align}
Based on Corollary \ref{corollary:optimalrevenue}, we can easily verify that the right-hand side equals ${\tilde H}\left(-\phi\right)+b\eta N\frac{U}{c_{\max}}$. That is to say, we have
\begin{align}
\nonumber
& R^{\rm app}\left(l_b,p_b\right)+ \Pi^{\rm venue}\left({r^*}\left(l_b,p_b\right),I^*\left(l_b,p_b\right),l_b,p_b\right)=\\
& {\tilde H}\left(-\phi\right)+b\eta N\frac{U}{c_{\max}}.
\end{align}
Note that when $p_b=-\phi$, the app's revenue $R^{\rm app}\left(l_b,p_b\right)=l_b$. Then, we can rewrite problem (\ref{SM:bargain:obj})-(\ref{SM:bargain:var}) as
\begin{align}
& \max l_b^{\gamma} \Bigl( {\tilde H}\left(-\phi\right)- l_b \Bigr)^{1-\gamma} \label{SM:bargain:new:a}\\
& {\rm var.~~~} 0\le l_b \le {\tilde H}\left(-\phi\right).\label{SM:bargain:new:b}
\end{align}
The objective function (\ref{SM:bargain:new:a}) is concave, and we can easily obtain the optimal solution $l_b^*=\gamma {\tilde H}\left(-\phi\right)$ by examining the first-order derivative of the objective function. 

According to the three steps, we can see that $p_b^*=-\phi$ and $l_b^*=\gamma {\tilde H}\left(-\phi\right)$.
\endproof

\section{Influences of $\eta$ and $N$ in Special Cases}

\subsection{Influence of $\eta$ When $\phi<{bV}/{U}$ and $\delta\le\delta_2$}\label{SM:numerical:popularity}

In this section, we numerically show that when $\phi<{bV}/{U}$ and $\delta\le\delta_2$, $R^{\rm app}\left(l^*,p^*\right)$ may decrease with $\eta$. 
We choose $N=200$, $c_{\max}=36$, $U=6$, $V=5$, $I_0=25$, $k=3$, $b=7$, $\theta=0.05$, $\phi=2$, and $\delta=9$. Based on the definition of $\delta_2$ in Lemma \ref{lemma:delta1}, we can compute $\delta_2=12.09$. Hence, we can verify that both $\phi<{bV}/{U}$ and $\delta\le\delta_2$ hold under the parameter setting. 

\begin{figure}[h]
  \centering
  \includegraphics[scale=0.36]{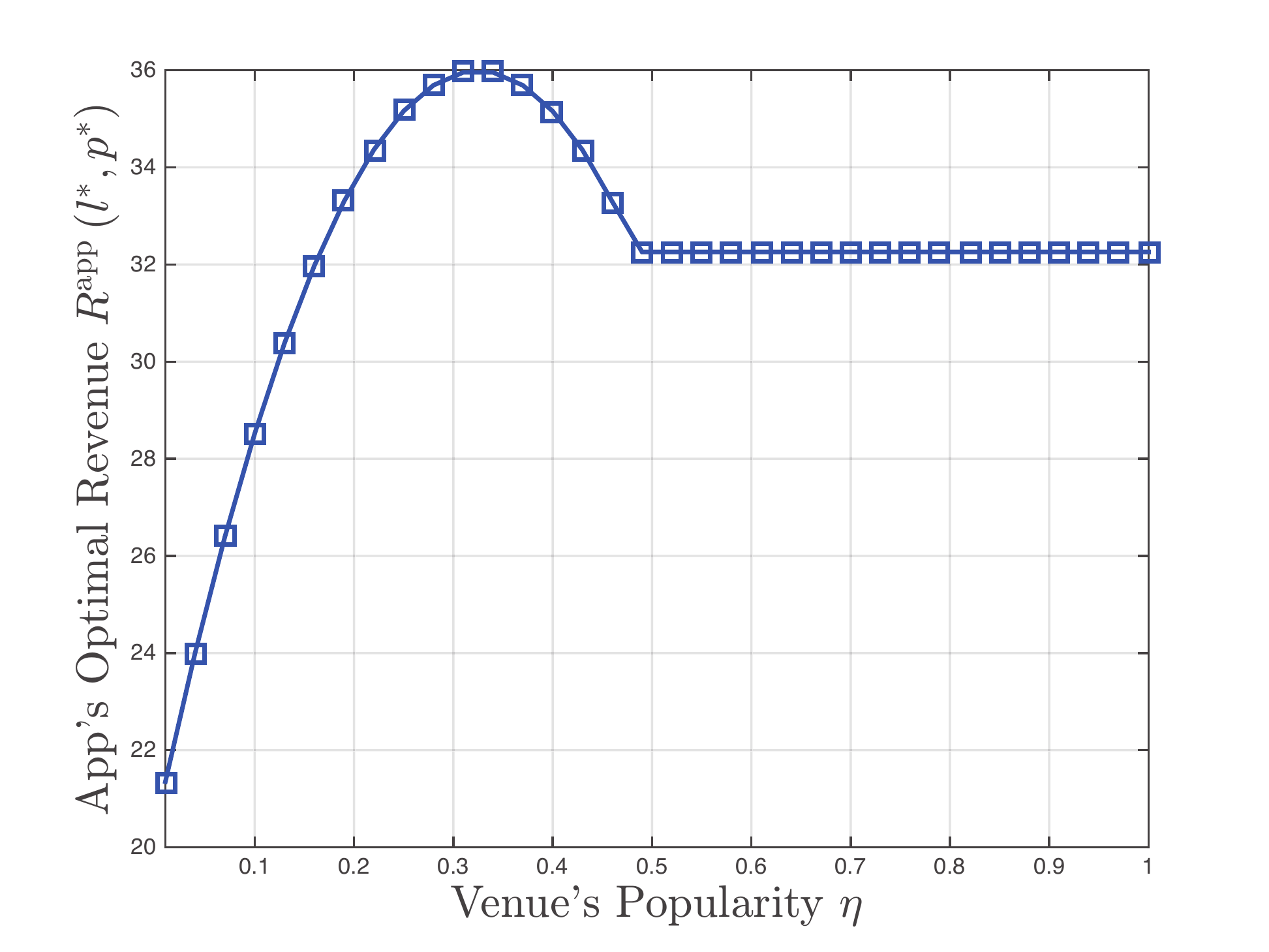}\\
  \caption{Impact of $\eta$ on $R^{\rm app}\left(l^*,p^*\right)$ When $\phi<{bV}/{U}$ and $\delta\le\delta_2$.}
  \label{SM:fig:numerical:popularity}
  \vspace{-0.4cm}
\end{figure}

We change $\eta$ from $0.01$ to $1$, and plot $R^{\rm app}\left(l^*,p^*\right)$ against $\eta$ in Fig. \ref{SM:fig:numerical:popularity}. We can see that $R^{\rm app}\left(l^*,p^*\right)$ decreases with $\eta$ when $\eta\in\left(0.34,0.49\right)$ and does not change with $\eta$ when $\eta\in\left(0.49,1\right)$. This numerical result is consistent with our conclusion in Table \ref{table:1}, i.e., if the unit advertising revenue $\phi$ is small, the app may avoid collaborating with a popular venue. 

\subsection{Influence of $N$ When $\phi<{bV}/{U}$ and $\delta\le\delta_3$}\label{SM:numerical:populationsize}

In this section, we numerically show that when $\phi<{bV}/{U}$ and $\delta\le\delta_3$, $R^{\rm app}\left(l^*,p^*\right)$ may decrease with $N$. 
We choose $c_{\max}=100$, $U=6$, $V=5$, $I_0=3$, $k=5.2$, $b=29$, $\eta=0.3$, $\theta=0.1$, $\phi=8$, and $\delta=12$. Based on the definition of $\delta_3$ in Lemma \ref{lemma:delta:2}, we can compute $\delta_3=12.40$. Hence, we can verify that both $\phi<{bV}/{U}$ and $\delta\le\delta_3$ hold under the parameter setting. 

\begin{figure}[h]
  \centering
  \includegraphics[scale=0.36]{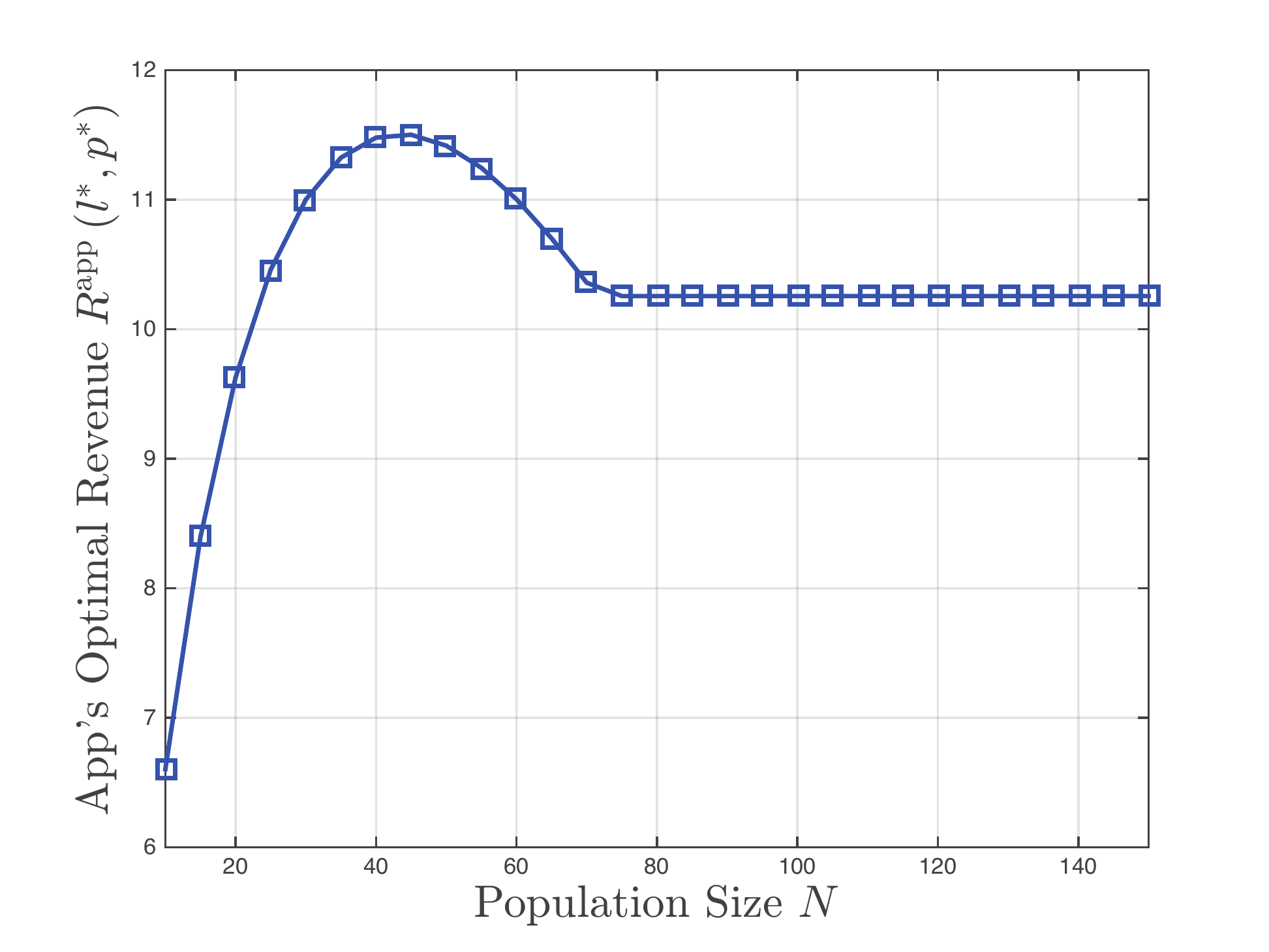}\\
  \caption{Impact of $N$ on $R^{\rm app}\left(l^*,p^*\right)$ When $\phi<{bV}/{U}$ and $\delta\le\delta_3$.}
  \label{SM:fig:numerical:populationsize}
  \vspace{-0.4cm}
\end{figure}

We change $N$ from $10$ to $150$, and plot $R^{\rm app}\left(l^*,p^*\right)$ against $N$ in Fig. \ref{SM:fig:numerical:populationsize}. We can see that $R^{\rm app}\left(l^*,p^*\right)$ decreases with $N$ when $N\in\left(45,75\right)$ and does not change with $N$ when $N\in\left(75,150\right)$. This numerical result is consistent with our conclusion in Table \ref{table:1}, i.e., if the unit advertising revenue $\phi$ is small, the app may avoid collaborating with a venue in a busy area.

\section{Numerical Computation of $p_u^*$ and $l_u^*$ and Impact of App's Risk Preference}\label{SM:riskpreference}
First, we briefly discuss the numerical approach to compute $p_u^*$ and $l_u^*$ for the risk-averse and risk-seeking apps (shown in the second and third bullets of Theorem \ref{theorem:uncertainty}). 
From Theorem \ref{theorem:uncertainty}, the optimal tariff $\left(l_u^*,p_u^*\right)$ satisfies the relation $l_u^*={\tilde H}\left(p_u^*\right)$. Hence, we can substitute $l_u={\tilde H}\left(p_u\right)$ into the objective function in (\ref{equ:uncertainty:obj}) to obtain a new objective function, which is only a function of $p_u$. 
Then, we can numerically show that the new objective function is always strictly unimodal in $p_u$ when $p_u$ is below a specific threshold, and is a constant function otherwise. 
In this case, we can directly compute $p_u^*$ via the Golden Section method. 
{{The computational complexity is $O\left(\log\frac{1}{\epsilon}\right)$, where $\epsilon>0$ captures the accuracy of the solution, i.e., the gap between the computed price and the optimal price is upper bounded by $\epsilon$.}} After obtaining $p_u^*$, we can use the relation $l_u^*={\tilde H}\left(p_u^*\right)$ to compute $l_u^*$.

Second, we investigate the impact of the degree of app's risk aversion on the $p_u^*$, $l_u^*$, and ${{\mathbb E}_{\phi}\left\{J\big(R^{\rm app}\left(l_u^*,p_u^*\right) \big)\right\}}$ through numerical experiments. 
We choose $N=100$, $c_{\max}=25$, $U=4$, $V=5$, $I_0=0.1$, $k=3$, $b=0.5$, $\eta=0.1$, $\theta=0.05$, and $\delta=0.3$. Moreover, we assume that $\phi$ follows a uniform distribution, i.e., $\phi\sim{\cal U}\left[\phi_{\min},\phi_{\max}\right]$. We fix $\frac{\phi_{\max}+\phi_{\min}}{2}=0.2$, and vary $\frac{\phi_{\max}-\phi_{\min}}{2}$ from $0$ to $0.18$. The value of $\frac{\phi_{\max}-\phi_{\min}}{2}$ reflects the randomness of $\phi$. We consider the following utility function for the app:
\begin{align}
J\left(z\right)=2-e^{-\alpha z},z\in\left(-\infty,\infty\right),
\end{align}
which is a concave function. Note that parameter $\alpha>0$ describes the degree of risk aversion of the app server. The constant $2$ does not affect the results' insights, and simply ensures that the app's expected utility ${{\mathbb E}_{\phi}\left\{J\big(R^{\rm app}\left(l_u^*,p_u^*\right) \big)\right\}}$ is always positive.

\begin{figure*}[t]
  \centering
  \subfigure[Per-Player Charge $p_u^*$.]{
  \label{SM:fig:simu:1}
    \includegraphics[scale=0.299]{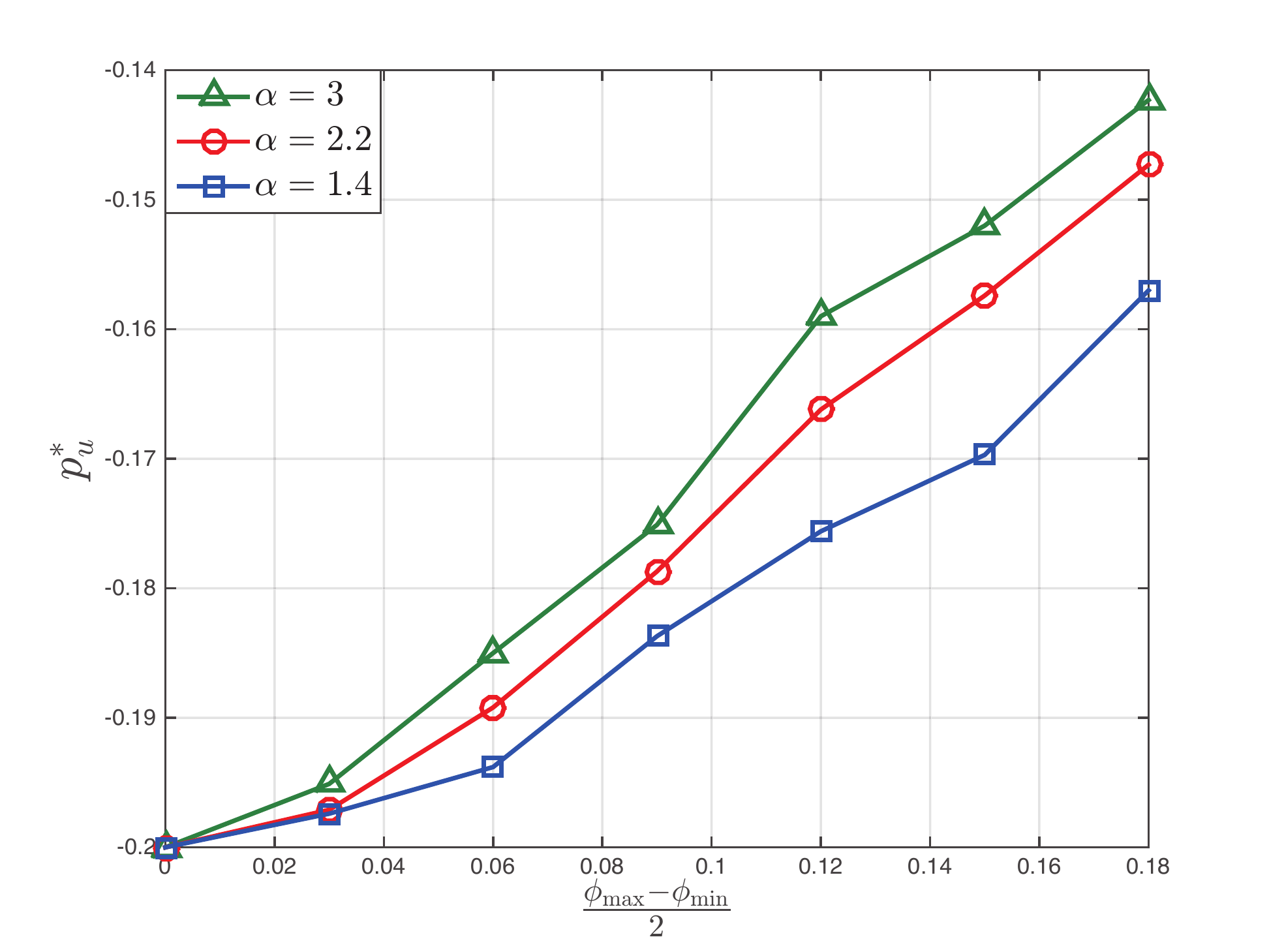}}
  \subfigure[Lump-Sum Fee $l_u^*$.]{
  \label{SM:fig:simu:2}
    \includegraphics[scale=0.299]{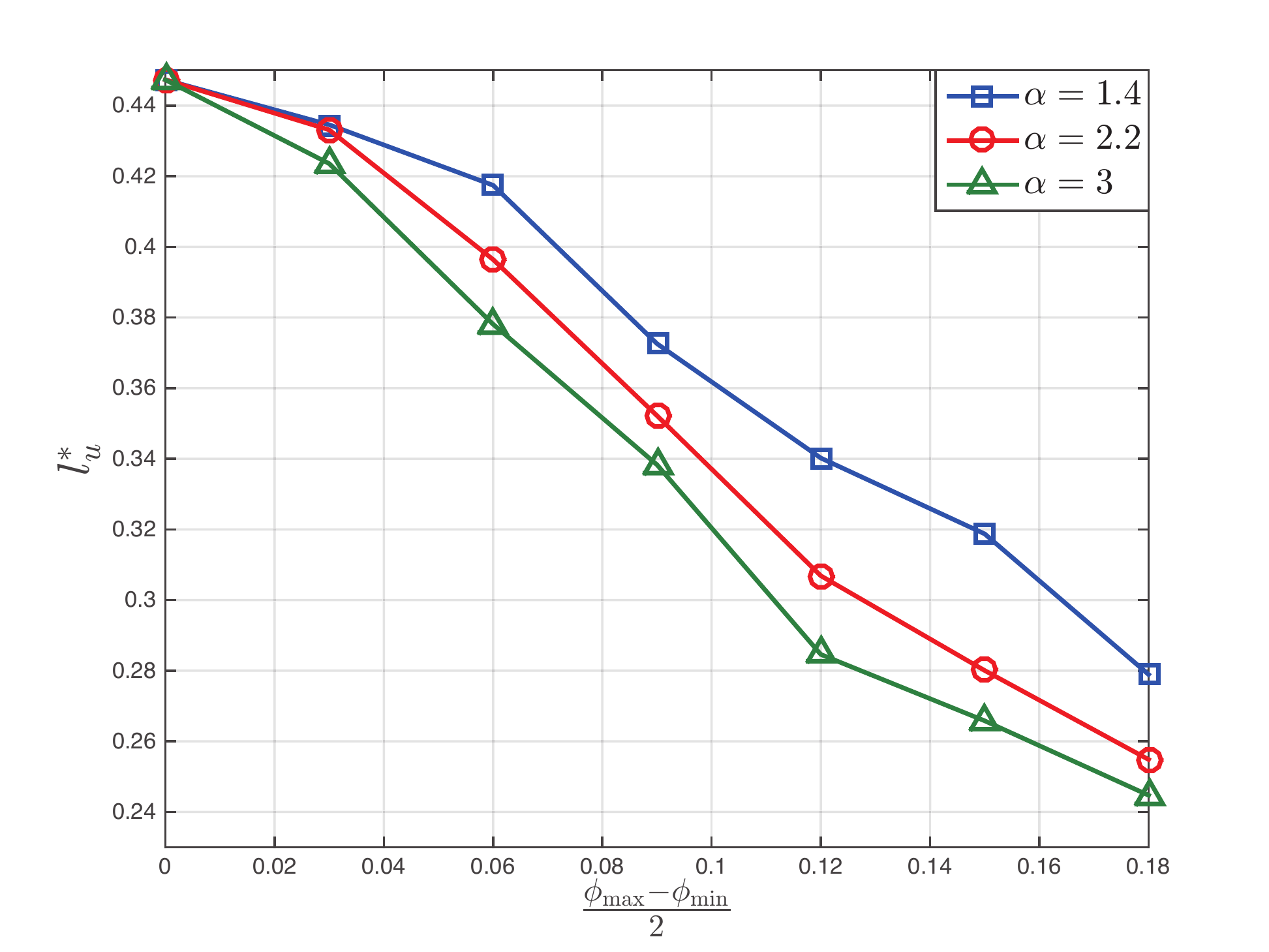}}\vspace{-0.2cm}
      \subfigure[App's Expected Utility.]{
  \label{SM:fig:simu:3}
    \includegraphics[scale=0.299]{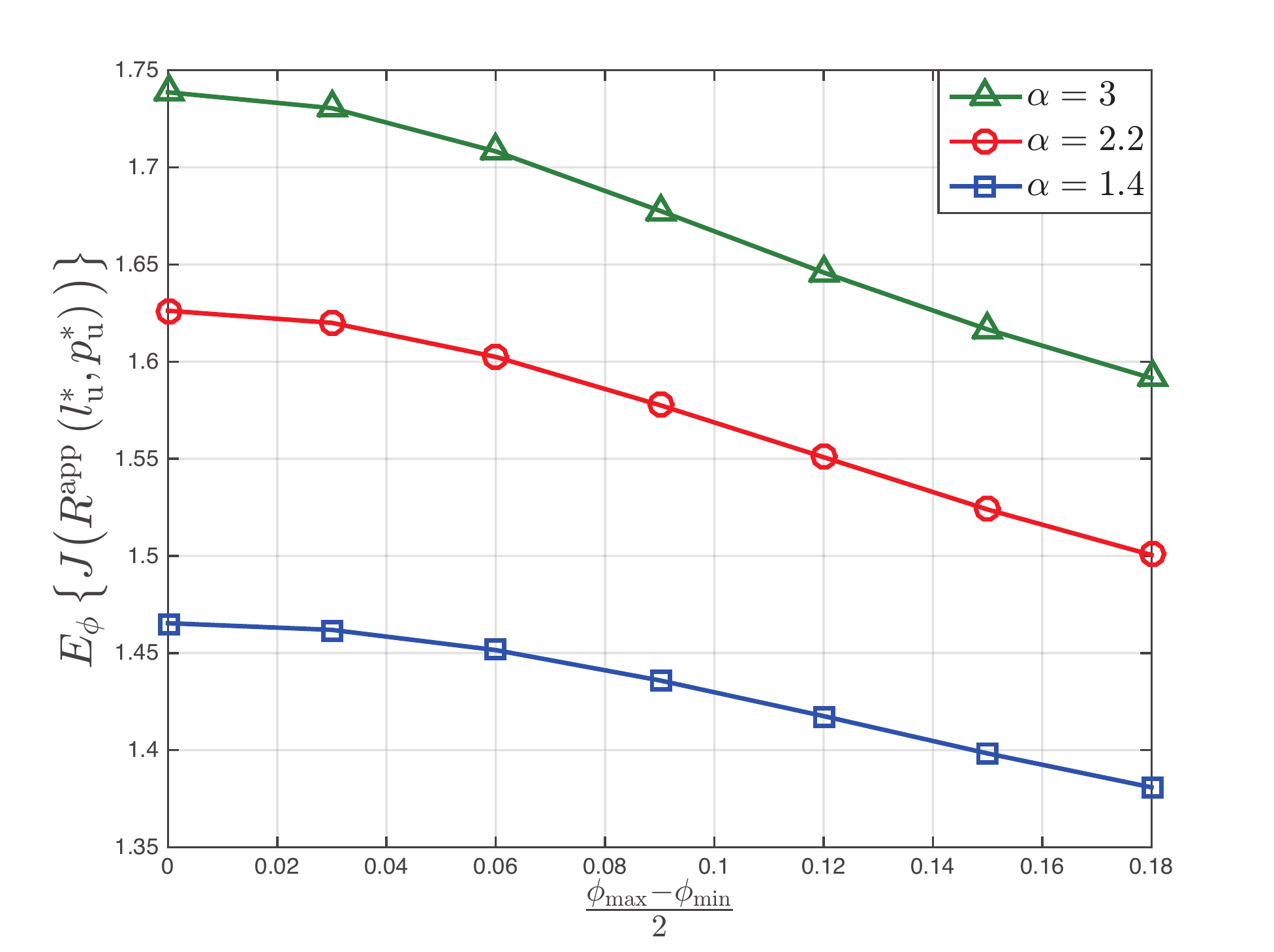}}
    \caption{Risk-Averse App.\vspace{-0.25cm}}
\end{figure*}

\begin{figure*}[t]
  \centering  
  \subfigure[Per-Player Charge $p_u^*$.]{
  \label{SM:fig:simu:4}
    \includegraphics[scale=0.299]{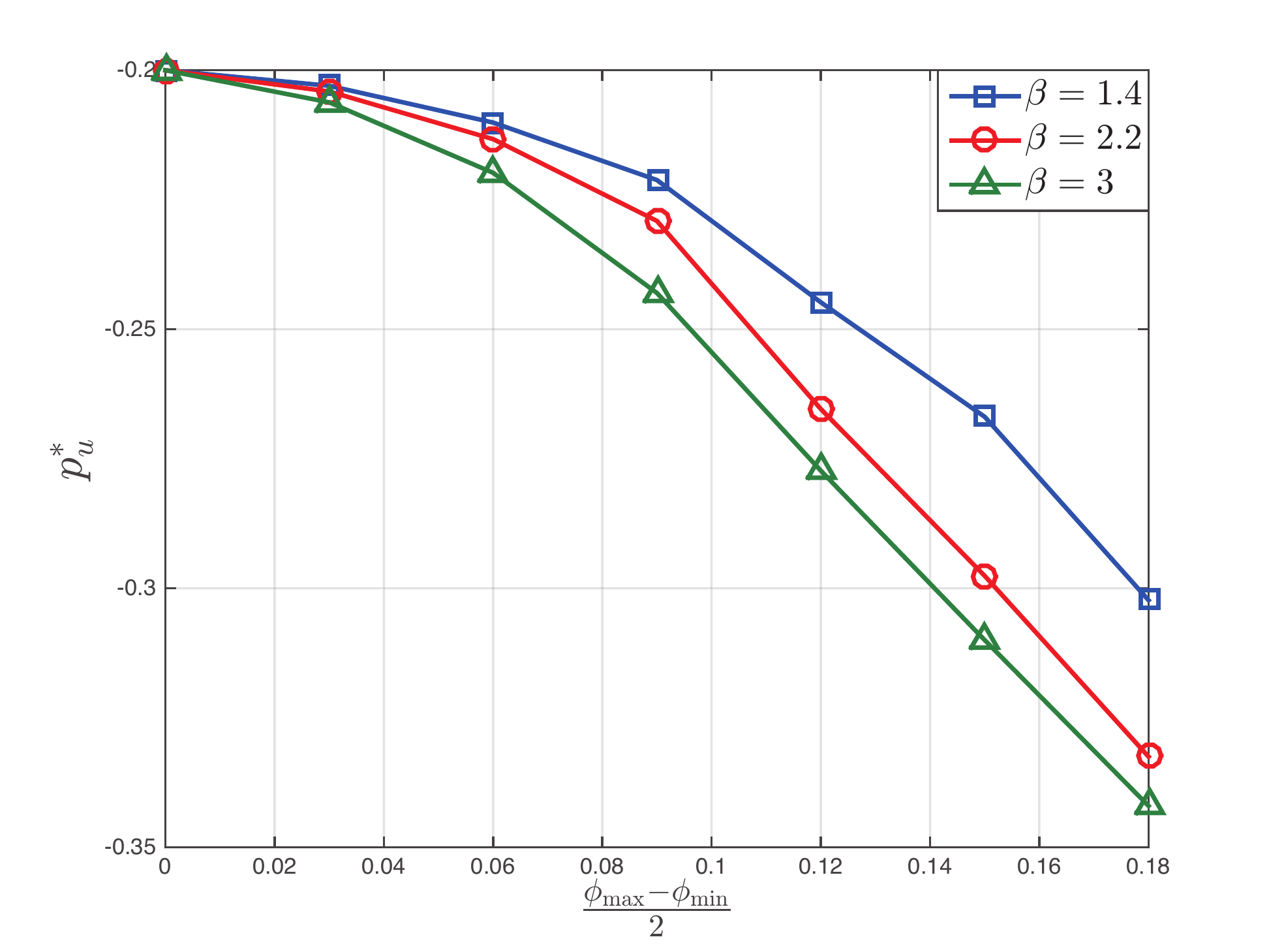}}
  \subfigure[Lump-Sum Fee $l_u^*$.]{
  \label{SM:fig:simu:5}
    \includegraphics[scale=0.299]{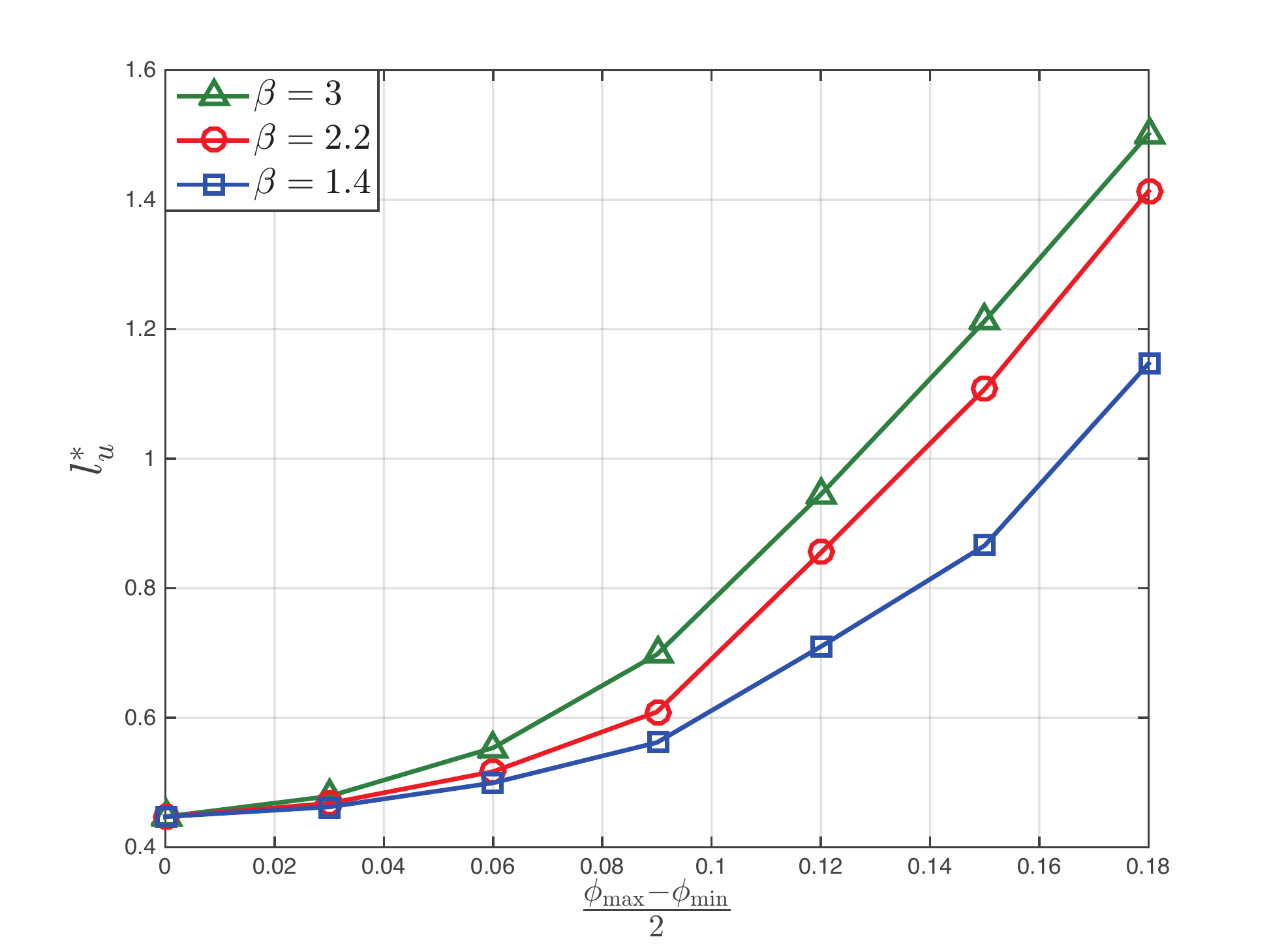}}\vspace{-0.2cm}
      \subfigure[App's Expected Utility.]{
  \label{SM:fig:simu:6}
    \includegraphics[scale=0.299]{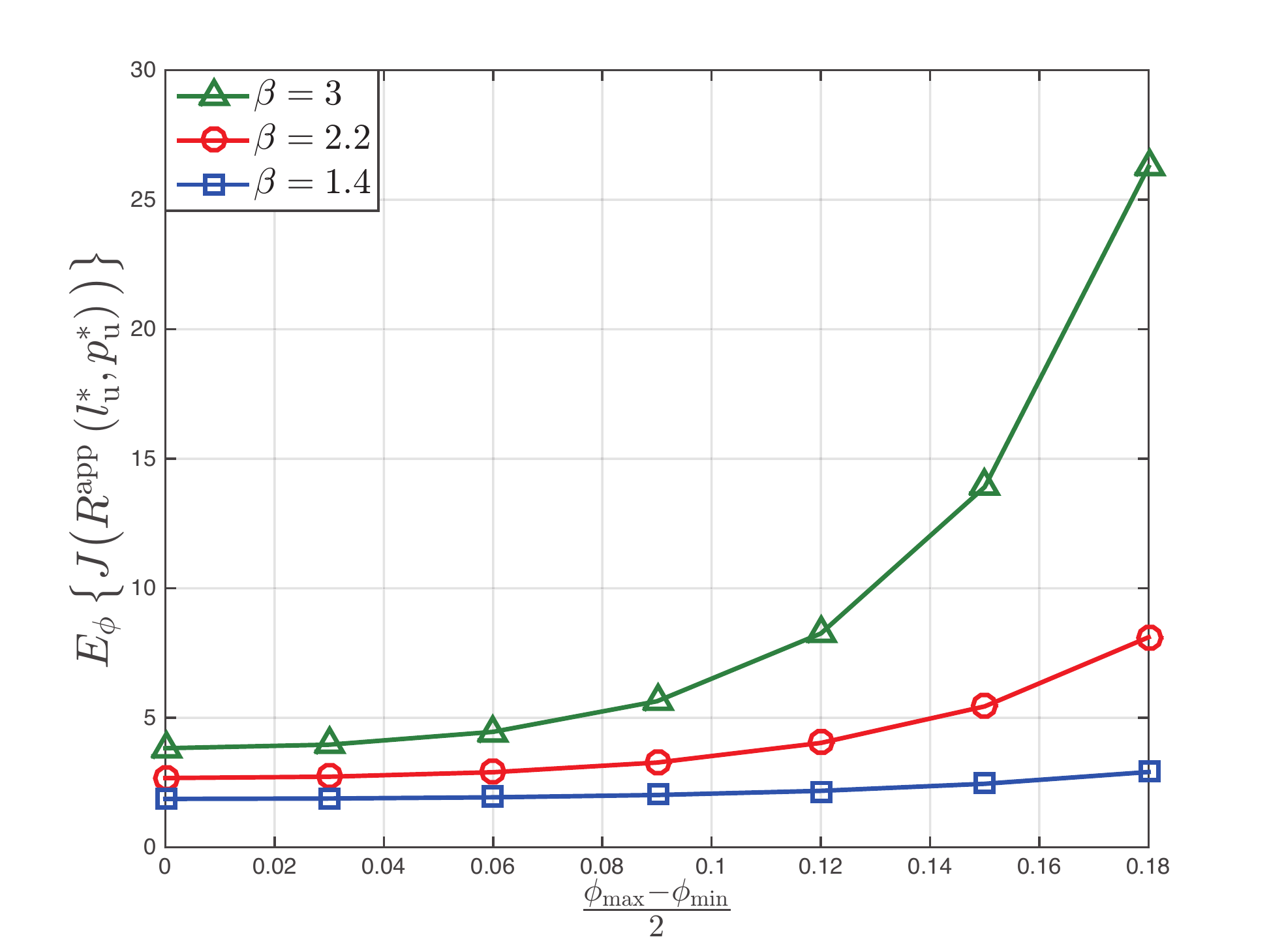}}
    \caption{Risk-Seeking App.\vspace{-0.25cm}}
\end{figure*}

We show the app's expected utility and optimal tariff under different $\frac{\phi_{\max}-\phi_{\min}}{2}$ and $\alpha$ in Fig. \ref{SM:fig:simu:1}, Fig. \ref{SM:fig:simu:2}, and Fig. \ref{SM:fig:simu:3}. Because $\frac{\phi_{\max}-\phi_{\min}}{2}$ reflects the randomness of $\phi$, and $\alpha$ describes the degree of risk aversion, we have the following observations.
\begin{observation}
If the app is risk-averse, the per-player charge $p_u^*$ is increasing in the randomness of $\phi$. Moreover, $p_u^*$ is increasing in the degree of risk aversion. 
\end{observation}
\begin{observation}
If the app is risk-averse, the lump-sum fee $l_u^*$ is decreasing in the randomness of $\phi$. Moreover, $l_u^*$ is decreasing in the degree of risk aversion. 
\end{observation}
\begin{observation}
If the app is risk-averse, its expected utility is decreasing in the randomness of $\phi$.
\end{observation}

Third, we investigate the impact of the degree of app's seeking risk on the $p_u^*$, $l_u^*$, and ${{\mathbb E}_{\phi}\left\{J\big(R^{\rm app}\left(l_u^*,p_u^*\right) \big)\right\}}$ through numerical experiments. The results are basically opposite to the results for a risk-averse app. In order to model a risk-seeking app, we consider the following utility function for the app:
\begin{align}
J\left(z\right)=e^{\beta z},z\in\left(-\infty,\infty\right),
\end{align}
which is a convex function. Parameter $\beta$ describes the app's degree of seeking risk. The remaining parameter settings are the same as the risk-averse app case. 

We show the app's expected utility and optimal tariff under different $\frac{\phi_{\max}-\phi_{\min}}{2}$ and $\beta$ in Fig. \ref{SM:fig:simu:4}, Fig. \ref{SM:fig:simu:5}, and Fig. \ref{SM:fig:simu:6}. Because $\frac{\phi_{\max}-\phi_{\min}}{2}$ reflects the randomness of $\phi$, and $\beta$ describes the app's degree of seeking risk, we have the following observations.
\begin{observation}
If the app is risk-seeking, the per-player charge $p_u^*$ is decreasing in the randomness of $\phi$. Moreover, $p_u^*$ is decreasing in the degree of seeking risk. 
\end{observation}
\begin{observation}
If the app is risk-seeking, the lump-sum fee $l_u^*$ is increasing in the randomness of $\phi$. Moreover, $l_u^*$ is increasing in the degree of seeking risk. 
\end{observation}
\begin{observation}
If the app is risk-seeking, the its expected utility is increasing in the randomness of $\phi$.\end{observation}


\section{More Numerical Results}\label{SM:numerical:result}

\begin{figure*}[t]
  \centering
  \subfigure[Example 1.]{
    \label{SM:fig:delta:1}
    \includegraphics[scale=0.299]{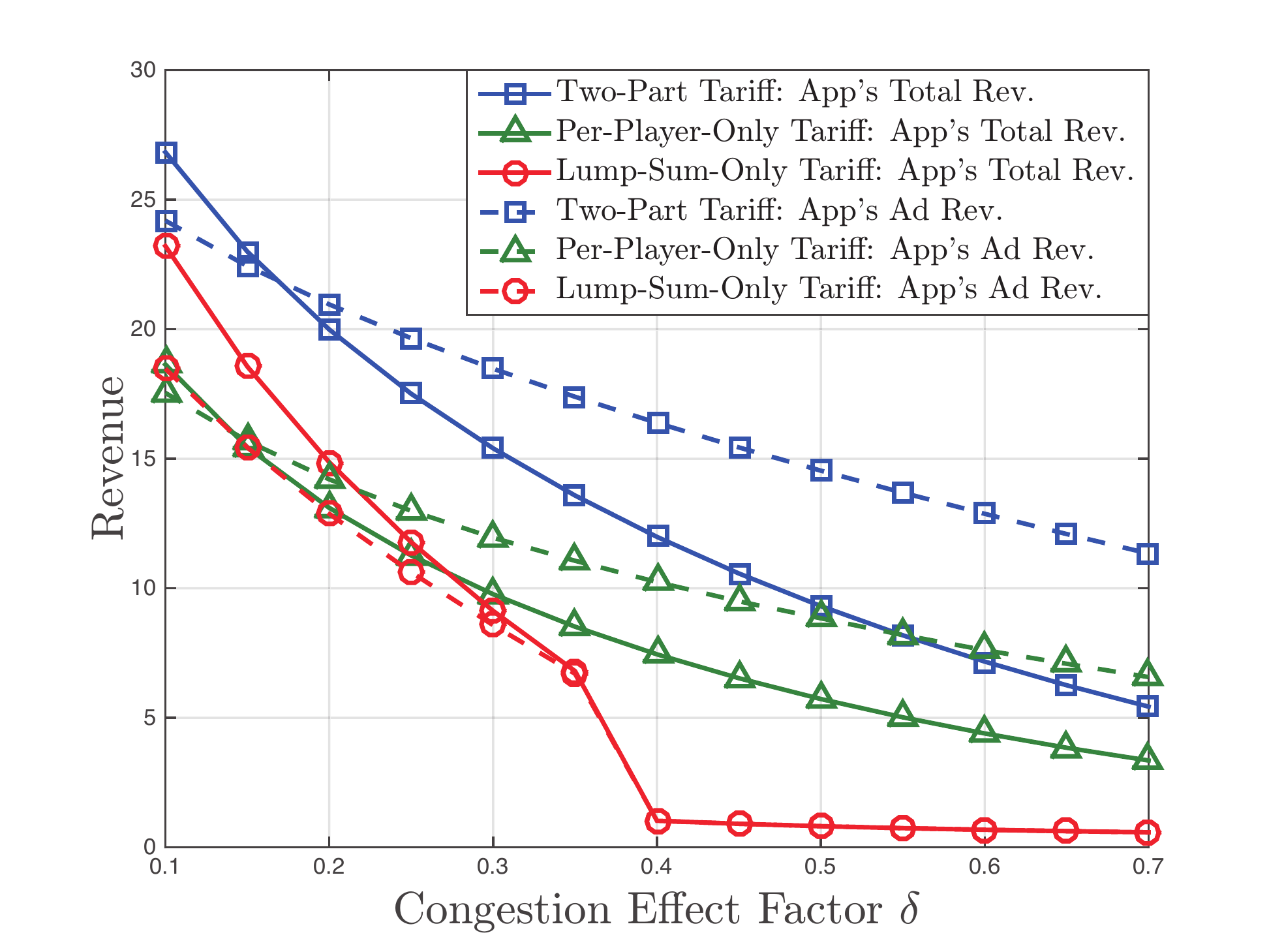}}
  \subfigure[Example 2.]{
    \label{SM:fig:delta:2}
    \includegraphics[scale=0.299]{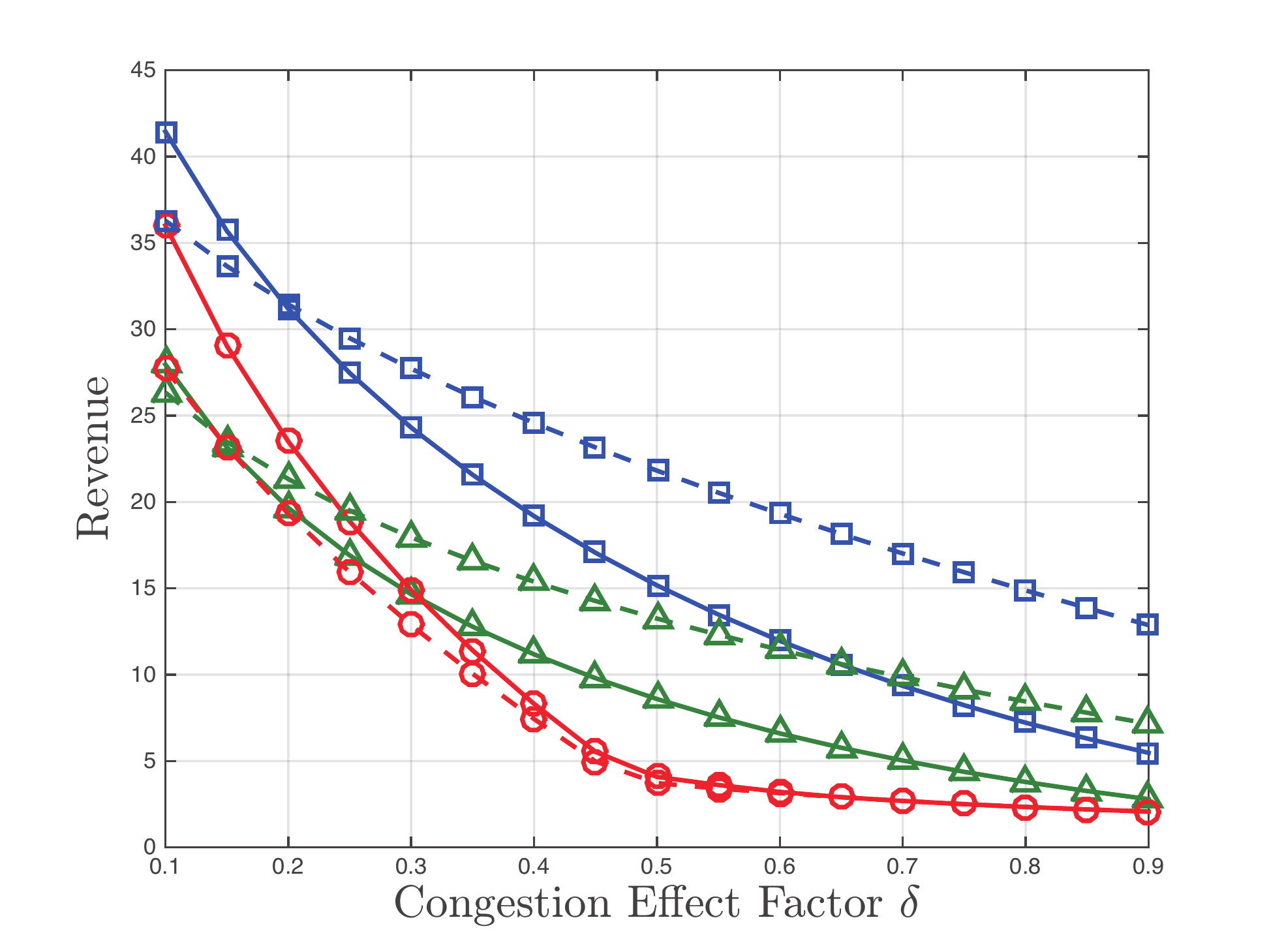}}
  \subfigure[Example 3.]{
    \label{SM:fig:delta:3}
    \includegraphics[scale=0.299]{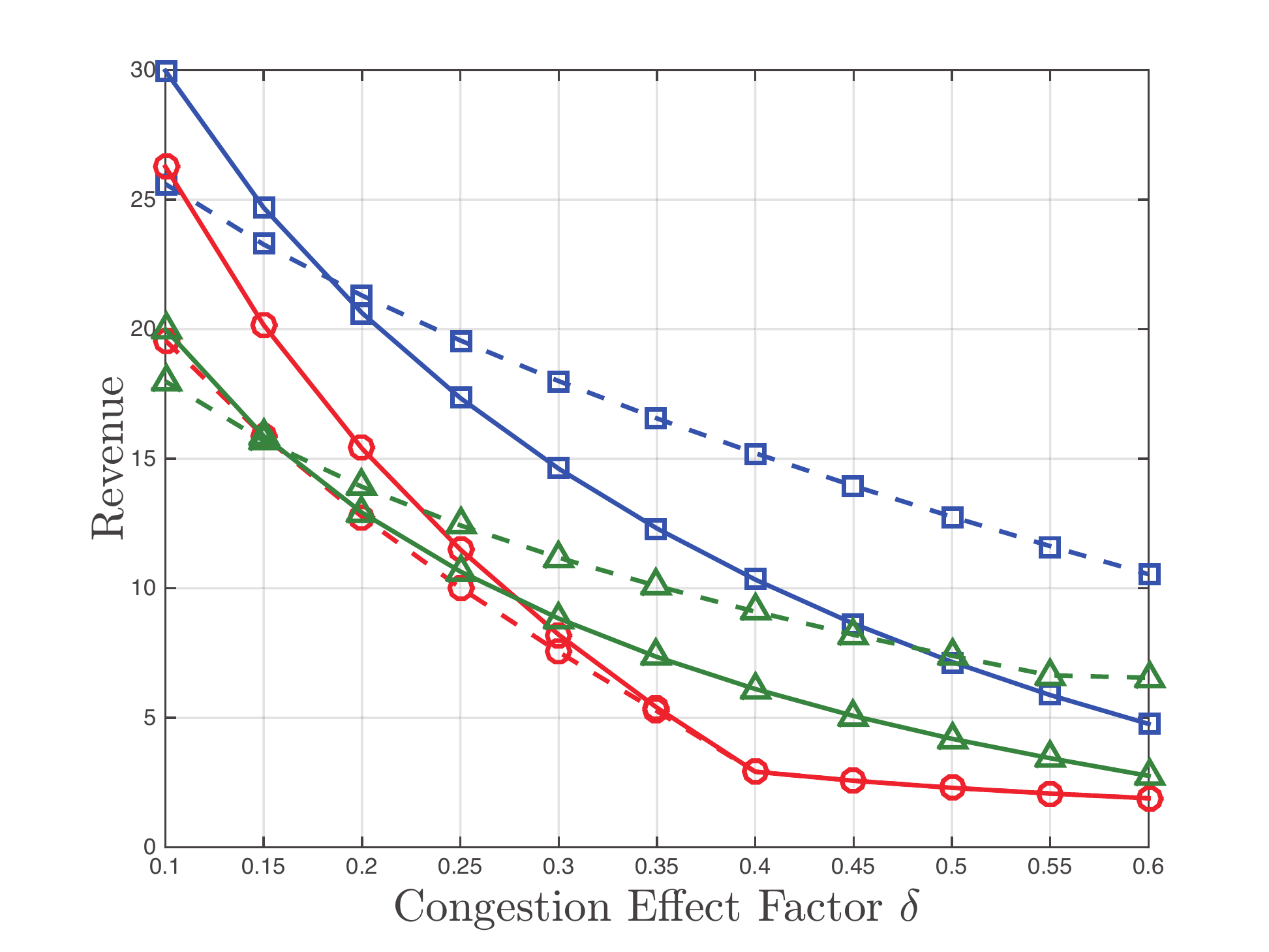}}    
  \caption{Comparison Under Different $\delta$.}
  \end{figure*}

\begin{figure*}[t]
  \centering
  \subfigure[Example 1.]{
    \label{SM:fig:theta:1}
    \includegraphics[scale=0.299]{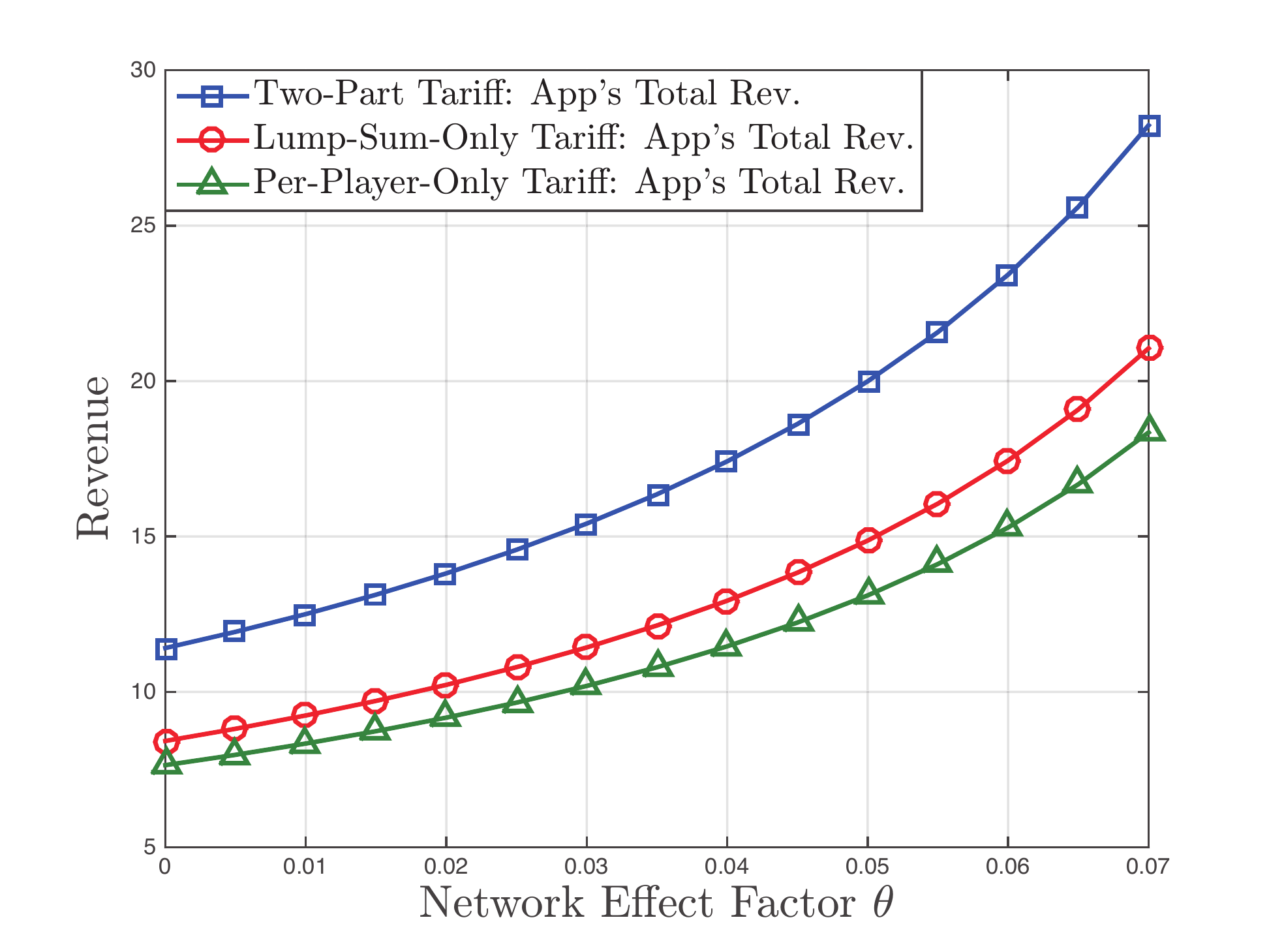}}
  \subfigure[Example 2.]{
    \label{SM:fig:theta:2}
    \includegraphics[scale=0.299]{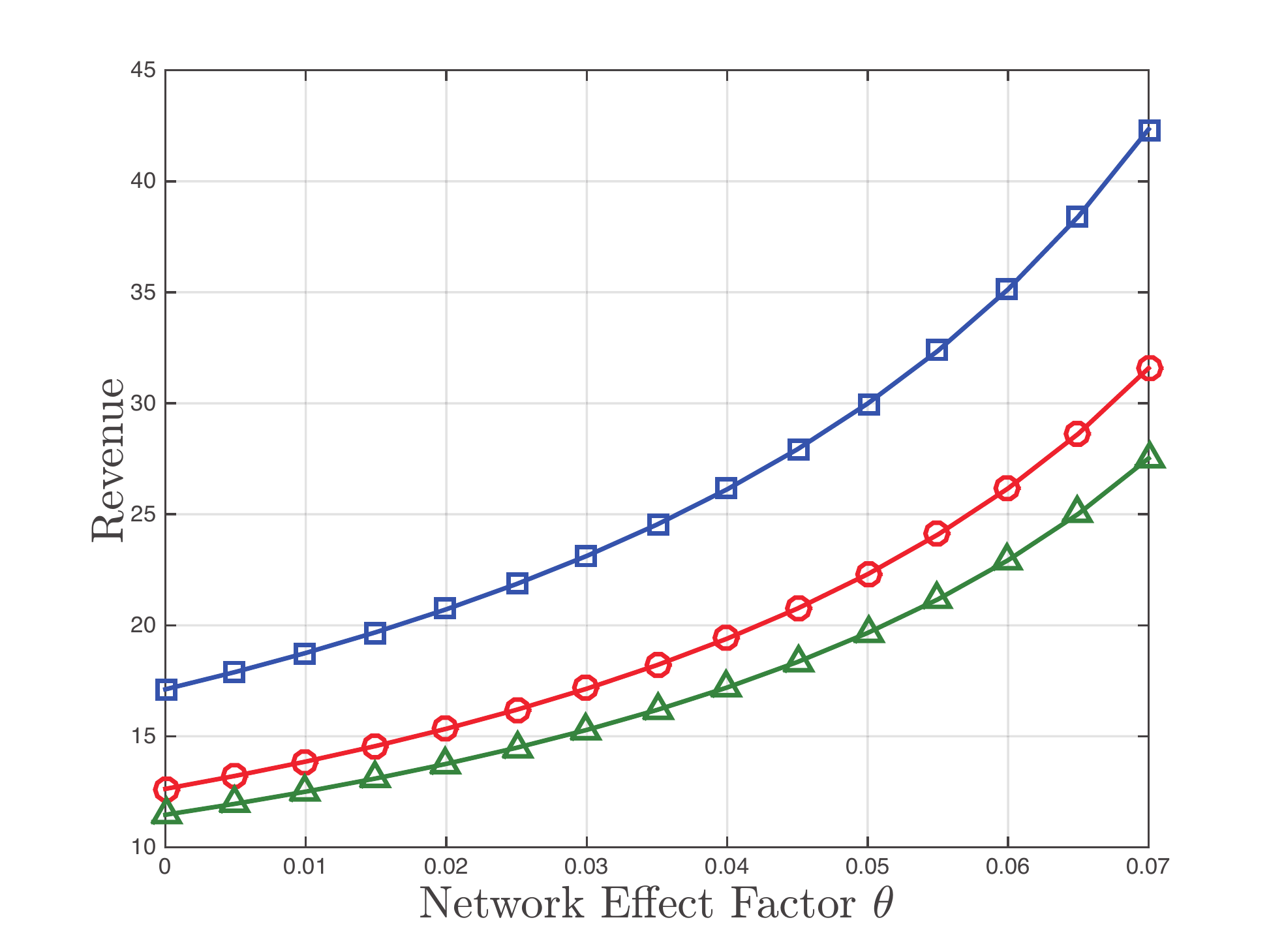}}
  \subfigure[Example 3.]{
    \label{SM:fig:theta:3}
    \includegraphics[scale=0.299]{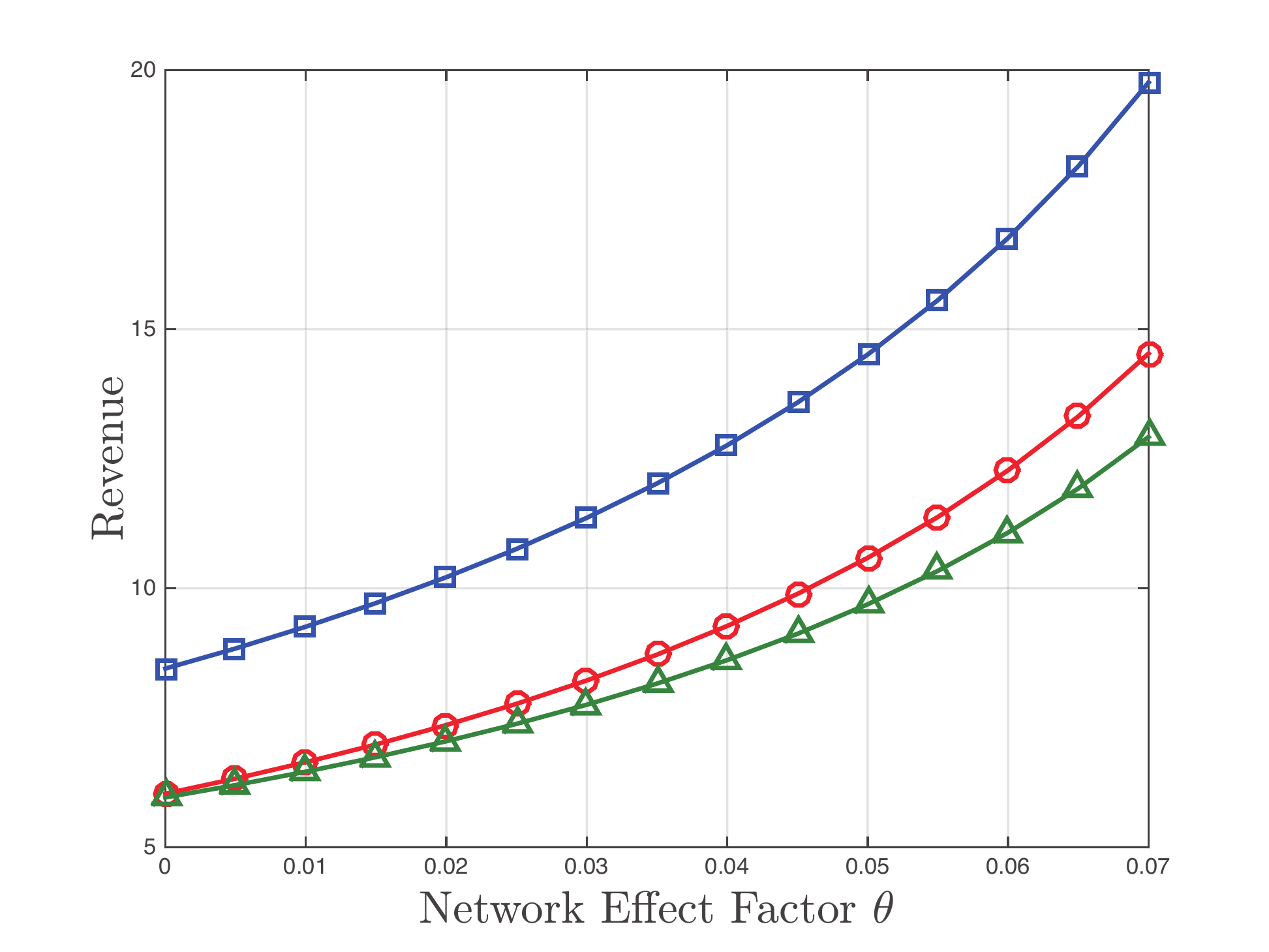}}    
  \caption{Comparison Under Different $\theta$.}
  \end{figure*}

\begin{figure*}[t]
  \centering
  \subfigure[Example 1.]{
    \label{SM:fig:phi:1}
    \includegraphics[scale=0.299]{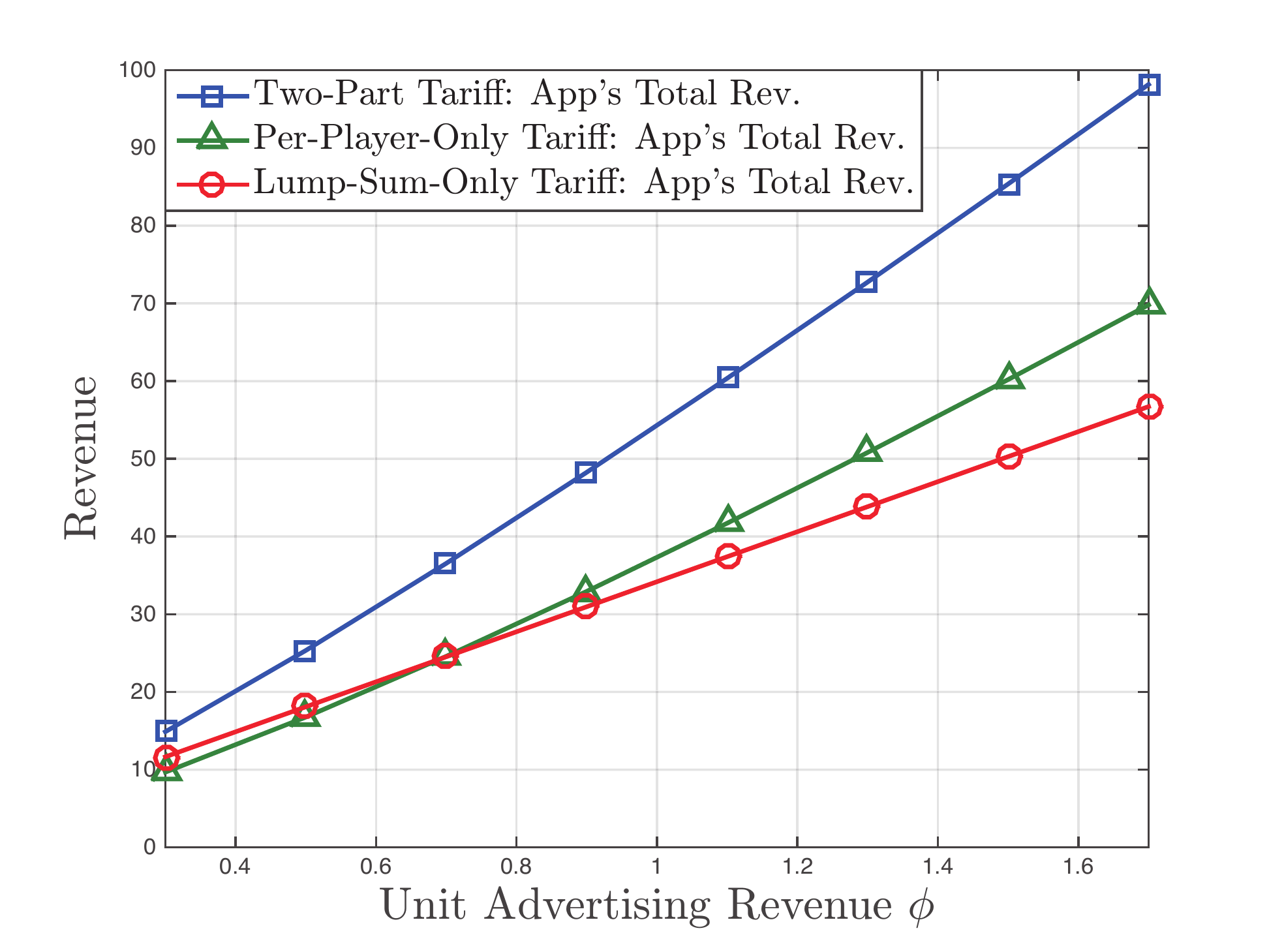}}
  \subfigure[Example 2.]{
    \label{SM:fig:phi:2}
    \includegraphics[scale=0.299]{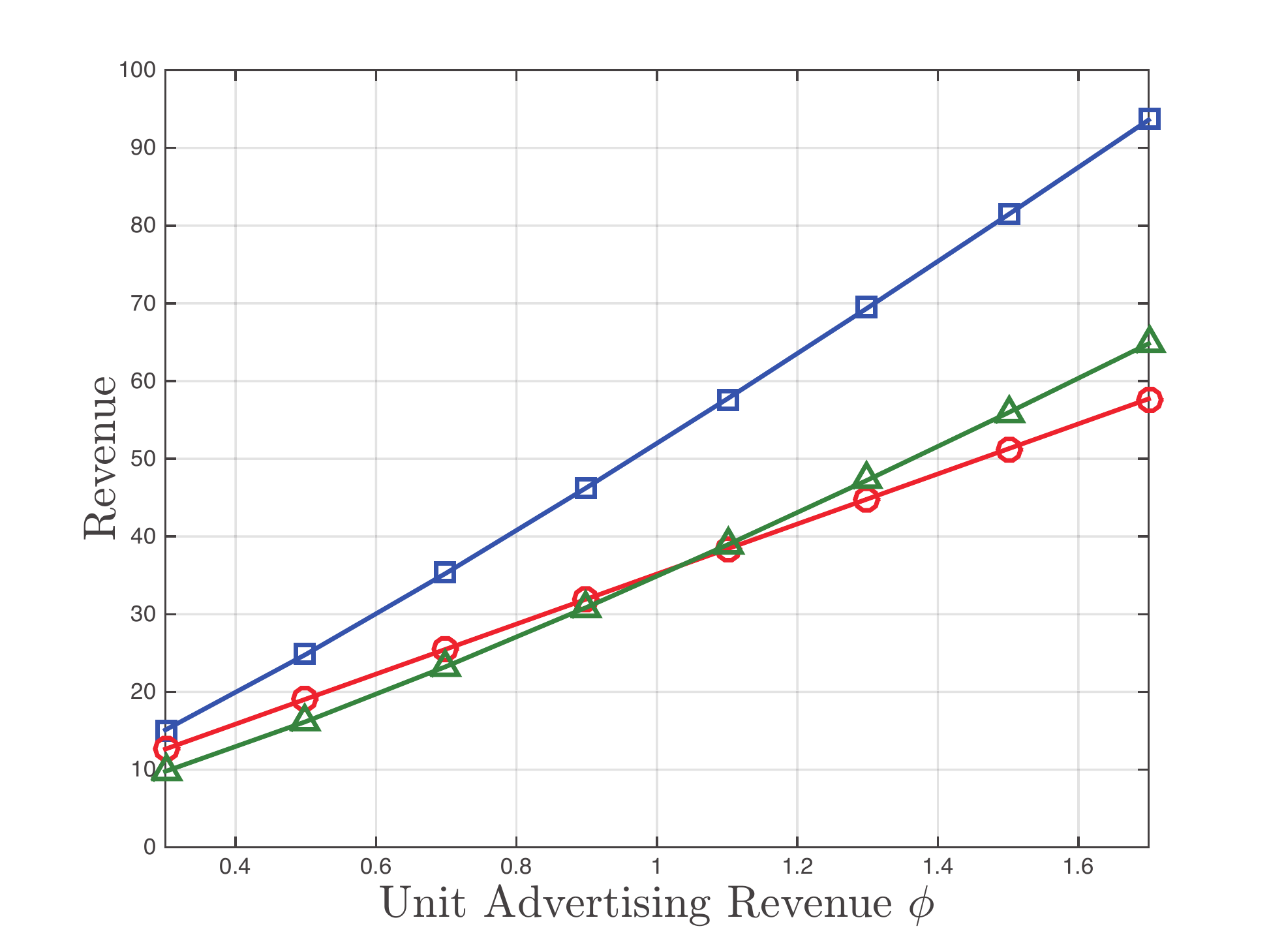}}
  \subfigure[Example 3.]{
    \label{SM:fig:phi:3}
    \includegraphics[scale=0.299]{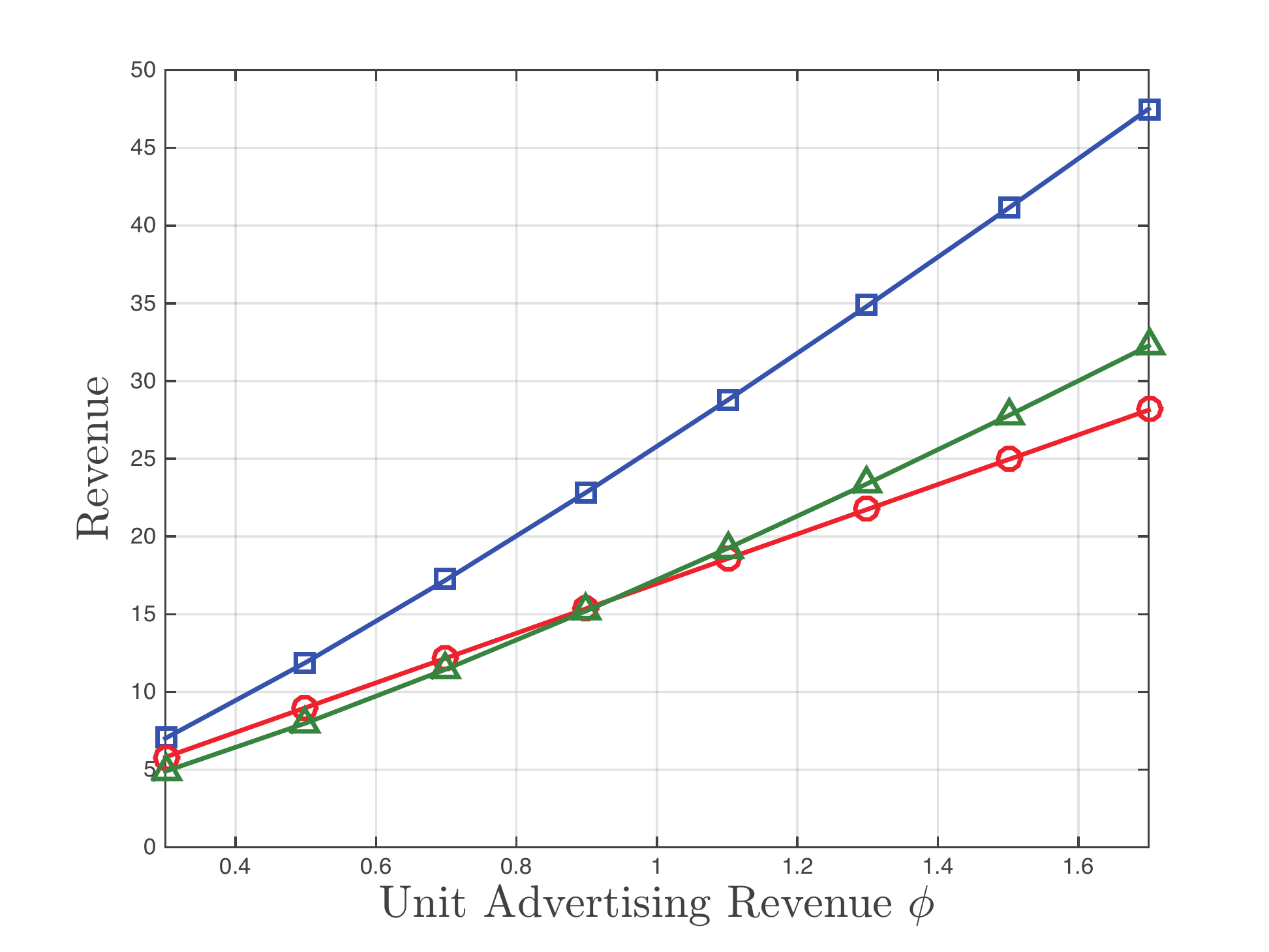}}    
  \caption{Comparison Under Different $\phi$.}
  \end{figure*}

{{
\subsection{Impact of $\delta$}
In this section, we provide more numerical results about the impact of $\delta$. We consider the following three parameter settings:
\begin{itemize}
\item $N=200$, $c_{\max}=24$, $U=3$, $V=5$, $I_0=0.2$, $k=2$, $b=1$, $\eta=0.2$, $\theta=0.05$, and $\phi=0.4$;
\item $N=200$, $c_{\max}=24$, $U=2$, $V=5$, $I_0=0.6$, $k=3$, $b=1$, $\eta=0.3$, $\theta=0.05$, and $\phi=0.6$;
\item $N=200$, $c_{\max}=22$, $U=3$, $V=3$, $I_0=0.6$, $k=3$, $b=2$, $\eta=0.2$, $\theta=0.05$, and $\phi=0.6$.
\end{itemize}
We show the corresponding numerical results in Fig. \ref{SM:fig:delta:1}, Fig. \ref{SM:fig:delta:2}, and Fig. \ref{SM:fig:delta:3}. We can see that the key insight still holds: \emph{if the congestion effect factor is medium, our tariff significantly outperforms the lump-sum-only tariff.}

\subsection{Impact of $\theta$}
In this section, we provide more numerical results about the impact of $\theta$. We consider the following three parameter settings:
\begin{itemize}
\item $N=200$, $c_{\max}=24$, $U=3$, $V=5$, $I_0=0.2$, $k=2$, $b=1$, $\eta=0.2$, $\phi=0.4$, and $\delta=0.2$;
\item $N=200$, $c_{\max}=24$, $U=2$, $V=5$, $I_0=0.2$, $k=3$, $b=1$, $\eta=0.3$, $\phi=0.6$, and $\delta=0.2$;
\item $N=200$, $c_{\max}=26$, $U=3$, $V=3$, $I_0=0.2$, $k=3$, $b=2$, $\eta=0.2$, $\phi=0.6$, and $\delta=0.2$.
\end{itemize}
We show the corresponding numerical results in Fig. \ref{SM:fig:theta:1}, Fig. \ref{SM:fig:theta:2}, and Fig. \ref{SM:fig:theta:3}. We can see that the key insight still holds: \emph{if the network effect factor is large, our tariff significantly outperforms the per-player-only tariff.}

\subsection{Impact of $\phi$}
In this section, we provide more numerical results about the impact of $\phi$. We consider the following three parameter settings:
\begin{itemize}
\item $N=200$, $c_{\max}=24$, $U=3$, $V=5$, $I_0=0.2$, $k=2$, $b=1$, $\eta=0.2$, $\theta=0.05$, and $\delta=0.2$;
\item $N=200$, $c_{\max}=24$, $U=2$, $V=5$, $I_0=0.2$, $k=3$, $b=1$, $\eta=0.3$, $\theta=0.05$, and $\delta=0.2$;
\item $N=200$, $c_{\max}=26$, $U=3$, $V=3$, $I_0=0.2$, $k=3$, $b=2$, $\eta=0.2$, $\theta=0.05$, and $\delta=0.2$.
\end{itemize}
We show the corresponding numerical results in Fig. \ref{SM:fig:phi:1}, Fig. \ref{SM:fig:phi:2}, and Fig. \ref{SM:fig:phi:3}. We can see that the key insight still holds: \emph{if the unit ad revenue is large, our tariff significantly outperforms the per-player-only tariff and lump-sum-only tariff.}

}}

\end{document}